\documentclass[onecolumn]{IEEE-Con-Sys-mag}

\usepackage[compress]{cite}
\usepackage{circledsteps}

\usepackage{pstricks}
\usepackage{psfrag}
\usepackage{multirow}
 \usepackage{dsfont}

\newcommand{\tras}{^{\mbox{\tiny T}}}

\newcommand{\muno}{^{\mbox{\tiny -1}}}

\newcommand{\ds}{\displaystyle}
\newcommand{\ts}{\textstyle}
\newcommand{\scr}{\scriptstyle}

 \newcommand{\mat}[2]{\left[\begin{array}{#1} #2 \end{array}\right]}

\def\ib{{\bf i}}   
 \def\n{{\bf n}}  
\def\q{{\bf q}}   \def\t{{\bf t}}
\def\u{{\bf u}} \def\v{{\bf v}}  \def\x{{\bf x}}
\def\y{{\bf y}} \def\z{{\bf z}}

\def\A{{\bf A}} \def\B{{\bf B}} \def\C{{\bf C}} \def\D{{\bf D}}
 \def\F{{\bf F}}  \def\H{{\bf H}}
\def\I{{\bf I}} \def\J{{\bf J}} \def\K{{\bf K}} \def\L{{\bf L}}
   \def\P{{\bf P}}
\def\Q{{\bf Q}} \def\R{{\bf R}} \def\S{{\bf S}} \def\T{{\bf T}}
 \def\V{{\bf V}}

   \def\cD{{\cal D}}

 \def\cR{{\cal R}}

  \def\tx{\tilde{\x}}

\def\tA{\tilde{\A}} \def\tB{\tilde{\B}} \def\tC{\tilde{\C}}

\def\tS{\tilde{\S}}

\def\ooA{\overline{\A}} \def\ooB{\overline{\B}}

\def\ooS{\overline{\S}}

  \def\hx{\hat{\x}}

\def\hA{\hat{\A}} \def\hB{\hat{\B}} \def\hC{\hat{\C}}
\def\hD{\hat{\D}}  
  
  \def\hL{\hat{\L}}

\def\hS{\hat{\S}}

 \psset{unit=1.0\unitlength}
 \SpecialCoor

\newcommand{\schemaEMR}[3]{
 \setlength{\unitlength}{#2}
 \psset{unit=1.0\unitlength}
 \SpecialCoor
 \begin{pspicture}#1
 \put(0,0){$#3$}
 \end{pspicture}
}

\newrgbcolor{greenSor}{0.8 1 0.8}
\newrgbcolor{rosaAcc}{1 0.8 0.6}
\newrgbcolor{azzurro}{0 0.9 1}

\newcommand{\EMRnrsezgiu}[2]{\begin{picture}(0,0)
 \rput[b](0,1){\small\cput[linewidth=0.3pt,framesep=1pt](0,0){\tiny #1}}
 \psline[linestyle=dashed,linewidth=0.2pt](0,-#2)
 \end{picture}}

\newcommand{\EMRsorin}[3]{\begin{picture}(7.5,5)
 \psellipse[linecolor=green,fillstyle=solid,fillcolor=greenSor](3,2.5)(3,2)
 \rput(3,2.5){\normalsize #1}
 \psline(5,4)(7.5,4)
 \psline{<-}(5,1)(7.5,1)
 \rput[b](7.5,4.5){$#2$}
 \rput[t](7.5,0.5){$#3$}
\end{picture}}

\newcommand{\EMRsorout}[3]{\begin{picture}(7.5,5)
 \psellipse[linecolor=green,fillstyle=solid,fillcolor=greenSor](4.5,2.5)(3,2)
 \rput(4.5,2.5){\normalsize #1}
 \psline{->}(0,4)(2.5,4)
 \psline(0,1)(2.5,1)
 \rput[b](0,4.5){$#2$}
 \rput[t](0,0.5){$#3$}
\end{picture}}

\newcommand{\EMRacc}[2]{\begin{picture}(7,5)
 \psline{->}(0,4)(2.5,4)
 \psline(0,1)(2.5,1)
 \psframe[linecolor=red,fillstyle=solid,fillcolor=rosaAcc](2.5,0)(4.5,5)
\psline[linecolor=red](2.5,0)(4.5,5)
 \psline(4.5,4)(7,4)
 \psline{<-}(4.5,1)(7,1)
 \rput[b](7,4.5){$#1$}
 \rput[t](7,0.5){$#2$}
\end{picture}}

\newcommand{\EMRelemecconuno}[2]{\begin{picture}(10,5)
 \psline{->}(0,4)(2.9,4)
 \psline(0,1)(2.9,1)
 \pscircle[linecolor=red,fillstyle=solid,fillcolor=rosaAcc](5,2.5){2.5}
 \psline(7.1,4)(10,4)
 \psline{<-}(7.1,1)(10,1)
 \rput[b](10,4.5){$#1$}
 \rput[t](10,0.5){$#2$}
\end{picture}}

 \psset{unit=1.0\unitlength}
 \SpecialCoor

\newcommand{\schemaPOG}[3]{
 \setlength{\unitlength}{#2}
 \psset{unit=1.0\unitlength}
 \SpecialCoor
 \begin{pspicture}#1
 \put(0,0){$#3$}
 \end{pspicture}
}

\newcommand{\punto}{\pscircle*{0.16}}

\newcommand{\tratteggio}{\psline[linestyle=dashed,linewidth=0.2pt](0,-0.75)(0,10.75)}

\newcommand{\piu}{\begin{pspicture}(0,0)
 \put(0,0){\pscircle[linewidth=0.4pt,fillstyle=none]{0.5}}
 \psline(0.5; 45)(0.5;225)
 \psline(0.5;315)(0.5;135)
\end{pspicture}}

\newcommand{\dst}{\begin{pspicture}(0,0)
 \put(0.293,0){\pscircle*{0.207}}
\end{pspicture}}

\newcommand{\giu}{\begin{pspicture}(0,0)
 \put(0      ,-0.293 ){\pscircle*{0.207}}
\end{pspicture}}

\newcommand{\su}{\begin{pspicture}(0,0)
 \put(0      ,0.293 ){\pscircle*{0.207}}
\end{pspicture}}

\newcommand{\sumdstsnt}[1]{\begin{pspicture}(0,1)
   \put(4,0){\piu}
   \put(4,0){#1}
   \put(0,0){\psline{->}(3.5,0)}
   \put(8,0){\psline{->}(-3.5,0)}
\end{pspicture}}

\newcommand{\sumdstdst}[1]{\begin{pspicture}(0,1)
   \put(4,0){\piu}
   \put(4,0){#1}
   \put(0,0){\psline{->}(3.5,0)}
   \put(4.5,0){\psline{->}(3.5,0)}
\end{pspicture}}

\newcommand{\sumsntsnt}[1]{\begin{pspicture}(0,1)
   \put(4,0){\piu}
   \put(4,0){#1}
   \put(3.5,0){\psline{->}(-3.5,0)}
   \put(8,0){\psline{->}(-3.5,0)}
\end{pspicture}}

\newcommand{\sumdstsntvi}[1]{\begin{pspicture}(0,1)
   \put(3,0){\piu}
   \put(3,0){#1}
   \put(0,0){\psline{->}(2.5,0)}
   \put(6,0){\psline{->}(-2.5,0)}
\end{pspicture}}

\newcommand{\sumdstdstvi}[1]{\begin{pspicture}(0,1)
   \put(3,0){\piu}
   \put(3,0){#1}
   \put(0,0){\psline{->}(2.5,0)}
   \put(3.5,0){\psline{->}(2.5,0)}
\end{pspicture}}

\newcommand{\sumsntsntvi}[1]{\begin{pspicture}(0,1)
   \put(3,0){\piu}
   \put(3,0){#1}
   \put(2.5,0){\psline{->}(-2.5,0)}
   \put(6,0){\psline{->}(-2.5,0)}
\end{pspicture}}

\newcommand{\sumdstsntiv}[1]{\begin{pspicture}(0,1)
   \put(2,0){\piu}
   \put(2,0){#1}
   \put(0,0){\psline{->}(1.5,0)}
   \put(4,0){\psline{->}(-1.5,0)}
\end{pspicture}}

\newcommand{\sumsntsntiv}[1]{\begin{pspicture}(0,1)
   \put(2,0){\piu}
   \put(2,0){#1}
   \put(1.5,0){\psline{->}(-1.5,0)}
   \put(4,0){\psline{->}(-1.5,0)}
\end{pspicture}}

\newcommand{\dotdstdst}{\begin{pspicture}(0,1)
   \put(4,0){\punto}
   \put(0,0){\psline{->}(4,0)}
   \put(4,0){\psline{->}(4,0)}
\end{pspicture}}

\newcommand{\dotsntsnt}{\begin{pspicture}(0,1)
   \put(4,0){\punto}
   \put(4,0){\psline{->}(-4,0)}
   \put(8,0){\psline{->}(-4,0)}
\end{pspicture}}

\newcommand{\dotsntdst}{\begin{pspicture}(0,1)
   \put(4,0){\punto}
   \put(4,0){\psline{->}(-4,0)}
   \put(4,0){\psline{->}(4,0)}
\end{pspicture}}

\newcommand{\dotdstdstvi}{\begin{pspicture}(0,1)
   \put(3,0){\punto}
   \put(0,0){\psline{->}(3,0)}
   \put(3,0){\psline{->}(3,0)}
\end{pspicture}}

\newcommand{\dotsntsntvi}{\begin{pspicture}(0,1)
   \put(3,0){\punto}
   \put(3,0){\psline{->}(-3,0)}
   \put(6,0){\psline{->}(-3,0)}
\end{pspicture}}

\newcommand{\dotsntdstvi}{\begin{pspicture}(0,1)
   \put(3,0){\punto}
   \put(3,0){\psline{->}(-3,0)}
   \put(3,0){\psline{->}(3,0)}
\end{pspicture}}

\newcommand{\dotdstdstiv}{\begin{pspicture}(0,1)
   \put(2,0){\punto}
   \put(0,0){\psline{->}(2,0)}
   \put(2,0){\psline{->}(2,0)}
\end{pspicture}}

\newcommand{\dotsntdstiv}{\begin{pspicture}(0,1)
   \put(2,0){\punto}
   \put(2,0){\psline{->}(-2,0)}
   \put(2,0){\psline{->}(2,0)}
\end{pspicture}}

\newcommand{\midgiuaddbase}[2]{\begin{pspicture}(0,1)
   \put( 0,0.5){#1}
   \put( 0.25,1.5){\makebox(0,0)[lb]{$#2$}}
\end{pspicture}}

\newcommand{\midsuaddbase}[2]{\begin{pspicture}(0,1)
   \put( 0,-0.5){#1}
   \put( 0.25,8.5){\makebox(0,0)[lt]{$#2$}}
\end{pspicture}}

\newcommand{\midgiuaddbasetop}[2]{\begin{pspicture}(0,1)
   \put( 0,0.5){#1}
   \put( 0.25,0.75){\makebox(0,0)[lb]{$#2$}}
\end{pspicture}}

\newcommand{\midsuaddbasetop}[2]{\begin{pspicture}(0,1)
   \put( 0,-0.5){#1}
   \put( 0.25,9.25){\makebox(0,0)[lt]{$#2$}}
\end{pspicture}}

\newcommand{\midgiu}[2]{\begin{pspicture}(0,1)
   \put(-2,3){\psframe(4,4)\rput(2,2){$#1$}}
   \put(0,3){\psline{->}(0,-3)}
   \put(0,9.5){\psline{->}(0,-2.5)}
   \put(0,-0.5){\makebox(0,0)[t]{$#2$}}
\end{pspicture}}

\newcommand{\midgiuadd}[2]{\midgiuaddbase{\midgiu{#1}{}}{#2}}

\newcommand{\midsu}[2]{\begin{pspicture}(0,1)
   \put(-2,3){\psframe(4,4)\rput(2,2){$#1$}}
   \put(0,7){\psline{->}(0,3)}
   \put(0,0.5){\psline{->}(0,2.5)}
   \put(0,10.5){\makebox(0,0)[b]{$#2$}}
\end{pspicture}}

\newcommand{\midsuadd}[2]{\midsuaddbase{\midsu{#1}{}}{#2}}

\newcommand{\midsuintm}[2]{\begin{pspicture}(0,1)
   \put(    0,0.5){\psline{->}(0,1)}
   \put(-1.25,1.5){\psframe(2.5,2.5)\rput(1.25,1.25){{\bf $\frac{1}{s}$}}}
   \put(    0,  4){\psline{->}(0,1)}
   \put(   -2,  5){\psframe(4,4)\rput(2,2){$#1$}}
   \put(    0,  9){\psline{->}(0,1)}
   \put(0,10.5){\makebox(0,0)[b]{$#2$}}
\end{pspicture}}

\newcommand{\midsuintmadd}[2]{\midsuaddbasetop{\midsuintm{#1}{}}{#2}}

\newcommand{\midgiumxm}[4]{\begin{pspicture}(0,1)
   \put(    0,9.5){\psline{->}(0,-1)}
   \put(-1.25,6.0){\psframe(2.5,2.5)\rput(1.25,1.25){{\bf $#3$}}}
   \put(    0,  6){\psline{->}(0,-1)}
   \put(  0.5,5.5){\makebox(0,0)[l]{$#4$}}
   \put(   -2,  1){\psframe(4,4)\rput(2,2){$#1$}}
   \put(    0,  1){\psline{->}(0,-1)}
   \put(0,-0.5){\makebox(0,0)[t]{$#2$}}
\end{pspicture}}

\newcommand{\midgiuivmxm}[4]{\begin{pspicture}(0,1)
   \put(    0,9.5){\psline{->}(0,-1.5)}
   \put(-1.25,5.5){\psframe(2.5,2.5)\rput(1.25,1.25){{\bf $#3$}}}
   \put(    0,5.5){\psline{->}(0,-1.5)}
   \put(  0.5,4.75){\makebox(0,0)[l]{$#4$}}
   \put(-1.25,1.5){\psframe(2.5,2.5)\rput(1.25,1.25){$#1$}}
   \put(    0,1.5){\psline{->}(0,-1.5)}
   \put(0,-0.5){\makebox(0,0)[t]{$#2$}}
\end{pspicture}}

\newcommand{\midsumxm}[4]{\begin{pspicture}(0,1)
   \put(    0,0.5){\psline{->}(0,1)}
   \put(-1.25,1.5){\psframe(2.5,2.5)\rput(1.25,1.25){{\bf $#3$}}}
   \put(    0,  4){\psline{->}(0,1)}
   \put(  0.5,4.5){\makebox(0,0)[l]{$#4$}}
   \put(   -2,  5){\psframe(4,4)\rput(2,2){$#1$}}
   \put(    0,  9){\psline{->}(0,1)}
   \put(0,10.5){\makebox(0,0)[b]{$#2$}}
\end{pspicture}}

\newcommand{\midsumxmadd}[4]{\midsuaddbasetop{\midsumxm{#1}{}{#3}{#4}}{#2}}

\newcommand{\midsuivmxm}[4]{\begin{pspicture}(0,1)
   \put(    0,0.5){\psline{->}(0,1.5)}
   \put(-1.25,  2){\psframe(2.5,2.5)\rput(1.25,1.25){{\bf $#3$}}}
   \put(    0,4.5){\psline{->}(0,1.5)}
   \put(  0.5,5.25){\makebox(0,0)[l]{$#4$}}
   \put(-1.25,  6){\psframe(2.5,2.5)\rput(1.25,1.25){$#1$}}
   \put(    0,8.5){\psline{->}(0,1.5)}
   \put(0,10.5){\makebox(0,0)[b]{$#2$}}
\end{pspicture}}

\newcommand{\midgiuintm}[2]{\begin{pspicture}(0,1)
   \put(    0,9.5){\psline{->}(0,-1)}
   \put(-1.25,  6){\psframe(2.5,2.5)\rput(1.25,1.25){{\bf $\frac{1}{s}$}}}
   \put(    0,  6){\psline{->}(0,-1)}
   \put(   -2,  1){\psframe(4,4)\rput(2,2){$#1$}}
   \put(    0,  1){\psline{->}(0,-1)}
   \put(0,-0.5  ){\makebox(0,0)[t]{$#2$}}
\end{pspicture}}

\newcommand{\midgiuintmadd}[2]{\midgiuaddbasetop{\midgiuintm{#1}{}}{#2}}

\newcommand{\midgiuiv}[2]{\begin{pspicture}(0,1)
   \put(-1.5,3.5){\psframe(3,3)\rput(1.5,1.5){$#1$}}
   \put(0,9.5){\psline{->}(0,-3)}
   \put(0,3.5){\psline{->}(0,-3.5)}
   \put(0,-0.5){\makebox(0,0)[t]{$#2$}}
\end{pspicture}}

\newcommand{\midgiuivadd}[2]{\midgiuaddbase{\midgiuiv{#1}{}}{#2}}

\newcommand{\midsuiv}[2]{\begin{pspicture}(0,1)
   \put(-1.5,3.5){\psframe(3,3)\rput(1.5,1.5){$#1$}}
   \put(0,6.5){\psline{->}(0,3.5)}
   \put(0,0.5){\psline{->}(0,3)}
   \put(0,10.5){\makebox(0,0)[b]{$#2$}}
\end{pspicture}}

\newcommand{\midsuivadd}[2]{\midsuaddbase{\midsuiv{#1}{}}{#2}}

\newcommand{\blosubase}[2]{\begin{pspicture}(8,10)
   \put(0, 0){\tratteggio}
   \put(0,10){\dotsntdst}
   \put(4, 0){#1}
   \put(0, 0){\sumdstsnt{#2}}
\end{pspicture}}

\newcommand{\blovisubase}[2]{\begin{pspicture}(6,10)
   \put(0, 0){\tratteggio}
   \put(0,10){\dotsntdstvi}
   \put(3, 0){#1}
   \put(0, 0){\sumdstsntvi{#2}}
\end{pspicture}}

\newcommand{\bloivsubase}[2]{\begin{pspicture}(4,10)
   \put(0, 0){\tratteggio}
   \put(0,10){\dotsntdstiv}
   \put(2, 0){#1}
   \put(0, 0){\sumdstsntiv{#2}}
\end{pspicture}}

\newcommand{\blosuorbase}[2]{\begin{pspicture}(8,10)
   \put(0, 0){\tratteggio}
   \put(0,10){\sumdstdst{#2}}
   \put(4, 0){#1}
   \put(0, 0){\dotsntsnt}
\end{pspicture}}

\newcommand{\blovisuorbase}[2]{\begin{pspicture}(6,10)
   \put(0, 0){\tratteggio}
   \put(0,10){\sumdstdstvi{#2}}
   \put(3, 0){#1}
   \put(0, 0){\dotsntsntvi}
\end{pspicture}}

\newcommand{\blosuaorbase}[2]{\begin{pspicture}(8,10)
   \put(0, 0){\tratteggio}
   \put(0,10){\sumsntsnt{#2}}
   \put(4, 0){#1}
   \put(0, 0){\dotdstdst}
\end{pspicture}}

\newcommand{\blovisuaorbase}[2]{\begin{pspicture}(6,10)
   \put(0, 0){\tratteggio}
   \put(0,10){\sumsntsntvi{#2}}
   \put(3, 0){#1}
   \put(0, 0){\dotdstdstvi}
\end{pspicture}}

\newcommand{\bloivsuaorbase}[2]{\begin{pspicture}(4,10)
   \put(0, 0){\tratteggio}
   \put(0,10){\sumsntsntiv{#2}}
   \put(2, 0){#1}
   \put(0, 0){\dotdstdstiv}
\end{pspicture}}

\newcommand{\blogiubase}[2]{\begin{pspicture}(8,10)
   \put(0, 0){\tratteggio}
   \put(0,10){\sumdstsnt{#2}}
   \put(4, 0){#1}
   \put(0, 0){\dotsntdst}
\end{pspicture}}

\newcommand{\blovigiubase}[2]{\begin{pspicture}(6,10)
   \put(0, 0){\tratteggio}
   \put(0,10){\sumdstsntvi{#2}}
   \put(3, 0){#1}
   \put(0, 0){\dotsntdstvi}
\end{pspicture}}

\newcommand{\bloivgiubase}[2]{\begin{pspicture}(4,10)
   \put(0, 0){\tratteggio}
   \put(0,10){\sumdstsntiv{#2}}
   \put(2, 0){#1}
   \put(0, 0){\dotsntdstiv}
\end{pspicture}}

\newcommand{\blogiuorbase}[2]{\begin{pspicture}(8,10)
   \put(0, 0){\tratteggio}
   \put(0,10){\dotdstdst}
   \put(4, 0){#1}
   \put(0, 0){\sumsntsnt{#2}}
\end{pspicture}}

\newcommand{\blovigiuorbase}[2]{\begin{pspicture}(6,10)
   \put(0, 0){\tratteggio}
   \put(0,10){\dotdstdstvi}
   \put(3, 0){#1}
   \put(0, 0){\sumsntsntvi{#2}}
\end{pspicture}}

\newcommand{\bloivgiuorbase}[2]{\begin{pspicture}(4,10)
   \put(0, 0){\tratteggio}
   \put(0,10){\dotdstdstiv}
   \put(2, 0){#1}
   \put(0, 0){\sumsntsntiv{#2}}
\end{pspicture}}

\newcommand{\blogiuaorbase}[2]{\begin{pspicture}(8,10)
   \put(0, 0){\tratteggio}
   \put(0,10){\dotsntsnt}
   \put(4, 0){#1}
   \put(0, 0){\sumdstdst{#2}}
\end{pspicture}}

\newcommand{\blovigiuaorbase}[2]{\begin{pspicture}(6,10)
   \put(0, 0){\tratteggio}
   \put(0,10){\dotsntsntvi}
   \put(3, 0){#1}
   \put(0, 0){\sumdstdstvi{#2}}
\end{pspicture}}

\newcommand{\bloin}[2]{\begin{pspicture}(0,0)
 \put(-0.5,10){\makebox(0,0)[r]{$#1$}}
 \put(-0.5, 0){\makebox(0,0)[r]{$#2$}}
 \put(0, 0){\punto}
 \put(0,10){\punto}
 \end{pspicture}}

\newcommand{\bloout}[2]{\begin{pspicture}(0,10)
   \put(0.5,10){\makebox(0,0)[l]{$#1$}}
   \put(0.5, 0){\makebox(0,0)[l]{$#2$}}
   \put(0, 0){\punto}
   \put(0,10){\punto}
   \put(0,0){\tratteggio}
\end{pspicture}}

\newcommand{\blosu}[3]{\blosubase{\midsu{#1}{#2}}{#3}}

\newcommand{\blosuor}[3]{\blosuorbase{\midsuadd{#1}{#2}}{#3}}

\newcommand{\blosuaor}[3]{\blosuaorbase{\midsuadd{#1}{#2}}{#3}}

\newcommand{\bloivsuaor}[3]{\bloivsuaorbase{\midsuivadd{#1}{#2}}{#3}}

\newcommand{\blovisuintm}[3]{\blovisubase{\midsuintm{#1}{#2}}{#3}}

\newcommand{\blovisuintmor}[3]{\blovisuorbase{\midsuintmadd{#1}{#2}}{#3}}

\newcommand{\blovisuintmaor}[3]{\blovisuaorbase{\midsuintmadd{#1}{#2}}{#3}}

\newcommand{\blovisumxm}[5]{\blovisubase{\midsumxm{#1}{#2}{#4}{#5}}{#3}}
\newcommand{\bloivsumxm}[5]{\bloivsubase{\midsuivmxm{#1}{#2}{#4}{#5}}{#3}}

\newcommand{\blovisumxmaor}[5]{\blovisuaorbase{\midsumxmadd{#1}{#2}{#4}{#5}}{#3}}

\newcommand{\blogiu}[3]{\blogiubase{\midgiu{#1}{#2}}{#3}}
\newcommand{\blovigiu}[3]{\blovigiubase{\midgiu{#1}{#2}}{#3}}
\newcommand{\bloivgiu}[3]{\bloivgiubase{\midgiuiv{#1}{#2}}{#3}}
\newcommand{\blogiuor}[3]{\blogiuorbase{\midgiuadd{#1}{#2}}{#3}}

\newcommand{\bloivgiuor}[3]{\bloivgiuorbase{\midgiuivadd{#1}{#2}}{#3}}
\newcommand{\blogiuaor}[3]{\blogiuaorbase{\midgiuadd{#1}{#2}}{#3}}

\newcommand{\blovigiuintm}[3]{\blovigiubase{\midgiuintm{#1}{#2}}{#3}}

\newcommand{\blovigiuintmor}[3]{\blovigiuorbase{\midgiuintmadd{#1}{#2}}{#3}}

\newcommand{\blovigiuintmaor}[3]{\blovigiuaorbase{\midgiuintmadd{#1}{#2}}{#3}}

\newcommand{\blovigiumxm}[5]{\blovigiubase{\midgiumxm{#1}{#2}{#4}{#5}}{#3}}
\newcommand{\bloivgiumxm}[5]{\bloivgiubase{\midgiuivmxm{#1}{#2}{#4}{#5}}{#3}}

\newcommand{\kxdst}[4]{\begin{pspicture}(0,1)
   \put(0,0){\psline{->}(1,0)}
   \psset{unit=0.5\unitlength}
   \rput[lb](2,-#1){\psset{unit=2.0\unitlength}
   \framebox(#1,#1){$#2$}}
   \psset{unit=2.0\unitlength}
   \put(#1,0){
   \put(1,0){\psline{->}(1,0)}
   \put(2.25, 0.5){\makebox(0,0)[bl]{$#3$}}
   \put(2.25,-0.5){\makebox(0,0)[tl]{$#4$}}}
\end{pspicture}}

\newcommand{\kxsnt}[4]{\begin{pspicture}(0,1)
   \put(1,0){\psline{->}(-1,0)}
   \put(-0.25, 0.5){\makebox(0,0)[br]{$#3$}}
   \put(-0.25,-0.5){\makebox(0,0)[tr]{$#4$}}
   \put(#1,0){
   \put(2,0){\psline{->}(-1,0)}
   \psset{unit=0.5\unitlength}
   \rput[rb](2,-#1){\psset{unit=2.0\unitlength}
   \framebox(#1,#1){$#2$}}
   \psset{unit=2.0\unitlength}}
\end{pspicture}}

\newcommand{\kxydst}[5]{\begin{pspicture}(0,1)
   \put(0,0){\psline{->}(1,0)}
   \psset{unit=0.5\unitlength}
   \rput[lb](2,-#2){\psset{unit=2.0\unitlength}
   \framebox(#1,#2){$#3$}}
   \psset{unit=2.0\unitlength}
   \put(#1,0){
   \put(1,0){\psline{->}(1,0)}
   \put(2.25, 0.5){\makebox(0,0)[bl]{$#4$}}
   \put(2.25,-0.5){\makebox(0,0)[tl]{$#5$}}}
\end{pspicture}}

\newcommand{\kxysnt}[5]{\begin{pspicture}(0,1)
   \put(1,0){\psline{->}(-1,0)}
   \put(-0.25, 0.5){\makebox(0,0)[br]{$#4$}}
   \put(-0.25,-0.5){\makebox(0,0)[tr]{$#5$}}
   \put(#1,0){
   \put(2,0){\psline{->}(-1,0)}
   \psset{unit=0.5\unitlength}
   \rput[rb](2,-#2){\psset{unit=2.0\unitlength}
   \framebox(#1,#2){$#3$}}
   \psset{unit=2.0\unitlength}}
\end{pspicture}}

\newcommand{\blokxbase}[3]{\begin{pspicture}(#1,10)
   \put(0, 0){\tratteggio}
   \put(0, 0){#2}
   \put(0,10){#3}
\end{pspicture}
\begin{pspicture}(2,10)\put(0, 0){}\end{pspicture}
}
\newcommand{\blokxor}[5]{\blokxbase{#1}{\kxsnt{#1}{#3}{}{#5}}{\kxdst{#1}{#2}{#4}{}}}
\newcommand{\blokxaor}[5]{\blokxbase{#1}{\kxdst{#1}{#3}{}{#5}}{\kxsnt{#1}{#2}{#4}{}}}
\newcommand{\blokxyor}[6]{\blokxbase{#1}{\kxysnt{#1}{#2}{#4}{}{#6}}{\kxydst{#1}{#2}{#3}{#5}{}}}
\newcommand{\blokxyaor}[6]{\blokxbase{#1}{\kxydst{#1}{#2}{#4}{}{#6}}{\kxysnt{#1}{#2}{#3}{#5}{}}}

\newcommand{\flowrightsu}[1]{\begin{pspicture}(0,10)
   \psline[linewidth=0.4pt]{->}(-0.5,11.5)(0.5,11.5)
   \put(0,11.75){\makebox(0,0)[b]{$#1$}}
\end{pspicture}}

\newcommand{\flowright}[1]{\begin{pspicture}(0,12)
   \put(0,-1){\makebox(0,0)[t]{$\Rightarrow$}}
\end{pspicture}}

\newcommand{\flowleft}[1]{\begin{pspicture}(0,12)
   \put(0,-1){\makebox(0,0)[t]{$\leftarrow$}}
\end{pspicture}}

\newcommand{\POGvigiu}[4]{\begin{pspicture}(6,10)
   \put(0, 0){\blovigiu{#1}{}{#4}}
   \rput[lb](3.25,8.00){$#2$}
   \rput[lb](3.25,1.00){$#3$}
\end{pspicture}}

  \newcommand{\loopxysnt}[5]{%
  \rput(1,4.5){\psline[#5]{#3}(-1,#2)(-#1,#2)}
  \rput(1,4.0){\psarc[#5](-#1,#2){0.5}{90}{180}}
  \rput(0.5,4){\psline[#5](-#1,#2)(-#1,1)}
  \rput(0.5,6){\psline[#5](-#1,-1)(-#1,-#2)}
  \rput(1,6.0){\psarc[#5](-#1,-#2){0.5}{180}{270}}
  \rput(1,5.5){\psline[#5]{#4}(-1,-#2)(-#1,-#2)}
  }
  \newcommand{\loopxydst}[5]{
  \rput(-1,4.5){\psline[#5]{#3}(1,#2)(#1,#2)}
  \rput(-1,4.0){\psarc[#5](#1,#2){0.5}{0}{90}}
  \rput(-0.5,4){\psline[#5](#1,#2)(#1,1)}
  \rput(-0.5,6){\psline[#5](#1,-1)(#1,-#2)}
  \rput(-1,6.0){\psarc[#5](#1,-#2){0.5}{270}{0}}
  \rput(-1,5.5){\psline[#5]{#4}(1,-#2)(#1,-#2)}
  }
  \newcommand{\loopxsnt}[4]{\loopxysnt{#1}{5}{#2}{#3}{#4}}

  \newcommand{\loopxdst}[4]{\loopxydst{#1}{5}{#2}{#3}{#4}}
  \newcommand{\loopvisntor}[1]{\loopxsnt{3}{<-}{-}{#1}}
  \newcommand{\loopvisntaor}[1]{\loopxsnt{3}{-}{<-}{#1}}
  
  \newcommand{\loopvidstor}[1]{\loopxdst{3}{-}{<-}{#1}}
  \newcommand{\loopvidstaor}[1]{\loopxdst{3}{<-}{-}{#1}}
\csname PSTplotLoaded\endcsname
\let\PSTplotLoaded 
\ifx\PSTricksLoaded \else\input pstricks.tex\fi
\ifx\PSTXKeyLoaded \else \input pst-xkey.tex\fi
\ifx\PSTtoolsloaded \else\input pst-tools.tex\fi
\ifx\PSTtoolsloaded \else\input pstricks-add.tex\fi
\ifx\PSTFPloaded \else   \input pst-fp.tex  \fi
\ifx\MultidoLoaded \else \input multido.tex \fi
\def\fileversion{1.92}
\def\filedate{2019/05/16}
\edef\TheAtCode{\the\catcode`\@}
\catcode`\@=11
\pst@addfams{pst-plot}
\def\pst@linetype{2}%
\SpecialCoor

\newdimen\pstRadUnit
\newdimen\pstRadUnitInv
\begingroup
\catcode`\{=13
\catcode`\}=13
\catcode`\(=13
\catcode`\)=13
\catcode`\,=13
\catcode`\!=1
\catcode`\*=2
\catcode`\ =13
\catcode`\_=13
\catcode`\^^M=13
\gdef\pst@datadelimiters!
\catcode`\{=13%
\catcode`\}=13%
\catcode`\(=13%
\catcode`\)=13%
\catcode`\,=13%
\catcode`\ =13%
\catcode`\^^M=13%
\def,##1!%
\ifcat\noexpand,\noexpand##1%
\expandafter##1%
\else\space%
D\space##1%
\fi*%
\let(,\let),\let{,\let},\let ,\let^^M,\let_\@empty*
\endgroup
\define@key[psset]{pst-plot}{ignoreLines}[0]{\def\psk@ignoreLines{#1}}
\psset[pst-plot]{ignoreLines=0}
\newcount\pst@linecnt
\begingroup
\catcode`\,=13
\catcode`\_=13
\gdef\savedata@#1[#2]{%
  \xdef\pst@tempg{#2_}%
  \endgroup
  \let#1\pst@tempg
  \global\let\pst@tempg\relax
  \ignorespaces}
\gdef\readdata@{%
  \read1 to \pst@tempA
  \ifnum\pst@linecnt=\psk@nStep
    \global\pst@linecnt=0
    \expandafter\readdata@@\pst@tempA_\@nil
  \fi
  \global\advance\pst@linecnt by 1
  \ifeof1\else\expandafter\readdata@\fi}
\gdef\pst@@readfile#1#2\@nil{\addto@pscode{,#1#2}}%
\gdef\readdata@@#1#2\@nil{\xdef\pst@tempg{\pst@tempg,#1#2}}%
\endgroup
\def\readdata{\@ifnextchar[{\readdata@i}{\readdata@i[]}}
\def\readdata@i[#1]#2#3{%
  \openin1=#3
  \begingroup
  \ifx#1\relax#1\else\psset{#1}\fi
  \def\pst@tempg{}%
  \ifeof1
    \@pstrickserr{Data file `#3' not found.}\@ehpa
  \else
    \pst@datadelimiters
    \catcode`\[=1
    \catcode`\]=2
    \pst@cnta=0
    \loop \ifnum\the\pst@cnta<\psk@ignoreLines
      \advance\pst@cnta by 1\relax
      \read1 to \pst@tempA
    \repeat
    \psDEBUG[pst-plot]{>>> ignored \the\pst@cnta\space data lines}%
    \global\pst@linecnt=\psk@nStep
    \readdata@
  \fi
  \endgroup
  \global\let#2\pst@tempg%
  \global\let\pst@tempg\relax%
\ignorespaces}
\def\pst@readfile#1{{\let\readdata@@\pst@@readfile\readdata\pst@tempg{#1}}}
\def\pst@altreadfile#1{%
  \openin1=#1
  \ifeof1
    \@pstrickserr{Data file `#1' not found.}\@ehpa
  \else
    \catcode`\{=10
    \catcode`\}=10
    \catcode`\(=10
    \catcode`\)=10
    \catcode`\,=10
    \catcode`\^^M=10
    \catcode`\[=1
    \catcode`\]=2
    \pst@@altreadfile
  \fi}
\def\pst@@altreadfile{%
  \read1 to \pst@tempg
  \expandafter\pst@@@altreadfile\pst@tempg\@empty\@nil
  \ifeof1\else\expandafter\pst@@@altreadfile\fi}
\def\pst@@@altreadfile#1#2\@nil{\addto@pscode{#1#2}}%
\def\savedata#1{\begingroup\pst@datadelimiters\savedata@{#1}}
\newread\RCD@file
\def\psreadDataColumn{\@ifnextchar[\psreadDataColumn@i{\psreadDataColumn@i[]}}
\def\psreadDataColumn@i[#1]{%
  \psset{#1}%
  \psreadDataColumn@ii
}
\def\psreadDataColumn@ii#1#2#3#4{
  \immediate\openin\RCD@file=#4\relax
  \global\let#3=\@empty
  \pst@cnta=0
  \loop \ifnum\the\pst@cnta<\psk@ignoreLines
      \advance\pst@cnta by 1\relax
      \read\RCD@file to \@tempa
  \repeat
  \loop
    \read\RCD@file to \@tempa
    \ifeof\RCD@file\else
      \edef\@tempa{\@tempa#2}%
      \def\reserved@b{}%
      \@tempswafalse
      \@tempcnta=#1\relax
    \expandafter\@tfor\expandafter\reserved@a
      \expandafter:\expandafter=\@tempa\do{
      \if\reserved@a#2\relax
        \advance\@tempcnta \m@ne
        \ifnum \@tempcnta=\z@
          \expandafter\g@addto@macro\expandafter#3\expandafter{\reserved@b\space}%
          \@tempswatrue
        \fi
        \def\reserved@b{}
      \else
        \edef\reserved@b{\reserved@b\reserved@a}
      \fi
      \if@tempswa\@break@tfor\fi
    }%
  \repeat
  \immediate\closein\RCD@file
}

\def\beginplot@line{\begin@OpenObj}
\def\endplot@line{\psline@ii}
\def\beginplot@polygon{\begin@ClosedObj}
\def\endplot@polygon{\pspolygon@ii}
\def\beginplot@curve{\begin@OpenObj}
\def\endplot@curve{\pscurve@ii}
\def\beginplot@ecurve{\begin@OpenObj}
\def\endplot@ecurve{\psecurve@ii}
\def\beginplot@ccurve{\begin@ClosedObj}
\def\endplot@ccurve{\psccurve@ii}
\def\beginplot@dots{\begin@SpecialObj}
\def\endplot@dots{\psdots@ii}
\define@key[psset]{pst-plot}{Hue}[180]{\pst@getint{#1}\pst@HueAngle}
\psset[pst-plot]{Hue=180}
\def\beginplot@colordots{\begin@SpecialObj}
\def\endplot@colordots{%
  \addto@pscode{%
    \psk@dotsize
    \@nameuse{psds@\psk@dotstyle}
    newpath
    /MaxValue 0 def
    /m n 2 mul def
    n { 
      dup MaxValue gt { dup /MaxValue ED } if
      m 2 roll
    } repeat
    n { dup MaxValue div  
      \pst@number\psyunit div abs 
      \pst@HueAngle\space 360 div exch dup sethsbcolor 
      transform floor .5 add exch floor
      .5 add exch itransform Dot stroke } repeat }%
  \end@SpecialObj%
}
\def\beginplot@bubble{\begin@SpecialObj}
\def\endplot@bubble{%
  \addto@pscode{%
    newpath
    n { dup 
      \pst@number\psyunit div abs 
      transform floor .5 add exch floor
      .5 add exch itransform 
      0 360 arc \psk@fill 
      stroke } repeat }%
  \end@SpecialObj%
}
\def\beginplot@bezier{\begin@OpenObj}
\def\endplot@bezier{\psbezier@ii}
\def\beginplot@cbezier{\begin@ClosedObj}
\def\endplot@cbezier{\pscbezier@ii}
\def\beginplot@cspline{\begin@OpenObj}
\def\endplot@cspline{\pscspline@ii}
\let\beginplot@LineToYAxis\beginplot@line  
\def\endplot@LineToYAxis{\psLineToYAxis@ii}
\let\beginqp@LineToYAxis\beginqp@line
\let\doqp@LineToYAxis\doqp@line
\let\endqp@LineToYAxis\endqp@line
\let\testqp@LineToYAxis\testqp@line
\let\beginplot@LineToXAxis\beginplot@line
\def\endplot@LineToXAxis{\psLineToXAxis@ii}
\let\beginqp@LineToXAxis\beginqp@line
\let\doqp@LineToXAxis\doqp@line
\let\endqp@LineToXAxis\endqp@line
\let\testqp@LineToXAxis\testqp@line
\newif\ifPst@interrupt \Pst@interruptfalse
\define@key[psset]{pst-plot}{barwidth}[0.25cm]{\pst@getlength{#1}\Add@barwidth}
\psset[pst-plot]{barwidth=0.25cm}
\define@key[psset]{pst-plot}{interrupt}[]{\expandafter\pst@interrupt#1,,,\@nil}
\def\pst@interrupt#1,#2,#3,#4\@nil{%
  \ifx\relax#1\relax \Pst@interruptfalse
  \else
    \Pst@interrupttrue
    \def\pst@interrupt@YMax{#1 }%
    \def\pst@interrupt@YMaxSep{#2 }%
    \def\pst@interrupt@YMaxDiff{#3 }%
  \fi
}
\def\psbar@ii{\addto@pscode{false \tx@NArray \psbar@iii}}

\def\psbar@iii{%
  \ifPst@interrupt
    /YMax \pst@interrupt@YMax \strip@pt\psyunit\space mul def
    /YMaxSep \pst@interrupt@YMaxSep \strip@pt\psyunit\space mul def
    /YMaxDiff \pst@interrupt@YMaxDiff \strip@pt\psyunit\space mul def
    /Tilde { 
      /Op ED 
      /DX ED
      currentpoint 2 copy
      /Y ED /X ED   
      X DX add Y YMaxSep 2 div Op   
      X DX dup add add Y           
      curveto
      currentpoint 2 copy pop /X ED 
      X DX add Y YMaxSep 2 div neg Op  
      X DX dup add add Y    
      curveto      
    } def  
    newpath
    n { 
      /Yval exch def /Xval exch def 
      Xval \number\Add@barwidth 0.5 mul sub 0 moveto 
      Yval YMax le {  
        0 Yval rlineto \number\Add@barwidth 0 rlineto 
        0 Yval neg rlineto \number\Add@barwidth neg 0 rlineto
      }{
        0 YMax rlineto 
        \number\Add@barwidth 4 div 
        { add } Tilde
        0 YMax neg rlineto 
        \number\Add@barwidth neg 0 rlineto
        closepath
        Xval \number\Add@barwidth 0.5 mul sub YMax YMaxSep add moveto 
        0 Yval YMax sub YMaxSep sub YMaxDiff sub rlineto 
        \number\Add@barwidth 0 rlineto 
        0 Yval YMax YMaxSep add sub YMaxDiff sub neg rlineto 
        \number\Add@barwidth 4 div neg 
        { sub } Tilde
      } ifelse
    } repeat
  \else
    newpath
    n { 
      /Yval exch def /Xval exch def 
      Xval \number\Add@barwidth 0.5 mul sub 0 moveto 
      0 Yval rlineto \number\Add@barwidth 0 rlineto 
      0 Yval neg rlineto \number\Add@barwidth neg 0 rlineto
    } repeat
  \fi
}%
\def\beginplot@bar{\begin@SpecialObj}
\def\endplot@bar{%
  \psbar@ii\psk@fillstyle\ifpsshadow\pst@closedshadow\fi%
  \pst@stroke
  \end@SpecialObj}
\def\psybar@ii{\addto@pscode{false \tx@NArray \psybar@iii}}
\def\psybar@iii{%
  newpath
  n { 
    /Yval exch def /Xval exch def 
    0 Yval \number\Add@barwidth 0.5 mul sub moveto 
    Xval 0 rlineto 0 \number\Add@barwidth rlineto 
    Xval neg 0 rlineto 0 \number\Add@barwidth neg rlineto
  } repeat
}%
\def\beginplot@ybar{\begin@SpecialObj}
\def\endplot@ybar{%
  \psybar@ii\psk@fillstyle\ifpsshadow\pst@closedshadow\fi%
  \pst@stroke
  \end@SpecialObj}
\def\psLSM@ii{\addto@pscode{ false \tx@NArray \psLSM@iii }}
\def\psLSM@iii{%
  /xiSquare 0 def				
  /xi 0 def					
  /fi 0 def					
  /xifi 0 def					
  exch dup dup /xEnd ED /xStart ED exch
  n { 						
    /Yval ED /Xval ED 				
    /xi xi Xval add def				
    /xiSquare xiSquare Xval dup mul add def	
    /xifi xifi Xval Yval mul add def		
    /fi fi Yval add def				
    Xval xStart lt { /xStart Xval def } if	
    Xval xEnd gt { /xEnd Xval def } if		
  } repeat
  /u xiSquare fi mul xi xifi mul sub n xiSquare mul xi dup mul sub div def
  /v n xifi mul xi fi mul sub n xiSquare mul xi dup mul sub div def
  \Pst@Debug\space 0 gt { 			
    /NimbusSanL-Regu findfont 12 scalefont setfont	
    0 -50 moveto (y=) show 			
    v \pst@number\psyunit \pst@number\psxunit div div 20 string cvs show ( x+) show		
    u \pst@number\psyunit div 20 string cvs show } if
  newpath
  (\psk@xStart) length 0 gt 			
    { \psk@xStart\space \pst@number\psxunit mul }
    { xStart } ifelse 
  dup v mul u add 				
  moveto		 			
  (\psk@xEnd) length 0 gt 			
    { \psk@xEnd\space \pst@number\psxunit mul }
    { xEnd } ifelse 
  dup v mul u add 				
  lineto					
}%
\def\beginplot@LSM{\begin@SpecialObj}
\def\endplot@LSM{%
  \psLSM@ii\psk@fillstyle\ifpsshadow\pst@closedshadow\fi%
  \pst@stroke
  \end@SpecialObj}
\define@key[psset]{pst-plot}{IQLfactor}{\pst@checknum{#1}\pst@IQLfactor}
\psset[pst-plot]{IQLfactor=1.5}
\define@key[psset]{pst-plot}{postAction}[]{\def\psk@postAction{%
  \ifx\relax#1\relax\else\pst@number\psyunit div #1 \pst@number\psyunit mul \fi }}
\psset[pst-plot]{postAction=}
\define@key[psset]{pst-plot}{mediancolor}[black]{\pst@getcolor{#1}\median@linecolor}
\psset[pst-plot]{mediancolor=black}
\define@boolkey[psset]{pst-plot}[Pst@]{markMedian}[true]{}
\psset[pst-plot]{markMedian=false}

\def\psBoxplot@ii{%
  \addto@pscode{
    /Barwidth \number\Add@barwidth 2 div def  
    /Endwidth Barwidth \psk@arrowlength\space mul def  
   NArray bubblesort
   /NArray ED 				
   [ NArray { yUnit mul } forall ] /NArray ED 
   NArray 0 get /MinVal ED		
   NArray m 1 sub get /MaxVal ED	
   m 2 div cvi /M ED 			
   NArray length 2 mod 0 eq {		
     M 1 sub NArray exch get 		
     NArray M get          		
     add 2 div /Median ED  		
   }{
     NArray M get /Median ED  		
   } ifelse
   m 4 mod 0 eq {	  		
     m 4 div cvi dup 1 sub NArray exch get
     exch NArray exch get
     add 2 div floor /LowerQuartil ED
   }{ 
     NArray M 2 div cvi get /LowerQuartil ED 
   } ifelse				
   m 0.75 mul dup dup cvi sub 0 eq {	
     cvi dup 1 sub NArray exch get exch NArray exch get
     add 2 div floor /UpperQuartil ED
   }{					
     NArray m 0.75 mul floor cvi get /UpperQuartil ED
   } ifelse 
   /IQL UpperQuartil LowerQuartil sub \pst@IQLfactor\space mul def
   0 1 m 1 sub { 
     dup /Index ED
     NArray exch get LowerQuartil sub abs IQL sub 0 gt { 
       \psk@dotsize
       \@nameuse{psds@\psk@dotstyle}
       0 NArray Index get \psk@postAction
       Dot
       NArray Index LowerQuartil UpperQuartil LowerQuartil sub \pst@IQLfactor\space mul sub 
       dup /MinVal ED put 
       NArray Index 1 add get /MinVal ED 
    } { exit } ifelse
   } for
   m 1 sub -1 0 {	
     dup /Index ED
     NArray exch get UpperQuartil sub abs IQL sub 0 gt { 
       \psk@dotsize
       \@nameuse{psds@\psk@dotstyle}
       0 NArray Index get \psk@postAction\space
       Dot
       NArray Index UpperQuartil LowerQuartil sub \pst@IQLfactor\space mul UpperQuartil add 
       dup /MaxVal ED put 
       NArray Index 1 sub get /MaxVal ED 
     }{ exit } ifelse
   } for
   Endwidth neg MaxVal \psk@postAction moveto			
   Endwidth dup add 0 rlineto 
   0 MaxVal \psk@postAction moveto 
   0 UpperQuartil \psk@postAction lineto			
   MinVal \psk@postAction MaxVal \psk@postAction lt {
     0 LowerQuartil \psk@postAction moveto			
     0 MinVal \psk@postAction lineto 
     Endwidth neg MinVal \psk@postAction moveto 
     Endwidth dup add 0 rlineto 
   } if
   gsave
   \pst@number\pslinewidth SLW
   \pst@usecolor\pslinecolor
   \tx@setStrokeTransparency 
   \@nameuse{psls@\pslinestyle}
   stroke
   grestore
   newpath
   Barwidth neg LowerQuartil \psk@postAction moveto	
   Barwidth neg UpperQuartil \psk@postAction lineto
   Barwidth dup add 0 rlineto
   Barwidth LowerQuartil \psk@postAction lineto
   closepath
   \pst@usecolor\psfillcolor
   gsave \pst@usecolor\psfillcolor \tx@setTransparency fill grestore
   \@nameuse{psls@solid}
   \ifPst@markMedian
     \pst@number\pslabelsep neg Median moveto currentpoint 
     /YMedian ED /XMedian ED 
      Barwidth neg Median \psk@postAction lineto  
   \else
      Barwidth neg Median \psk@postAction moveto  
   \fi
   Barwidth dup add 0 rlineto 
   \pst@number\pslinewidth SLW
   \pst@usecolor\median@linecolor
   \tx@setStrokeTransparency
   stroke
  }
}%
\def\beginplot@Boxplot{\init@pscode}
\def\endplot@Boxplot{%
  \psBoxplot@ii\psk@fillstyle\ifpsshadow\pst@closedshadow\fi%
  \pst@stroke
  \end@SpecialObj}
\def\psBoxplot{\def\pst@par{}\pst@object{psBoxplot}}
\def\psBoxplot@i#1{%
  \leavevmode
  \pst@killglue
  \begingroup
  \addbefore@par{barwidth=40pt,arrowlength=0.75}%
  \addto@par{plotstyle=Boxplot}%
  \use@par
  \@nameuse{beginplot@\psplotstyle}%
  \addto@pscode{
    /D {} def
    [ #1 ] /NArray ED 
    NArray aload length /m ED
    /xUnit \pst@number\psxunit def
    /yUnit \pst@number\psyunit def
  }%
  \@nameuse{endplot@\psplotstyle}%
  \ignorespaces%
}
\define@key[psset]{pst-plot}{plotstyle}[line]{%
  \@ifundefined{beginplot@#1}%
    {\@pstrickserr{Plot style `#1' not defined}\@eha}%
    {\def\psplotstyle{#1}}}
\psset[pst-plot]{plotstyle=line}
\define@key[psset]{pst-plot}{plotpoints}[50]{%
  \pst@cntg=#1\relax
  \ifnum\pst@cntg<2
    \@pstrickserr{plotpoints parameter must be at least 2}\@ehpa
  \else
    \advance\pst@cntg-1
    \edef\psk@plotpoints{\the\pst@cntg\space}%
  \fi}
\psset[pst-plot]{plotpoints=50}
\define@key[psset]{pst-plot}{yMaxValue}[1.e30]{\def\psk@yMaxValue{#1 }\def\psk@yMinValue{#1 neg }}
\psset{yMaxValue=1.e30}
\define@key[psset]{pst-plot}{yMinValue}[-1.e30]{\def\psk@yMinValue{#1 }}
\psset{yMinValue=-1.e30}
\def\beginqp@line{\pst@oplineto}
\def\doqp@line{ 
  dup
  \psk@yMaxValue \pst@number\psyunit mul gt 
    { moveto }
    { dup \psk@yMinValue \pst@number\psyunit mul lt 
      { moveto }
      { L } ifelse 
    } ifelse
}
\def\endqp@line{%
  \ifPst@variableLW \addto@pscode{ \pst@flattenpath }\fi%
  \end@OpenObj}%

\def\testqp@line{%
  \ifdim\pslinearc>\z@\else
    \ifshowpoints\else
      \ifx\psk@arrowA\@empty
        \ifx\psk@arrowB\@empty
          \@psttrue
        \fi
      \fi
    \fi
  \fi}
\def\beginqp@polygon{moveto }
\def\doqp@polygon{ 
      dup
      \psk@yMaxValue \pst@number\psyunit mul gt 
      { moveto }{ 
          dup
          \psk@yMinValue \pst@number\psyunit mul lt 
          { moveto }{ L } ifelse 
      } ifelse
}
\def\endqp@polygon{%
  \addto@pscode{closepath}%
  \end@ClosedObj}
\def\testqp@polygon{%
  \ifdim\pslinearc>\z@\else
    \ifshowpoints\else
      \@psttrue
    \fi
  \fi}
\def\beginqp@dots{%
  \psk@dotsize
  \@nameuse{psds@\psk@dotstyle}
  Dot }
\def\doqp@dots{Dot }
\def\endqp@dots{\end@SpecialObj}
\def\testqp@dots{\@psttrue}
\def\beginqp@bezier{/n 0 def \pst@oplineto}
\def\doqp@bezier{/n n 1 add def n 3 mod 0 eq { 
    dup \psk@yMaxValue\space \pst@number\psyunit mul gt 
    { moveto pop pop pop pop}
    { dup \psk@yMinValue\space \pst@number\psyunit mul lt 
      { moveto pop pop pop pop}{ curveto } ifelse 
    } ifelse 
  } if
}
\def\endqp@bezier{%
  \addto@pscode{n 3 mod { pop pop } repeat}
  \end@OpenObj}%
\def\testqp@bezier{%
  \ifshowpoints\else
    \ifx\psk@arrowA\@empty
      \ifx\psk@arrowB\@empty
        \@psttrue
      \fi
    \fi
  \fi}
\def\beginqp@cbezier{/n 0 def moveto }
\def\doqp@cbezier{\doqp@bezier}
\def\endqp@cbezier{%
  \addto@pscode{n 3 mod { pop pop } repeat closepath}
  \end@ClosedObj}%
\def\testqp@cbezier{\ifshowpoints\else\@psttrue\fi}
\def\tx@LineToYAxis{LineToYAxis }
\def\psLineToYAxis@ii{%
\addto@pscode{\pst@cp \psline@iii \psk@Ox\space \pst@number\psxunit mul \tx@LineToYAxis}%
\end@OpenObj}
\def\tx@LineToXAxis{LineToXAxis }
\def\psLineToXAxis@ii{%
\addto@pscode{\pst@cp \psline@iii \psk@Oy\space \pst@number\psyunit mul \tx@LineToXAxis}%
\end@OpenObj}
\define@key[psset]{pst-plot}{PSfont}[NimbusRomNo9L-Regu]{\def\psk@PSfont{/#1 }}
\define@key[psset]{pst-plot}{valuewidth}[10]{\pst@getint{#1}\psk@valuewidth }
\define@key[psset]{pst-plot}{fontscale}[10]{\pst@checknum{#1}\psk@fontscale }
\define@key[psset]{pst-plot}{decimals}[-1]{\pst@getint{#1}\psk@decimals }
\psset[pst-plot]{PSfont=NimbusRomNo9L-Regu,fontscale=10,valuewidth=10,decimals=-1}
\newdimen\psxlabelsep
\newdimen\psylabelsep
\define@key[psset]{pst-plot}{xlabelsep}[5pt]{\pssetlength\psxlabelsep{#1}}
\define@key[psset]{pst-plot}{ylabelsep}[5pt]{\pssetlength\psylabelsep{#1}}
\psset[pst-plot]{xlabelsep=5pt,ylabelsep=5pt}

\newif\ifPst@valuesStar\Pst@valuesStarfalse
\newif\ifPst@xvalues\Pst@xvaluesfalse
\def\psvalues@ii{\addto@pscode{ false \tx@NArray \psvalues@iii }}
\def\psvalues@iii{
  \psk@PSfont findfont \psk@fontscale scalefont setfont 
  newpath 
  n { /yO ED /xO ED
      gsave
      \ifPst@xvalues
        xO \pst@number\psxunit div
      \else
        yO \pst@number\psyunit div
      \fi
      \psk@decimals 0 eq { cvi } if
      \psk@decimals 0 gt { 10 \psk@decimals exp dup 3 1 roll mul cvi exch div } if
      \psk@valuewidth string cvs /Str ED
      \ifPst@valuesStar
      Str stringwidth pop /yS \psk@fontscale def /xS ED 
      gsave newpath 
        xO \ifPst@xvalues \pst@number\pslabelsep add \fi 
        yO \ifPst@xvalues \psk@fontscale 4 div sub \else \pst@number\pslabelsep add \fi 
        moveto \ifx\psk@rot\@empty\else\psk@rot rotate \fi
        xS 0 rlineto 0 yS rlineto xS neg 0 rlineto 0 yS neg rlineto 
        closepath  1 setgray fill stroke 
      grestore 
      \fi
      xO \ifPst@xvalues \pst@number\pslabelsep add \fi
      yO \ifPst@xvalues \psk@fontscale 4 div sub \else \pst@number\pslabelsep add \fi 
      moveto \ifx\psk@rot\@empty\else\psk@rot rotate \fi 
      Str show 
      grestore } repeat 
}%
\def\beginplot@values{\Pst@valuesStarfalse\begin@SpecialObj}
\expandafter\def\csname beginplot@values*\endcsname{\Pst@valuesStartrue\begin@SpecialObj}
\def\beginplot@xvalues{\Pst@valuesStarfalse\begin@SpecialObj}
\expandafter\def\csname beginplot@xvalues*\endcsname{\Pst@valuesStartrue\begin@SpecialObj}
\def\endplot@values{%
  \Pst@xvaluesfalse%
  \psvalues@ii%
  \pst@stroke
  \end@SpecialObj}
\@namedef{endplot@values*}{\endplot@values}
\def\endplot@xvalues{%
  \Pst@xvaluestrue%
  \psvalues@ii%
  \pst@stroke
  \end@SpecialObj}
\@namedef{endplot@xvalues*}{\endplot@xvalues}
\def\psdataplot{\def\pst@par{}\pst@object{dataplot}}
\def\dataplot{\def\pst@par{}\pst@object{dataplot}}
\def\dataplot@i#1{%
  \pst@killglue
  \begingroup
    \use@par
    \@pstfalse
    \@nameuse{testqp@\psplotstyle}%
    \if@pst
      \dataplot@ii{\addto@pscode{#1}}%
    \else
      \listplot@ii{\addto@pscode{#1}}%
    \fi
  \endgroup
  \ignorespaces}
\def\dataplot@ii#1{%
  \@nameuse{beginplot@\psplotstyle}%
    \addto@pscode{%
      /Dx { \pst@number\psxunit mul /D { Dy } def } def
      /Dy { \pst@number\psyunit mul Do /D { Dx } def } def
      /D { /D { Dx } def } def
      /Do {
        \@nameuse{beginqp@\psplotstyle}%
        /Do { \@nameuse{doqp@\psplotstyle}} def
      } def}%
    #1
    \addto@pscode{ D }%
  \@nameuse{endqp@\psplotstyle}}
\def\psfileplot{\def\pst@par{}\pst@object{fileplot}}
\def\fileplot{\def\pst@par{}\pst@object{fileplot}}
\def\fileplot@i#1{%
  \pst@killglue%
  \begingroup%
    \use@par%
    \@pstfalse%
    \@nameuse{testqp@\psplotstyle}%
    \if@pst\dataplot@ii{\pst@readfile{#1}}\else\listplot@ii{\pst@altreadfile{#1}}\fi%
  \endgroup%
  \ignorespaces}
\def\pslistplot{\pst@object{listplot}}
\def\listplot{\pst@object{listplot}}
\def\listplot@i#1{\listplot@ii{\addto@pscode{#1}}}
\def\listplot@ii#1{%
  \@nameuse{beginplot@\psplotstyle}%
  \addto@pscode{/D {} def mark}%
  #1%
  \addto@pscode{
    \tx@PreparePoints
    \pst@number\psxunit
    \pst@number\psyunit
    \tx@ScalePoints
  }%
  \@nameuse{endplot@\psplotstyle}%
}
\define@boolkey[psset]{pst-plot}[Psk@]{xyValues}[true]{}
\define@boolkey[psset]{pst-plot}[Pst@]{ChangeOrder}[true]{}
\psset[pst-plot]{xyValues,ChangeOrder=false}
\pst@def{PreparePoints}<{%
  counttomark /m exch def
  /maxYValues \psk@plotNoMax\space def
  /YValuePos \psk@plotNo\space def
  /XValuePos \psk@plotNoX\space def
  /maxYvalue \ifx\empty\psk@plotYMax false \else true \fi def
  maxYvalue { /YMaxValue \psk@plotYMax\space def } if
  \ifPsk@xyValues\else 
    /mm m def
    /M m 1 add def
    m { mm exch M 2 roll /M M 1 add def /mm mm 1 sub def } repeat
    /m m dup add def
  \fi
  \ifPst@ChangeOrder
    /m0 m def
    m maxYValues 1 add div 1 sub cvi {
      m0 maxYValues 1 add roll /m0 m0 maxYValues 1 add sub def
    } repeat
  \fi
  /n m maxYValues 1 add div cvi def
  XValuePos 1 gt {
    n {
      maxYValues 1 add XValuePos neg roll    
      dup /XValue ED
      maxYValues 1 add XValuePos 1 sub  roll 
      pop XValue                             
      maxYValues 1 add 1 roll                
      m maxYValues 1 add  roll               
    } repeat
  } if 
  maxYValues 1 gt {
    n {
      maxYValues YValuePos 1 sub neg roll 
      maxYValues 1 sub { pop } repeat     
      /m m maxYValues 1 sub sub def
      m 2 roll
    } repeat
  } if 
  /xMax -99999 def /yMax -99999 def
  /xP 0 def /yP 0 def
  m copy
  n {
    /y exch def /x exch def
    xMax x lt { /xMax x def } if
    yMax y lt {/yMax y def } if
    xP x gt { /xP x def } if
    yP y gt { /yP y def } if
  } repeat
  \psk@xStep\space 0 gt \psk@yStep\space 0 gt or (\psk@xStart) length 0 gt or
  (\psk@yStart) length 0 gt or (\psk@xEnd) length 0 gt or (\psk@yEnd) length 0 gt or {
    (\psk@xStart) length 0 gt {\psk@xStart\space }{ xP } ifelse /xStart exch def
    (\psk@yStart) length 0 gt {\psk@yStart\space }{ yP } ifelse /yStart exch def
    (\psk@xEnd) length 0 gt { \psk@xEnd\space }{ xMax } ifelse /xEnd exch def
    (\psk@yEnd) length 0 gt { \psk@yEnd\space }{ yMax } ifelse /yEnd exch def
    n {
      m -2 roll
      2 copy /yVal exch def /xVal exch def
      xVal xP ge
      yVal yP ge and
      xVal xEnd le and
      yVal yEnd le and
      xVal xStart ge and
      yVal yStart ge and {
        /xP xP \psk@xStep\space add def
        /yP yP \psk@yStep\space add def
      }{%
        pop pop
        /m m 2 sub def
      } ifelse
    } repeat
  }{%
    /ncount 1 def
    (\psk@nEnd) length 0 gt { \psk@nEnd\space }{ m } ifelse 
    /nEnd exch def
    n {
      m -2 roll
      \psk@nStep\space 1 gt { ncount \psk@nStart\space sub \psk@nStep\space mod 0 eq }{ true } ifelse
        ncount nEnd le and
        ncount \psk@nStart\space ge and not {
          pop pop
          /m m 2 sub def
        } if
        /ncount ncount 1 add def
      } repeat
  } ifelse
  maxYvalue { 
    /m0 m def
    n {
      dup YMaxValue gt { pop pop /m m 2 sub def } if
      m -2 roll 
    } repeat
  } if
}>
\define@boolkey[psset]{pst-plot}[Pst@]{polarplot}[true]{}
\define@boolkey[psset]{pst-plot}[Pst@]{VarStep}[true]{}
\define@key[psset]{pst-plot}{PlotDerivative}{\def\psk@PlotDerivative{#1}}%
\define@key[psset]{pst-plot}{VarStepEpsilon}{\def\psk@VarStepEpsilon{#1}}%
\define@key[psset]{pst-plot}{method}{\def\psk@method{#1}}
\def\@rkiv{rk4}
\def\@varrkiv{varrkiv}
\def\@adams{adams}
\def\@default{default}
\psset[pst-plot]{VarStep=false,PlotDerivative=none,VarStepEpsilon=default,polarplot=false,method={}}
\def\psplotinit#1{\xdef\psplot@init{#1 }}
\def\psplot@init{}
\def\psplot{\def\pst@par{}\pst@object{psplot}}
\def\psplot@i#1#2{\@ifnextchar[{\psplot@x{#1}{#2}}{\psplot@x{#1}{#2}[]}}
\def\psplot@x#1#2[#3]#4{%
  \pst@killglue
  \begingroup
    \use@par
    \@nameuse{beginplot@\psplotstyle}%
    \ifPst@polarplot
      \addto@pscode{
        \psplot@init
        #3 
        /x #1 def
        /x1 #2 def
        /dx x1 x sub \psk@plotpoints div def
        /F@pstplot \ifPst@algebraic (#4)
                    \ifx\psk@PlotDerivative\@none\else
                      \psk@PlotDerivative\space { (x) tx@Derive begin Derive end } repeat
                    \fi\space
                    tx@AlgToPs begin AlgToPs end cvx
                 \else { #4 } \fi  def
        \ifPst@VarStep
          /StillZero 0 def /LastNonZeroStep dx def
          /F2@pstplot tx@Derive begin (#4) (x) Derive (x) Derive end
                     \ifx\psk@PlotDerivative\@none\else
                       \psk@PlotDerivative\space { (x) tx@Derive begin Derive end } repeat
                     \fi\space
                    tx@AlgToPs begin AlgToPs end cvx def
          /epsilon12 \ifx\psk@VarStepEpsilon\@default tx@Derive begin F2@pstplot end dx 3 exp abs mul abs
                    \else\psk@VarStepEpsilon\space 12 mul \fi def
          /ComputeStep {
            dup 1e-4 lt
            { pop StillZero 2 ge { LastNonZeroStep 2 mul } { LastNonZeroStep } ifelse /StillZero StillZero 1 add def }
            { epsilon12 exch div 1 3 div exp /StillZero 0 def }
            ifelse } bind def
        \fi
        /xy {
          F@pstplot x \ifPst@algebraic RadtoDeg \fi PtoC
          \pst@number\psyunit mul exch
          \pst@number\psxunit mul exch
        } def}%
    \else
    \addto@pscode{
      \psplot@init
      #3 
      /x #1 def
      /x1 #2 def
      /dx x1 x sub \psk@plotpoints div def
      /F@pstplot \ifPst@algebraic (#4)
                    \ifx\psk@PlotDerivative\@none\else
                      \psk@PlotDerivative\space { (x) tx@Derive begin Derive end } repeat
                    \fi\space
                    tx@AlgToPs begin AlgToPs end cvx
                 \else { #4 } \fi  def
      \ifPst@VarStep
         /StillZero 0 def /LastNonZeroStep dx def
         /F2@pstplot tx@Derive begin (#4) (x) Derive (x) Derive end
                     \ifx\psk@PlotDerivative\@none\else
                       \psk@PlotDerivative\space { (x) tx@Derive begin Derive end } repeat
                     \fi\space
                    tx@AlgToPs begin AlgToPs end cvx def
         /epsilon12 \ifx\psk@VarStepEpsilon\@default tx@Derive begin F2@pstplot end dx 3 exp abs mul abs
                    \else\psk@VarStepEpsilon\space 12 mul \fi def
         /ComputeStep {
           dup 1e-4 lt
           { pop StillZero 2 ge { LastNonZeroStep 2 mul } { LastNonZeroStep } ifelse /StillZero StillZero 1 add def }
           { epsilon12 exch div 1 3 div exp /StillZero 0 def }
           ifelse } bind def
      \fi
      /xy { x \pst@number\psxunit mul F@pstplot \pst@number\psyunit mul
      } def}%
    \fi
    \gdef\psplot@init{}%
    \ifx\pslinestyle\psls@@symbol
      \psplot@iii
    \else
      \@pstfalse
      \@nameuse{testqp@\psplotstyle}%
      \if@pst\psplot@ii\else\psplot@iii\fi
    \fi
  \endgroup
  \ignorespaces}
\def\psplot@ii{%
  \ifPst@VarStep%
    \addto@pscode{%
      mark xy \@nameuse{beginqp@\psplotstyle}
      { F2@pstplot abs ComputeStep
        x 2 copy add dup x1 gt {pop x1} if /x exch def F2@pstplot abs ComputeStep
        /x 3 -1 roll def 2 copy gt { exch } if pop
        /x x 3 -1 roll add dup x1 gt {pop x1} if def
        xy \@nameuse{doqp@\psplotstyle}
        x x1 eq { exit } if} loop}%
  \else
    \pst@killglue%
    \addto@pscode{
      /ps@Exit false def
      xy \@nameuse{beginqp@\psplotstyle}
      \ifx\psk@method\@varrkiv\else\psk@plotpoints 1 sub \fi {
        /x x dx add \ifx\psk@method\@varrkiv  dup x1 gt { pop x1 } if \fi def
        xy \@nameuse{doqp@\psplotstyle}
        \ifx\psk@method\@varrkiv  x x1 eq { exit } if \fi
      } 
      ps@Exit { exit } if
      \ifx\psk@method\@varrkiv loop \else repeat \fi
      ps@Exit not {
        /x x1 def
        xy \@nameuse{doqp@\psplotstyle}
      } if }%
  \fi%
  \@nameuse{endqp@\psplotstyle}}
\def\psplot@iii{%
  \ifPst@VarStep%
    \addto@pscode{
      /n 2 def
      mark
      { xy n 2 roll F2@pstplot abs
        ComputeStep x 2 copy add dup x1 gt {pop x1} if
        /x exch def F2@pstplot abs ComputeStep
        /x 3 -1 roll def 2 copy gt { exch } if pop
        /x x 3 -1 roll dup /LastNonZeroStep exch def add dup x1 gt {pop x1} if def /n n 2 add def
        x x1 eq { exit } if } loop
      xy 
      n 2 roll}%
  \else\pst@killglue%
    \addto@pscode{
      mark
      /n 2 def
      \ifx\psk@method\@varrkiv\else\psk@plotpoints\fi {
        xy
        n 2 roll
        /n n 2 add def
        /x x dx add \ifx\psk@method\@varrkiv  dup x1 gt { pop x1 } if \fi def
        \ifx\psk@method\@varrkiv  x x1 eq { exit } if \fi
      } \ifx\psk@method\@varrkiv loop\else repeat \fi \space
      /x x1 def
      xy 
      2 copy \tx@UserCoor 2 array astore /FinalState ED
      n 2 roll}%
  \fi%
  \@nameuse{endplot@\psplotstyle}}
\def\psparametricplot{\pst@object{parametricplot}}
\def\parametricplot{\pst@object{parametricplot}}
\def\parametricplot@i#1#2{\@ifnextchar[{\parametricplot@x{#1}{#2}}{\parametricplot@x{#1}{#2}[]}}
\def\parametricplot@x#1#2[#3]{\@ifnextchar[{\parametricplot@xi{#1}{#2}[#3]}{\parametricplot@xi{#1}{#2}[#3][]}}
\def\parametricplot@xi#1#2[#3][#4]#5{%
  \pst@killglue%
  \begingroup%
    \use@par%
    \@nameuse{beginplot@\psplotstyle}%
    \addto@pscode{%
      #3 
      \psplot@init
      /t #1 def
      /t1 #2 def
      /dt t1 t sub \psk@plotpoints div def
      /F@pstplot \ifPst@algebraic (#5)
                    \ifx\psk@PlotDerivative\@none\else
                      \psk@PlotDerivative\space { (t) tx@Derive begin Derive end } repeat
                    \fi\space
                    tx@AlgToPs begin AlgToPs end cvx
                 \else { #5 } \fi  def
      \ifPst@VarStep
         /StillZero 0 def /LastNonZeroStep dt def
         /F2@pstplot tx@Derive begin (#5) (t) Derive (t) Derive end
                     \ifx\psk@PlotDerivative\@none\else
                       \psk@PlotDerivative\space { (t) tx@Derive begin Derive end } repeat
                     \fi\space
                    tx@AlgToPs begin AlgToPs end cvx def
         /epsilon12 \ifx\psk@VarStepEpsilon\@default
                       tx@Derive begin F2@pstplot end Pyth
                       dt 3 exp abs mul
                    \else\psk@VarStepEpsilon\space 12 mul \fi def
         /ComputeStep {
           dup 1e-4 lt
           { pop StillZero 2 ge { LastNonZeroStep 2 mul } { LastNonZeroStep } ifelse /StillZero StillZero 1 add def }
           { epsilon12 exch div 1 3 div exp /StillZero 0 def }
           ifelse } bind def
      \fi
      /xy {
        \ifPst@algebraic F@pstplot \else #5 \fi
        \pst@number\psyunit mul exch
        \pst@number\psxunit mul exch
      } def
      }%
    \gdef\psplot@init{}%
    \@pstfalse
    \@nameuse{testqp@\psplotstyle}%
    \if@pst\parametricplot@ii{#4}\else\parametricplot@iii{#4}\fi
  \endgroup%
  \ignorespaces}
\def\parametricplot@ii#1{
  \ifPst@VarStep%
    \addto@pscode{%
      mark xy \@nameuse{beginqp@\psplotstyle}
      { F2@pstplot Pyth ComputeStep
        t 2 copy add dup t1 gt {pop t1} if /t exch def F2@pstplot Pyth ComputeStep
        /t 3 -1 roll def 2 copy gt { exch } if pop
        /t t 3 -1 roll add dup t1 gt {pop t1} if def
        xy \@nameuse{doqp@\psplotstyle}
        t t1 eq { exit } if } loop}%
  \else\pst@killglue%
    \addto@pscode{%
      /ps@Exit false def
      xy \@nameuse{beginqp@\psplotstyle}
      \psk@plotpoints 1 sub {
        /t t dt add def
        xy \@nameuse{doqp@\psplotstyle}
        ps@Exit { exit } if 
      } repeat
      ps@Exit not {
        /t t1 def
        xy \@nameuse{doqp@\psplotstyle}
      } if 
    }%
  \fi%
  \addto@pscode{ #1 }%
  \@nameuse{endqp@\psplotstyle}}
\def\parametricplot@iii#1{%
  \ifPst@VarStep%
    \addto@pscode{%
      /n 2 def
      mark
      { xy n 2 roll F2@pstplot Pyth
        ComputeStep t 2 copy add dup t1 gt {pop t1} if
        /t exch def F2@pstplot Pyth ComputeStep
        /t 3 -1 roll def 2 copy gt { exch } if pop
        /t t 3 -1 roll dup /LastNonZeroStep exch def add dup t1 gt {pop t1} if def /n n 2 add def
        t t1 eq { exit } if } loop
      xy 
      2 copy \tx@UserCoor 2 array astore /FinalState ED
      n 2 roll}%
  \else\pst@killglue%
    \addto@pscode{
      mark
      /n 2 def
      \psk@plotpoints {
        xy
        n 2 roll
        /n n 2 add def
        /t t dt add def
      } repeat
      /t t1 def
      xy
      n 2 roll}%
  \fi%
  \addto@pscode{ #1 }%
  \@nameuse{endplot@\psplotstyle}}
\newdimen\psk@subticksize\psk@subticksize=\z@
\newdimen\pst@xticksizeA
\newdimen\pst@xticksizeB
\newdimen\pst@xticksizeC
\newdimen\pst@yticksizeA
\newdimen\pst@yticksizeB
\newdimen\pst@yticksizeC
\define@key[psset]{pst-plot}{ticks}[all]{\pst@expandafter\psset@@ticks{#1}\@nil\psk@ticks}
\def\psset@@ticks#1#2\@nil#3{%
  \ifx#1a\let#3\z@\else
    \ifx#1x\let#3\@ne\else
      \ifx#1y\let#3\tw@\else
        \ifx#1n\let#3\thr@@\else
          \@pstrickserr{Bad argument: `#1#2'}\@ehpa
  \fi\fi\fi\fi}
\psset[pst-plot]{ticks=all}
\define@key[psset]{pst-plot}{labels}[all]{\pst@expandafter\psset@@ticks{#1}\@nil\psk@labels}
\psset[pst-plot]{labels=all}
\define@key[psset]{pst-plot}{Ox}[0]{\def\psk@Ox{#1}}
\psset[pst-plot]{Ox=0}
\define@key[psset]{pst-plot}{Dx}[1]{\def\psk@Dx{#1}}
\psset[pst-plot]{Dx=1}
\define@key[psset]{pst-plot}{dx}[0]{%
  \pssetxlength\pst@dimg{#1}%
  \edef\psk@dx{\number\pst@dimg}}
\psset[pst-plot]{dx=0}
\define@key[psset]{pst-plot}{Oy}[0]{\def\psk@Oy{#1}}
\psset[pst-plot]{Oy=0}
\define@key[psset]{pst-plot}{Dy}[1]{\def\psk@Dy{#1}}
\psset[pst-plot]{Dy=1}
\define@key[psset]{pst-plot}{dy}[0]{%
  \pssetylength\pst@dimg{#1}%
  \edef\psk@dy{\number\pst@dimg}}
\psset[pst-plot]{dy=0}
\define@boolkey[psset]{pst-plot}[]{showXorigin}[true]{}
\define@boolkey[psset]{pst-plot}[]{showYorigin}[true]{}
\define@boolkey[psset]{pst-plot}[]{showorigin}[true]{%
  \ifshoworigin
    \showXorigintrue\showYorigintrue
  \else
    \showXoriginfalse\showYoriginfalse
  \fi
}
\psset[pst-plot]{showorigin=true}
\long\def\psrotatebox#1#2{%
  \leavevmode
  \Grot@setangle{#1}%
  \setbox\z@\hbox{{#2}}%
  \Grot@x\z@
  \Grot@y\z@
  \Grot@box}
\def\Grot@setangle#1{\edef\Grot@angle{#1}}
\def\Grot@Px#1#2#3{%
        #1\Grot@cos#2%
        \advance#1-\Grot@sin#3}
\def\Grot@Py#1#2#3{%
        #1\Grot@sin#2%
        \advance#1\Grot@cos#3}
\def\Grot@box{%
  \begingroup
  \CalculateSin\Grot@angle
  \CalculateCos\Grot@angle
  \edef\Grot@sin{\UseSin\Grot@angle}%
  \edef\Grot@cos{\UseCos\Grot@angle}%
  \Grot@r\wd\z@  \advance\Grot@r-\Grot@x
  \Grot@l\z@     \advance\Grot@l-\Grot@x
  \Grot@h\ht\z@  \advance\Grot@h-\Grot@y
  \Grot@d-\dp\z@ \advance\Grot@d-\Grot@y
  \ifdim\Grot@sin\p@>\z@
    \ifdim\Grot@cos\p@>\z@
      \Grot@Py\Grot@height \Grot@r\Grot@h
      \Grot@Px\Grot@right  \Grot@r\Grot@d
      \Grot@Px\Grot@left   \Grot@l\Grot@h
      \Grot@Py\Grot@depth  \Grot@l\Grot@d
    \else
      \Grot@Py\Grot@height \Grot@r\Grot@d
      \Grot@Px\Grot@right  \Grot@l\Grot@d
      \Grot@Px\Grot@left   \Grot@r\Grot@h
      \Grot@Py\Grot@depth  \Grot@l\Grot@h
    \fi
  \else
    \ifdim\Grot@cos\p@<\z@
      \Grot@Py\Grot@height \Grot@l\Grot@d
      \Grot@Px\Grot@right  \Grot@l\Grot@h
      \Grot@Px\Grot@left   \Grot@r\Grot@d
      \Grot@Py\Grot@depth  \Grot@r\Grot@h
    \else
      \Grot@Py\Grot@height \Grot@l\Grot@h
      \Grot@Px\Grot@right  \Grot@r\Grot@h
      \Grot@Px\Grot@left   \Grot@l\Grot@d
      \Grot@Py\Grot@depth  \Grot@r\Grot@d
    \fi
  \fi
  \advance\Grot@height\Grot@y
  \advance\Grot@depth\Grot@y
  \Grot@Px\dimen@  \Grot@x\Grot@y
  \Grot@Py\dimen@ii \Grot@x\Grot@y
  \dimen@-\dimen@     \advance\dimen@-\Grot@left
  \dimen@ii-\dimen@ii \advance\dimen@ii\Grot@y
  \setbox\z@\hbox{%
    \kern\dimen@
    \raise\dimen@ii\hbox{\Grot@start\box\z@\Grot@end}}%
  \ht\z@\Grot@height
  \dp\z@-\Grot@depth
  \advance\Grot@right-\Grot@left\wd\z@\Grot@right
  \leavevmode\box\z@
  \endgroup}
\define@key[psset]{pst-plot}{labelFontSize}[{}]{\def\psk@xlabelFontSize{#1}\def\psk@ylabelFontSize{#1}}%
\define@key[psset]{pst-plot}{xlabelFontSize}[{}]{\def\psk@xlabelFontSize{#1}}%
\define@key[psset]{pst-plot}{ylabelFontSize}[{}]{\def\psk@ylabelFontSize{#1}}%
\define@boolkey[psset]{pst-plot}[Pst@]{mathLabel}[true]{%
  \ifPst@mathLabel%
    \Pst@xmathLabeltrue \Pst@ymathLabeltrue
    \def\pshlabel##1{$\psk@xlabelFontSize##1$}%
    \def\psvlabel##1{$\psk@ylabelFontSize##1$}%
  \else%
    \Pst@xmathLabelfalse \Pst@ymathLabelfalse
    \def\pshlabel##1{\psk@xlabelFontSize##1}%
    \def\psvlabel##1{\psk@ylabelFontSize##1}%
  \fi}
\define@boolkey[psset]{pst-plot}[Pst@]{xmathLabel}[true]{%
  \ifPst@xmathLabel%
    \def\pshlabel##1{$\psk@xlabelFontSize##1$}\else\def\pshlabel##1{\psk@xlabelFontSize##1}\fi}
\define@boolkey[psset]{pst-plot}[Pst@]{ymathLabel}[true]{%
  \ifPst@ymathLabel%
    \def\psvlabel##1{$\psk@ylabelFontSize##1$}\else\def\psvlabel##1{\psk@ylabelFontSize##1}\fi}
\psset[pst-plot]{labelFontSize={},mathLabel}
\define@boolkey[psset]{pst-plot}[Pst@]{xAxis}[true]{}
\define@boolkey[psset]{pst-plot}[Pst@]{yAxis}[true]{}
\define@boolkey[psset]{pst-plot}[Pst@]{xyAxes}[true]{%
    \@nameuse{Pst@xAxis#1}\@nameuse{Pst@yAxis#1}}%
\psset[pst-plot]{xAxis,yAxis}%
\define@key[psset]{pst-plot}{xlabelPos}[b]{\pst@expandafter\psset@@xlabelPos#1\@nil}
\define@key[psset]{pst-plot}{ylabelPos}[l]{\pst@expandafter\psset@@ylabelPos#1\@nil}
\def\psset@@xlabelPos#1#2\@nil{%
  \ifx#1t\relax
    \let\psk@xlabelPos\tw@
    \pst@xticksizeC=\pst@xticksizeB
  \else
    \ifx#1a\relax
      \let\psk@xlabelPos\@ne 
      \pst@xticksizeC=\z@
    \else
      \def\psk@xlabelPos{\z@}
      \pst@xticksizeC=\pst@xticksizeA
  \fi\fi
}
\def\psset@@ylabelPos#1#2\@nil{%
  \ifx#1r\relax
    \def\psk@ylabelPos{\tw@}
    \pst@yticksizeC=\pst@yticksizeB
  \else
    \ifx#1a\relax
      \def\psk@ylabelPos{\@ne}
      \pst@yticksizeC=\z@
    \else 
      \def\psk@ylabelPos{\z@}
      \pst@yticksizeC=\pst@yticksizeA
  \fi\fi
}
\psset[pst-plot]{xlabelPos=b, ylabelPos=l}%
\define@key[psset]{pst-plot}{xyDecimals}[{}]{\def\psk@xDecimals{#1}\def\psk@yDecimals{#1}}
\define@key[psset]{pst-plot}{xDecimals}[{}]{\def\psk@xDecimals{#1}}
\define@key[psset]{pst-plot}{yDecimals}[{}]{\def\psk@yDecimals{#1}}
\psset[pst-plot]{xyDecimals={}}%
\define@key[psset]{pst-plot}{xlogBase}[{}]{\def\psk@xlogBase{#1}}
\define@key[psset]{pst-plot}{ylogBase}[{}]{\def\psk@ylogBase{#1}}
\define@key[psset]{pst-plot}{xylogBase}[{}]{\def\psk@xlogBase{#1}\def\psk@ylogBase{#1}}%
\psset[pst-plot]{xylogBase={}}%
\define@key[psset]{pst-plot}{trigLabelBase}[0]{\pst@getint{#1}{\psk@xtrigLabelBase}\let\psk@ytrigLabelBase\psk@xtrigLabelBase}
\define@key[psset]{pst-plot}{xtrigLabelBase}[0]{%
  \pst@getint{#1}{\psk@xtrigLabelBase}\psset{xtrigLabels}}
\define@key[psset]{pst-plot}{ytrigLabelBase}[0]{%
  \pst@getint{#1}{\psk@ytrigLabelBase}\psset{ytrigLabels}}
\psset[pst-plot]{trigLabelBase=0}
\define@key[psset]{pst-plot}{fractionLabelBase}[0]{\pst@getint{#1}{\psk@xfractionLabelBase}\let\psk@yfractionLabelBase\psk@xfractionLabelBase}
\define@key[psset]{pst-plot}{xfractionLabelBase}[0]{%
  \pst@getint{#1}{\psk@xfractionLabelBase}\psset{xfractionLabels}}
\define@key[psset]{pst-plot}{yfractionLabelBase}[0]{%
  \pst@getint{#1}{\psk@yfractionLabelBase}\psset{yfractionLabels}}
\psset[pst-plot]{fractionLabelBase=0}
\def\setDefaulthLabels{%
  \ifPst@xmathLabel\def\pshlabel##1{$\psk@xlabelFontSize##1$}\else\def\pshlabel##1{\psk@xlabelFontSize##1}\fi
  \def\pst@@@hlabel##1{%
      \edef\@xyDecimals{\psk@xDecimals}%
      \ifnum\psk@labels<\tw@\relax
        \ifx\psk@xlogBase\@empty
          \pshlabel{\psk@xlabelFontSize\expandafter\@LabelComma##1..\@nil\psk@xlabelFactor}%
        \else
          \ifPst@xmathLabel
            \pshlabel{\psk@xlabelFontSize\psk@xlogBase^{\expandafter\@stripDecimals##1..\@nil}}%
          \else
            \pshlabel{\psk@xlabelFontSize\psk@xlogBase\textsuperscript{\expandafter\@stripDecimals##1..\@nil}}%
          \fi
        \fi
      \fi
    }%
    \ifPst@xmathLabel\def\pshlabel##1{$\psk@xlabelFontSize##1$}\else\def\pshlabel##1{\psk@xlabelFontSize##1}\fi
}
\def\setTrighLabels{%
    \def\pst@@@hlabel##1{\pshlabel{##1}}%
    \def\pshlabel##1{%
      \ifnum\psk@xtrigLabelBase<2
        \def\de@nominator{\@ne}\else\def\de@nominator{\psk@xtrigLabelBase}\fi
      \def\pst@tempA{##1}%
      \pst@abs{\pst@tempA}\pst@cntm 
      \pst@mod{\pst@cntm}{\de@nominator}\pst@cntp 
      \ifnum\@ne>\pst@cntp                  
        \pst@cnto=\pst@cntm \divide\pst@cnto by \de@nominator  
	\ifPst@xmathLabel
          $\psk@xlabelFontSize
  	  \ifnum\pst@tempA<0 -\fi
          \ifnum\pst@cnto=\@ne                
            \pi                 	      
          \else
            \ifnum\pst@cnto=\z@ 0\else
            \the\pst@cnto\pi 	              
          \fi\fi$%
	\else%
          \psk@xlabelFontSize
  	  \ifnum\pst@tempA<0 -\fi
          \ifnum\pst@cnto=\@ne
            $\pi$
          \else%
            \the\pst@cnto$\pi$
          \fi%
	\fi%
      \else%
	\ifPst@xmathLabel%
          $\psk@xlabelFontSize%
          \ifnum\pst@cntp=\@ne
            \if\pst@cntm=\@ne%
              \frac{\pi}{\de@nominator}
            \else\ifnum\pst@tempA=-1 \frac{-\pi}{\de@nominator}%
              \else \ifnum\pst@tempA=1 \frac{\pi}{\de@nominator}%
                \else\frac{\pst@tempA\pi}{\de@nominator}
            \fi\fi\fi%
          \else%
            \ifnum\pst@tempA=1 \frac{\pi}{\de@nominator}%
            \else\ifnum\pst@tempA=\de@nominator \pi%
              \else\frac{\pst@tempA\pi}{\de@nominator}%
          \fi\fi\fi$%
	\else%
          \psk@xlabelFontSize%
          \ifnum\pst@cntp=\@ne
            \if\pst@cntm=\@ne%
              $\frac{\pi}{\de@nominator}$
            \else\ifnum\pst@tempA=-1 $\frac{-\pi}{\de@nominator}$%
              \else \ifnum\pst@tempA=1 $\frac{\pi}{\de@nominator}$%
                \else$\frac{\pst@tempA\pi}{\de@nominator}$
            \fi\fi\fi
          \else
            \ifnum\pst@tempA=1 $\frac{\pi}{\de@nominator}$%
            \else\ifnum\pst@tempA=\de@nominator $\pi$%
              \else$\frac{\pst@tempA\pi}{\de@nominator}$%
          \fi\fi\fi
	\fi
      \fi
    }%
}
\def\setDefaultvLabels{%
  \ifPst@ymathLabel\def\psvlabel##1{$\psk@ylabelFontSize##1$}\else\def\psvlabel##1{\psk@ylabelFontSize##1}\fi
    \def\pst@@@vlabel##1{%
      \edef\@xyDecimals{\psk@yDecimals}%
      \ifodd\psk@labels 
      \else%
        \ifx\psk@ylogBase\@empty
          \psvlabel{\expandafter\@LabelComma##1..\@nil\psk@ylabelFactor}%
        \else%
          \ifPst@ymathLabel%
            \psvlabel{\psk@ylogBase^{\expandafter\@stripDecimals##1..\@nil }}%
	  \else
            \psvlabel{\psk@ylogBase\textsuperscript{\expandafter\@stripDecimals##1..\@nil }}%
          \fi%
        \fi%
      \fi%
    }%
}%
\def\setTrigvLabels{%
  \def\pst@@@vlabel##1{\psvlabel{##1}}%
    \def\psvlabel##1{%
      \ifnum\psk@ytrigLabelBase<2 \def\de@nominator{\@ne}\else\def\de@nominator{\psk@ytrigLabelBase}\fi
      \def\pst@tempA{##1} 
      \pst@abs{\pst@tempA}\pst@cntm 
      \pst@mod{\pst@cntm}{\de@nominator}\pst@cntp 
      \ifnum\@ne>\pst@cntp                  
        \pst@cnto=\pst@cntm \divide\pst@cnto by \de@nominator  
	\ifPst@ymathLabel%
          $\psk@ylabelFontSize
  	  \ifnum\pst@tempA<0 -\fi
          \ifnum\pst@cnto=\@ne                
            \pi                 	      
          \else
            \the\pst@cnto\pi 	              
          \fi$%
	\else%
          \psk@ylabelFontSize%
  	  \ifnum\pst@tempA<0 -\fi
          \ifnum\pst@cnto=\@ne
            $\pi$
          \else
            \the\pst@cnto$\pi$
          \fi
	\fi
      \else
	\ifPst@ymathLabel%
          $\psk@ylabelFontSize
          \ifnum\pst@cntp=\@ne
            \if\pst@cntm=\@ne%
              \frac{\pi}{\de@nominator}
            \else\ifnum\pst@tempA=-1 \frac{-\pi}{\de@nominator}%
              \else \ifnum\pst@tempA=1 \frac{\pi}{\de@nominator}%
                \else\frac{\pst@tempA\pi}{\de@nominator}
            \fi\fi\fi%
          \else%
            \ifnum\pst@tempA=1 \frac{\pi}{\de@nominator}%
            \else\ifnum\pst@tempA=\de@nominator \pi%
              \else\frac{\pst@tempA\pi}{\de@nominator}%
          \fi\fi\fi$%
	\else
          \psk@ylabelFontSize
          \ifnum\pst@cntp=\@ne
            \if\pst@cntm=\@ne
              $\frac{\pi}{\de@nominator}$
            \else\ifnum\pst@tempA=-1 $\frac{-\pi}{\de@nominator}$%
              \else \ifnum\pst@tempA=1 $\frac{\pi}{\de@nominator}$%
                \else$\frac{\pst@tempA\pi}{\de@nominator}$
            \fi\fi\fi
          \else
            \ifnum\pst@tempA=1 $\frac{\pi}{\de@nominator}$%
            \else\ifnum\pst@tempA=\de@nominator $\pi$%
              \else$\frac{\pst@tempA\pi}{\de@nominator}$%
          \fi\fi\fi
	\fi
      \fi
    }%
}
\def\setFractionvLabels{%
  \def\pst@@@vlabel##1{\psvlabel{##1}}%
  \def\psvlabel##1{%
      \ifnum\psk@yfractionLabelBase<2 \def\de@nominator{\@ne}\else\def\de@nominator{\psk@yfractionLabelBase}\fi
      \def\pst@tempA{##1}%
      \pst@abs{\pst@tempA}\pst@cntm 
      \pst@mod{\pst@cntm}{\de@nominator}\pst@cntp 
      \ifnum\@ne>\pst@cntp                  
        \pst@cnto=\pst@cntm \divide\pst@cnto by \de@nominator  
	\ifPst@ymathLabel$\psk@ylabelFontSize\ifnum\pst@tempA<0 -\fi\the\pst@cnto\psk@ylabelFactor$%
	\else             \psk@ylabelFontSize\ifnum\pst@tempA<0 -\fi\the\pst@cnto\psk@ylabelFactor
	\fi
      \else
	\ifPst@ymathLabel
          $\psk@ylabelFontSize
          \ifnum\pst@cntp=\@ne                
            \if\pst@cntm=\@ne
              \frac{1}{\de@nominator}\psk@ylabelFactor
            \else\ifnum\pst@tempA=-1 \frac{-1}{\de@nominator}\psk@ylabelFactor%
              \else \ifnum\pst@tempA=1 \frac{1}{\de@nominator}\psk@ylabelFactor%
                \else\frac{\pst@tempA}{\de@nominator}\psk@ylabelFactor
            \fi\fi\fi
          \else
            \ifnum\pst@tempA=1 \frac{1}{\de@nominator}\psk@ylabelFactor%
            \else\ifnum\pst@tempA=\de@nominator 1\psk@xlabelFactor \else\frac{\pst@tempA}{\de@nominator}\psk@ylabelFactor%
          \fi\fi\fi$
	\else
          \psk@ylabelFontSize
          \ifnum\pst@cntp=\@ne
            \if\pst@cntm=\@ne
              $\frac{1}{\de@nominator}\psk@ylabelFactor$
            \else\ifnum\pst@tempA=-1 $\frac{-1}{\de@nominator}\psk@ylabelFactor$%
              \else \ifnum\pst@tempA=1 $\frac{1}{\de@nominator}\psk@ylabelFactor$%
                \else$\frac{\pst@tempA}{\de@nominator}\psk@ylabelFactor$
            \fi\fi\fi%
          \else%
            \ifnum\pst@tempA=1 $\frac{1}{\de@nominator}\psk@ylabelFactor$%
            \else\ifnum\pst@tempA=\de@nominator 1\psk@ylabelFactor
              \else$\frac{\pst@tempA}{\de@nominator}\psk@ylabelFactor$
          \fi\fi\fi
	\fi
      \fi
    }%
}
\def\setFractionhLabels{%
  \def\pst@@@hlabel##1{\pshlabel{##1}}%
  \def\pshlabel##1{%
      \ifnum\psk@xfractionLabelBase<2 \def\de@nominator{\@ne}\else\def\de@nominator{\psk@xfractionLabelBase}\fi
      \def\pst@tempA{##1}%
      \pst@abs{\pst@tempA}\pst@cntm 
      \pst@mod{\pst@cntm}{\de@nominator}\pst@cntp
      \ifnum\@ne>\pst@cntp                  
        \pst@cnto=\pst@cntm \divide\pst@cnto by \de@nominator  
	\ifPst@xmathLabel$\psk@xlabelFontSize\ifnum\pst@tempA<0 -\fi\the\pst@cnto\psk@xlabelFactor$%
	\else             \psk@xlabelFontSize\ifnum\pst@tempA<0 -\fi\the\pst@cnto\psk@xlabelFactor
	\fi
      \else
	\ifPst@xmathLabel
          $\psk@xlabelFontSize
          \ifnum\pst@cntp=\@ne
            \if\pst@cntm=\@ne \frac{1}{\de@nominator}\psk@xlabelFactor
            \else\ifnum\pst@tempA=-1 \frac{-1}{\de@nominator}\psk@xlabelFactor%
              \else\ifnum\pst@tempA=1 \frac{1}{\de@nominator}\psk@xlabelFactor%
                \else\frac{\pst@tempA}{\de@nominator}\psk@xlabelFactor
            \fi\fi\fi%
          \else%
            \ifnum\pst@tempA=1 \frac{1}{\de@nominator}\psk@xlabelFactor%
            \else\ifnum\pst@tempA=\de@nominator 1\psk@xlabelFactor\else\frac{\pst@tempA}{\de@nominator}\psk@xlabelFactor%
          \fi\fi\fi$
	\else
          \psk@xlabelFontSize
          \ifnum\pst@cntp=\@ne
            \if\pst@cntm=\@ne $\frac{1}{\de@nominator}\psk@xlabelFactor$
            \else\ifnum\pst@tempA=-1 $\frac{-1}{\de@nominator}\psk@xlabelFactor$%
              \else \ifnum\pst@tempA=1 $\frac{1}{\de@nominator}\psk@xlabelFactor$%
                \else$\frac{\pst@tempA}{\de@nominator}\psk@xlabelFactor$
            \fi\fi\fi
          \else
            \ifnum\pst@tempA=1 $\frac{1}{\de@nominator}\psk@xlabelFactor$%
            \else\ifnum\pst@tempA=\de@nominator 1\psk@xlabelFactor%
              \else$\frac{\pst@tempA}{\de@nominator}\psk@xlabelFactor$
          \fi\fi\fi
	\fi
      \fi
    }%
}
\define@boolkey[psset]{pst-plot}[Pst@]{xtrigLabels}[true]{%
  \ifPst@xtrigLabels \setTrighLabels \else \setDefaulthLabels\fi}
\define@boolkey[psset]{pst-plot}[Pst@]{ytrigLabels}[true]{%
  \ifPst@ytrigLabels \setTrigvLabels \else \setDefaultvLabels \fi}
\define@boolkey[psset]{pst-plot}[Pst@]{trigLabels}[true]{%
  \ifPst@trigLabels\psset[pst-plot]{xtrigLabels,ytrigLabels=false}
  \else            \psset[pst-plot]{xtrigLabels=false,ytrigLabels=false}%
  \fi}
\psset[pst-plot]{trigLabels=false}
\define@boolkey[psset]{pst-plot}[Pst@]{xfractionLabels}[true]{%
  \ifPst@xfractionLabels \setFractionhLabels \else \setDefaulthLabels \fi}
\define@boolkey[psset]{pst-plot}[Pst@]{yfractionLabels}[true]{%
  \ifPst@yfractionLabels \setFractionvLabels \else \setDefaultvLabels \fi}
\define@boolkey[psset]{pst-plot}[Pst@]{fractionLabels}[true]{%
  \ifPst@fractionLabels \setFractionhLabels\setFractionvLabels\Pst@xfractionLabelstrue\Pst@yfractionLabelstrue\fi}
\psset[pst-plot]{fractionLabels=false}
\def\psk@logLines{3}
\define@key[psset]{pst-plot}{logLines}[none]{\pst@expandafter\psset@@logLines#1\@nil\psk@logLines}
\def\psset@@logLines#1#2\@nil#3{%
  \ifx#1a\relax
    \let#3\z@
    \Pst@maxxTickstrue\Pst@maxyTickstrue
    \set@xticksize{0 4pt}\set@yticksize{0 4pt}%
    \def\psk@xsubticksize{1}\def\psk@ysubticksize{1}%
  \else
    \ifx#1x\relax
      \let#3\@ne
      \Pst@maxxTickstrue\Pst@maxyTicksfalse
      \set@xticksize{0 4pt}\def\psk@xsubticksize{1}%
    \else
      \ifx#1y\relax
        \let#3\tw@
	\Pst@maxyTickstrue\Pst@maxxTicksfalse
	\set@yticksize{0 4pt}\def\psk@ysubticksize{1}%
      \else
        \ifx#1n\let#3\thr@@\else
          \@pstrickserr{Bad argument: `#1#2'}\@ehpa
  \fi\fi\fi\fi}
\psset[pst-plot]{logLines=none}%
\define@key[psset]{pst-plot}{ylabelFactor}[\relax]{\def\psk@ylabelFactor{#1}}
\define@key[psset]{pst-plot}{xlabelFactor}[\relax]{\def\psk@xlabelFactor{#1}}
\define@boolkey[psset]{pst-plot}[Pst@]{showOriginTick}[true]{}%
\psset[pst-plot]{xlabelFactor=\relax,ylabelFactor=\relax,showOriginTick}%

\def\psxTick{\pst@object{psxTick}}
\def\psxTick@i{\@ifnextchar({\psxTick@ii{0}}\psxTick@ii}
\def\psxTick@ii#1(#2)#3{{%
  \pst@killglue
  \addbefore@par{arrows=-,linewidth=\psk@xtickwidth\pslinewidth}
  \ifPst@xtrigLabels\addto@par{xtrigLabels=false}\fi 
  \use@par
  \edef\temp@coor{(!#2 \pst@number\pst@xticksizeB \pst@number\psyunit div)(!#2 \pst@number\pst@xticksizeA \pst@number\psyunit div)}%
  \expandafter\psline\temp@coor
  \rput[t]{#1}(! \psk@origin 
                 #2 \pst@number\psxlabelsep \pst@number\pst@xticksizeB add
                 \pst@number\psyunit div neg ){\pshlabel{#3\vphantom{1}}}%
  }\ignorespaces}
\def\psyTick{\pst@object{psyTick}}
\def\psyTick@i{\@ifnextchar({\psyTick@ii{0}}\psyTick@ii}
\def\psyTick@ii#1(#2)#3{{%
  \pst@killglue
  \addbefore@par{arrows=-,linewidth=\psk@ytickwidth\pslinewidth}
  \ifPst@ytrigLabels \setDefaultvLabels \fi
  \use@par
  \edef\temp@coor{(!\pst@number\pst@yticksizeB \pst@number\psxunit div #2)(!\pst@number\pst@yticksizeA \pst@number\psxunit div #2)}%
  \expandafter\psline\temp@coor
    \rput[r]{#1}(!\psk@origin
                  \pst@number\pst@yticksizeB \pst@number\psylabelsep add
                  \pst@number\psxunit div neg #2){\psvlabel{#3}}}\ignorespaces}
\define@boolkey[psset]{pst-plot}[Pst@]{markPoint}[true]{}%
\psset[pst-plot]{markPoint}
\def\psCoordinates{\pst@object{psCoordinates}}
\def\psCoordinates@i(#1){%
  \pst@killglue%
  \begingroup
  \addbefore@par{showpoints=false,markPoint}
  \use@par
  \psline(#1|0,0)(#1)
  \psline(#1)(0,0|#1)%
  \ifPst@markPoint\psdot(#1)\fi%
  \endgroup
  \ignorespaces
}
\def\stripDecimals#1{\expandafter\@stripDecimals#1..\@nil}
\def\@stripDecimals#1.#2.#3\@nil{%
  \def\pst@dummy{#1}%
  \ifx\pst@dummy\@empty\the\@zero\else#1\fi
}
\newcount\@digitcounter\@digitcounter=0\relax
\def\@inc@digitcounter{\global\advance\@digitcounter by 1\relax}
\def\@get@digitcounter{\the\@digitcounter\relax}
\def\@Reset@digitcounter{\global\@digitcounter=0\relax}
\def\@zeroFill{%
  \ifnum \@xyDecimals>\@get@digitcounter
    \bgroup
      0\@inc@digitcounter\@zeroFill
    \egroup
  \fi
}
\def\@process@digits#1#2;{%
  \ifx *#1\@zeroFill\else#1\@inc@digitcounter 
  \ifnum\@xyDecimals>\@get@digitcounter\expandafter\@process@digits#2;\fi\fi%
}
\def\@writeDecimals#1{%
  \ifx\@xyDecimals\@empty
    \def\@tempa{#1}
    \ifx\@tempa\@empty
    \else\ifmmode\expandafter\mathord\expandafter{\psk@decimalSeparator}\else\psk@decimalSeparator\fi#1\fi%
  \else
    \ifnum\@xyDecimals>\@zero
      \ifmmode\expandafter\mathord\expandafter{\psk@decimalSeparator}\else\psk@decimalSeparator\fi%
        \@Reset@digitcounter
        \expandafter\@process@digits#1*;%
      \fi%
  \fi%
}
\def\@LabelComma#1.#2.#3\@nil{%
  \def\pst@tempA{#1}%
  \ifx\pst@tempA\@empty\the\@zero\else#1\fi
  \def\pst@tempA{#2}%
  \ifx\pst@tempA\@empty\@writeDecimals{}\else\@writeDecimals{#2}\fi
}
\def\set@xticksize#1{%
  \pst@expandafter\pst@getydimdim{#1} {} {}\@nil
  \ifdim\pst@dimm>\pst@dimn
    \pst@xticksizeA=\the\pst@dimn%
    \pst@xticksizeB=\the\pst@dimm%
  \else%
    \pst@xticksizeA=\the\pst@dimm%
    \pst@xticksizeB=\the\pst@dimn
  \fi%
  \edef\psk@xticksize{\pst@number\pst@xticksizeA \pst@number\pst@xticksizeB}%
  \ifnum\psk@xlabelPos<\z@\relax
    \pst@xticksizeC=\pst@dimn
  \else
    \pst@xticksizeC=\pst@dimm
  \fi
}
\def\set@yticksize#1{%
  \pst@expandafter\pst@getxdimdim{#1} {} {}\@nil
  \ifdim\pst@dimm>\pst@dimn\relax
    \pst@yticksizeA=\the\pst@dimn%
    \pst@yticksizeB=\the\pst@dimm%
  \else%
    \pst@yticksizeA=\the\pst@dimm%
    \pst@yticksizeB=\the\pst@dimn
  \fi%
  \edef\psk@yticksize{\pst@number\pst@yticksizeA \pst@number\pst@yticksizeB}%
  \ifnum\psk@ylabelPos<\z@	
    \pst@yticksizeC=\pst@dimn%
  \else%
      \pst@yticksizeC=\pst@dimo
  \fi%
}
\newif\ifPst@maxxTicks
\newif\ifPst@maxyTicks
\define@key[psset]{pst-plot}{ticksize}[-4pt 4pt]{%
  \def\pst@tempA{max}%
  \def\pst@tempB{#1}%
  \ifx\pst@tempA\pst@tempB%
    \Pst@maxxTickstrue\Pst@maxyTickstrue%
    \set@xticksize{0 4pt}\set@yticksize{0 4pt}%
  \else%
    \Pst@maxxTicksfalse\Pst@maxyTicksfalse%
    \set@xticksize{#1}\set@yticksize{#1}%
  \fi}
\define@key[psset]{pst-plot}{xticksize}{%
  \def\pst@tempA{max}%
  \def\pst@tempB{#1}%
  \ifx\pst@tempA\pst@tempB
    \Pst@maxxTickstrue\set@xticksize{0 4pt}%
  \else\set@xticksize{#1}\Pst@maxxTicksfalse\fi}
\define@key[psset]{pst-plot}{yticksize}{%
  \def\pst@tempA{max}%
  \def\pst@tempB{#1}%
  \ifx\pst@tempA\pst@tempB%
    \Pst@maxyTickstrue\set@yticksize{0 4pt}%
  \else\set@yticksize{#1}\Pst@maxyTicksfalse\fi}%
\psset[pst-plot]{ticksize=-4pt 4pt}
\define@key[psset]{pst-plot}{tickstyle}[full]{\pst@expandafter\psset@@tickstyle{#1}\@nil}
\def\psset@@tickstyle#1#2\@nil{%
  \ifx#1f\let\psk@tickstyle\z@\else			
    \ifx#1t\let\psk@tickstyle\@ne			
      \edef\psk@xticksize{0 \pst@number\pst@xticksizeB}%
      \edef\psk@yticksize{0 \pst@number\pst@yticksizeB}%
    \else\ifx#1b\let\psk@tickstyle\m@ne			
      \edef\psk@xticksize{\pst@number\pst@xticksizeA 0}%
      \edef\psk@yticksize{\pst@number\pst@yticksizeA 0}%
      \else\ifx#1i\let\psk@tickstyle\tw@
        \else\@pstrickserr{Bad tick style: `#1#2'}\@ehpa
  \fi\fi\fi\fi}
\psset[pst-plot]{tickstyle=full}
\define@key[psset]{pst-plot}{subticks}[1]{\def\psk@xsubticks{#1}\def\psk@ysubticks{#1}}
\define@key[psset]{pst-plot}{xsubticks}[1]{\def\psk@xsubticks{#1}}
\define@key[psset]{pst-plot}{ysubticks}[1]{\def\psk@ysubticks{#1}}
\define@key[psset]{pst-plot}{subticksize}[0.75]{\def\psk@xsubticksize{#1}\def\psk@ysubticksize{#1}}
\define@key[psset]{pst-plot}{xsubticksize}[0.75]{\def\psk@xsubticksize{#1}}
\define@key[psset]{pst-plot}{ysubticksize}[0.75]{\def\psk@ysubticksize{#1}}
\define@key[psset]{pst-plot}{tickwidth}[0.5\pslinewidth]{%
  \pst@getlength{#1}\psk@xtickwidth%
  \pst@getlength{#1}\psk@ytickwidth}
\define@key[psset]{pst-plot}{xtickwidth}[0.5\pslinewidth]{\pst@getlength{#1}\psk@xtickwidth}
\define@key[psset]{pst-plot}{ytickwidth}[0.5\pslinewidth]{\pst@getlength{#1}\psk@ytickwidth}
\define@key[psset]{pst-plot}{subtickwidth}[0.25\pslinewidth]{%
  \pst@getlength{#1}\psk@xsubtickwidth%
  \pst@getlength{#1}\psk@ysubtickwidth}
\define@key[psset]{pst-plot}{xsubtickwidth}[0.25\pslinewidth]{\pst@getlength{#1}\psk@xsubtickwidth}
\define@key[psset]{pst-plot}{ysubtickwidth}[0.25\pslinewidth]{\pst@getlength{#1}\psk@ysubtickwidth}
\define@key[psset]{pst-plot}{labelOffset}[0pt]{%
  \pst@getlength{#1}\psk@xlabelOffset%
  \pst@getlength{#1}\psk@ylabelOffset}
\define@key[psset]{pst-plot}{xlabelOffset}[0pt]{\pst@getlength{#1}\psk@xlabelOffset}
\define@key[psset]{pst-plot}{ylabelOffset}[0pt]{\pst@getlength{#1}\psk@ylabelOffset}
\define@key[psset]{pst-plot}{frameOffset}[0pt]{\pst@getlength{#1}\psk@frameOffset}
\define@key[psset]{pst-plot}{tickcolor}[black]{%
    \pst@getcolor{#1}\psk@xtickcolor%
    \pst@getcolor{#1}\psk@ytickcolor}
\define@key[psset]{pst-plot}{xtickcolor}[black]{\pst@getcolor{#1}\psk@xtickcolor}
\define@key[psset]{pst-plot}{ytickcolor}[black]{\pst@getcolor{#1}\psk@ytickcolor}
\define@key[psset]{pst-plot}{subtickcolor}[gray]{%
  \pst@getcolor{#1}\psk@xsubtickcolor%
  \pst@getcolor{#1}\psk@ysubtickcolor}
\define@key[psset]{pst-plot}{xsubtickcolor}[gray]{\pst@getcolor{#1}\psk@xsubtickcolor}
\define@key[psset]{pst-plot}{ysubtickcolor}[gray]{\pst@getcolor{#1}\psk@ysubtickcolor}
\define@key[psset]{pst-plot}{xticklinestyle}[solid]{%
  \@ifundefined{psls@#1}%
    {\@pstrickserr{Line style `#1' not defined}\@eha}%
    {\def\psxticklinestyle{#1}}}
\define@key[psset]{pst-plot}{xsubticklinestyle}[solid]{%
  \@ifundefined{psls@#1}%
    {\@pstrickserr{Line style `#1' not defined}\@eha}%
    {\def\psxsubticklinestyle{#1}}}
\define@key[psset]{pst-plot}{yticklinestyle}[solid]{%
  \@ifundefined{psls@#1}%
    {\@pstrickserr{Line style `#1' not defined}\@eha}%
    {\def\psyticklinestyle{#1}}}
\define@key[psset]{pst-plot}{ysubticklinestyle}[solid]{%
  \@ifundefined{psls@#1}%
    {\@pstrickserr{Line style `#1' not defined}\@eha}%
    {\def\psysubticklinestyle{#1}}}
\define@key[psset]{pst-plot}{ticklinestyle}[solid]{%
  \@ifundefined{psls@#1}%
    {\@pstrickserr{Line style `#1' not defined}\@eha}%
    {\def\psxticklinestyle{#1}\def\psyticklinestyle{#1}}}
\define@key[psset]{pst-plot}{subticklinestyle}[solid]{%
  \@ifundefined{psls@#1}%
    {\@pstrickserr{Line style `#1' not defined}\@eha}%
    {\def\psxsubticklinestyle{#1}\def\psysubticklinestyle{#1}}}
\psset[pst-plot]{subticksize=0.75,subticks=1,tickcolor=black,ticklinestyle=solid,
  subticklinestyle=solid,subtickcolor=gray,tickwidth=0.5\pslinewidth,
  subtickwidth=0.25\pslinewidth,labelOffset=0pt,frameOffset=0pt}
\define@key[psset]{pst-plot}{nStep}[1]{\def\psk@nStep{#1}}
\define@key[psset]{pst-plot}{nStart}[0]{\def\psk@nStart{#1}}
\define@key[psset]{pst-plot}{nEnd}[{}]{\def\psk@nEnd{#1}}
\define@key[psset]{pst-plot}{xStep}[0]{\def\psk@xStep{#1}}
\define@key[psset]{pst-plot}{yStep}[0]{\def\psk@yStep{#1}}
\define@key[psset]{pst-plot}{xStart}[{}]{\def\psk@xStart{#1}}
\define@key[psset]{pst-plot}{xEnd}[{}]{\def\psk@xEnd{#1}}
\define@key[psset]{pst-plot}{yStart}[{}]{\def\psk@yStart{#1}}
\define@key[psset]{pst-plot}{yEnd}[{}]{\def\psk@yEnd{#1}}
\define@key[psset]{pst-plot}{plotNoX}[1]{\def\psk@plotNoX{#1}}
\define@key[psset]{pst-plot}{plotNo}[1]{\def\psk@plotNo{#1}}
\define@key[psset]{pst-plot}{plotNoMax}[1]{\def\psk@plotNoMax{#1}}
\define@key[psset]{pst-plot}{plotYMax}[1]{\def\psk@plotYMax{#1}}
\psset[pst-plot]{nStep=1, nStart=0, nEnd={},%
  xStep=0, yStep=0, xStart={}, xEnd={},  yStart={}, yEnd={}, 
  plotNo=1,plotNoMax=1,plotNoX=1,plotYMax={}}%
\def\pstScalePoints(#1,#2)#3#4{%
  \def\pstXScale{#1 }%
  \def\pstYScale{#2 }%
  \def\pstXPSScale{#3 }%
  \def\pstYPSScale{#4 }%
  \pst@def{ScalePoints}<%
    /yVal ED /xVal ED
    /yPSOp { #4 yVal mul #2 mul } def
    /xPSOp { #3 xVal mul #1 mul } def
    counttomark dup dup cvi eq not { exch pop } if
    /m exch def /n m 2 div cvi def
    n {
      \ifPst@polarplot exch cvi 360 mod PtoC \fi  
      yPSOp m 1 roll xPSOp m 1 roll 
      /m m 2 sub
      def } repeat>%
}
\pstScalePoints(1,1){}{}
\def\psxs@none{\let\psk@arrowA\@empty\let\psk@arrowB\@empty\psxs@axes}
\def\psxs@axes{{%
  \ifPst@xAxis\psxs@@axes\pst@dima\pst@dimb\pst@dimc\pst@dimd{}{x}\fi%
  \ifPst@yAxis\psxs@@axes\pst@dima\pst@dimb\pst@dimc\pst@dimd{exch}{y}\fi%
}}
\newif\ifSpecialLabelsDone
\def\psaxes{\pst@object{psaxes}}
\def\psaxes@i{%
  \let\pst@par@save\pst@par
  \pst@getarrows\psaxes@ii}
\def\psaxes@ii(#1){\@ifnextchar({\psaxes@iii(#1)}{\psaxes@iv(0,0)(0,0)(#1)}}
\def\psaxes@iii(#1)(#2){\@ifnextchar({\psaxes@iv(#1)(#2)}{\psaxes@iv(#1)(#1)(#2)}}
\def\psaxes@iv(#1)(#2)(#3){\@ifnextchar[{\psaxes@v(#1)(#2)(#3)}{\psaxes@vii(#1)(#2)(#3)}}%
\def\psaxes@v(#1)(#2)(#3)[#4]{\@ifnextchar[{\psaxes@vi(#1)(#2)(#3)[#4]}{\psaxes@vi(#1)(#2)(#3)[#4][]}}%
\def\psaxes@vi(#1)(#2)(#3)[#4,#5][#6,#7]{%
  \psaxes@vii(#1)(#2)(#3)%
  \let\pst@par\pst@par@save
  \begingroup
  \SpecialCoor
  \use@par
  \ifshowgrid\psgrid[style=gridstyleA]\fi
  \uput{\psxlabelsep}[#5](#3|#1){#4}\uput{\psylabelsep}[#7](#1|#3){#6}%
  \endgroup
  \ignorespaces
}
\def\psaxes@vii(#1,#2)(#3,#4)(#5,#6){%
  \pst@killglue
  \begingroup
  \ifdim\pst@dimc<\z@\relax 
    \ifdim\pst@dimd<\z@\relax 
      \addbefore@par{xlabelPos=t,ylabelPos=r}%
  \fi\fi
  \use@par
  \pssetxlength\pst@dimc{#5}
  \pssetylength\pst@dimd{#6}
    \pssetxlength\pst@dimg{#1}
    \pssetylength\pst@dimh{#2}
    \pssetxlength\pst@dima{#3}
    \pssetylength\pst@dimb{#4}
    \pst@dima=\dimexpr\pst@dima-\pst@dimg\relax
    \pst@dimb=\dimexpr\pst@dimb-\pst@dimh\relax
    \pst@dimc=\dimexpr\pst@dimc-\pst@dimg\relax
    \pst@dimd=\dimexpr\pst@dimd-\pst@dimh\relax
   \setbox\pst@hbox=\hbox\bgroup
    \ifshowgrid\psgrid[style=gridstyleA]\fi
    \@nameuse{psxs@\psk@axesstyle}
    \ifPst@xAxis
      \SpecialLabelsDonefalse
      \begingroup
      \ifnum\psk@dx=\z@
        \pst@dimg=\psk@Dx\psxunit
        \ifdim\pst@dimg<\p@ 
          \pst@cnta=\psk@Dx
          \edef\psk@Dx{\the\numexpr-1*\pst@cnta}%
        \fi
        \edef\psk@dx{\number\pst@dimg}%
      \fi
      \pst@hlabels{\pst@dimc}{\psk@arrowB}{#3}{#5}
      \ifPst@yAxis\showoriginfalse\fi
      \pst@hlabels{\pst@dima}{\psk@arrowA}{#3}{#5}
      \endgroup
    \fi
    \ifPst@yAxis
      \SpecialLabelsDonefalse
      \begingroup
      \ifdim\pst@dima=\z@ \else\ifPst@xtrigLabels\showoriginfalse\fi\fi
      \ifnum\psk@dy=\z@
        \pst@dimg=\psk@Dy\psyunit
        \ifdim\pst@dimg<\p@ 
          \pst@cnta=\psk@Dy
          \edef\psk@Dy{\the\numexpr-1*\pst@cnta}%
        \fi
        \edef\psk@dy{\number\pst@dimg}%
      \fi
      \pst@vlabels{\pst@dimb}{\psk@arrowA}{#4}{#6}%
      \ifPst@xAxis\ifdim\pst@dima<\z@ \showoriginfalse\fi\fi 
      \pst@vlabels{\pst@dimd}{\psk@arrowB}{#4}{#6}%
      \endgroup
    \fi
  \egroup%
  \pssetxlength\pst@dimg{#1}%
  \pssetylength\pst@dimh{#2}%
  \leavevmode
  \psput@cartesian\pst@hbox
  \endgroup
  \ignorespaces
}
\newif\ifis@yAxis%
\def\psxs@@axes#1#2#3#4#5#6{
  \pst@killglue
  \begin@SpecialObj
    \ifx#6x\relax
      \is@yAxisfalse
      \ifnum\psk@dx=\z@
        \pst@dimg=\psk@Dx\psxunit
        \def\psk@dx{\number\pst@dimg}%
      \fi
    \else
      \is@yAxistrue
      \ifnum\psk@dy=\z@
        \pst@dimg=\psk@Dy\psyunit
        \def\psk@dy{\number\pst@dimg}%
      \fi
    \fi
    \let\pst@linetype\pst@arrowtype
    \def\pst@axes{axes}%
    \pst@addarrowdef
    \addto@pscode{
      /showOrigin \ifPst@showOriginTick true \else false \fi def 	
      \ifis@yAxis 0 \pst@number#4 \else \pst@number#3 0 \fi
      \ifis@yAxis 0 \pst@number#2 \else \pst@number#1 0 \fi
      ArrowA
      CP 4 2 roll
      ArrowB 
      2 copy
      /yEnd exch def /xEnd exch def
      \ifx\psk@axesstyle\@none   
        pop pop 
      \else
        gsave                              		
        L                                  		
        \@nameuse{psls@\pslinestyle}                 	
        stroke                                       	
        grestore
      \fi
      /yStart exch def
      /xStart exch def
      \number\psk@ticks\space dup 2 mod 0 eq \ifis@yAxis true \else false \fi and 
      exch 2 lt \ifis@yAxis false \else true \fi and or {
      /viceversa 
        \ifis@yAxis\pst@number#2 \pst@number#4 \else\pst@number#1 \pst@number#3 \fi
         gt { true }{ false } ifelse def           
      /epsilon 0.01 def                            
      /minTickline \ifis@yAxis \pst@number#1 \else \pst@number#2 \fi def
      /maxTickline \ifis@yAxis \pst@number#3 \else \pst@number#4 \fi def
      /dT \ifis@yAxis \psk@dy \else \psk@dx \fi\space abs  
        65536 div viceversa { neg } if def                 
      /DT \ifis@yAxis \psk@Dy \else \psk@Dx \fi\space abs viceversa { neg } if def  
      /subTNo \ifis@yAxis\psk@ysubticks\else\psk@xsubticks\fi \space def
      subTNo 0 gt { /dsubT dT subTNo div def}{ /dsubT 0 def } ifelse  
      \ifis@yAxis \psk@yticksize \else \psk@xticksize \fi
      /tickend exch def /tickstart exch def
      /Twidth \ifis@yAxis \psk@ytickwidth \else \psk@xtickwidth \fi\space def
      /subTwidth \ifis@yAxis \psk@ysubtickwidth \else \psk@xsubtickwidth \fi\space def
      /STsize \ifis@yAxis \psk@ysubticksize \else \psk@xsubticksize \fi\space def
      /TColor {
        \ifis@yAxis\pst@usecolor\psk@ytickcolor
        \else\pst@usecolor\psk@xtickcolor\fi\space } def
      /subTColor {
        \ifis@yAxis\pst@usecolor\psk@ysubtickcolor
        \else\pst@usecolor\psk@xsubtickcolor\fi\space } def
      /MinValue { \ifis@yAxis yStart \else xStart \fi
        \ifx\psk@arrowA\@empty\else 
          \psk@arrowsize\space CLW mul add \psk@arrowlength\space mul 
           viceversa { sub epsilon add }{ add epsilon sub } ifelse \fi } def
      /MaxValue { \ifis@yAxis yEnd \else xEnd \fi 
        \ifx\psk@arrowB\@empty\else
          \psk@arrowsize\space CLW mul add \psk@arrowlength\space mul 
           viceversa { add epsilon sub }{ sub epsilon add } ifelse \fi } def
      /logLines {
        \ifnum\psk@logLines=\z@ true \else         
          \ifnum\psk@logLines<\tw@                 
            \ifis@yAxis false \else true \fi       
          \else
            \ifnum\psk@logLines<\thr@@             
              \ifis@yAxis true \else false \fi     
            \else 
              false                                
            \fi
          \fi
        \fi
      } def
      /LSstroke {                                  
        \ifis@yAxis \@nameuse{psls@\psyticklinestyle}
        \else       \@nameuse{psls@\psxticklinestyle}\fi 
        stroke} def
      /subLSstroke {                               
        \ifis@yAxis \@nameuse{psls@\psysubticklinestyle}
        \else       \@nameuse{psls@\psxsubticklinestyle}\fi 
        stroke} def
      0 dT MaxValue 1 add {                        
        /cntTick exch def                          
        logLines {                                 
          gsave
          1 1 DT {
           1 sub /OffSet exch def
          -10 subTNo 1 add div dup 10 add exch dup -0.1 mul 1 add {                   
            /dx exch def                           
            /x dx log OffSet add \ifis@yAxis\pst@number\psyunit\else\pst@number\psxunit\fi\space mul cntTick add def       %
            x abs MaxValue abs le {                
	      \ifis@yAxis
	        \ifPst@maxyTicks true \else false \fi
	      \else
	        \ifPst@maxxTicks true \else false \fi
	      \fi
                { x minTickline #5 moveto
                  x maxTickline #5 lineto }
                { x tickstart STsize mul #5 moveto
                  x tickend STsize mul #5 lineto } ifelse
            } if
          } for } for
          subTwidth SLW subTColor                  
          subLSstroke
          grestore                                 
          stroke
          /dsubT 0 def                             
        } if 					   
        dsubT abs 0 gt {                           
          gsave                                    
          /cntsubTick cntTick dsubT add def
          subTNo 1 sub {
            cntsubTick abs MaxValue abs le {       
    	    \ifis@yAxis
              \ifPst@maxyTicks true \else false \fi
    	    \else
              \ifPst@maxxTicks true \else false \fi
    	    \fi
              { cntsubTick minTickline STsize mul #5 moveto
                cntsubTick maxTickline STsize mul #5 lineto }
              { cntsubTick tickstart STsize mul #5 moveto
                cntsubTick tickend STsize mul #5 lineto } ifelse
            }{ exit }  ifelse
            /cntsubTick cntsubTick dsubT add def
          } repeat 
          subTwidth SLW subTColor               
          subLSstroke
          grestore                              
        } if
        showOrigin {
          gsave
          \ifis@yAxis
            \ifPst@maxyTicks true \else false \fi
          \else
            \ifPst@maxxTicks true \else false \fi
          \fi
            { cntTick minTickline #5 moveto
              cntTick maxTickline #5 lineto }
            { cntTick tickstart #5 moveto        
              cntTick tickend #5 lineto } ifelse 
          Twidth SLW TColor                      
          LSstroke
          grestore
        }{ /showOrigin true def } ifelse         
      } for
      /showOrigin \ifPst@showOriginTick true \else false \fi def 
      /dT dT neg def                               
      /dsubT dsubT neg def
      0 dT MinValue epsilon viceversa { add }{ sub } ifelse {
        /cntTick exch def
        logLines {                                 
          gsave
          1 1 DT cvi {
            1 sub /OffSet exch def
          -10 subTNo 1 add div dup 10 add exch dup -0.1 mul 1 add {                   
            /dx exch def                           
            /x dx log OffSet add \ifis@yAxis\pst@number\psyunit\else\pst@number\psxunit\fi\space mul cntTick add def
            x abs MinValue abs le {                
	      \ifis@yAxis
	        \ifPst@maxyTicks true \else false \fi
	      \else
	        \ifPst@maxxTicks true \else false \fi
	      \fi
                { x minTickline #5 moveto
                  x maxTickline #5 lineto }
                { x tickstart STsize mul #5 moveto
                  x tickend STsize mul #5 lineto } ifelse
            } if
          } for } for
          /dsubT 0 def 
          subTwidth SLW subTColor                  
          subLSstroke
          grestore
        }                                          
        dsubT abs 0 gt {                           
          gsave                                    
          /cntsubTick cntTick dsubT add def
          subTNo 1 sub {
            cntsubTick abs MinValue abs le {       
              cntsubTick tickstart STsize mul #5 moveto
              cntsubTick tickend STsize mul #5 lineto
            }{ exit } ifelse
            /cntsubTick cntsubTick dsubT add def
          } repeat 
          subTwidth SLW subTColor                  
          subLSstroke
          grestore                                 
        } if
        showOrigin {
          gsave
          cntTick tickstart #5 moveto         	
          cntTick tickend #5 lineto    	       	
          Twidth SLW TColor                         
          LSstroke
          grestore
        }{ /showOrigin true def } ifelse         
      } for
    } if
   }
  \end@SpecialObj%
  \ifx\psk@axesstyle\@none\else
    \ifPst@yAxis\psline[linecolor=\pslinecolor](0,#2)(0,#4)\fi
    \ifPst@xAxis\psline[linecolor=\pslinecolor](#1,0)(#3,0)\fi
  \fi
  \ignorespaces
}%
\def\psxs@frame{%
  \psset{axesstyle=none}%
  \begin@SpecialObj%
    \addto@pscode{					
      \pst@number\pst@dima \psk@frameOffset sub \pst@number\pst@dimb \psk@frameOffset sub moveto 	
      \pst@number\pst@dimc \psk@frameOffset add \pst@number\pst@dimb \psk@frameOffset sub L	
      \pst@number\pst@dimc \psk@frameOffset add \pst@number\pst@dimd \psk@frameOffset add L 	
      \pst@number\pst@dima \psk@frameOffset sub \pst@number\pst@dimd \psk@frameOffset add L 	
      closepath 
      }%
    \pst@stroke%
    \psk@fillstyle%
  \end@SpecialObj%
  \let\psk@arrowA\@empty%
  \let\psk@arrowB\@empty%
  \pst@xticksizeC=\z@\pst@yticksizeC=\z@  
  \ifPst@xAxis\psxs@@axes\pst@dima\pst@dimb\pst@dimc\pst@dimd{}{x}\fi
  \ifPst@yAxis\psxs@@axes\pst@dima\pst@dimb\pst@dimc\pst@dimd{ exch }{y}\fi
  \ifnum\psk@tickstyle=\tw@	
    \psDEBUG[psxs@frame]{psk@tickstyle=2 (inner)}%
    \psDEBUG[psxs@frame]{pst@dima=\pst@number\pst@dima}%
    \psDEBUG[psxs@frame]{pst@dimb=\pst@number\pst@dimb}%
    \psDEBUG[psxs@frame]{pst@dimc=\pst@number\pst@dimc}%
    \psDEBUG[psxs@frame]{pst@dimd=\pst@number\pst@dimd}%
    \ifPst@xAxis\psxs@@axes\pst@dima\pst@dimb\pst@dimc\pst@dimd{ neg \pst@number\pst@dimd add }{x}\fi
    \ifPst@yAxis\psxs@@axes\pst@dima\pst@dimb\pst@dimc\pst@dimd{ neg \pst@number\pst@dimc add exch }{y}\fi
  \fi%
}
\def\psxs@polar{
  \pst@killglue
  \begingroup
  \edef\pst@dimC{\strip@pt\pst@dimc}
  \pstFPDiv\pstR@dius{\pst@dimC}{\strip@pt\psxunit}
  \edef\pst@dimD{\strip@pt\pst@dimd}
  \pstFPDiv\psk@EndAngle{\pst@dimD}{\strip@pt\psyunit}
  \ifnum\psk@EndAngle=0 \def\psk@EndAngle{360}\fi
  \use@keep@par
  \pstFPDiv\pstN@lpha{\psk@EndAngle}{\psk@Dy}
  \pstFPdiv\pstd@lpha{\psk@Dy}{\psk@ysubticks}
  \pstFPdiv\pstdR@dius{1}{\psk@xsubticks}
  \pst@cntm=\psk@xsubticks\advance\pst@cntm by \m@ne
  \multido{\iA=\psk@Dx+\psk@Dx,\rB=\pstdR@dius+\psk@Dx,\iB=0+1}{\pstR@dius}{%
    \multido{\rA=\rB+\pstdR@dius}{\the\pst@cntm}{%
      \psarc[linestyle=\psxsubticklinestyle,
         linecolor=\psk@xsubtickcolor,linewidth=\psk@xsubtickwidth pt](0,0){\rA}{0}{\psk@EndAngle}}    
    \psarc[linestyle=\psxticklinestyle,linecolor=\psk@xtickcolor,
		linewidth=\psk@xtickwidth pt](0,0){\iA}{0}{\psk@EndAngle}%
    \ifnum\psk@labels<2\relax
      \uput[-45](\iB,0){\pshlabel{\iB}}\uput[45](0,\iB){\pshlabel{\iB}}%
    \fi%
  }%
  \pst@cntm=\psk@ysubticks\advance\pst@cntm by \m@ne
  \multido{\iA=\psk@Oy+\psk@Dy,\rB=\pstd@lpha+\psk@Dy}{\pstN@lpha}{%
    \multido{\rA=\rB+\pstd@lpha}{\the\pst@cntm}{\psline[linestyle=\psysubticklinestyle,
      linecolor=\psk@ysubtickcolor,linewidth=\psk@ysubtickwidth pt](\pstR@dius;\rA)} 
    \psline[linestyle=\psyticklinestyle,
      linecolor=\psk@ytickcolor,linewidth=\psk@ytickwidth pt](\pstR@dius;\iA)%
    \ifodd\psk@labels\else
      \uput[\iA](\pstR@dius;\iA){\psvlabel{\iA\psk@ylabelFactor}}%
    \fi%
  }%
  \ifnum\psk@EndAngle<360 \psline[linestyle=\psyticklinestyle,
      linecolor=\psk@ytickcolor,linewidth=\psk@ytickwidth pt](\pstR@dius;0)\fi
  \endgroup\ignorespaces%
  \Pst@xAxisfalse\Pst@yAxisfalse%
}
\def\@polar{polar}
\define@key[psset]{pst-plot}{axesstyle}[axes]{%
  \@ifundefined{psxs@#1}%
    {\@pstrickserr{Axes style `#1' not defined}\@eha}%
    {\def\psk@axesstyle{#1}%
     \ifx\psk@axesstyle\@polar\psset{Dy=30}\fi}}
\psset[pst-plot]{axesstyle=axes}
\define@key[psset]{pst-plot}{xLabels}[]{\def\psk@xLabels{#1}}
\define@key[psset]{pst-plot}{xLabelsRot}[0]{\pst@getangle{#1}\pst@xLabelsRot}
\psset[pst-plot]{xLabels=,xLabelsRot=0}
\define@key[psset]{pst-plot}{yLabels}[]{\def\psk@yLabels{#1}}
\define@key[psset]{pst-plot}{yLabelsRot}[0]{\pst@getangle{#1}\pst@yLabelsRot}
\psset[pst-plot]{yLabels=,yLabelsRot=0}
\def\pst@hlabels#1#2#3#4{%
  \ifSpecialLabelsDone
  \else
    \kern\psk@xlabelOffset pt            
    \ifx\empty\psk@xLabels
      \ifdim#1=\z@
      \else                   
        \ifx#2\empty
        \else
          \advance#1\ifdim#1>\z@-\fi7\pslinewidth
        \fi
        \pst@cnta=#1\relax                
        \divide\pst@cnta\psk@dx\relax     
        \ifnum\pst@cnta=\z@
        \else
          \pst@dimb=\psk@dx sp            
          \ifnum\psk@labels<\tw@ \ifPst@xAxis\pst@@hlabels\fi\fi
          \showoriginfalse
        \fi
      \fi
   \else
     \ifnum\psk@xlabelPos=\tw@ \def\pst@tempC{90}\else\def\pst@tempC{-90}\fi
       \pstFPsub\pst@pmtempa{#4}{#3}%
       \pstFPDiv\pst@pmtempb{\pst@pmtempa}{\psk@Dx}%
       \pstFPadd\pst@pmtempc{\pst@pmtempb}{-1}%
       \pstFPadd\pst@pmtempd{\pst@pmtempb}{1}%
       \ifdim\pst@pmtempb pt < \z@ 
         \def\pst@pmtempe{\pst@int{\pst@pmtempc}}%
       \else
         \def\pst@pmtempe{\pst@int{\pst@pmtempd}}%
       \fi
       \multido{\nA=0+1,\rA=#3+\psk@Dx}{\pst@pmtempe}{%
         \ifdim \nA pt < \z@ \def\nB{-\nA} \else \def\nB{\nA} \fi
         \uput{\psxlabelsep}[\pst@tempC]{\pst@xLabelsRot}(\rA,0){%
              \strut\expandafter\pshlabel\expandafter{\psPutXLabel{\nB}}}}%
       \SpecialLabelsDonetrue
    \fi
  \fi
}
\def\pst@@hlabels{%
  \setbox\z@=\vbox{
    \ifcase\psk@xlabelPos
      \vskip-\pst@xticksizeA\vskip\psxlabelsep\or
      \vskip-1ex\vskip-\pslabelsep\or
      \vskip-\pst@xticksizeB\vskip-\psxlabelsep\vskip-1ex
    \fi
    \ifnum\pst@cnta<\z@ \pst@dimb=-\pst@dimb\fi
    \hbox to \z@{%
      \ifshoworigin\hbox to \z@{\hss\pst@@@hlabel{\psk@Ox}\hss}\fi
      \mmultido{\nA=\psk@Ox+\psk@Dx}{\pst@cnta}{%
        \hskip\pst@dimb \hbox to \z@{\hss
          \ifdim\nA pt=\z@\relax\ifshoworigin\pst@@@hlabel{0}\fi
          \else\expandafter\pst@@@hlabel{\nA}%
          \fi
        \hss}%
      }\hss
    }%
  }\ht\z@\z@ \dp\z@\z@ \box\z@}
\def\pst@vlabels#1#2#3#4{%
  \ifSpecialLabelsDone\else
      \ifx\empty\psk@yLabels
        \ifdim#1=\z@\else
          \ifx#2\empty\else\ifdim#1>\z@ \advance#1 by -7\pslinewidth\else\advance#1 by 7\pslinewidth\fi\fi
          \pst@cnta=#1\relax           
          \divide\pst@cnta\psk@dy\relax
          \ifnum\pst@cnta=\z@\else
            \pst@dima=\psk@dy sp
            \ifodd\number\psk@labels\else\ifPst@yAxis\pst@@vlabels\fi
          \fi
          \showoriginfalse
        \fi
      \fi
    \else
	\pstFPsub\pst@pmtempa{#4}{#3}%
	\pstFPDiv\pst@pmtempb{\pst@pmtempa}{\psk@Dy}%
	\pstFPadd\pst@pmtempc{\pst@pmtempb}{-1}%
	\pstFPadd\pst@pmtempd{\pst@pmtempb}{1}%
	\ifdim\pst@pmtempb pt < \z@ \def\pst@pmtempe{\pst@int{\pst@pmtempc}}\else\def\pst@pmtempe{\pst@int{\pst@pmtempd}}\fi
	\multido{\nA=0+1,\rA=#3+\psk@Dy}{\pst@pmtempe}{%
	  \ifdim \nA pt < \z@ \def\nB{-\nA}\else \def\nB{\nA}\fi
	  \ifnum\psk@ylabelPos=0
            \uput{\psylabelsep}[180]{\pst@yLabelsRot}(0,\rA){%
              \strut\expandafter\psvlabel\expandafter{\psPutYLabel{\nB}}}%
          \else
            \uput{\psylabelsep}[0]{\pst@yLabelsRot}(0,\rA){%
              \strut\expandafter\psvlabel\expandafter{\psPutYLabel{\nB}}}%
          \fi
        }%
      \SpecialLabelsDonetrue
    \fi
  \fi
}
\def\pst@@vlabels{%
  \vbox to\z@{%
   \vbox to -\psk@ylabelOffset pt{}
    \ifnum\pst@cnta>\z@ \pst@dima=-\pst@dima\fi
    \offinterlineskip
    \ifshoworigin
      \vbox to \z@{\vss\hbox to\z@{%
        \ifcase\psk@ylabelPos
	  \hss\pst@@@vlabel{\psk@Oy}\hskip\psylabelsep\hskip-\pst@yticksizeA\or%
	  \hskip\pslabelsep\hss\pst@@@vlabel{\psk@Oy}\hss\or
	  \hskip\pst@yticksizeB\hskip\psylabelsep\pst@@@vlabel{\psk@Oy}%
	\fi}\vss}%
    \fi
    \mmultido{\nA=\psk@Oy+\psk@Dy}{\pst@cnta}{%
      \vbox to\pst@dima{\vss}%
      \vbox to \z@{%
        \vss\hbox to\z@{%
        \ifcase\psk@ylabelPos 
	  \hss\ifdim\nA pt=\z@ \ifshoworigin\pst@@@vlabel{0}\fi\else\pst@@@vlabel{\nA}\fi
	    \hskip\psylabelsep\hskip-\pst@yticksizeA\or
	  \hss\ifdim\nA pt=\z@ \ifshoworigin\pst@@@vlabel{0}\fi\else\pst@@@vlabel{\nA}\fi
	  \ifdim\psylabelsep=\z@\hss\else\kern-\psylabelsep\fi\or
	  \hskip\pst@yticksizeB\hskip\psylabelsep
	  \ifdim\nA pt=\z@ \ifshoworigin\pst@@@vlabel{0}\fi\else\pst@@@vlabel{\nA}\fi
	\fi}\vss}%
    }\vss}%
}
\define@key[psset]{pst-plot}{xAxisLabel}[x]{\def\psk@xAxisLabel{#1}}
\define@key[psset]{pst-plot}{yAxisLabel}[y]{\def\psk@yAxisLabel{#1}}
\psset[pst-plot]{xAxisLabel=x,yAxisLabel=y}
\define@key[psset]{pst-plot}{xAxisLabelPos}[{}]{\def\psk@xAxisLabelPos{#1}}
\define@key[psset]{pst-plot}{yAxisLabelPos}[{}]{\def\psk@yAxisLabelPos{#1}}
\psset[pst-plot]{yAxisLabelPos={},xAxisLabelPos={}}
\newdimen\psk@llx
\newdimen\psk@lly
\newdimen\psk@urx
\newdimen\psk@ury
\define@key[psset]{pst-plot}{llx}[\z@]{\pssetxlength\psk@llx{#1}}
\define@key[psset]{pst-plot}{lly}[\z@]{\pssetylength\psk@lly{#1}}
\define@key[psset]{pst-plot}{urx}[\z@]{\pssetxlength\psk@urx{#1}}
\define@key[psset]{pst-plot}{ury}[\z@]{\pssetylength\psk@ury{#1}}
\psset[pst-plot]{llx=\z@, lly=\z@, urx=\z@, ury=\z@}
\define@boolkey[psset]{pst-plot}[Pst@]{psgrid}[true]{}
\define@key[psset]{pst-plot}{gridpara}[{}]{\def\psk@gridpara{#1}}
\define@key[psset]{pst-plot}{gridcoor}[\relax]{\def\psk@gridcoor{#1}}
\psset[pst-plot]{psgrid=false,gridpara={gridlabels=0pt,gridcolor=red!30,subgridcolor=green!30,subgridwidth=0.5\pslinewidth,
  subgriddiv=5},gridcoor=\relax}
\define@key[psset]{pst-plot}{axespos}[bottom]{\pst@expandafter\psset@@axespos{#1}\@nil}
\def\psset@@axespos#1#2\@nil{%
  \ifx#1b\let\psk@axespos\z@\else		
    \ifx#1t\let\psk@axespos\@ne			
      \else\@pstrickserr{Bad axes position: `#1#2'}\@ehpa
  \fi\fi}
\psset[pst-plot]{axespos=b}
\newdimen\pst@xunit
\newdimen\pst@yunit
\def\pslegend{\@ifnextchar[\pslegend@i{\pslegend@i[rt]}}
\def\pslegend@i[#1]{\@ifnextchar({\pslegend@ii[#1]}{\pslegend@ii[#1](\pst@number\pslabelsep,\pst@number\pslabelsep)}}
\def\pslegend@ii[#1](#2,#3)#4{%
  \gdef\pslegend@ref{#1}%
  \xdef\pslegend@sepx{#2 }%
  \xdef\pslegend@sepy{#3 }%
  \gdef\pslegend@text{#4}}
\newpsstyle{legendstyle}{fillstyle=solid,fillcolor=white,linewidth=0.5pt}
\def\pslegend@iii[#1](#2){\rput[#1](#2){\psframebox[style=legendstyle]{%
  \footnotesize\tabcolsep=2pt%
  \tabular[t]{@{}ll@{}}\pslegend@text\endtabular}}\global\let\pslegend@text\relax}
\let\pslegend@text\relax
\def\psgraph{\pst@object{psgraph}}
\def\psgraph@i{%
  \let\psgraph@para\pst@par
  \let\psk@save@arrowA\psk@arrowA
  \let\psk@save@arrowB\psk@arrowB
  \pst@getarrows\psgraph@ii}
\def\psgraph@ii(#1,#2){\catcode`\!=12\relax
  \@ifnextchar({\psgraph@iii(#1,#2)}{\psgraph@iv(0,0)(#1,#2)}}
\def\psgraph@iii(#1,#2)(#3,#4){\@ifnextchar({\psgraph@v(#1,#2)(#3,#4)}{\psgraph@iv(#1,#2)(#3,#4)}}
\def\psgraph@iv(#1,#2)(#3,#4)#5#6{
  \pst@killglue%
  \begingroup
  \use@keep@par
  \pstFPsub\pst@tempA{#3}{#1}%
  \pst@dimm=#5
  \pst@dimo=\pst@tempA pt
  \pstFPdiv\pst@@dx{\strip@pt\pst@dimm}{\pst@tempA}%
  \pst@xunit=\pst@@dx\p@
  \ifx!#6\let\pst@yunit=\pst@xunit\else
    \pst@dimm=#6
    \pstFPsub\pst@tempA{#4}{#2}%
    \pstFPdiv\pst@@dy{\strip@pt\pst@dimm}{\pst@tempA}%
    \pst@yunit=\pst@@dy\p@
  \fi
  \pst@dimm=#1\pst@xunit\advance\pst@dimm by \psk@llx
  \pst@dimn=#2\pst@yunit\advance\pst@dimn by \psk@lly
  \pst@dimo=#3\pst@xunit\advance\pst@dimo by \psk@urx
  \pst@dimp=#4\pst@yunit\advance\pst@dimp by \psk@ury
  \if@star\pspicture*(\pst@dimm,\pst@dimn)(\pst@dimo,\pst@dimp)\else
  \pspicture(\pst@dimm,\pst@dimn)(\pst@dimo,\pst@dimp)\fi
  \let\psxunit\pst@xunit \let\psyunit\pst@yunit
  \ifdim\pst@xunit=\pst@yunit\relax\psset{runit=\pst@xunit}\fi%
  \bgroup
    \use@par
  \ifPst@psgrid
     \expandafter\psset\expandafter{\psk@gridpara}%
      \rput[lb](0,0){\expandafter\psgrid\psk@gridcoor}  
  \fi
    \ifnum\psk@axespos=0
      \expandafter\psaxes\expandafter[\psgraph@para](#1,#2)(#3,#4)%
    \else
      \xdef\psgraph@coor{(#1,#2)(#3,#4)(#5,#6)}%
    \fi
  \egroup
  \psgraph@vi(#1,#2)(#1,#2)(#3,#4)%
}
\def\psgraph@v(#1,#2)(#3,#4)(#5,#6)#7#8{
  \pst@killglue%
  \let\psgraph@para\pst@par
  \begingroup%
  \use@keep@par
  \pstFPsub\pst@tempA{#5}{#3}%
  \pst@dimm=#7%
  \pst@dimo=\pst@tempA pt%
  \pstFPdiv\pst@@dx{\strip@pt\pst@dimm}\pst@tempA%
  \pst@xunit=\pst@@dx\p@%
  \ifx!#8\let\pst@yunit=\pst@xunit\else
    \pst@dimm=#8%
    \pstFPsub\pst@tempA{#6}{#4}%
    \pstFPdiv\pst@@dy{\strip@pt\pst@dimm}\pst@tempA%
    \pst@yunit=\pst@@dy\p@%
  \fi%
  \pst@dima=#3\pst@xunit \advance\pst@dima by \psk@llx%
  \pst@dimb=#4\pst@yunit \advance\pst@dimb by \psk@lly%
  \pst@dimc=#5\pst@xunit \advance\pst@dimc by \psk@urx%
  \pst@dimd=#6\pst@yunit \advance\pst@dimd by \psk@ury%
  \if@star\pspicture*(\pst@dima,\pst@dimb)(\pst@dimc,\pst@dimd)\else%
          \pspicture(\pst@dima,\pst@dimb)(\pst@dimc,\pst@dimd)\fi%
  \psset{xunit=\pst@xunit,yunit=\pst@yunit}
  \ifdim\pst@xunit=\pst@yunit \psset{runit=\pst@xunit}\fi%
  \bgroup%
    \use@par%
  \ifPst@psgrid
     \expandafter\psset\expandafter{\psk@gridpara}%
      \rput[lb](0,0){\expandafter\psgrid\psk@gridcoor}
  \fi%
    \ifnum\psk@axespos=0
      \psaxes(#1,#2)(#3,#4)(#5,#6)%
    \else
      \xdef\psgraph@coor{(#1,#2)(#3,#4)(#5,#6)}%
    \fi
  \egroup
  \psgraph@vi(#1,#2)(#3,#4)(#5,#6)%
}
\def\setxLabelC@@r#1,#2(#3,#4)(#5){%
  \pst@getcoor{#5}\pst@tempB%
  \ifx c#1 
    \pssetylength\pst@dimm{#2}%
    \rput(! #4 #3 add 2 div \pst@number\pst@dimm \pst@tempB\space exch pop add 
      \pst@number\psyunit div ){\psk@xAxisLabel}%
  \else%
    \pst@getcoor{\psk@xAxisLabelPos}\pst@tempA%
    \rput(! \pst@tempA\space \pst@tempB\space exch pop add \tx@UserCoor ){\psk@xAxisLabel}%
  \fi}
\def\setyLabelC@@r#1,#2(#3,#4)(#5){%
  \pst@getcoor{#5}\pst@tempB%
  \ifx c#2
    \pssetxlength\pst@dimm{#1}%
    \rput{90}(! \pst@number\pst@dimm \pst@tempB\space pop add \pst@number\psxunit div #4 #3 add 2 div ){\psk@yAxisLabel}%
  \else%
    \pst@getcoor{\psk@yAxisLabelPos}\pst@tempA%
    \rput{90}(! \pst@tempB\space pop \pst@tempA\space 3 1 roll add exch \tx@UserCoor ){\psk@yAxisLabel}%
  \fi}
\def\psgraph@vi(#1,#2)(#3,#4)(#5,#6){%
  \ifx\psk@xAxisLabel\@empty\else%
    \ifx\psk@xAxisLabelPos\@empty\uput[0](#5,#2){\psk@xAxisLabel}%
    \else\expandafter\setxLabelC@@r\psk@xAxisLabelPos(#3,#5)(#1,#2)\fi%
  \fi%
  \ifx\psk@yAxisLabel\@empty\else%
    \ifx\psk@yAxisLabelPos\@empty\uput[90](#1,#6){\psk@yAxisLabel}%
    \else\expandafter\setyLabelC@@r\psk@yAxisLabelPos(#4,#6)(#1,#2)\fi%
  \fi%
  \def\lt@@{lt}\def\lb@@{lb}\def\rb@@{rb}%
  \ifx\pslegend@ref\lb@@    \gdef\pslegend@coor{#3 \pslegend@sepx \pst@number\psxunit div add 
                                                   \pslegend@sepy \pst@number\psyunit div}%
  \else%
    \ifx\pslegend@ref\lt@@  \gdef\pslegend@coor{#3 \pslegend@sepx \pst@number\psxunit div add 
                                                #6 \pslegend@sepy \pst@number\psyunit div sub}%
    \else%
      \ifx\pslegend@ref\rb@@\gdef\pslegend@coor{#5 \pslegend@sepx \pst@number\psxunit div sub 
                                                   \pslegend@sepy \pst@number\psyunit div}%
      \else                 \gdef\pslegend@coor{#5 \pslegend@sepx \pst@number\psxunit div sub 
                                                #6 \pslegend@sepy \pst@number\psyunit div sub}%
      \fi%
    \fi%
  \fi%
  \xdef\psgraphLLx{#3}\xdef\psgraphLLy{#4}\xdef\psgraphURx{#5}\xdef\psgraphURy{#6}%
  \global\let\psk@arrowA\psk@save@arrowA
  \global\let\psk@arrowB\psk@save@arrowB
  \ignorespaces
}
\def\endpsgraph{%
  \ifx\relax\pslegend@text\relax \else\pslegend@iii[\pslegend@ref](!\pslegend@coor)\fi
  \expandafter\psset\expandafter{\psgraph@para}%
  \ifnum\psk@axespos>0
    \expandafter\psaxes\psgraph@coor
  \fi
  \endpspicture
  \endgroup\ignorespaces}
\@namedef{psgraph*}{\psgraph*}
\@namedef{endpsgraph*}{\endpsgraph}
\def\psPutXLabel#1{%
  \global\pst@cnto=0\relax
  \global\pst@cntp=#1\relax
  \expandafter\get@Label\psk@xLabels,,\@nil%
}
\def\psPutYLabel#1{%
  \global\pst@cnto=0\relax
  \global\pst@cntp=#1\relax
  \expandafter\get@Label\psk@yLabels,,\@nil%
}
\def\get@Label#1,#2,#3\@nil{%
    \ifnum\the\pst@cnto<\the\pst@cntp
      \global\advance\pst@cnto by \@ne 
      \ifx\relax#3\relax\else\expandafter\get@Label#2,#3\@nil\fi%
    \else #1\fi%
}
\def\psVectorfield{\pst@object{psVectorfield}}
\def\psVectorfield@i(#1,#2)(#3,#4)#5{{%
  \addbefore@par{Dx=0.1,Dy=0.1,Ox=3,arrows=->,linewidth=0.2pt}%
  \begin@SpecialObj
  \SpecialCoor
  \pstFPsub\pst@tempA{#3}{#1}%
  \pstFPsub\pst@tempB{#4}{#2}%
  \pstFPDiv{\pst@tempC}{\pst@tempA}{\psk@Dx}%
  \pstFPDiv{\pst@tempD}{\pst@tempB}{\psk@Dy}%
  \pstVerb{ /yStrich \ifPst@algebraic (#5) tx@AlgToPs begin AlgToPs end cvx
                \else { #5 } \fi def }%
  \multido{\rX=#1+\psk@Dx}{\numexpr\pst@tempC+1}{%
    \multido{\rY=#2+\psk@Dy}{\numexpr\pst@tempD+1}{%
       \psline%
         (! /x \rX\space def 
            /y \rY\space def 
            /yTemp yStrich \psk@Dx\space \psk@Ox\space div mul def 
            \rX\space \psk@Dx\space \psk@Ox\space div sub \rY\space yTemp sub)%
         (! /x \rX\space def 
            /y \rY\space def 
            /yTemp yStrich \psk@Dx\space \psk@Ox\space div mul def 
            \rX\space \psk@Dx\space \psk@Ox\space div add \rY\space yTemp add)%
   }}%
  \end@SpecialObj
}\ignorespaces}  
\def\psFixpoint{\pst@object{psFixpoint}}
\def\psFixpoint@i#1#2#3{
  \pst@killglue%
  \begingroup%
  \use@par%
  \@nameuse{beginplot@\psplotstyle}%
  \addto@pscode{
    \psplot@init
      /x #1 def
      /F@pstplot \ifPst@algebraic (#2) tx@AlgToPs begin AlgToPs end cvx
                 \else { #2 } \fi  def
      /xy { x \pst@number\psxunit mul F@pstplot dup /x ED \pst@number\psyunit mul } def 
  }%
  \gdef\psplot@init{}%
  \@pstfalse%
  \@nameuse{testqp@\psplotstyle}%
  \addto@pscode{
      mark
      x \pst@number\psxunit mul 0
      /n 2 def
      #3 {
        xy 
        dup dup 
        /n n 4 add def
      } repeat 
  }%
  \@nameuse{endplot@\psplotstyle}%
  \endgroup%
  \ignorespaces}
\define@boolkey[psset]{pst-plot}[Pst@]{showDerivation}[true]{}
\psset{showDerivation}
\def\psNewton{\pst@object{psNewton}}
\def\psNewton@i#1#2{\@ifnextchar[{\psNewton@ii{#1}{#2}}{\psNewton@iii{#1}{#2}}}
\def\psNewton@ii#1#2[#3]#4{
  \pst@killglue%
  \begingroup%
  \addbefore@par{showDerivation}%
  \use@par%
  \@nameuse{beginplot@\psplotstyle}%
  \addto@pscode{
    \psplot@init
      /x #1 def
      /F@pstplot \ifPst@algebraic (#2) tx@AlgToPs begin AlgToPs end cvx \else { #2 } \fi  def
      /F@pstplotDerive \ifPst@algebraic (#3) tx@AlgToPs begin AlgToPs end cvx \else { #3 } \fi  def
      /newxVal { 
        F@pstplotDerive 
        div neg 
      } def
  }%
  \gdef\psplot@init{}%
  \@pstfalse%
  \@nameuse{testqp@\psplotstyle}%
  \addto@pscode{
      mark
      x 0 \tx@ScreenCoor 
      /n 2 def
      #4 {
        F@pstplot /yVal ED
        x yVal \tx@ScreenCoor
        /n n 2 add def
        yVal newxVal x add /x ED
        x 0 \tx@ScreenCoor 
        \ifPst@showDerivation /n n 4 add def \else moveto /n n 2 add def\fi
      } repeat 
      pstack
  }%
  \@nameuse{endplot@\psplotstyle}%
  \endgroup%
  \ignorespaces}
\def\psNewton@iii#1#2#3{
  \pst@killglue%
  \begingroup%
  \addbefore@par{VarStepEpsilon=0.01,showDerivation}%
  \use@par%
  \@nameuse{beginplot@\psplotstyle}%
  \addto@pscode{
    \psplot@init
      /epsilon \psk@VarStepEpsilon\space def
      /x #1 def
      /F@pstplot \ifPst@algebraic (#2) tx@AlgToPs begin AlgToPs end cvx \else { #2 } \fi  def
      /newxVal { 
        /saveX x def
        saveX epsilon add /x ED F@pstplot saveX epsilon sub /x ED F@pstplot sub epsilon dup add div 
        div neg 
        /x saveX def
      } def
  }%
  \gdef\psplot@init{}%
  \@pstfalse%
  \@nameuse{testqp@\psplotstyle}%
  \addto@pscode{
      mark
      x 0 \tx@ScreenCoor 
      /n 2 def
      #3 {
        F@pstplot /yVal ED
        x yVal \tx@ScreenCoor
        yVal newxVal x add /x ED
        x 0 \tx@ScreenCoor 
        \ifPst@showDerivation /n n 4 add def \else moveto /n n 2 add def\fi
      } repeat 
  }%
  \@nameuse{endplot@\psplotstyle}%
  \endgroup%
  \ignorespaces}
\def\psResetPlotValues{%
  \psset{method={}}%
}%
\catcode`\@=\TheAtCode\relax
 
\csname PSTnodesLoaded\endcsname
\let\PSTnodesLoaded 
\ifx\PSTricksLoaded \else\input pstricks.tex \fi\relax
\ifx\PSTXKeyLoaded \else \input pst-xkey \fi
\def\fileversion{1.42}
\def\filedate{2019/03/03}
\edef\TheAtCode{\the\catcode`\@}
\catcode`\@=11
\pstheader{pst-node.pro}
\pst@addfams{pst-node}
\SpecialCoor
\define@boolkey[psset]{pst-node}[Pst@]{trueAngle}[true]{}
\psset[pst-node]{trueAngle=false}
\define@boolkey[psset]{pst-node}[Pst@]{storeNodeInfo}[true]{}
\psset[pst-node]{storeNodeInfo=false}
\def\pst@nodedict{tx@NodeDict begin }
\def\pst@zapspace#1 #2{%
#1%
\ifx#2\@empty\else\expandafter\pst@zapspace\fi
#2}
\def\pst@getnode#1#2{\pst@expandafter\pst@@getnode{#1},,\@nil#2}
\def\pst@@getnode#1,#2,#3\@nil#4{%
  \ifx\@empty#3\@empty
    \edef#4{/N@\pst@zapspace#1 \@empty\space}%
  \else
    \pst@cntg=#1\relax
    \pst@cnth=#2\relax
    \edef#4{/N@M-\ifnum\psmatrixcnt=\z@ 1\else\the\psmatrixcnt\fi
    -\the\pst@cntg-\the\pst@cnth\space}%
  \fi}
\def\tx@NewNode{/NodeScale {\ifx\pstnodescale\@undefined  \else\pstnodescale \fi} def NewNode }
\def\psopenNodeFile{%
  \pst@Verb{ 
    (\jobname.nodes) (w) file /NodeFile exch def 
  }}
\def\pscloseNodeFile{\pstVerb{ tx@NodeDict begin NodeFile closefile end }}
\define@boolkey[psset]{pst-node}[Pst@]{showNode}[true]{\ifPst@showNode\psopenNodeFile\fi}
\define@boolkey[psset]{pst-node}[Pst@]{markNode}[true]{}
\define@boolkey[psset]{pst-node}[Pst@]{saveNodeCoors}[true]{}
\define@key[psset]{pst-node}{NodeCoorPrefix}[]{\def\psk@NodeCoorPrefix{#1}}
\psset[pst-node]{saveNodeCoors=false,showNode=false,markNode=false,NodeCoorPrefix=}
\def\pst@newnode#1#2#3#4{%
\pst@killglue
\leavevmode
\pst@getnode{#1}\pst@thenode
\pst@Verb{
  \ifPst@saveNodeCoors
    \ifx\relax#3\relax 0 0 \else gsave \pst@dict STV CP T end #3 \tx@UserCoor grestore \fi 
    \if$\psk@NodeCoorPrefix$
      /N-#1.y exch def
      /N-#1.x exch def
    \else
      /\psk@NodeCoorPrefix#1y exch def
      /\psk@NodeCoorPrefix#1x exch def
    \fi
  \fi
  \pst@nodedict
  {#3}
  \ifx\psk@name\relax false \else \psk@name true \fi
  \pst@thenode
  #2
  {#4}
  \ifPst@showNode 
  exch dup /NodeType ED 
  exch
   NodeType 10 eq {  
    5 copy 
    cvlit aload pop
    20 string cvs (; )   6 2 roll 
    20 string cvs (; )   7 2 roll 
    20 string cvs (; )   8 2 roll 
    20 string cvs (; )   9 2 roll 
    cvlit dup length 2 eq 
      { aload pop exch 
        20 string cvs (; ) 11 2 roll 
        20 string cvs (, ) 12 2 roll  
        (\string\n)                   
        13 array astore concatstringarray 
      }
      { 255 string cvs (; ) 10 2 roll 
        (\string\n)                   
        11 array astore concatstringarray 
      } ifelse 
    NodeFile exch writestring 
  } if
  NodeType 14 eq {  
    5 copy 
    /@@temp ED 
    @@temp  
    4 -1 roll cvlit pop
    ( OvalNodePos ) (; )  5 2 roll
    20 string cvs (; )   6 2 roll 
    20 string cvs (; )   7 2 roll 
    20 string cvs (; )   8 2 roll 
    Y 20 string cvs (; ) 10 2 roll
    X 20 string cvs (, ) 12 2 roll
    (\string\n)                   
    13 array astore concatstringarray 
    tx@NodeDict begin NodeFile exch writestring end
  } if
  \fi
  \tx@NewNode
  end 
}%
\global\let\psk@name\relax%
\pstree@nodehook%
\global\let\pstree@nodehook\relax}
\let\pstree@nodehook\relax
\define@boolkey[psset]{pst-node}[Pst@]{nodealign}[true]{}
\psset[pst-node]{nodealign=false}

\def\pst@nodealign{%
\pst@dimg=\ht\pst@hbox
\advance\pst@dimg by -\dp\pst@hbox
\divide\pst@dimg by \tw@
\lower\pst@dimg}
\def\tx@InitPnode{InitPnode }
\def\pnode{\@ifnextchar[{\pnode@i}{\pnode@iii}}
\def\pnode@i[#1]{\@ifnextchar({\pnode@ii[#1]}{\pnode@ii[#1](0,0)}}
\def\pnode@ii[#1](#2)#3{%
  \pst@getcoor{#1}\pst@tempA%
  \pst@getcoor{#2}\pst@tempB%
  \pst@newnode{#3}{10}{\pst@tempA \pst@tempB 3 -1 roll add 3 1 roll add exch }{\tx@InitPnode}%
  \ifPst@showNode\psdot(#3)\uput[\ifx\psk@rot\@empty0\else\psk@rot\fi]{0}(#3){#3}\fi
  \ignorespaces}
\def\pnode@iii{\@ifnextchar({\pnode@}{\pnode@(0,0)}}
\def\pnode@(#1)#2{%
  \pst@@getcoor{#1}%
  \pst@newnode{#2}{10}{\pst@coor}{\tx@InitPnode}%
  \ifPst@showNode\psdot(#2)\uput[\ifx\psk@rot\@empty0\else\psk@rot\fi]{0}(#2){#2}\fi
  \ignorespaces}
\def\pnodes{\@ifnextchar[{\pnodes@i}{\pnodes@i[0,0]}}
\def\pnodes@i[#1]{\@ifnextchar({\psnodes@ii[#1]}{\pnodes@ii}}
\def\psnodes@ii[#1](#2)#3{%
  \pnode[#1](#2){#3}%
  \@ifnextchar({\psnodes@ii[#1]}{}%
}
\def\tx@InitCnode{InitCnode }
\def\cnode{\pst@object{cnode}}
\def\cnode@i{\@ifnextchar({\cnode@ii}{\cnode@ii(0,0)}}
\def\cnode@ii(#1)#2#3{%
  \leavevmode
  \hbox{%
    \use@par
    \pst@@getcoor{#1}%
    \pssetlength\pst@dimc{#2}%
    \pst@dimg=\psk@dimen\pslinewidth
    \advance\pst@dimc-\pst@dimg
    \advance\pst@dimc.5\pslinewidth
    \ifPst@nodealign
      \kern\pst@dimc
      \vrule width\z@ height \pst@dimc depth \pst@dimc
    \fi
    \pscircle@do(#1){#2}%
    \pst@newnode{#3}{11}{\pst@coor \pst@number\pst@dimc}{\tx@InitCnode}%
    \ifPst@nodealign\kern\pst@dimc\fi%
  }%
  \ignorespaces}
\def\Cnode{\pst@object{Cnode}}
\def\Cnode@i{\@ifnextchar({\Cnode@ii}{\Cnode@ii(0,0)}}
\def\Cnode@ii(#1)#2{\cnode@ii(#1){\psk@radius}{#2}}%
\def\cnodeput{\pst@object{cnodeput}}
\def\cnodeput@i{\@ifnextchar({\cnodeput@iii}{\cnodeput@ii}}
\def\cnodeput@ii#1{%
  \addto@par{rot={#1}}%
  \@ifnextchar({\cnodeput@iii}{\cnodeput@iii(\z@,\z@)}%
}
\def\cnodeput@iii(#1)#2{%
  \pst@killglue
  \@fixedradiusfalse
  \def\pst@nodehook{\cnodeput@iv{#2}}%
  \pst@makebox{\cput@v{#1}}%
}
\def\cnodeput@iv#1{%
  \pst@newnode{#1}{11}{\pscirclebox@iv \pst@number\pslinewidth add}{\tx@InitCnode}%
  \global\let\pst@nodehook\relax
  \ignorespaces
}
\def\Cnodeput{\pst@object{Cnodeput}}
\def\Cnodeput@i{\@ifnextchar({\Cnodeput@iii}{\Cnodeput@ii}}
\def\Cnodeput@ii#1{%
  \addto@par{rot={#1}}%
  \@ifnextchar({\Cnodeput@iii}{\Cnodeput@iii(\z@,\z@)}}
\def\Cnodeput@iii(#1)#2{%
  \pst@killglue
  \@fixedradiustrue
  \def\pst@nodehook{\Cnodeput@iv{#2}}%
  \pst@makebox{\cput@v{#1}}%
}
\def\Cnodeput@iv#1{%
  \pst@newnode{#1}{11}{%
    \pst@number{\wd\pst@hbox} 2 div \pst@number\pst@dima 
    \pst@number\pst@dimb \pst@number\pslinewidth \psk@dimen .5 sub mul sub }
       {\tx@InitCnode}%
  \global\let\pst@nodehook\relax}
\def\circlenode{\pst@object{circlenode}}
\def\circlenode@i#1{\pst@makebox{\circlenode@ii{#1}}}
\def\circlenode@ii#1{%
  \begingroup
  \pst@useboxpar
  \setbox\pst@hbox=\hbox{%
    \cnodeput@iv{#1}%
    \pscirclebox@iii
    \box\pst@hbox}%
  \ifPst@nodealign \psboxseptrue \fi
  \ifpsboxsep \pscirclebox@sep \fi
  \leavevmode
  \ifPst@nodealign\pst@nodealign\fi
  \box\pst@hbox
  \endgroup}
\def\Circlenode{\pst@object{Circlenode}}
\def\Circlenode@i#1{\pst@makebox{\Circlenode@ii{#1}}}
\def\Circlenode@ii#1{%
\begingroup
  \pst@useboxpar
  \pst@dima=\ht\pst@hbox
  \advance\pst@dima by -\dp\pst@hbox
  \divide\pst@dima by \tw@
  \pssetlength\pst@dimb\psk@radius
  \setbox\pst@hbox=\hbox{%
  \Cnodeput@iv{#1}%
  \pscircle(.5\wd\pst@hbox,\pst@dima){\pst@dimb}%
  \box\pst@hbox}%
  \ifPst@nodealign \psboxseptrue \fi
  \ifpsboxsep \psCirclebox@sep \fi
  \leavevmode
  \ifPst@nodealign\pst@nodealign\fi
  \box\pst@hbox
  \endgroup}
\def\tx@GetRnodePos{GetRnodePos }
\def\tx@InitRnode{InitRnode }
\def\psnode{\pst@object{psnode}}
\def\psnode@i{\@ifnextchar(\psnode@ii{\psnode@ii(0,0)}}
\def\psnode@ii(#1)#2#3{
  \rput(#1){\rnode{#2}{#3}}}
\def\rnode{\@ifnextchar[{\rnode@i}{\def\pst@par{}\rnode@ii}}
\def\rnode@i[#1]{\def\pst@par{ref=#1}\rnode@ii}
\def\rnode@ii#1{\pst@makebox{\rnode@iii\rnode@iv{#1}}}
\def\rnode@iii#1#2{%
\leavevmode
\begingroup
\pst@useboxpar
#1%
\ifPst@nodealign\lower\pst@dimb\fi
\hbox{%
\pst@newnode{#2}{16}{%
\pst@number{\ht\pst@hbox}%
\pst@number{\dp\pst@hbox}%
\pst@number{\wd\pst@hbox}%
\pst@number\pst@dima%
\pst@number\pst@dimb}%
{\tx@InitRnode}%
\box\pst@hbox}%
\endgroup}
\def\rnode@iv{%
\pst@dima=\psk@xref\wd\pst@hbox
\ifx\psk@yref\relax
\pst@dimb=\z@
\else
\pst@dimb=\ht\pst@hbox
\advance\pst@dimb\dp\pst@hbox
\pst@dimb=\psk@yref\pst@dimb
\advance\pst@dimb-\dp\pst@hbox
\fi}
\define@key[psset]{pst-node}{href}[0]{\pst@checknum{#1}\psk@href}
\psset[pst-node]{href=0}
\define@key[psset]{pst-node}{vref}[0.7ex]{\def\psk@vref{#1}}
\psset[pst-node]{vref=0.7ex}
\def\Rnode{\pst@object{Rnode}}
\def\Rnode@i#1{\pst@makebox{\rnode@iii\Rnode@ii{#1}}}
\def\Rnode@ii{%
\use@par
\pst@dima=\psk@href\wd\pst@hbox
\advance\pst@dima\wd\pst@hbox
\divide\pst@dima 2
\pssetlength\pst@dimb{\psk@vref}}
\def\tx@DiaNodePos{DiaNodePos }
\def\dianode{\pst@object{dianode}}
\def\dianode@i#1{\pst@makebox{\dianode@ii{#1}}}
\def\dianode@ii#1{%
\begingroup
\pst@useboxpar
\psdiabox@iii
\setbox\pst@hbox=\hbox{%
\pst@newnode{#1}{14}{}{%
/X \pst@number\pst@dima def
/Y \pst@number\pst@dimb def
/w \pst@number\pst@dimc 2 mul def
/h \pst@number\pst@dimd 2 mul def
/NodePos { \tx@DiaNodePos } def}%
\box\pst@hbox}%
\ifPst@nodealign\psboxseptrue\fi
\ifpsboxsep\psdiabox@sep\fi
\leavevmode
\ifPst@nodealign\lower\pst@dimb\fi
\box\pst@hbox
\endgroup}
\def\tx@TriNodePos{TriNodePos }
\def\tx@InitTriNode{InitTriNode }
\def\trinode{\pst@object{trinode}}
\def\trinode@i#1{\pst@makebox{\trinode@ii{#1}}}
\def\trinode@ii#1{%
  \begingroup%
  \pst@useboxpar%
  \pstribox@iii
  \setbox\pst@hbox=\hbox{%
    \pst@newnode{#1}{14}{}{
      \pst@number\pst@dimc
      \pst@number\pst@dimd
      \ifodd\psk@trimode
        exch
        \pst@number\pst@dima
      \else
        \pst@number\pst@dimb
      \fi
      \psk@trimode
      \pst@number{\wd\pst@hbox}
      \pst@number{\ht\pst@hbox}
      \pst@number{\dp\pst@hbox}
      \tx@InitTriNode
    }%
    \box\pst@hbox%
  }%
  \ifPst@nodealign\psboxseptrue\fi
  \ifpsboxsep\pstribox@sep\fi
  \leavevmode
  \ifPst@nodealign\lower\pst@tempa\fi
  \box\pst@hbox%
  \endgroup}
\def\tx@OvalNodePos{OvalNodePos }
\def\ovalnode{\pst@object{ovalnode}}
\def\ovalnode@i#1{\pst@makebox{\ovalnode@ii{#1}}}
\def\ovalnode@ii#1{%
\begingroup
\pst@useboxpar
\psovalbox@iii
\setbox\pst@hbox=\hbox{%
\pst@newnode{#1}{14}{}{%
/X \pst@number\pst@dima def
/Y \pst@number\pst@dimb def
/w \pst@number\pst@dimc def
/h \pst@number\pst@dimd def
/NodePos { \tx@OvalNodePos } def}%
\unhbox\pst@hbox}%
\ifPst@nodealign\psboxseptrue\fi
\ifpsboxsep\psovalbox@sep\fi
\leavevmode
\ifPst@nodealign\lower\pst@dimb\fi
\box\pst@hbox
\endgroup}
\def\dotnode{\pst@object{dotnode}}
\def\dotnode@i{\@ifnextchar({\dotnode@ii}{\dotnode@ii(\z@,\z@)}}
\def\dotnode@ii(#1)#2{%
  \leavevmode
  \hbox{%
    \use@par
    \pst@@getcoor{#1}%
    \pst@getdotsize
    \pstree@nodehook
    \ifPst@nodealign
      \pst@dima=\pst@dimg
      \kern\pst@dima
      \vrule width\z@ height \pst@dimh depth \pst@dimh
    \fi
    \pst@newnode{#2}{14}{}{
      \pst@coor
      /Y exch def /X exch def
      /w \pst@number\pst@dimg def
      /h \pst@number\pst@dimh def
      /NodePos { \tx@OvalNodePos } def}%
    \psdot@ii(#1)%
    \ifPst@nodealign\kern\pst@dima\fi}%
  \ifPst@markNode\uput[\ifx\psk@rot\@empty0\else\psk@rot\fi]{0}(#2){#2}\fi
  \ignorespaces}
\def\dotnodes{\pst@object{dotnodes}}
\def\dotnodes@i{\use@par\dotnodes@ii}
\def\dotnodes@ii(#1)#2{%
  \dotnode(#1){#2}%
  \@ifnextchar(\dotnodes@ii{\def\pst@par{}}}
\define@key[psset]{pst-node}{framesize}{\pst@expandafter\psset@@framesize{#1} \@nil}
\def\psset@@framesize#1 #2\@nil{%
  \pssetlength\pst@dimg{#1}%
  \divide\pst@dimg2
  \edef\psk@framewidth{\pst@number\pst@dimg}%
  \ifx\@empty#2\@empty
    \let\psk@frameheight\psk@framewidth
  \else
    \pssetlength\pst@dimg{#2}%
    \divide\pst@dimg2
    \edef\psk@frameheight{\pst@number\pst@dimg}%
  \fi}
\psset[pst-node]{framesize=10pt}
\def\fnode{\pst@object{fnode}}
\def\fnode@i{\@ifnextchar({\fnode@ii}{\fnode@ii(\z@,\z@)}}
\def\fnode@ii(#1)#2{%
  \leavevmode
  \pst@killglue
  \hbox{%
    \use@par%
    \begin@ClosedObj%
    \ifPst@nodealign
      \kern\psk@framewidth\p@
      \vrule width\z@ height \psk@frameheight\p@ depth \psk@frameheight\p@
      \edef\pst@coor{0 0 }%
    \else\pst@@getcoor{#1}\fi
    \pst@newnode{#2}{14}{}{
      \pst@coor
      /Y exch def /X exch def
      /d \psk@dimen .5 sub CLW mul neg def
      /r \psk@framewidth d add def
      /l r neg def
      /u \psk@frameheight d add def
      /d u neg def
      /NodePos { \tx@GetRnodePos } def}%
    \addto@pscode{
      /x2 \psk@framewidth CLW \psk@dimen mul sub def
      /y2 \psk@frameheight CLW \psk@dimen mul sub def
      \pst@coor 2 copy
      y2 sub /y1 ED
      x2 sub /x1 exch def
      y2 add /y2 exch def
      x2 add /x2 exch def
      \psk@cornersize
      1 index 0 eq { pop pop \tx@Rect } { \tx@OvalFrame } ifelse}%
    \def\pst@linetype{2}%
    \showpointsfalse%
    \end@ClosedObj%
    \ifPst@nodealign\kern\psk@framewidth\p@\fi}
  \ignorespaces}
\define@key[psset]{}{XnodesepA}{\pst@getlength{#1}\psk@nodesepA\def\psk@nodeseptypeA{2 }}
\define@key[psset]{}{XnodesepB}{\pst@getlength{#1}\psk@nodesepB\def\psk@nodeseptypeB{2 }}
\define@key[psset]{}{Xnodesep}{%
    \pst@getlength{#1}\psk@nodesepA
    \let\psk@nodesepB\psk@nodesepA
    \def\psk@nodeseptypeA{2 }%
    \def\psk@nodeseptypeB{2 }}
\define@key[psset]{}{YnodesepA}{\pst@getlength{#1}\psk@nodesepA\def\psk@nodeseptypeA{1 }}
\define@key[psset]{}{YnodesepB}{\pst@getlength{#1}\psk@nodesepB\def\psk@nodeseptypeB{1 }}
\define@key[psset]{}{Ynodesep}{%
    \pst@getlength{#1}\psk@nodesepA
    \let\psk@nodesepB\psk@nodesepA
    \def\psk@nodeseptypeA{1 }%
    \def\psk@nodeseptypeB{1 }}
\define@key[psset]{pst-node}{nodesepA}[0pt]{\pst@getlength{#1}\psk@nodesepA \def\psk@nodeseptypeA{0 }}
\define@key[psset]{pst-node}{nodesepB}[0pt]{\pst@getlength{#1}\psk@nodesepB \def\psk@nodeseptypeB{0 }}
\define@key[psset]{pst-node}{nodesep}[0pt]{%
  \pst@getlength{#1}\psk@nodesepA
  \let\psk@nodesepB\psk@nodesepA
  \def\psk@nodeseptypeA{0 }%
  \def\psk@nodeseptypeB{0 }}
\psset[pst-node]{nodesep=0pt}
\define@key[psset]{pst-node}{armA}[10pt]{\pst@getlength{#1}\psk@armA \def\psk@armtypeA{0 }}
\define@key[psset]{pst-node}{armB}[10pt]{\pst@getlength{#1}\psk@armB \def\psk@armtypeB{0 }}
\define@key[psset]{pst-node}{arm}[10pt]{%
  \pst@getlength{#1}\psk@armA
  \let\psk@armB\psk@armA
  \def\psk@armtypeA{0 }%
  \def\psk@armtypeB{0 }}
\psset[pst-node]{arm=10pt}
\define@key[psset]{pst-node}{XarmA}[]{\pst@getlength{#1}\psk@armA \def\psk@armtypeA{1 }}
\define@key[psset]{pst-node}{XarmB}[]{\pst@getlength{#1}\psk@armB \def\psk@armtypeB{1 }}
\define@key[psset]{pst-node}{Xarm}{%
  \pst@getlength{#1}\psk@armA
  \let\psk@armB\psk@armA
  \def\psk@armtypeA{1 }%
  \def\psk@armtypeB{1 }}
\define@key[psset]{pst-node}{YarmA}[]{\pst@getlength{#1}\psk@armA \def\psk@armtypeA{2 }}
\define@key[psset]{pst-node}{YarmB}[]{\pst@getlength{#1}\psk@armB \def\psk@armtypeB{2 }}
\define@key[psset]{pst-node}{Yarm}[]{%
  \pst@getlength{#1}\psk@armA
  \let\psk@armB\psk@armA
  \def\psk@armtypeA{2 }%
  \def\psk@armtypeB{2 }}
\define@key[psset]{pst-node}{offsetA}[0pt]{\pst@getlength{#1}\psk@offsetA}
\define@key[psset]{pst-node}{offsetB}[0pt]{\pst@getlength{#1}\psk@offsetB}
\define@key[psset]{pst-node}{offset}[0pt]{\pst@getlength{#1}\psk@offsetA\let\psk@offsetB\psk@offsetA}
\psset[pst-node]{offset=0pt}
\define@key[psset]{pst-node}{angleA}[0]{\pst@getangle{#1}\psk@angleA}
\define@key[psset]{pst-node}{angleB}[0]{\pst@getangle{#1}\psk@angleB}%
\define@key[psset]{pst-node}{angle}[0]{%
  \pst@getangle{#1}\psk@angleA
  \let\psk@angleB\psk@angleA}
\psset[pst-node]{angle=0}
\define@key[psset]{pst-node}{arcangleA}[8]{\pst@getangle{#1}\psk@arcangleA}
\define@key[psset]{pst-node}{arcangleB}[8]{\pst@getangle{#1}\psk@arcangleB}%
\define@key[psset]{pst-node}{arcangle}[8]{%
  \pst@getangle{#1}\psk@arcangleA
  \let\psk@arcangleB\psk@arcangleA}
\psset[pst-node]{arcangle=8}
\define@key[psset]{pst-node}{ncurvA}[0.67]{\pst@checknum{#1}\psk@ncurvA}
\define@key[psset]{pst-node}{ncurvB}[0.67]{\pst@checknum{#1}\psk@ncurvB}%
\define@key[psset]{pst-node}{ncurv}[0.67]{\pst@checknum{#1}\psk@ncurvA\let\psk@ncurvB\psk@ncurvA}
\psset[pst-node]{ncurv=0.67}
\def\tx@GetCenter{GetCenter }
\def\tx@XYPos{XYPos }
\def\tx@GetEdge{GetEdge }
\def\tx@AddOffset{AddOffset }
\def\tx@GetEdgeA{GetEdgeA }
\def\tx@GetEdgeB{GetEdgeB }
\def\tx@GetArmA{GetArmA }
\def\tx@GetArmB{GetArmB }
\def\check@arrow#1#2{%
  \check@@arrow#2-\@nil
  \if@pst\addto@par{arrows=#2}\def\next{#1}%
  \else\def\next{#1{#2}}\fi
  \next}
\def\check@@arrow#1-#2\@nil{%
\ifx\@nil#2\@nil\@pstfalse\else\@psttrue\fi}
\def\tx@InitNC{InitNC }
\def\nc@object#1#2#3#4#5{%
  \csname begin@#1Obj\endcsname
  \showpointsfalse
  \pst@getnode{#2}\pst@tempa
  \pst@getnode{#3}\pst@tempb
  \gdef\npos@default{#4 }%
  \addto@pscode{%
    /NCLW CLW def
    \pst@nodedict
    \psk@offsetA
    \psk@offsetB neg
    \psk@nodesepA
    \psk@nodesepB
    \psk@nodeseptypeA
    \psk@nodeseptypeB
    \pst@tempa
    \pst@tempb
    \tx@InitNC { #5 } if
    end }%
  \def\use@pscode{%
    \pst@Verb{gsave \tx@STV newpath \pst@code\space grestore}%
    \gdef\pst@code{}}%
  \csname end@#1Obj\endcsname
  \pst@shortput}
\def\npos@default{.5 }
\def\pc@object#1{%
  \@ifnextchar({\pc@@object#1}{\pst@getarrows{\pc@@object#1}}}
\def\pc@@object#1(#2)(#3){%
  \pnode(#2){@@A}\pnode(#3){@@B}%
  #1{@@A}{@@B}}
\def\tx@LPutLine{LPutLine }
\def\tx@LPutLines{LPutLines }
\def\tx@BezierMidpoint{BezierMidpoint }
\def\tx@HPosBegin{HPosBegin }
\def\tx@HPosEnd{HPosEnd }
\def\tx@HPutLine{HPutLine }
\def\tx@HPutLines{HPutLines }
\def\tx@VPosBegin{VPosBegin }
\def\tx@VPosEnd{VPosEnd }
\def\tx@VPutLine{VPutLine }
\def\tx@VPutLines{VPutLines }
\def\tx@HPutCurve{HPutCurve }
\def\tx@NCCoor{NCCoor }
\def\tx@NCLine{NCLine }
\def\ncline{\pst@object{ncline}}
\def\ncline@i{\check@arrow{\ncline@ii}}
\def\ncline@ii#1#2{\nc@object{Open}{#1}{#2}{.5}{\tx@NCLine}}
\def\pcline{\pst@object{pcline}}
\def\pcline@i{\pc@object\ncline@ii}
\def\ncLine{\pst@object{ncLine}}
\def\ncLine@i{\check@arrow{\ncLine@ii}}
\def\ncLine@ii#1#2{\nc@object{Open}{#1}{#2}{.5}%
{\tx@NCLine /LPutPos { xB yB xA yA \tx@LPutLine } def}}
\def\tx@NCLines{NCLines }
\def\nclines{\pst@object{nclines}}
\def\nclines@i{\check@arrow\nclines@ii}
\def\nclines@ii#1#2{%
\begingroup
\use@par
\def\pst@aftercoors{\nclines@iii{#1}{#2}}%
\def\pst@coors{}%
\pst@@getcoors}
\def\nclines@iii#1#2{%
\nc@object{Open}{#1}{#2}{.5}{%
tx@Dict begin \psline@iii pop end
mark \pst@coors \tx@NCLines}%
\endgroup
\ignorespaces}
\def\tx@NCCurve{NCCurve }
\def\nccurve{\pst@object{nccurve}}
\def\nccurve@i{\check@arrow{\nccurve@ii}}
\def\nccurve@ii#1#2{\nc@object{Open}{#1}{#2}{.5}{%
  /AngleA \psk@angleA\space def /AngleB \psk@angleB\space def
  \psk@ncurvB\space \psk@ncurvA\space
  \tx@NCCurve}}
\def\pccurve{\pst@object{pccurve}}
\def\pccurve@i{\pc@object\nccurve@ii}
\def\ncarc{\pst@object{ncarc}}
\def\ncarc@i{\check@arrow{\ncarc@ii}}
\def\ncarc@ii#1#2{\nc@object{Open}{#1}{#2}{.5}{%
  yB yA sub xB xA sub \tx@Atan dup
  \psk@arcangleA\space add /AngleA exch def
  \psk@arcangleB\space sub 180 add /AngleB exch def
  \psk@ncurvB\space \psk@ncurvA\space
  \tx@NCCurve}}
\def\pcarc{\pst@object{pcarc}}
\def\pcarc@i{\pc@object\ncarc@ii}
\def\tx@NCAngles{NCAngles }
\define@boolkey[psset]{pst-node}[Pst@]{pcRef}[true]{}
\psset[pst-node]{pcRef=false}
\def\ncangles{\pst@object{ncangles}}
\def\ncangles@i{\check@arrow{\ncangles@ii}}
\def\ncangles@ii#1#2{%
  \nc@object{Open}{#1}{#2}{1.5}{\ncangles@iii \tx@NCAngles}}
\def\ncangles@iii{
  tx@Dict begin \psline@iii pop end
  /AngleA \psk@angleA def
  /AngleB \psk@angleB def
  /ArmA \psk@armA \ifPst@pcRef 
    GetEdgeA yA yA1 sub dup mul xA xA1 sub dup mul add sqrt sub \fi def
  /ArmB \psk@armB def
  /ArmTypeA \psk@armtypeA def
  /ArmTypeB \psk@armtypeB def }
\def\pcangles{\pst@object{pcangles}}
\def\pcangles@i{\pc@object\ncangles@ii}
\def\tx@NCAngle{NCAngle }
\def\ncangle{\pst@object{ncangle}}
\def\ncangle@i{\check@arrow{\ncangle@ii}}
\def\ncangle@ii#1#2{%
\nc@object{Open}{#1}{#2}{1.5}{\ncangles@iii \tx@NCAngle}}
\def\pcangle{\pst@object{pcangle}}
\def\pcangle@i{\pc@object\ncangle@ii}
\def\tx@NCBar{NCBar }
\def\ncbar{\pst@object{ncbar}}
\def\ncbar@i{\check@arrow{\ncbar@ii}}
\def\ncbar@ii#1#2{\nc@object{Open}{#1}{#2}{1.5}{%
\ncangles@iii /AngleB \psk@angleA def \tx@NCBar}}
\def\pcbar{\pst@object{pcbar}}
\def\pcbar@i{\pc@object\ncbar@ii}
\define@key[psset]{pst-node}{lineAngle}[0]{%
  \ifdim#1pt=\z@\else\psset{armB=0.5}\fi
  \def\psk@lineAngle{#1}}%
\psset[pst-node]{lineAngle=0}%
\def\tx@NCDiag{NCDiag }
\def\ncdiag{\pst@object{ncdiag}}
\def\ncdiag@i{\check@arrow{\ncdiag@ii}}
\def\ncdiag@ii#1#2{%
  \nc@object{Open}{#1}{#2}{1.5}{\ncangles@iii \psk@lineAngle\space \tx@NCDiag}}
\def\pcdiag{\pst@object{pcdiag}}
\def\pcdiag@i{\pc@object\ncdiag@ii}
\def\tx@NCDiagg{NCDiagg }
\def\ncdiagg{\pst@object{ncdiagg}}
\def\ncdiagg@i{\check@arrow{\ncdiagg@ii}}
\def\ncdiagg@ii#1#2{%
  \nc@object{Open}{#1}{#2}{.5}{\ncangles@iii \psk@lineAngle\space \tx@NCDiagg}}
\def\pcdiagg{\pst@object{pcdiagg}}
\def\pcdiagg@i{\pc@object\ncdiagg@ii}
\def\tx@NCLoop{NCLoop }
\define@key[psset]{pst-node}{loopsize}{\pst@getlength{#1}\psk@loopsize}
\psset[pst-node]{loopsize=1cm}
\def\ncloop{\pst@object{ncloop}}
\def\ncloop@i{\check@arrow{\ncloop@ii}}
\def\ncloop@ii#1#2{%
\nc@object{Open}{#1}{#2}{2.5}%
{\ncangles@iii /loopsize \psk@loopsize def \tx@NCLoop}}
\def\pcloop{\pst@object{pcloop}}
\def\pcloop@i{\pc@object\ncloop@ii}
\def\tx@NCCircle{NCCircle }
\def\nccircle{\pst@object{nccircle}}
\def\nccircle@i{\check@arrow{\nccircle@ii}}
\def\nccircle@ii#1#2{%
\pssetlength\pst@dima{#2}%
\nc@object{Open}{#1}{#1}{.5}{%
/AngleA \psk@angleA def
/r \pst@number\pst@dima def
\tx@NCCircle \psarc@v end}}
\def\tx@NCBox{NCBox }
\def\ncbox{\pst@object{ncbox}}
\def\ncbox@i{\check@arrow{\ncbox@ii}}
\def\ncbox@ii#1#2{%
\def\pst@linetype{2}%
\nc@object{Closed}{#1}{#2}{.5}{%
tx@Dict begin \psline@iii pop end
\psk@boxheight \psk@boxdepth
\tx@NCBox}}
\def\pcbox{\pst@object{pcbox}}
\def\pcbox@i{\pc@object\ncbox@ii}
\def\tx@NCArcBox{NCArcBox }
\define@key[psset]{pst-node}{boxheight}[0.4cm]{\pst@getlength{#1}\psk@boxheight}
\define@key[psset]{pst-node}{boxdepth}[0.4cm]{\pst@getlength{#1}\psk@boxdepth}
\define@key[psset]{pst-node}{boxsize}[0.4cm]{%
  \pst@getlength{#1}\psk@boxheight%
  \let\psk@boxdepth\psk@boxheight}
\psset[pst-node]{boxsize=0.4cm}
\def\ncarcbox{\pst@object{ncarcbox}}
\def\ncarcbox@i{\check@arrow{\ncarcbox@ii}}
\def\ncarcbox@ii#1#2{%
\def\pst@linetype{1}%
\nc@object{Closed}{#1}{#2}{.5}{%
\psk@arcangleA \psk@boxheight \psk@boxdepth \pst@number\pslinearc
\tx@NCArcBox}}
\def\pcarcbox{\pst@object{pcarcbox}}
\def\pcarcbox@i{\pc@object\ncarcbox@ii}
\def\tx@Tfan{Tfan }
\begingroup
\catcode`\:=13
\gdef\pst@activerot{\def:{\string:}}
\endgroup
\define@key[psset]{pst-node}{nrot}[0]{%
  \begingroup
  \pst@activerot
  \pst@expandafter{\@ifnextchar:{\psset@@nrot}{\psset@@rot}}{#1}\@nil
  \global\let\pst@tempg\psk@rot
  \endgroup
  \let\psk@nrot\pst@tempg}
\def\psset@@nrot:#1\@nil{%
  \psset@@rot#1\@nil
  \edef\psk@rot{NAngle \ifx\psk@rot\@empty\else\psk@rot add \fi}}
\psset[pst-node]{nrot=0}
\def\tx@LPutCoor{LPutCoor }
\def\tx@LPut{LPut }
\define@key[psset]{pst-node}{npos}[{}]{%
  \def\pst@tempa{#1}%
  \ifx\pst@tempa\@empty\def\psk@npos{\npos@default}\else\pst@checknum{#1}\psk@npos\fi}
\psset[pst-node]{npos=}
\def\ncput{\pst@object{ncput}}
\def\ncput@i{\pst@killglue\pst@makebox{\ncput@ii}}
\def\ncput@ii{%
  \begingroup%
  \use@par%
  \if@star\pst@starbox\fi%
  \pst@makesmall\pst@hbox%
  \pst@rotate\psk@nrot\pst@hbox%
  \ncput@iii%
  \endgroup%
  \pst@shortput}
\def\ncput@iii{%
  \leavevmode%
  \hbox{%
    \pst@Verb{
      \pst@nodedict
      /t \psk@npos def
      \tx@LPut
      end
      \tx@PutBegin}%
    \box\pst@hbox%
    \pst@Verb{\tx@PutEnd}}}
\def\naput{\pst@object{naput}}
\def\naput@i{\pst@killglue\pst@makebox{\naput@ii{NAngle 90 add}}}
\def\naput@ii#1{%
  \begingroup
  \use@par
  \if@star\pst@starbox\fi
  \def\psk@refangle{#1 }%
  \let\psk@rot\psk@nrot
  \pst@Verb{ 
    gsave  STV CP T /ps@refangle {#1 } def 
    /ps@rot { \psk@rot } def grestore }
  \uput@vii
  {exch pop add a \tx@PtoC h1 add exch w1 add exch }%
  {tx@Dict /NCLW known { NCLW add } if }%
  \ncput@iii
  \endgroup
  \pst@shortput}
\def\nbput{\pst@object{nbput}}
\def\nbput@i{\pst@killglue\pst@makebox{\naput@ii{NAngle 90 sub}}}
\define@key[psset]{pst-node}{tpos}[0.5]{%
  \pst@checknum{#1}\psk@tpos
  \ifdim\psk@tpos \p@<\z@
    \def\psk@tpos{.5}%
    \@pstrickserr{Bad `tpos' value: `#1'. Must be 0<tpos<1}\@ehpa
  \else
    \ifdim\psk@tpos \p@>\p@
      \def\psk@tpos{.5}%
      \@pstrickserr{Bad `tpos' value: `#1'. Must be 0<tpos<1}\@ehpa%
    \fi%
  \fi}
\psset[pst-node]{tpos=0.5}
\def\nlput{\pst@object{nlput}}
\def\nlput@i(#1)(#2)#3#4{%
  \begin@SpecialObj
  \psLDNode(#1)(#2){#3}{temp@lnput}
  \pcline[linestyle=none](#1)(temp@lnput)%
  \ncput[npos=1]{#4}%
  \end@SpecialObj}
\def\tvput{\pst@object{tvput}}
\def\tvput@i{\pst@makebox{\psput@tput{H}{1}}}
\def\tlput{\pst@object{tlput}}
\def\tlput@i{\pst@makebox{\psput@tput{H}{true}}}
\def\trput{\pst@object{trput}}
\def\trput@i{\pst@makebox{\psput@tput{H}{false}}}
\def\thput{\pst@object{thput}}
\def\thput@i{\pst@makebox{\psput@tput{V}{1}}}
\def\taput{\pst@object{taput}}
\def\taput@i{\pst@makebox{\psput@tput{V}{true}}}
\def\tbput{\pst@object{tbput}}
\def\tbput@i{\pst@makebox{\psput@tput{V}{false}}}
\def\tx@HPutAdjust{HPutAdjust }
\def\tx@VPutAdjust{VPutAdjust }
\def\psput@tput#1#2{%
  \begingroup
  \use@par
  \pst@tputmakesmall
  \leavevmode
  \hbox{%
    \pst@Verb{%
      \pst@nodedict
      /t \psk@tpos \pst@tposflip def
      tx@NodeDict /HPutPos known
        { #1PutPos }
        { CP /Y exch def /X exch def /NAngle 0 def /NCLW 0 def }
      ifelse
      /Sin NAngle sin def
      /Cos NAngle cos def
      /s \pst@number\pslabelsep NCLW add def
      /l \pst@number\pst@dima def
      /r \pst@number\pst@dimb def
      /h \pst@number\pst@dimc def
      /d \pst@number\pst@dimd def
      \ifnum1=0#2 \else
        /flag #2 def
        \csname tx@#1PutAdjust\endcsname
      \fi
      \tx@LPutCoor
      end
      \tx@PutBegin}%
    \box\pst@hbox
    \pst@Verb{\tx@PutEnd}}%
  \endgroup
  \pst@shortput}
\def\pst@tposflip{}
\def\pst@tputmakesmall{%
\pst@dima=\wd\pst@hbox
\divide\pst@dima 2
\pst@dimg=\psk@href\pst@dimg
\pst@dimb\pst@dima
\advance\pst@dima\pst@dimg 
\advance\pst@dimb-\pst@dimg 
\pst@dimd=\psk@vref\relax
\pst@dimc=\ht\pst@hbox
\advance\pst@dimc-\pst@dimd 
\advance\pst@dimd\dp\pst@hbox 
\setbox\pst@hbox=\hbox to\z@{%
\kern-\pst@dima\vbox to\z@{\vss\box\pst@hbox\vskip-\pst@dimd}\hss}}
\def\MakeShortNab#1#2{%
  \def\pst@shortput@nab{%
    \def\pst@tempg{\next}%
    \ifx#1\next
      \let\pst@tempg\naput
    \else
      \ifx#2\next
        \let\pst@tempg\nbput
      \else
        \ifx\@sptoken\next
          \let\pst@tempg\pst@shortput
        \fi
      \fi
    \fi
    \pst@tempg}}
\MakeShortNab{^}{_}
\def\MakeShortTablr#1#2#3#4{%
  \def\pst@shortput@tablr{%
    \def\pst@tempg{\next}%
    \ifx#1\next
      \let\pst@tempg\taput
    \else
      \ifx#2\next
        \let\pst@tempg\tbput
      \else
        \ifx#3\next
          \let\pst@tempg\tlput
        \else
          \ifx#4\next
            \let\pst@tempg\trput
          \else
            \ifx\@sptoken\next
              \let\pst@tempg\pst@shortput
            \fi
          \fi
        \fi
      \fi
    \fi
    \pst@tempg}}
\MakeShortTablr{^}{_}{<}{>}
\def\MakeShortTab#1#2{%
  \def\pst@shortput@tab{%
    \def\pst@tempg{\next}%
    \ifx#1\next
      \def\pst@tempg{%
        \@nameuse{%
          t\ifodd\psk@treemode\ifpstreeflip b\else a\fi
          \else\ifpstreeflip r\else l\fi\fi put}}%
    \else
      \ifx#2\next
        \def\pst@tempg{%
          \@nameuse{%
            t\ifodd\psk@treemode\ifpstreeflip a\else b\fi
            \else\ifpstreeflip l\else r\fi\fi put}}%
      \else
        \ifx\@sptoken\next
          \let\pst@tempg\pst@shortput
        \fi
      \fi
    \fi
    \pst@tempg}}
\MakeShortTab{^}{_}
\define@key[psset]{pst-node}{shortput}[none]{%
  \def\pst@tempg{#1}%
  \ifx\pst@tempg\@none
    \let\pst@shortput\ignorespaces
  \else
    \@ifundefined{pst@shortput@#1}%
     {\@pstrickserr{Bad short put: `#1'}\@ehpa}%
     {\edef\pst@shortput{\noexpand\afterassignment\expandafter\noexpand
      \csname pst@shortput@#1\endcsname\noexpand\let\noexpand\next}}%
  \fi}
\psset[pst-node]{shortput=none}
\def\lput{\def\pst@par{}\pst@ifstar{\@ifnextchar[{\lput@i}{\lput@ii}}}
\def\lput@i[#1]{\addto@par{ref=#1}\lput@ii}
\def\lput@ii{\@ifnextchar({\lput@iv}{\lput@iii}}
\def\lput@iii#1{\addto@par{nrot=#1}\@ifnextchar({\lput@iv}{\ncput@i}}
\def\lput@iv(#1){\addto@par{npos=#1}\ncput@i}
\def\mput{\def\pst@par{}\pst@ifstar{\@ifnextchar[{\mput@i}{\ncput@i}}}
\def\mput@i[#1]{\addto@par{ref=#1}\ncput@i}
\def\Lput{\def\pst@par{}\pst@ifstar{\@ifnextchar[{\Lput@ii}{\Lput@i}}}
\def\Lput@i#1{\addto@par{labelsep=#1}\Lput@ii}
\def\Lput@ii[#1]{\addto@par{ref={#1}}\@ifnextchar({\Lput@iv}{\Lput@iii}}
\def\Lput@iii#1{\addto@par{nrot={#1}}\@ifnextchar({\Lput@iv}{\Lput@v}}
\def\Lput@iv(#1){\addto@par{npos=#1}\Lput@v}
\def\Lput@v{\pst@killglue\pst@makebox{\Lput@vi}}
\def\Lput@vi{%
\begingroup
\use@par
\if@star\pst@starbox\fi
\Rput@vi
\pst@makesmall\pst@hbox
\ifx\psk@rot\@empty\else\pst@rotate{ps@rot }\pst@hbox\fi
\ncput@iii
\endgroup
\pst@shortput}
\def\Mput{\def\pst@par{}\pst@ifstar{\@ifnextchar[{\Mput@ii}{\Mput@i}}}
\def\Mput@i#1{\addto@par{labelsep=#1}\Mput@ii}
\def\Mput@ii[#1]{\addto@par{ref={#1}}\Lput@v}
\def\aput@#1{\def\pst@par{}\pst@ifstar{\@ifnextchar[{\aput@i#1}{\aput@ii#1}}}
\def\aput@i#1[#2]{\addto@par{labelsep=#2}\aput@ii#1}
\def\aput@ii#1{\@ifnextchar({\aput@iv#1}{\aput@iii#1}}
\def\aput@iii#1#2{\addto@par{nrot=#2}\@ifnextchar({\aput@iv#1}{#1}}
\def\aput@iv#1(#2){\addto@par{npos=#2}#1}
\def\aput{\aput@\naput@i}
\def\bput{\aput@\nbput@i}
\def\Aput{\def\pst@par{}\pst@ifstar{\@ifnextchar[{\Aput@i}{\naput@i}}}
\def\Aput@i[#1]{\addto@par{labelsep=#1}\naput@i}
\def\Bput{\def\pst@par{}\pst@ifstar{\@ifnextchar[{\Bput@i}{\nbput@i}}}
\def\Bput@i[#1]{\addto@par{labelsep=#1}\nbput@i}
\def\node@coor#1;#2\@nil{
  \pst@getnode{#1}\pst@tempg
  \edef\pst@coor{%
    \pst@nodedict
    tx@NodeDict \pst@tempg known
    \pslbrace \pst@tempg load \tx@GetCenter \psrbrace
    \pslbrace 0 0 \psrbrace ifelse
    end }}
\def\Node@coor[#1]#2;#3\@nil{
\begingroup
\psset{angle=0,#1}
\@ifnextchar\bgroup{\Node@@@coor}
                   {\Node@@coor}#2\@nil
\endgroup
\let\pst@coor\pst@tempg}
\def\Node@@coor#1\@nil{%
\pst@getnode{#1}\pst@tempg
\xdef\pst@tempg{%
\pst@nodedict
tx@NodeDict \pst@tempg known
  { \psk@nodesepA \psk@angleA 
    \pst@tempg load \psk@nodeseptypeA \tx@GetEdge
    \psk@offsetA \psk@angleA \tx@AddOffset
    \pst@tempg load \tx@GetCenter
    3 -1 roll add 3 1 roll add exch }
  { CP } ifelse end }}
\def\Node@@@coor#1{
\pst@@getcoor{#1}%
\def\psk@angleA{%
  \pst@tempg load \tx@GetCenter \pst@coor
  3 -1 roll sub 3 1 roll sub neg \tx@Atan \psk@angleB add
  }%
\Node@@coor}
\def\nput{\pst@object{nput}}
\def\nput@i#1#2{\pst@killglue\pst@makebox{\nput@ii{#1}{#2}}}
\def\nput@ii#1#2{%
  \begingroup
  \use@par
  \if@star\pst@starbox\fi%
  \psset[pstricks]{refangle=#1}%
  \let\psk@angleA\psk@refangle
  \edef\psk@nodesepA{\pst@number\pslabelsep}%
  \def\psk@nodeseptypeA{0 }%
  \pslabelsep\z@
  \uput@vi
  \Node@@coor#2\@nil
  \let\pst@coor\pst@tempg
  \leavevmode
  \psput@special\pst@hbox
  \endgroup
  \ignorespaces}
\newcount\psrow
\newcount\pscol
\newcount\psmatrixcnt
\newskip\psrowsep
\newskip\pscolsep
\define@key[psset]{pst-node}{colsep}[1.5cm]{\pssetlength\pscolsep{#1}}
\define@key[psset]{pst-node}{rowsep}[1.5cm]{\pssetlength\psrowsep{#1}}
\psset[pst-node]{colsep=1.5cm}
\psset[pst-node]{rowsep=1.5cm}
\newif\ifpsmatrix
\ifx\mscount\@undefined\let\mscount\@multicnt\fi
\def\psmatrix{\begingroup{\ifnum0=`}\fi 
  \@ifnextchar[{\psmatrix@i}{\ifnum0=`{\fi}{}\psmatrix@ii}}
\def\psmatrix@i[#1]{%
  \ifnum0=`{\fi}{}%
  \psset{#1}%
  \psmatrix@ii}
\def\psmatrix@ii{%
  \KillGlue
  \edef\psm@beginmath{%
    \ifmmode$\m@th\ifinner\textstyle\else\displaystyle\fi\fi}%
  \edef\psm@endmath{\ifmmode$\fi}%
  \let\\\psm@cr
  \advance\psmatrixcnt by \@ne
  \def\psm@thenode{M-\the\psmatrixcnt-\the\psrow-\the\pscol}%
  \tabskip\z@
  \psrow=\@ne
  \pscol\z@
  \psset{shortput=tablr}%
  \leavevmode
  \vbox\bgroup\halign\bgroup&%
  \begingroup
  \global\advance\pscol by \@ne
  \csname psrowhook\romannumeral\psrow\endcsname
  \csname pscolhook\romannumeral\pscol\endcsname
  \psm@beginnode##\psm@endnode\endgroup
  \cr}
\def\endpsmatrix{%
  \crcr\egroup\unskip\egroup
  \endgroup}
\def\psm@cr{{\ifnum0=`}\fi\ps@ifnextchar[{\psm@@cr}{\psm@@@cr{}}}
\def\psm@@cr[#1]{\psm@@@cr{\vskip#1\relax}}
\def\psm@@@cr#1{%
  \ifnum0=`{\fi}{}\cr
  \noalign{%
  \global\advance\psrow 1
  \global\pscol\z@
  \vskip\psrowsep
  #1}}
\def\psm@beginnode{%
  \@ifnextchar\psm@endnode
    {\let\psm@endnode@i\relax\setbox\pst@hbox=\hbox{}}%
    {\pst@object{psm@beginnode}}}
\def\psm@beginnode@i{%
  \setbox\pst@hbox=\hbox\bgroup
  \psm@beginmath
  \begingroup
  \ignorespaces}
\def\psm@endnode@i{%
  \unskip
  \endgroup
  \psm@endmath
  \egroup
  \use@par
  \@psttrue}
\def\psm@endnode{%
  \@pstfalse
  \psm@endnode@i
  \ifnum\pscol>1\relax \pshskip\pscolsep \fi
  \psk@mnodesize
  \hfil
  \Pst@nodealigntrue
  \if@pst\csname mnode@\psk@mnode\endcsname
  \else\csname mnode@\psk@emnode\endcsname\fi
  \psk@mcol
  \psk@@mnodesize}
\def\psspan#1{\global\mscount#1\relax\pstloop\ifnum\mscount>\@ne\sp@n\repeat}
\def\pstloop#1\repeat{\gdef\pstiterate{#1\relax\expandafter\pstiterate\fi}%
  \pstiterate
  \let\pstiterate\relax}
\define@key[psset]{pst-node}{name}[\relax]{\pst@getnode{#1}\psk@name}
\let\psk@name\relax
\define@key[psset]{pst-node}{mcol}[c]{%
  \ifx r#1\relax\let\psk@mcol\relax\else
    \ifx l#1\relax\let\psk@mcol\hfill\else
    \let\psk@mcol\hfil\fi\fi}
\psset[pst-node]{mcol=c}
\define@key[psset]{pst-node}{mnodesize}[-1pt]{%
  \pssetlength\pst@dimg{#1}%
  \ifdim\pst@dimg<\z@
    \let\psk@mnodesize\relax
    \let\psk@@mnodesize\relax
  \else
    \edef\psk@mnodesize{\noexpand\hbox to\number\pst@dimg sp\noexpand\bgroup}%
    \let\psk@@mnodesize\egroup
  \fi}
\psset[pst-node]{mnodesize=-1pt}
\def\mnode@R{\rnode@iii\Rnode@ii{\psm@thenode}}
\def\mnode@r{\rnode@iii\rnode@iv{\psm@thenode}}
\def\mnode@oval{\ovalnode@ii{\psm@thenode}}
\def\mnode@tri{\trinode@ii{\psm@thenode}}
\def\mnode@dia{\dianode@ii{\psm@thenode}}
\def\mnode@C{{\Pst@nodealigntrue\cnode@ii(\z@,\z@){\psk@radius}{\psm@thenode}}}
\def\mnode@f{{\Pst@nodealigntrue\fnode@ii(\z@,\z@){\psm@thenode}}}
\def\mnode@circle{\circlenode@ii{\psm@thenode}}
\def\mnode@Circle{\Circlenode@ii{\psm@thenode}}
\def\mnode@p{\pnode(\z@,\z@){\psm@thenode}}
\def\mnode@dot{\dotnode@ii(\z@,\z@){\psm@thenode}}
\def\mnode@none{\box\pst@hbox}
\define@key[psset]{pst-node}{mnode}[R]{%
  \@ifundefined{mnode@#1}%
    {\@pstrickserr{\string\psmatrix\space node `#1' not defined.}\@ehpa}%
    {\edef\psk@mnode{#1}}}
\define@key[psset]{pst-node}{emnode}[none]{%
  \@ifundefined{mnode@#1}%
    {\@pstrickserr{\string\psmatrix\space node `#1' not defined.}\@ehpa}%
    {\edef\psk@emnode{#1}}}
\psset[pst-node]{mnode=R,emnode=none}
\def\nccoil{\pst@object{nccoil}}
\def\nccoil@i{\check@arrow{\nccoil@ii}}
\def\nccoil@ii#1#2{\nc@object{Open}{#1}{#2}{.5}{
  \tx@NCCoor
  tx@Dict begin
  4 2 roll
  \psk@coilwidth \pscoilheight
  \psk@coilarmA \psk@coilarmB
  \psk@coilaspect \psk@coilinc
  \pst@coildict \tx@Coil end
  end}%
}
\def\nczigzag{\pst@object{nczigzag}}
\def\nczigzag@i{\check@arrow{\nczigzag@ii}}
\def\nczigzag@ii#1#2{\nc@object{Open}{#1}{#2}{.5}{
  \tx@NCCoor
  tx@Dict begin
  4 2 roll
  \pscoilheight
  \psk@coilwidth
  \psk@coilarmA
  \psk@coilarmB
  \pst@coildict \tx@ZigZag end
  \psline@iii
  \tx@Line
  end}%
}
\def\psGetNodeCenter#1{ tx@NodeDict begin /N@#1 load GetCenter end 
  \pst@number\psyunit div /#1.y exch def 	
  \pst@number\psxunit div /#1.x exch def }	
\def\psGetEdgeA#1#2{
  tx@NodeDict begin \psk@offsetA \psk@offsetB neg 
    \psk@nodesepA \psk@nodesepB 0 0 
    /N@#1 /N@#2 InitNC { NCCoor } if pop pop \tx@UserCoor end}
\def\psGetEdgeB#1#2{
  tx@NodeDict begin \psk@offsetA \psk@offsetB neg 
    \psk@nodesepA \psk@nodesepB 0 0 
    /N@#1 /N@#2 InitNC { NCCoor } if 4 2 roll pop pop \tx@UserCoor end}
\def\ncbarr{\pst@object{ncbarr}}
\def\ncbarr@i#1#2{%
  \begingroup
  \use@par%
  \psLNode(#1)(#2){0.5}{barr@tempNode}%
  \pst@dimc=\psk@angleA pt
  \pst@dimd=180pt
  \ifdim\pst@dimc=\z@\else\ifdim\pst@dimc=\pst@dimd\else\psset{angleA=0}\fi\fi
  \ncbar[arrows=-]{#1}{barr@tempNode}
  \ifdim\psk@angleA pt=\z@\relax
    \ncbar[angleA=180,angleB=180]{barr@tempNode}{#2}
  \else\ncbar[angleA=0,angleB=0]{barr@tempNode}{#2}\fi%
  \endgroup%
}
\def\psLNode(#1)(#2)#3#4{%
  \pst@getcoor{#1}\pst@tempA%
  \pst@getcoor{#2}\pst@tempB%
  \pnode(!
    \pst@tempA /YA exch \pst@number\psyunit div def
    /XA exch \pst@number\psxunit div def
    \pst@tempB /YB exch \pst@number\psyunit div def
    /XB exch \pst@number\psxunit div def
    /dx XB XA sub def
    /dy YB YA sub def
    XA dx #3\space mul add YA dy #3\space mul add){#4}}
\def\psLCNode(#1)#2(#3)#4#5{%
  \pst@getcoor{#1}\pst@tempA%
  \pst@getcoor{#3}\pst@tempB%
  \pnode(!
    \pst@tempA /YA exch \pst@number\psyunit div def
    /XA exch \pst@number\psxunit div def
    \pst@tempB /YB exch \pst@number\psyunit div def
    /XB exch \pst@number\psxunit div def
    XA #2\space mul XB #4\space mul add
    YA #2\space mul YB #4\space mul add){#5}}
\def\psLDNode(#1)(#2)#3#4{%
  \pst@getcoor{#1}\pst@tempA%
  \pst@getcoor{#2}\pst@tempB%
  \pssetlength\pst@dimb{#3}%
  \pnode(!%
    \pst@tempA /YA exch \pst@number\psyunit div def
    /XA exch \pst@number\psxunit div def
    \pst@tempB /YB exch \pst@number\psyunit div def
    /XB exch \pst@number\psxunit div def
    /dx XB XA sub def
    /dy YB YA sub def
    /angle dy dx Atan def
    /linelength \pst@number\pst@dimb \pst@number\psunit div def
    XA linelength angle cos mul add YA linelength angle sin mul add ){#4}%
}
\def\psRelNode{\pst@object{psRelNode}}
\def\psRelNode@i(#1)(#2)#3#4{{
  \use@par
  \pst@getcoor{#1}\pst@tempA%
  \pst@getcoor{#2}\pst@tempB%
  \pnode(!
    \pst@tempA /YA exch \pst@number\psyunit div def
    /XA exch \pst@number\psxunit div def
    \pst@tempB /YB exch \pst@number\psyunit div def
    /XB exch \pst@number\psxunit div def
    /AlphaStrich \psk@angleA\space def
    /unit \pst@number\psyunit \pst@number\psxunit div def 
    /dx XB XA sub  def
    /dy YB YA sub \ifPst@trueAngle\space unit mul \fi\space def
    /laenge dy dup mul dx dup mul add sqrt #3 mul def
    /Alpha dy dx atan def 
    /beta Alpha AlphaStrich add def
    laenge beta cos mul XA add
    laenge beta sin mul \ifPst@trueAngle\space unit div \fi\space YA add ){#4}%
}\ignorespaces}
\define@cmdkeys[psset]{pstricks-add}[PSTPSPNk@]{
  blName,bcName,brName,
  clName,ccName,crName,
  tlName,tcName,trName}[]{}%
\psset[pstricks-add]{%
  blName=PSPbl,bcName=PSPbc,brName=PSPbr,
  clName=PSPcl,ccName=PSPcc,crName=PSPcr,
  tlName=PSPtl,tcName=PSPtc,trName=PSPtr}
\def\psDefPSPNodes{\def\pst@par{}\pst@object{psDefPSPNodes}}
\def\psDefPSPNodes@i{%
  \pst@killglue
  \begingroup
  \use@par
  \expandafter\psDefPSPNodes@ii\pic@coor}
\def\psDefPSPNodes@ii(#1)(#2)(#3){%
    \pnode(#1){PSPN@temp}\pnode([angle=45]PSPN@temp){\PSTPSPNk@blName}
    \pnode(#3){PSPN@temp}\pnode([angle=-135]PSPN@temp){\PSTPSPNk@trName}
    \pnode(\PSTPSPNk@blName|\PSTPSPNk@trName){\PSTPSPNk@tlName}
    \pnode(\PSTPSPNk@trName|\PSTPSPNk@blName){\PSTPSPNk@brName}
    \ncline[linestyle=none]{\PSTPSPNk@blName}{\PSTPSPNk@tlName}
    \ncput[npos=.5]{\pnode{\PSTPSPNk@clName}}
    \ncline[linestyle=none]{\PSTPSPNk@blName}{\PSTPSPNk@brName}
    \ncput[npos=.5]{\pnode{\PSTPSPNk@bcName}}
    \pnode(\PSTPSPNk@brName|\PSTPSPNk@clName){\PSTPSPNk@crName}
    \pnode(\PSTPSPNk@bcName|\PSTPSPNk@trName){\PSTPSPNk@tcName}
    \pnode(\PSTPSPNk@bcName|\PSTPSPNk@clName){\PSTPSPNk@ccName}
  \endgroup
  \ignorespaces}
\def\psDefBoxNodes#1#2{\rnode[tl]{#1:tl}{\rnode[Bl]{#1:Bl}{\rnode[tr]{#1:tr}{%
\rnode[bl]{#1:bl}{\rnode[Br]{#1:Br}{\rnode[br]{#1:br}{#2}}}}}}%
\pnode(!\psGetNodeCenter{#1:bl}
          \psGetNodeCenter{#1:tl} 
          #1:bl.x #1:tl.x add 2 div #1:bl.y #1:tl.y add 2 div ){#1:Cl}%
\pnode(!\psGetNodeCenter{#1:tr}
          \psGetNodeCenter{#1:br} 
          #1:tr.x #1:br.x add 2 div #1:tr.y #1:br.y add 2 div ){#1:Cr}%
\pnode(!\psGetNodeCenter{#1:Cl}
          \psGetNodeCenter{#1:Cr} 
          #1:Cl.x #1:Cr.x add 2 div #1:Cl.y #1:Cr.y add 2 div ){#1:C}%
\pnode(!\psGetNodeCenter{#1:Br}
          \psGetNodeCenter{#1:Bl} 
          #1:Br.x #1:Bl.x add 2 div #1:Br.y #1:Bl.y add 2 div ){#1:BC}%
\pnode(!\psGetNodeCenter{#1:tr}
          \psGetNodeCenter{#1:tl} 
          #1:tr.x #1:tl.x add 2 div #1:tr.y #1:tl.y add 2 div ){#1:tC}%
\pnode(!\psGetNodeCenter{#1:br}
          \psGetNodeCenter{#1:bl} 
          #1:br.x #1:bl.x add 2 div #1:br.y #1:bl.y add 2 div ){#1:bC}}%
\newcount\pst@args
\newcount\num@pts
\newcount\pst@argcnt
\def\PST@root{}
\let\pst@next\relax
\def\my@tempA{}
\def\my@tempB{}
\def\my@tempC{}
\def\my@tempD{}
\def\my@next{}
\newif\if@paren%
\newif\if@equal%
\newif\if@colon%
\newif\ifshow
\def\plussign{+}\def\minussign{-}
\def\defaultvalue#1#2{
  \ifdefined#1\ifx#1\@empty\xdef#1{#2}\fi\else\xdef#1{#2}\fi}%
\def\testAlg#1|#2\@nil{%
\ifx\relax#2\relax%
   \let\my@next\psparnode\xdef\my@tempD{}%
\else%
   \let\my@next\algparnode\xdef\my@tempD{A}
\fi}%
\def\trim #1{\expandafter\trim@\expandafter{#1 }#1}%
\def\trim@ #1{\trim@@ @#1 @ #1 @ @@}%
\def\trim@@ #1@ #2@ #3@@{\trim@@@\empty #2 @}%
\def\unbrace#1{#1}%
\unbrace{\def\trim@@@ #1 } #2@#3{\expandafter\def%
  \expandafter #3\expandafter {#1}}%
\def\hasparen#1(#2\@nil{
  \ifx\relax#2\relax \@parenfalse \else \@parentrue\fi}%
\def\hasequal#1=#2\@nil{
  \ifx\relax#2\relax \@equalfalse \else \@equaltrue\fi
  \hascolon#2:\@nil}%
\def\hascolon#1:#2\@nil{
\ifx\relax#2\relax \@colonfalse \else \@colontrue\fi}%
\def\equalwhat#1=#2:#3\@nil{{#2}{#3}}%
\def\parsenodexn#1(#2)#3\@nil{%
  \def\coeffA{#1}\edef\nodeA{#2}%
  \trim\coeffA%
  \ifx\nodeA\@empty\else%
    \pnode(#2){@@TMP}%
    \ifx\coeffA\@empty\def\coeffA{1}\else%
      \ifx\coeffA\plussign\def\coeffA{1}\else\ifx\coeffA\minussign\def\coeffA{-1}\fi\fi\fi%
  \edef\cmd{\noexpand\psLCNode(@TMP\the\pst@argcnt){1}(@@TMP){\coeffA}{@TMP}}%
  \cmd%
  \advance\pst@argcnt by \@ne%
  \pnode(@TMP){@TMP\the\pst@argcnt}%
  \parsenodexn#3\@nil%
  \fi}%
\def\normalvec(#1)#2{%
  \psRelNodeVar(0,0)(#1)(0,1){#2}}%
\def\curvepnode#1#2#3{%
  \edef\my@tempA{#2}
  \expandafter\testAlg\my@tempA|\@nil\my@next {#1}{#2}{#3}}
\def\psparnode#1#2#3{%
  \pnode(!/t #1 def #2){#3}%
  \pnode(!/t #1 .001 sub def #2 
          /t #1 .001 add def 
           #2 3 -1 roll sub 3 1 roll sub neg 
           2 copy Pyth dup 3 1 roll div 3 1 roll div ){#3tang}}
\def\algparnode#1#2#3{%
  \pstVerb{tx@Dict begin /Func (#2) AlgParser cvx def end }
  \pnode(!/t #1 def Func){#3}
  \pnode(!/t #1 .001 sub def Func 
          /t #1 .001 add def 
          Func 3 -1 roll sub 3 1 roll sub neg 
          2 copy Pyth dup 3 1 roll div 3 1 roll div ){#3tang}
}%
\def\nodex#1{%
\expandafter\hasparen#1(\@nil%
\if@paren
  \pnode(0,0){@TMP0}%
  \pst@argcnt=0%
  \expandafter\parsenodexn#1()\@nil%
\else%
  \def\my@tempC{#1}%
  \ifx\my@tempC\@empty\pnode(0,0){@TMP}\else\pnode(#1){@TMP}\fi%
\fi}
\def\nodexn#1#2{%
\expandafter\hasparen#1(\@nil
\if@paren
  \pnode(0,0){@TMP0}%
  \pst@argcnt=0%
  \parsenodexn#1()\@nil%
  \pnode(@TMP){#2}%
\else%
  \def\my@tempC{#1}%
  \ifx\my@tempC\@empty\pnode(0,0){#2}\else\pnode(#1){#2}\fi%
\fi}
\def\psxline{\pst@object{psxline}}%
\def\psxline@i{\@ifnextchar({\psxline@iii}{\psxline@ii}}%
\def\psxline@ii#1{%
\addto@par{arrows=#1}%
\psxline@iii}%
\def\psxline@iii(#1)#2#3{{
\pst@killglue%
\use@par%
\nodexn{#2}{@TMP@a}%
\AplusB(#1)(@TMP@a){@TMP@A}%
\nodexn{#3}{@TMP@a}%
\AplusB(#1)(@TMP@a){@TMP@B}%
\psline(@TMP@A)(@TMP@B)%
}%
\ignorespaces}%
\def\curvepnodes{\pst@object{curvepnodes}}
\def\curvepnodes@i#1#2#3#4{{
  \pst@killglue
  \use@par
  \edef\my@tempA{#3}
  \expandafter\testAlg\my@tempA|\@nil %
  \pstVerb{%
	tx@Dict begin 
	/t0 #1 def
	/t1 #2 def  
	 t1 t0 sub end \psk@plotpoints div /dt exch def }%
  \pst@cntc=\psk@plotpoints\relax
  \advance\pst@cntc by \@ne\relax 
  \ifx\my@tempD\@empty\pstVerb{tx@Dict begin /Func (#3) cvx def end }
  \else\pstVerb{tx@Dict begin /Func (#3 ) AlgParser cvx def end }%
  \fi%
    \multido{\i=0+1}{\pst@cntc}{%
      \pnode(! /t #1 dt \i\space mul add def Func ){#4\i}}
    \expandafter\xdef \csname #4nodecount\endcsname {\psk@plotpoints}%
    \ifnum\Pst@Debug>0 \typeout{Created nodes #40 .. #4\psk@plotpoints}\fi%
}\ignorespaces}%
\def\fnpnode{\pst@object{fnpnode}}
\def\fnpnode@i#1#2#3{{
  \pst@killglue
  \use@par
  \ifPst@algebraic\pnode(*#1 {#2}){#3}\else\pnode(! /x #1 def x #2){#3}\fi
}\ignorespaces}%
\def\fnpnodes{\pst@object{fnpnodes}}
\def\fnpnodes@i#1#2#3#4{{
\pst@killglue
\use@par
\pst@dima=#1pt \pst@dimb=#2pt \advance\pst@dimb -\pst@dima%
\pst@cnta=\psk@plotpoints \relax 
\def\PST@root{#4}
\divide\pst@dimb by \pst@cnta
\pst@cntc=\pst@cnta %
\advance\pst@cntc by 1 \relax 
\ifPst@algebraic 
  \multido{\i=0+1}{\pst@cntc}{\pnode(*{\pst@number\pst@dima} {#3}){#4\i}
  \advance\pst@dima \pst@dimb}%
\else
    \multido{\i=0+1}{\pst@cntc}{\pnode(!/x \pst@number\pst@dima\space def x #3){#4\i}%
  \advance\pst@dima \pst@dimb}%
\fi%
  \expandafter\xdef \csname \PST@root nodecount\endcsname {\the\pst@cnta}%
  \ifnum\Pst@Debug>0 \typeout{Created nodes #40 .. #4\the\pst@cnta}\fi%
}\ignorespaces}%
\def\AtoB(#1)(#2)#3{\psLCNodeVar(#1)(#2)(-1,1){#3}}
\def\AplusB(#1)(#2)#3{\psLCNodeVar(#1)(#2)(1,1){#3}}
\def\midAB(#1)(#2)#3{\psLCNodeVar(#1)(#2)(.5,.5){#3}}
\def\psnline{\pst@object{psnline}}
\def\psnline@i{\pst@getarrows{\psnline@ii}}
\def\psnline@ii(#1,#2)#3{{%
\pst@killglue%
\use@par%
\pst@cnta=#2 \relax\advance\pst@cnta by 1
\edef\@tmp{}%
\multido{\i=#1+1}{\pst@cnta}{\xdef\@tmp{\@tmp(#3\i)}}%
\expandafter\psline\@tmp}%
\ignorespaces}%
\def\psnpolygon{\pst@object{psnpolygon}}
\def\psnpolygon@i{\pst@getarrows{\psnpolygon@ii}}
\def\psnpolygon@ii(#1,#2)#3{{%
\pst@killglue%
\use@par%
\pst@cnta=#2 \relax\advance\pst@cnta by 1
\edef\@tmp{}%
\multido{\i=#1+1}{\pst@cnta}{\xdef\@tmp{\@tmp(#3\i)}}%
\expandafter\pspolygon\@tmp}%
\ignorespaces}%
\def\psncurve{\pst@object{psncurve}}
\def\psncurve@i{\pst@getarrows{\psncurve@ii}}
\def\psncurve@ii(#1,#2)#3{{%
\pst@killglue%
\use@par%
\pst@cnta=#2 \relax\advance\pst@cnta by 1
\edef\@tmp{}%
\multido{\i=#1+1}{\pst@cnta}{\xdef\@tmp{\@tmp(#3\i)}}%
\expandafter\pscurve\@tmp}%
\ignorespaces}%
\def\psnccurve{\pst@object{psnccurve}}
\def\psnccurve@i{\pst@getarrows{\psnccurve@ii}}
\def\psnccurve@ii(#1,#2)#3{{%
\pst@killglue%
\use@par%
\pst@cnta=#2 \relax\advance\pst@cnta by 1
\xdef\@tmp{}%
\multido{\i=#1+1}{\pst@cnta}{\xdef\@tmp{\@tmp(#3\i)}}%
\expandafter\psccurve\@tmp}%
\ignorespaces}%
\def\shownode(#1){
  \pst@killglue%
  \pstVerb{%
    gsave tx@Dict begin %
    tx@NodeDict /N@#1 known { 
      /tmpar [(Node #1: ) <28> () (, ) () <29>] def %
      /str 12 string def 
      STV CP T \psGetNodeCenter{#1}\space 
      tmpar 2 #1.x str cvs put 
      /str 12 string def 
      tmpar 4 #1.y str cvs put 
      tmpar concatstringarray = }%
    {
      (Node #1: (NOT KNOWN)) = %
    } ifelse %
    end grestore }%
  \ignorespaces}%
\def\pnodes@ii#1{\getnodelist{#1}{}}
\def\getnodelist#1#2{%
\pst@args=0 \relax%
\def\PST@root{#1}%
\def\pst@next{#2}
\getnext@Node}%
\def\getnext@Node{\@ifnextchar({\getnext@Node@i}%
  {\advance\pst@args by \m@ne \expandafter\xdef \csname \PST@root nodecount\endcsname {\the\pst@args}
  \ifnum\Pst@Debug>0 \typeout{Created nodes \PST@root0 .. \PST@root\the\pst@args}\fi%
  \pst@next}%
}%
\def\getnext@Node@i(#1){%
\pnode(#1){\PST@root\the\pst@args}%
\advance\pst@args by \@ne\relax%
\getnext@Node}%
\def\psLCNodeVar(#1)(#2)(#3)#4{%
\pst@getcoor{#1}\my@tempA%
\pst@getcoor{#2}\my@tempB%
\pnode(#3){tmpLCn@de}%
\pnode(!%
  \my@tempA /YA exch \pst@number\psyunit div def
  /XA exch \pst@number\psxunit div def
  \my@tempB /YB exch \pst@number\psyunit div def
  /XB exch \pst@number\psxunit div def 
  \psGetNodeCenter{tmpLCn@de}\space
  XA tmpLCn@de.x mul XB tmpLCn@de.y mul add
  YA tmpLCn@de.x mul YB tmpLCn@de.y mul add){tmpLCn@deA}%
\pnode(tmpLCn@deA){#4}%
}%
\def\psRelNodeVar{\pst@object{psRelNodeVar}}
\def\psRelNodeVar@i(#1)(#2)(#3)#4{{
  \use@par
  \pst@getcoor{#1}\my@tempA%
  \pst@getcoor{#2}\my@tempB%
   \pnode(#3){tmpn@de}%
\pnode(!
  /unit \pst@number\psyunit \pst@number\psxunit div def 
    \my@tempA /YA exch \pst@number\psyunit div def
    /XA exch \pst@number\psxunit div def
    \my@tempB /YB exch \pst@number\psyunit div YA sub 
    \ifPst@trueAngle\space unit mul \fi\space def
    /XB exch \pst@number\psxunit div XA sub def
    \psGetNodeCenter{tmpn@de}
    XB tmpn@de.x mul YB tmpn@de.y mul sub
    YB tmpn@de.x mul XB tmpn@de.y mul add
    \ifPst@trueAngle\space unit div \fi\space 
   YA add exch XA add exch 
    ){#4}%
}}
\def\psRelLineVar{\pst@object{psRelLineVar}}
\def\psRelLineVar@i{\@ifnextchar({\psRelLineVar@iii}{\psRelLineVar@ii}}
\def\psRelLineVar@ii#1{%
  \addto@par{arrows=#1}%
  \psRelLineVar@iii}
\def\psRelLineVar@iii(#1)(#2)(#3)#4{{%
  \pst@killglue
  \use@par
  \psRelNodeVar(#1)(#2)(#3){#4}%
  \psline(#1)(#4)%
}\ignorespaces}
\def\rhombus#1(#2)(#3)#4#5{
\AtoB(#2)(#3){node@P}
\pnode(! 
/tmp \psGetNodeCenter{node@P} node@P.x node@P.y 
Pyth 2 div def 
/ang tmp #1\space div Acos def 
#1\space tmp 2 mul div 
dup ang cos mul exch ang sin mul ){node@A1}
\pnode(! \psGetNodeCenter{node@A1} node@A1.x node@A1.y neg ){node@A2}
\psRelNodeVar(#2)(#3)(node@A1){#4}%
\psRelNodeVar(#2)(#3)(node@A2){#5}%
}%
\def\psrline{\pst@object{psrline}}
\def\psrline@i{\@ifnextchar({\psrline@iii}{\psrline@ii}}%
\def\psrline@ii#1{%
\addto@par{arrows=#1}%
\psrline@iii}%
\def\psrline@iii{%
\getnodelist{@tmpnode}{\psrline@iv}%
}%
\def\psrline@iv{%
   \ifnum\pst@args<0\else
      \pnode(@tmpnode0){@tmpnodeB0}%
      \multido{\iA=1+1,\iB=0+1}{\pst@args}{%
      \AplusB(@tmpnodeB\iB)(@tmpnode\iA){@tmpnodeB\iA}}%
      \psrline@v%
   \fi%
}%
\def\psrline@v{{
  \pst@killglue%
  \use@par%
  \xdef\tmp{(@tmpnodeB0)}%
  \multido{\i=1+1}{\pst@args}%
{\xdef\tmp{\tmp(@tmpnodeB\i)}}%
\expandafter\psline\tmp%
}\ignorespaces}%
\def\polyIntersections#1#2(#3)(#4){%
\def\nodenameA{#1}\def\nodenameB{#2}%
\pnode(#3){P@A}\pnode(#4){P@B}%
\@ifnextchar({\polyIntersections@next}{\polyIntersections@ii}%
}%
\def\polyIntersections@ii#1#2{%
\def\root@node{#1}\num@pts=#2 \relax%
\polyIntersections@iii}%
\def\polyIntersections@next{
\def\root@node{P@}\getnodelist{P@}{\num@pts=\pst@args \relax\polyIntersections@iii}%
}%
\def\polyIntersections@iii{
\pst@cnta=\num@pts \relax\advance\pst@cnta by 1 \relax%
\pstVerb{%
 /xarray \the\pst@cnta\space array def
 /yarray \the\pst@cnta\space array def  tx@Dict begin }%
\multido{\i=0+1}{\the\pst@cnta}{\pstVerb{ \psGetNodeCenter{\root@node\i} xarray \i\space \root@node\i.x put yarray \i\space \root@node\i.y put }}%
\pstVerb{ /tposmin 100 def /tnegmax -100 def 
\psGetNodeCenter{P@B} \psGetNodeCenter{P@A} 
/dx P@B.x P@A.x sub def 
/dy P@B.y P@A.y sub def 
/lenAB dx dy Pyth def
/oldx xarray 0 get def /oldy yarray 0 get def 
1 1 \the\num@pts\space {/k exch def /newx xarray k get def /newy yarray k get def 
/ddx newx oldx sub def /ddy newy oldy sub def 
/det ddy dx mul ddx dy mul sub def
det abs lenAB ddx ddy Pyth mul .001 mul gt 
{/ac oldx P@A.x sub def /bd oldy P@A.y sub def 
 /tt  ac ddy mul bd ddx mul sub det div def 
 /ss ac  dy mul bd dx mul sub det div def 
ss 0 ge 
   {ss 1 le 
        {tt 0 lt {tt tnegmax gt {/tnegmax tt def} if } {tt tposmin lt {/tposmin tt def} if } ifelse }
    if } 
if }
 if 
 /oldx newx def /oldy newy def} for end }%
\pnode(! \psGetNodeCenter{P@A} \psGetNodeCenter{P@B} P@B.x P@A.x sub  tposmin mul P@A.x add  P@B.y P@A.y sub tposmin  mul P@A.y add ){\nodenameA}%
\pnode(! \psGetNodeCenter{P@A} \psGetNodeCenter{P@B} P@B.x P@A.x sub tnegmax mul P@A.x add P@B.y P@A.y sub tnegmax mul P@A.y add){\nodenameB}%
}%
\def\actualscale#1 #2 scale{
#1}
\def\psGetCenter#1{ tx@NodeDict begin /N@#1 load GetCenter end }
\def\ArrowNotch{\pst@object{ArrowNotch}}
\def\ArrowNotch@i#1#2#3#4{{%
\pst@killglue%
\use@par%
\def\inc{-1}%
\ifx#3<\def\inc{1}\fi
\pstVerb{ 
    1 \psk@arrowinset\space sub \psk@arrowlength\space \psk@arrowsize\space  
    \pst@number\pslinewidth \space mul add  mul mul 
    \expandafter\actualscale\psk@arrowscale \space  mul 
    /hh exch def /hh1 hh .05 sub def }
\def\root@node{#1}\num@pts=\csname\root@node nodecount\endcsname %
\pst@cntb=\num@pts \advance\pst@cntb by \@ne
\pst@cnta=\num@pts \advance\pst@cnta by \thr@@
\pst@cntc=#2 \relax
\ifnum\pst@cntc>\num@pts \pnode(0,0){#4}\else
\pstVerb{%
/PythSq { dup mul exch dup mul add } def
/PtSub {					
  3 -1 roll 		
  sub neg		
  3 1 roll 		
  sub			
  exch                     
} def
  /xarray \the\pst@cnta\space array def
  /yarray \the\pst@cnta\space array def  
  tx@Dict begin }
\multido{\i=0+1,\ib=1+1}{\the\pst@cntb}{\pnode(! \psGetCenter{\root@node\i}\space  
yarray \ib\space 3 -1 roll put xarray \ib\space 3 -1 roll put 0 0 ){@tmp}}
\pnode(! xarray 1 get dup yarray 1 get dup 3 1 roll 
xarray 2 get yarray 2 get PtSub  
2 copy Pyth hh div 2 div dup 
3 1 roll 
div 3 1 roll div 
3 1 roll 
add 3 1 roll add 
 xarray 0 3 -1 roll put yarray 0 3 -1 roll put 
 xarray length 2 sub /topnum exch def 
 xarray topnum get dup yarray topnum get dup 3 1 roll 
topnum 1 sub /topnum exch def xarray topnum get yarray topnum get 
3 -1 roll sub  neg 3 1 roll sub exch 
2 copy Pyth hh div 2 div dup 
3 1 roll div 3 1 roll div 
3 -1 roll add 3 1 roll 
topnum 2 add /topnum exch def xarray topnum 3 -1 roll put yarray topnum 3 -1 roll put 
 /oldcindex \the\pst@cntc\space 1 add def 
 xarray oldcindex get /xc exch def yarray oldcindex get /yc exch def
/inc \inc\space def 
/cindex oldcindex def 
{cindex inc add /cindex exch def xarray cindex get xc sub yarray cindex get yc sub Pyth dup hh1 gt 
{ exit } if } loop 
 hh1 .1 add lt { xarray cindex get yarray cindex get } 
{ xarray cindex inc sub get dup yarray cindex inc sub get dup 4 -1 roll exch 
xarray cindex get yarray cindex get PtSub /dy1 exch def /dx1 exch def dx1 dy1 PythSq /Aterm exch def 
 2 copy xc yc PtSub 
 2 copy 2 copy 3 -1 roll mul 3 1 roll mul add hh dup mul sub 
 Aterm div /Cterm exch def  
 dx1 dy1 
 4 1 roll mul 3 1 roll mul add Aterm div /Bterm exch def 
 Bterm abs neg dup dup mul Cterm sub sqrt add dup /tval exch def
 dup dx1 dy1 4 1 roll mul 3 1 roll mul  
 PtSub } ifelse 
 \pst@number\psyunit div exch \pst@number\psxunit div exch  
){#4}\fi%
\pstVerb { end } 
}\ignorespaces}%
\def\saveDataAsNodes#1#2{
  \psLoopIndex=0\relax
  \typeout{Open file #1}%
  \openin7=#1
  \loop
    \read7 to \@Data
    \ifeof7\else
      \ifx\@Data\@empty
      \else
        \pnode(!\@Data){#2\the\psLoopIndex}%
        \typeout{#2\the\psLoopIndex -> \@Data}%
	\advance\psLoopIndex by 1
        \let\@oldData\@Data
      \fi
  \repeat
  \closein7
  \advance\psLoopIndex by -1
  \pnode(!\@oldData){#2Last}%
}
\catcode`\@=\TheAtCode\relax
 
\csname PSTcoilsLoaded\endcsname
\let\PSTcoilsLoaded 
\ifx\PSTricksLoaded   \else\input pstricks.tex\fi
\ifx\PSTnodeLoaded    \else\fi
\ifx\PSTXKeyLoaded    \else\input pst-xkey \fi
\def\fileversion{1.07}
\def\filedate{2015/05/13}
\edef\TheAtCode{\the\catcode`\@}
\catcode`\@=11
\pst@addfams{pst-coil}
\pstheader{pst-coil.pro}
\edef\pst@theheaders{\pst@theheaders,pst-coil.pro}
\def\pst@CoilDict{tx@CoilDict begin }
\def\tx@CoilLoop  {\pst@CoilDict CoilLoop   end }
\def\tx@Coil      {\pst@CoilDict Coil       end }
\def\tx@AltCoil   {\pst@CoilDict AltCoil    end }
\def\tx@ZigZag    {\pst@CoilDict ZigZag     end }
\def\tx@ZigZagCirc{\pst@CoilDict ZigZagCirc end }
\def\tx@Sin       {\pst@CoilDict Sin        end }
\define@key[psset]{pst-coil}{coilwidth}[1cm]{\pst@getlength{#1}\psk@coilwidth}
\define@key[psset]{pst-coil}{coilheight}[1]{\pst@checknum{#1}\pscoilheight}
\define@key[psset]{pst-coil}{coilarmA}[0.5cm]{\pst@getlength{#1}\psk@coilarmA}
\define@key[psset]{pst-coil}{coilarmB}[0.5cm]{\pst@getlength{#1}\psk@coilarmB}
\define@key[psset]{pst-coil}{coilarm}[0.5cm]{%
  \pst@getlength{#1}\psk@coilarmA%
  \let\psk@coilarmB\psk@coilarmA}
\define@key[psset]{pst-coil}{coilaspect}[45]{\pst@getangle{#1}\psk@coilaspect}
\define@key[psset]{pst-coil}{coilinc}[10]{\pst@getangle{#1}\psk@coilinc}
\psset[pst-coil]{coilaspect=45,coilarm=.5cm,coilheight=1,coilwidth=1cm,coilinc=10}
\def\pscoil{\def\pst@par{}\pst@object{pscoil}}
\def\pscoil@i{\pst@getarrows\pscoil@ii}
\def\pscoil@ii(#1){\@ifnextchar({\pscoil@iii{1}(#1)}{\pscoil@iii{\z@}(0,0)(#1)}}
\def\pscoil@iii#1(#2)(#3){%
  \begin@OpenObj
  \pst@getcoor{#2}\pst@tempa
  \pst@getcoor{#3}\pst@tempb
  \pst@optcp{#1}\pst@tempa
  \addto@pscode{%
    \pst@tempa \pst@tempb
    \psk@coilwidth \pscoilheight
    \psk@coilarmA \psk@coilarmB
    \psk@coilaspect \psk@coilinc
    \tx@Coil }%
    \showpointsfalse
  \end@OpenObj}
\def\psCoil{\def\pst@par{}\pst@object{psCoil}}
\def\psCoil@i#1#2{%
  \begin@AltOpenObj
  \showpointsfalse
  \pst@getangle{#1}\pst@tempa
  \pst@getangle{#2}\pst@tempb
  \addto@pscode{%
    \pst@tempa
    \pst@tempb
    \psk@coilwidth
    \pscoilheight
    \psk@coilaspect
    \psk@coilinc
    \tx@AltCoil  
    \@nameuse{psls@\pslinestyle} }%
  \end@OpenObj}
\define@key[psset]{pst-coil}{bow}[0]{%
  \pst@getlength{#1}\psk@bow%
  \pst@dimm=\psk@bow pt%
  \pst@absdim{\pst@dimm}{\pst@dimn}%
  \ifdim\pst@dimn<1pt \def\psk@bow{0}\fi}%
\psset[pst-coil]{bow=0}
\def\pszigzag{\def\pst@par{}\pst@object{pszigzag}}
\def\pszigzag@i{\pst@getarrows\pszigzag@ii}
\def\pszigzag@ii(#1){\@ifnextchar({\pszigzag@iii{1}(#1)}{\pszigzag@iii{\z@}(0,0)(#1)}}
\def\pszigzag@iii#1(#2)(#3){%
  \addbefore@par{bow=0}%
  \begin@OpenObj%
  \pst@getcoor{#2}\pst@tempA%
  \pst@getcoor{#3}\pst@tempB%
  \pst@optcp{#1}\pst@tempA%
  \addto@pscode{%
    \pst@tempA
    \pst@tempB
    \pscoilheight
    \psk@coilwidth
    \psk@coilarmA
    \psk@coilarmB 
    \ifdim\psk@bow pt=\z@ \tx@ZigZag \else \psk@bow\space \tx@ZigZagCirc \fi
    \psline@iii
    \tx@Line }%
  \end@OpenObj}
\def\nccoil{\pst@object{nccoil}}
\def\nccoil@i{\check@arrow{\nccoil@ii}}
\def\nccoil@ii#1#2{\nc@object{Open}{#1}{#2}{.5}{%
  \tx@NCCoor
  tx@Dict begin
  4 2 roll
  \psk@coilwidth \pscoilheight
  \psk@coilarmA \psk@coilarmB
  \psk@coilaspect \psk@coilinc
  \tx@Coil 
  end }}
\def\pccoil{\def\pst@par{}\pst@object{pccoil}}
\def\pccoil@i{\pc@object\nccoil@ii}
\def\nczigzag{\pst@object{nczigzag}}
\def\nczigzag@i{\check@arrow{\nczigzag@ii}}
\def\nczigzag@ii#1#2{\nc@object{Open}{#1}{#2}{.5}{%
  \tx@NCCoor
  tx@Dict begin
  4 2 roll
  \pscoilheight
  \psk@coilwidth
  \psk@coilarmA
  \psk@coilarmB
  \ifdim\psk@bow pt=\z@\tx@ZigZag\else\psk@bow\space\tx@ZigZagCirc\fi 
  \psline@iii
  \tx@Line
  end }}
\def\pczigzag{\def\pst@par{}\pst@object{pczigzag}}
\def\pczigzag@i{\pc@object\nczigzag@ii}
\def\pst@checkUnit#1#2{\expandafter\pst@checkUnit@i#1!!#2}
\def\pst@checkUnit@i{\@ifnextchar*%
  {\def\pst@roundValue{0 }\pst@checkUnit@ii}%
  {\def\pst@roundValue{-1 }\pst@checkUnit@iii**}}
\def\pst@checkUnit@ii*{\@ifnextchar*%
  {\def\pst@roundValue{1 }\pst@checkUnit@iii*}%
  {\pst@checkUnit@iii**}}
\def\pst@checkUnit@iii**#1!!#2{%
  \edef\ps@next{#1}%
  \ifx\ps@next\@empty\let\pst@num\z@%
  \else\expandafter\pst@@checknum\ps@next..\@nil%
  \fi%
  \ifnum\pst@num=\z@\pst@getlength{#1}{#2}\def\pst@relativePeriod{false }%
  \else%
    \def\pst@relativePeriod{true }%
    \edef#2{\ifnum\pst@num=\tw@-\fi\the\pst@cntg.%
    \expandafter\@gobble\the\pst@cnth\space}%
  \fi}
\define@key[psset]{pst-coil}{periods}[1]{\pst@checkUnit{#1}{\psk@periods}}
\define@key[psset]{pst-coil}{amplitude}[1]{\def\psk@amplitude{#1 }}
\define@key[psset]{pst-coil}{ppoints}[360]{\def\psk@ppoints{#1 }}
\define@key[psset]{pst-coil}{function}[sin]{\def\psk@function{#1 }}
\psset[pst-coil]{periods=1,amplitude=1,ppoints=360,function=sin}
\def\pssin{\pst@object{pssin}}
\def\pssin@i{\pst@getarrows\pssin@ii}
\def\pssin@ii(#1){\@ifnextchar({\pssin@iii{1}(#1)}{\pssin@iii{\z@}(0,0)(#1)}}
\def\pssin@iii#1(#2)(#3){%
  \begin@OpenObj
  \pst@getcoor{#2}\pst@tempa
  \pst@getcoor{#3}\pst@tempb
  \pst@optcp{#1}\pst@tempa
  \addto@pscode{%
    \pst@tempa \pst@tempb
    \psk@periods 
    \pst@relativePeriod 
    \pst@roundValue
    \psk@amplitude \pst@number\psyunit mul
    \psk@coilarmA \psk@coilarmB 
    \psk@ppoints
    { \psk@function }
    \tx@Sin
  }%
  \showpointsfalse%
  \end@OpenObj}
\def\ncsin{\pst@object{ncsin}}
\def\ncsin@i{\check@arrow{\ncsin@ii}}
\def\ncsin@ii#1#2{\nc@object{Open}{#1}{#2}{.5}{%
  \tx@NCCoor
  tx@Dict begin
  4 2 roll
  \psk@periods 
  \pst@relativePeriod 
  \pst@roundValue
  \psk@amplitude \pst@number\psyunit mul
  \psk@coilarmA \psk@coilarmB 
  \psk@ppoints
  { \psk@function }
  \tx@Sin 
  end }}
\def\pcsin{\def\pst@par{}\pst@object{pcsin}}
\def\pcsin@i{\pc@object\ncsin@ii}
\catcode`\@=\TheAtCode\relax

\makeatletter
\newcommand\notsotiny{\@setfontsize\notsotiny\@vipt\@viipt}
\makeatother

\definecolor{celestialblue}{rgb}{0.29, 0.59, 0.82}
\definecolor{orange(colorwheel)}{rgb}{1.0, 0.5, 0.0}
\definecolor{persimmon}{rgb}{0.93, 0.35, 0.0}

 \definecolor{mygreen}{RGB}{28,172,0}

 \definecolor{selettrico}{RGB}{39,90,183}

 \definecolor{myorange}{rgb}{0 0.68 0} 

 \def\bPhi{{\bf \Phi}} \def\bphi{{\bf \phi}}

\newcommand{\sgn}{\text{sgn}}
\newcommand{\largearrow}{\psset{arrowsize=3pt 2}\psset{arrowinset=0.2}}

\newcommand{\aggiungi}[1]{\begin{picture}(0,0) #1 \end{picture}}
 \newcommand{\molver}[2]{\begin{picture}(6,2)
    \pszigzag[coilwidth=1,coilarm=1,linearc=0](0,0)(0,6)
    \put(-1.05,3){\makebox(0,0)[r]{$#1$}}
    \put( 1.05,3){\makebox(0,0)[l]{$#2$}}
    \end{picture}}          
\newcommand{\indver}[2]{\begin{picture}(6,2)
  \pscoil[coilarm=1.2,coilwidth=1](0,0)(0,6)
  \put(-1.15,3){\makebox(0,0)[r]{$#1$}}
  \put( 0.9,3){\makebox(0,0)[l]{$#2$}}
\end{picture}
}

 \newcommand{\nrsec}[1]{
 \setlength{\unitlength}{0.85mm}
 \psset{unit=1.0\unitlength}
 \begin{pspicture}(0,0)(2.25,0)\cput[framesep=1pt,linewidth=0.5pt](0,1.3){$\scr #1$} \end{pspicture}}

 \newrgbcolor{mygreen}{0.0 0.7 0.0}
 \newrgbcolor{myred}{0.7 0.0 0.0}

 \newtheorem{Prop}{\bf Property}
\newtheorem{Remar}{\bf Remark}

\newcommand{\Eff}{{\mygreen v_e}}
\newcommand{\Effi}{{\mygreen v_{e1}}}
\newcommand{\Effii}{{\mygreen v_{e2}}}
\newcommand{\Effiii}{{\mygreen v_{e}}}
\newcommand{\Effiv}{{\mygreen v_{e4}}}
\newcommand{\Effv}{{\mygreen v_{e5}}}
\newcommand{\Flow}{{\myred v_f}}
\newcommand{\Floi}{{\myred v_{f1}}}
\newcommand{\Floii}{{\myred v_{f2}}}
\newcommand{\Floiii}{{\myred v_{f}}}
\newcommand{\Floiv}{{\myred v_{f4}}}
\newcommand{\Flov}{{\myred v_{f5}}}
\newcommand{\Kn}{{K}}

\jname{IEEE CONTROL SYSTEMS}
\pubyear{2024}

\begin{document}


\title{The Power-Oriented Graphs Modeling Technique\stitle{From the fundamental principles to the systematic, step-by-step modeling of complex physical systems
}}

\author{DAVIDE TEBALDI$^*$, and ROBERTO ZANASI}
\affil{University of Modena and Reggio Emilia, Department of Engineering Enzo Ferrari, Via Pietro Vivarelli 10, Modena, 41125, Italy \\
$^*$ Corresponding Author \\
Email addresses: davide.tebaldi@unimore.it, roberto.zanasi@unimore.it\\
This work has been submitted to the IEEE for possible publication. Copyright may be transferred without notice, after which this version may no longer be accessible.
}

\maketitle

\dois{}{}
\chapterinitial{U}nderstanding the physical laws governing physical systems has always been 
a subject of study for many world-renowned scientists throughout history. A significant contribution was given by James Clerk Maxwell~\cite{maxwell1861}, who elaborated on the relations between the mechanical motion of a medium and the observed phenomena of magnetism and electricity. The existence of a unified framework for the study of physical systems in different energetic domains started to become popular in the last century thanks to the work of Gabriel Kron, who showed that the modeling and simulation of mechanical and hydraulic systems can be effectively performed using analogous electrical systems based on tensor analysis~\cite{kron1939tensor}. It is from this unified framework that graphical modeling techniques such as Bond Graph (BG)~\cite{paynter1961analysis,Gawthrop2007}, Power-Oriented Graphs (POG)~\cite{zanasi1991,zanasi2010,Nostro_5}, and Energetic Macroscopic Representation (EMR)~\cite{Bouscayrol_2000,BOUSCAYROL2003} were first inspired and developed, and have subsequently evolved over the years. Independently of the chosen modeling approach, comprehending the dynamics governing the evolution of the system trajectories is a
first step for control engineers in order to develop suitable control laws for the considered system~\cite{SCL_New_1}.
For this reason, the ablity to develop compact models that accurately represent the dynamics of the systems under consideration is of great interest in the field of control and systems engineering~\cite{bequette2003process}. Together with the graphical modeling techniques POG, BG, and EMR, the modeling of physical systems can be addressed using a variety of approaches and methodologies available in the literature, where the chosen approach typically depends on the energetic domain of the system under consideration.

Historically, one of the most well-known approaches for modeling mechanical systems has been the Lagrangian approach~\cite{Altro_1,Altro_2}. Among the areas of application of this approach, robotics stands out, since the Lagrangian equations are used to derive the classical robot manipulator dynamics~\cite{Abdallah1991CSM}. This applies to many different types of robotic manipulators; for instance, in~\cite{Lee2003CSM}, the Lagrangian equations are used to derive the dynamics of wheeled mobile robots. Another area of application is represented by vehicle dynamics. As an example, a comparison between Lagrangian mechanics and the POG technique when modeling a full toroidal variator~\cite{Altro_5} is given in \cite{Nostro_1}. Depending on the specific mechanical system under consideration, the Lagrangian approach may be replaced by other approaches, as in~\cite{Altro_3}, where vehicle differentials are modeled using a set of algebraic equations, in~\cite{Altro_4}, where planetary gear sets are modeled using the Lever Analogy, and in~\cite{Altro_5}, where full toroidal variators are modeled  using a Newtonian approach. 

As far as the electrical and electromechanical domains are concerned, reference can be made to power converters~\cite{Nostro_2} and  to Permanent Magnet Synchronous Motors (PMSMs) \cite{Nostro_4}, which are used in applications such as power grids~\cite{Nathabhat2013262} and hybrid or plug-in electric vehicles~\cite{Carli2013,Zhou2023,Harnefors2013first}.
Power converters are typically modeled in the form of equations \cite{Steckler22First,BenBrahim2016First} and can be simulated using dedicated platforms such as the Piecewise Linear Electrical Circuit
Simulation (PLECS) simulator~\cite{PLECSDocum}. Permanent magnet synchronous motors are typically modeled in the form of equations by giving the system equations in the d-q reference frame~\cite{Khalil2017CSM,Owen2006CSM} or in the static reference frame~\cite{CSL_New_2bis,Zhang2020stfr}. 

The modeling problem becomes more complex when physical systems composed of elements belonging to different energetic domains need to be addressed. 
Devices such as Electro Continuously Variable Transmission (ECVT) \cite{Altro_8,Tebaldi2021ECVT} or hydraulic CVT \cite{TebaldiZanasiCDC23,Liu2018HCVT}, which can be found in hybrid vehicles,
include physical elements in the electromechanical and hydromechanical domains, respectively.
Mechatronic machines~\cite{DeBaerdemaeker2001CSM,Kapila2004CSM} generally include physical elements in different energetic domains interacting with each other, and the same observation applies to robotic systems as well, which nowadays find applications in many fields including manufacturing, medical, and social sectors~\cite{Garcia2007RAM}.
Many other examples of multi-physics systems~\cite{CSL_New_last}, composed of physical elements from various energetic domains, can be found. This is because engineering applications typically involve the interaction of different physical systems. 
In these cases, it can be more appropriate to use port-based modeling approaches~\cite{CSL_New_last}. The main dedicated graphical modeling techniques which are suitable for modeling physical systems in different energetic domains are the aforementioned POG, BG, and EMR. These techniques have been widely employed in a range of applications over the years for the study of multi-physics systems. They are based on the same unified concepts of energy and power, but offer different pros and cons relative to each other~\cite{Geitner15POGBGEMR}. In this article, we dedicate a sidebar to comparing POG, BG and EMR, with reference to an example used as a case study, as shown in ``Comparison With Other Graphical Modeling Techniques''.
Graphical modeling techniques introduce a unified modeling framework~\cite{CSL_3_neww_bis} which, unlike purely analytical tools, 
also offers effective and energy-based graphical descriptions of physical systems.

At the beginning of its history~\cite{paynter1961analysis}, BG was mainly applied to the modeling of mechanical systems~\cite{Karnopp69BG}, while variants of BG were proposed in order to model systems such as transmission lines~\cite{Auslander68BG}. 
\begin{summary}
\summaryinitial{M}odeling physical systems is an essential skill for a control engineer, since it enables to achieve a deep understanding of their dynamic behavior and, consequently, the development of effective control strategies. The first part of this article provides a tutorial description of the fundamental principles and properties of the Power-Oriented Graphs (POG) modeling technique. Various case studies in different energetic domains are then presented to consolidate the fundamental principles, each highlighting different features of the POG modeling technique. The latter is then compared with the other two main graphical modeling techniques available in the literature, namely Bond Graph (BG) and Energetic Macroscopic Representation (EMR).
%
%
The second part of this article assumes once again a tutorial nature, in order to introduce the new Fast Modeling POG (FMPOG) procedure. The FMPOG, which operates in the POG framework, is a methodical step-by-step procedure that enables the readers 
to quickly derive the power-oriented graphical model of physical systems starting from their schematics. From the power-oriented graphical model, the state-space model can then be directly determined. 
To ensure the FMPOG procedure is easily usable by the entire community, we apply it to three examples in different energetic domains in this article, guiding the reader step-by-step through the derivation of the physical systems models. 
 A freely available Matlab/Simulink program is provided in a repository, allowing the users to automatically apply the FMPOG procedure to various classes of physical systems. This program allows to convert the physical systems schematics into the corresponding POG block schemes and, ultimately, into the state-space mathematical models.
\end{summary}
BG has also found academic applications to provide students with a graphical tool for modeling systems in different energetic domains, such as the dynamics of a sink or simple electrical networks~\cite{Rosenberg70BG}. Later, BG was applied to electrical machines and power engineering~\cite{Sirivadha83BG}, as well as to magnetic circuits~\cite{Fraisse95BG} and hydraulic systems~\cite{Kuang99BG}. Nowadays, BG is also used with interconnections~\cite{LCSS_New_5}, in order to model multi-physics systems such as vehicle dynamics~\cite{Benmoussa14BG}, robotic systems~\cite{Kumar23BG} or even biomolecular systems~\cite{Gawthrop17BG}. In the 90s, POG came into play~\cite{zanasi1991}. Over the years, it has been employed for modeling a large variety of physical systems providing a valid alternative to BG thanks to the more intuitive graphical construction. Among the multi-physics systems modeled using the POG technique are automotive systems~\cite{TebaldiZanasiCDC23,Nostro_1}, electrohydraulic 
\begin{pullquote}
Graphical modeling techniques introduce a unified modeling framework which, unlike purely analytical tools, 
also offers effective and energy-based graphical descriptions of physical systems.
\end{pullquote}
systems~\cite{zanasi2010}, electrical machines such as PMSMs~\cite{Nostro_4}, power converters~\cite{Nostro_2}, and hybrid electric vehicles~\cite{Nostro_3,Tebaldi2021ECVT}, showing the versatility of the POG technique. EMR was lastly invented at the beginning of this millennium~\cite{Bouscayrol_2000}. It adopts a different symbolism with respect to BG and POG, and can be employed for the modeling of multi-physic systems as well, although they have been mostly employed for electromechanical systems such as electric machines~\cite{Chen10EMR}, plenetary gear sets~\cite{Lhomme17EMR} or entire hybrid electric vehicles~\cite{Li24EMR}.

In this article, we propose a new Fast Modeling Power-Oriented Graphs (FMPOG) step-by-step modeling procedure in the framework of the POG technique.
The latter offers all the functionalities of the standard POG technique, including its applicability to different physical systems in different energetic domains. Additionally, FMPOG has the functionality of being methodical, since it consists in the application of a series of well-defined modeling rules following a step-by-step guided procedure. 

The fundamental concepts of the POG technique are first described in this article, by also providing several modeling case studies of different complexity. These case studies are designed to cover all the main advantages of the POG technique, which include POG block schemes directly implementable in the Simulink environment, the functionality of reading the POG state-space model directly from the POG block scheme, and the 
potential to apply
\begin{pullquote}
The POG is a graphical modeling technique adopting an energetic approach. Physical systems are modeled using a modular structure based on two main blocks: Elaboration Blocks, modeling dynamic and static elements, and Connection Blocks, modeling energy conversions. 
\end{pullquote}
the so-called congruent transformations for model reduction. The mathematical model calculation is a functionality that simulators do not generally provide: the user can simulate the system schematic, which is typically obtained by dragging and connecting the different physical elements together, but this does not provide the user with the system dynamic model/equations. For simple systems, the state-space equations may be read from the schematic; however, as the number of physical elements increases,
reading the state-space equations directly from the schematic may become very complex and easily subject to mistakes. Nevertheless, having the system dynamic model is highly desired, since it enables a control engineer to develop a proper control strategy for the system under consideration. Furthermore, the direct attainment of the state-space model starting from the graphical model is a functionality of the POG technique that is not available for the BG and EMR techniques. 
In this article, the new FMPOG procedure is proposed and applied to three examples in different energetic domains. We provide the general rules and the step-by-step procedure to be followed by the readers interested in modeling multi-physics systems, and we also provide a Matlab/Simulink program to convert the physical systems schematic into the corresponding POG block scheme and, ultimately, into the state-space mathematical model.

\section{Fundamentals}\label{Fundamentals_sect}
The POG is a graphical modeling technique adopting an energetic approach. Physical systems are modeled using a modular structure based on two main blocks: Elaboration Blocks (EBs), modeling dynamic and static elements, and Connection Blocks (CBs), modeling energy conversions. 
The main concepts which are employed are those of energy and power variables, whose description in the different energetic domains is given in the Section ``Energetic Domains and Physical Elements'' together with the description of the different physical elements. 
%
%
The elaboration and the connection blocks interact with each other and with the external world through the system energetic ports, also called power sections, as described in the Section ``Modular Structure of the POG''. Depending on the type of connection characterizing the Physical Elements (PEs), which can either be of the series or parallel type, a proper generalization of the Kirchhoff’s laws can be applied, as further discussed in the Section ``Series and Parallel Connections''. By combining together elaboration and connection blocks,
the so-called POG block scheme is obtained. 
Once the POG block scheme is build, it is always possible to determine the so-called POG state-space model 
in the linear case~\cite{DescriptorStSp2024}, as illustrated in Setion ``POG State-Space Model''. One of the most appealing features of such POG state-space model, as opposed to the classical POG state-space representation, is its suitability for the application of state-space transformation for model reduction, as further discussed in Section ``Congruent State-Space Transformations and
Model Reduction''.

\subsection{Energetic Domains and Physical Elements}\label{Energetic_Domains_sect}
When modeling physical systems, the main energetic domains are the electrical domain, the mechanical translational domain, the mechanical rotational domain, and the hydraulic domain, as enlisted in Table~\ref{Ambiti_Energetici}.
\begin{table*}[t!]
  \centering \scriptsize
  \caption{The dynamic elements $\mygreen D_{e}$ and $\myred D_{f}$, the static element $\R$, the energy variables $\mygreen q_{e}$ and $\myred
q_{f}$, and the power variables $\mygreen v_{e}$ and $\myred v_{f}$ in the electrical, mechanical translational, mechanical rotational, and hydraulic energetic domains.\\[-3mm]}\label{Ambiti_Energetici}
\normalsize
\centering
\begin{tabular}{|c|c|c|c|c|} \hline
& \bf{Electrical}  & \bf{Mechanical Translational} &
\bf{Mechanical Rotational} &
 \bf{Hydraulic} \\ 
  \hline
  \hline
 \mygreen $D_e$ & Capacitor $C$ & Mass $M$ & Inertia $J$ &
 Hyrdaulic Capacitor  $C_I$ \\
\hline
 \mygreen $q_e$ & Charge \mygreen $Q$ & Momentum \mygreen $p$ & Angular Momentum \mygreen $p$ & Volume \mygreen $V$
  \\   \hline

 \mygreen $v_e$ & \mygreen Voltage $V$ & \mygreen Speed $v$ & \mygreen Angular Speed $\omega$
 & \mygreen Pressure $P$
  \\   \hline
  \hline
 \myred $D_f$ & Inductor $L$ & Spring $E$ &
 Rotational Spring $E_r$ & Hydraulic Inductor $L_I$
  \\   \hline
  \myred $q_f$ & Flux \myred $\phi$ & Displacement \myred $x$ & Angular Displacement \myred $\theta$ & Hydraulic Flux \myred $\phi_I$
  \\   \hline
 \myred $v_f$ & \myred Current $I$ & \myred Force $F$ & \myred Torque $\tau$
 & \myred Volume Flow Rate $Q$
  \\   \hline
  \hline
 $R$ & Resistor $R$ & Friction $b$ &
 Angular Friction $d$ & Hydraulic Resistor $R_I$
  \\   \hline
\end{tabular}
\vspace{-2.6mm}
\end{table*}
 Each energetic domain is characterized by three different types of physical elements: two dynamic elements $\mygreen D_{e}$ and $\myred D_{f}$, storing the energy in the system, and a static element $\cR$, dissipating the energy. 
 %
 
\begin{sidebar}{First Case Study: A DC Motor Driving an Hydraulic Pump}

 \setcounter{sequation}{0}
\renewcommand{\thesequation}{S\arabic{sequation}}
\setcounter{stable}{0}
\renewcommand{\thestable}{S\arabic{stable}}
\setcounter{sfigure}{0}
\renewcommand{\thesfigure}{S\arabic{sfigure}}

Figure~\ref{pompa_elettro_idraulica} provides a schematic representation of the first case study, consisting in a DC motor driving an hydraulic pump~\cite{zanasi2010}. 

\sdbarfig{  
 \centering \footnotesize
 \setlength{\unitlength}{2.5mm}
 \psset{unit=\unitlength}
\begin{pspicture}(-14,-10)(14,7) 
  \psfrag{Vl}[rt][rt]{}
  \psfrag{Y}[rt][rt]{$\ts\omega_{2}$}
  \psfrag{DC}[c]{\footnotesize }
  \psfrag{Mot}[b][b]{\scriptsize DC Motor}
  \psfrag{or}[b][b]{\scriptsize }
  \psfrag{Pp}[c]{\scriptsize Pump}
  \psfrag{Ft}[r][r]{\scriptsize Filter}
  \psfrag{Tk}[bc]{\scriptsize Tank}
  \psfrag{Ac}[r][r]{\scriptsize Accumulator }
  \psfrag{R}{}
  \psfrag{T}{}
  \psfrag{Va}{$V_a$}
  \psfrag{La}{$L_1$}
  \psfrag{Ra}{$R_1$}
  \psfrag{Ia}{$I_a$}
  \psfrag{Jm}[r][r]{\footnotesize $J_2$}
  \psfrag{Jp}{$ $}
  \psfrag{bp}{$b_2$}
  \psfrag{Ar}{}
  \psfrag{Pl}{$P_b$ $\hspace{0.025cm}C_3$}
  \psfrag{Ql}{$Q_b$}
  \psfrag{Qu}{}
  \psfrag{At}{}
  \psfrag{Qp}{$ $}
  \psfrag{Qa}{$ $}
  \psfrag{V0}{$ $}
\rput(-1,0){
  \includegraphics[width=7.25cm]{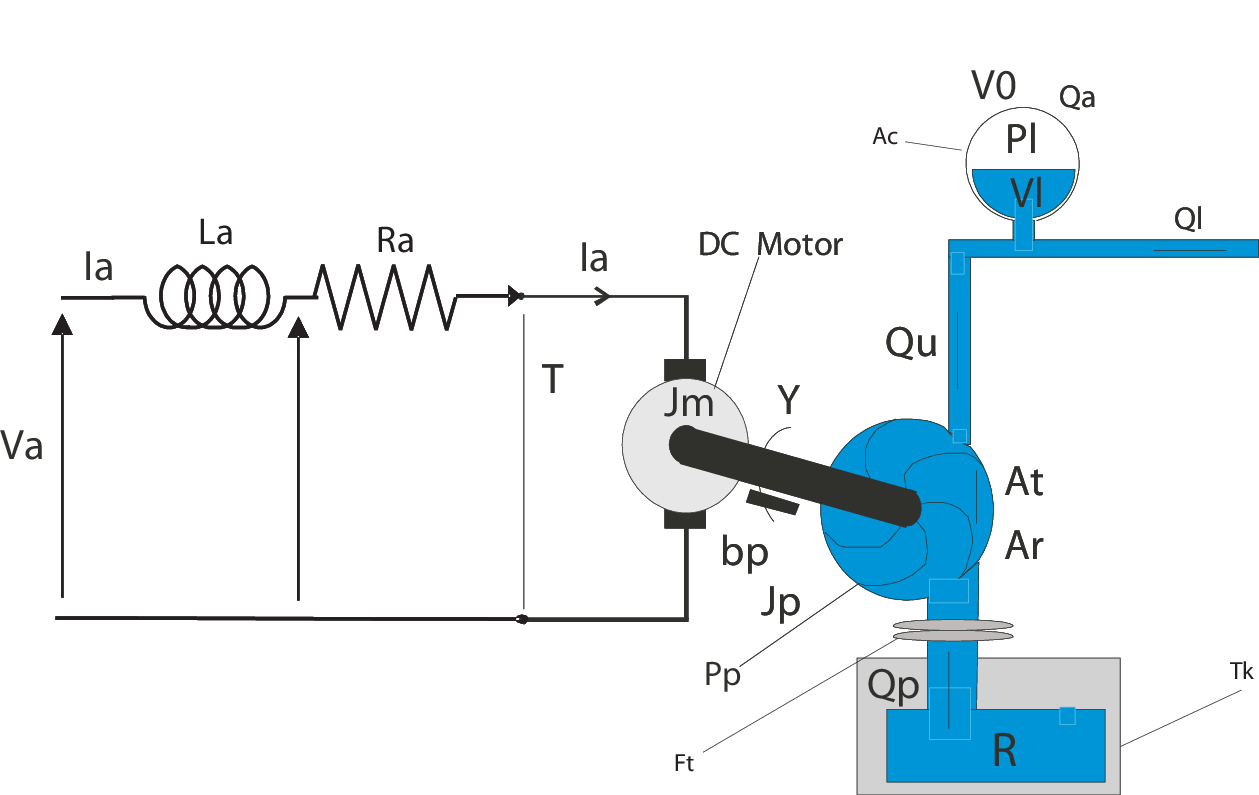}}  
  \psline[linestyle=dashed,linewidth=0.4pt,linecolor=red](-13.9,-6)(-13.9,3)
  \psline[linestyle=dashed,linewidth=0.4pt,linecolor=red](-8.5,-6)(-8.5,3)
  \psline[linestyle=dashed,linewidth=0.4pt,linecolor=red](-3.3,-6)(-3.3,3)
  \psline[linestyle=dashed,linewidth=0.4pt,linecolor=red](1.25,-4.5)(1.25,1)
  \psline[linestyle=dashed,linewidth=0.4pt,linecolor=red](6.6,-8)(6.6,0)
  \psline[linestyle=dashed,linewidth=0.4pt,linecolor=red](5.5,1.5)(8,1.5)
  \psline[linestyle=dashed,linewidth=0.4pt,linecolor=red](10.75,4.5)(10.75,2)
  \pscircle[linewidth=0.4pt](-14,-7){1.25mm}
  \rput(-14,-7){\scriptsize $1$}
  \pscircle[linewidth=0.4pt](-8.5,-7){1.25mm}
  \rput(-8.5,-7){\scriptsize $2$}
  \pscircle[linewidth=0.4pt](-4,-7){1.25mm}
  \rput(-4,-7){\scriptsize $3$}
  \pscircle[linewidth=0.4pt](-2.6,-7){1.25mm}
  \rput(-2.6,-7){\scriptsize $4$}
  \pscircle[linewidth=0.4pt](1.25,-5.5){1.25mm}
  \rput(1.25,-5.5){\scriptsize $5$}
  \pscircle[linewidth=0.4pt](5.95,-6.5){1.25mm}
  \rput(5.95,-6.5){\scriptsize $6$}
  \pscircle[linewidth=0.4pt](7.2,-6.5){1.25mm}
  \rput(7.2,-6.5){\scriptsize $7$}
  \pscircle[linewidth=0.4pt](9,1.5){1.25mm}
  \rput(9,1.5){\scriptsize $8$}
  \pscircle[linewidth=0.4pt](10.75,1){1.25mm}
  \rput(10.75,1){\scriptsize $9$}
  %
\end{pspicture}
   }{
    First Case Study: A DC motor driving an hydraulic pump. The electrical input power of the DC motor is given by $V_a\,I_a$. After some dissipation associated with the electrical resistance $R_1$, the input power is converted into 
    mechanical power, in order to drive the rotor inertia $J_2$ subject to the friction coefficient $b_2$. The rotor of the DC motor is connected to an hydraulic pump, which pumps a volume flow rate $Q_3$ 
    from a tank into an hydrulic capacitor $C_3$ (accumulator). The hydraulic output power flux is given by $P_b\,Q_b$.
    \label{pompa_elettro_idraulica}
     }


The electrial part of the DC motor is composed of the stator inductor $L_1$ and stator resistance $R_1$, respectively. The torque constant 
is then responsible for the power conversion from the electrical to the mechanical rotational domains, and viceversa. The electrical input power $V_a\,I_a$ is therefore converted to the mechanical domain, after some dissipation associated with the electrical resistance $R_1$, while the mechanical dissipations are represented by the rotor friction coefficient $b_2$. The hydraulic pump is 
responsible for converting the rotor movement, at angular speed $\omega_2$, into a volume flow rate $Q_3$. Accounting for some dissipation which occurs in the hydraulic domain due to the hydraulic resistance $R_3$, the volume flow rate 
is pumped from a tank into an hydrulic capacitor $C_3$, which is denoted as Accumulator in Figure~\ref{pompa_elettro_idraulica}. The hydraulic output power is given by $P_b\,Q_b$, where $P_b$ is the pressure within the hydraulic capacitor $C_3$. The physical system in Figure~\ref{pompa_elettro_idraulica} can be modeled using the POG block scheme of Figure~\ref{pompa_elettro_idraulica_POG}. The first two elaboration blocks between power sections  \Circled{\scriptsize 1} - \Circled{\scriptsize 2} and \Circled{\scriptsize 2} - \Circled{\scriptsize 3}
in the POG block scheme, describing the stator inductor $L_1$ and the resistor $R_1$, are connected in series. Indeed, the respective summation nodes apply the following generalized VKL: 
${V''_a}\black={V_a}\black-{V'_a}$ and ${V'_a}\black={V_1}\black+{V_{R_1}}$.
The first connection block between power sections
\Circled{\scriptsize 3} - \Circled{\scriptsize 4}
describes the electrical/mechanical rotational power conversion associated with the torque constant $K_{12}$ of the DC motor.
The third and fourth elaboration blocks between power sections 
\Circled{\scriptsize 4} - \Circled{\scriptsize 5} and \Circled{\scriptsize 5} - \Circled{\scriptsize 6}
in the POG block scheme of Figure~\ref{pompa_elettro_idraulica_POG}, describing the rotor inertia $J_2$ and the friction coefficient $b_2$, are connected in parallel. Indeed, the respective summation nodes apply the following generalized CKL: 
${\tau''_2}\black={\tau_2}\black-{\tau'_2}$ and ${\tau'_2}\black={\tau_3}\black+{\tau_{b_2}}$. The second connection block between power sections 
\Circled{\scriptsize 6} - \Circled{\scriptsize 7} 
describes the mechanical rotational/hydraulic power conversion associated with the hydraulic pump coefficient $K_{23}$. The fifth and sixth elaboration blocks between power sections 
\Circled{\scriptsize 7} - \Circled{\scriptsize 8} and \Circled{\scriptsize 8} - \Circled{\scriptsize 9}
in the POG block scheme, describing the hydraulic capacitor $C_3$ and the hydraulic resistor $R_3$, are also connected in parallel. Indeed, the respective summation nodes apply the following generalized { CKL}: 
${Q''_3}\black={Q_3}\black-{Q'_3}$ and ${Q'_3}\black={Q_b}\black+{Q_{R_3}}$.

\sdbarfig{  
\begin{pspicture}(0,0)(5,31)
\centering 
\schemaPOG{(-3,-3)(34,16)}{2.1mm}{
  \thicklines
 \bloin{ V_a}{ I_a}
 \bloivgiumxm{
  {\frac{1}{L_{1}}}}{
    { I_{1}}}{\dst}{\frac{1}{s}}{\scr\phi_{I_1}}
 \bloivsuaor{R_1}{}{}
 \blokxaor{2}{K_{12}\tras}{K_{12}}{ }{ \!\!\!\!\!\tau_2 }
 \bloivsumxm{
   {\frac{1}{J_{2}}}}{ 
   {\omega_{2}}}{\dst}{\frac{1}{s}}{\scr  p_2}
 \bloivgiuor{b_{2}}{}{}
 \blokxor{2}{K_{23}\tras}{K_{23}}{ \!\!\!\!\! Q_3}{ }
 \bloivgiumxm{
   {\frac{1}{C_{3}}}}{
   { P_3}}{\dst}{\frac{1}{s}}{\scr  V_3}
 \bloivsuaor{R_3}{ }{}
 \bloout{ Q_b}{ P_b}
 } 
\pscircle[linewidth=0.4pt](-34,0.56){1.25mm}
\rput(-34,0.56){\scriptsize $1$}
\pscircle[linewidth=0.4pt](-30,0.56){1.25mm}
\rput(-30,0.56){\scriptsize $2$}
\pscircle[linewidth=0.4pt](-26,0.56){1.25mm}
\rput(-26,0.56){\scriptsize $3$}
\pscircle[linewidth=0.4pt](-22,0.56){1.25mm}
\rput(-22,0.56){\scriptsize $4$}
\pscircle[linewidth=0.4pt](-18,0.56){1.25mm}
\rput(-18,0.56){\scriptsize $5$}
\pscircle[linewidth=0.4pt](-14,0.56){1.25mm}
\rput(-14,0.56){\scriptsize $6$}
\pscircle[linewidth=0.4pt](-10,0.56){1.25mm}
\rput(-10,0.56){\scriptsize $7$}
\pscircle[linewidth=0.4pt](-6,0.56){1.25mm}
\rput(-6,0.56){\scriptsize $8$}
\pscircle[linewidth=0.4pt](-2,0.56){1.25mm}
\rput(-2,0.56){\scriptsize $9$}
%
\rput(-31,12){\scriptsize  $V''_{a}$}
\rput(-30,14){\scriptsize  $V'_{a}$}
\rput(-26,14){\scriptsize  $V_1$}
\rput(-26.75,11){\scriptsize  $V_{R_1}$}
\rput(-19,4){\scriptsize  $\tau''_{2}$}
\rput(-18,2){\scriptsize  $\tau'_{2}$}
\rput(-14,2){\scriptsize  $\tau_{3}$}
\rput(-15,5){\scriptsize  $\tau_{b_2}$}
\rput(-7,12){\scriptsize  $Q''_{3}$}
\rput(-6,14){\scriptsize  $Q'_{3}$}
\rput(-2.5,11){\scriptsize $Q_{R_3}$}
%
%
\end{pspicture}
}{POG block scheme of the system in Figure~\ref{pompa_elettro_idraulica}. Starting from the left-hand side, the first two series-connected elaboration blocks describe the dynamics of the stator inductor $L_1$ and of the stator resistor $R_1$. The first connection block characterized by the torque constant $K_{12}$ describes the energy conversion between the electrical and mechanical rotational domains. The third and fourth parallel-connected elaboration blocks describe the dynamics of the rotor inertia $J_2$ and of its friction coefficient $b_2$. The second connection block characterized by the hydraulic pump coefficient $K_{23}$ describes the energy conversion between mechanical rotational and hydraulic domains. Finally, the fifth and sixth parallel-connected elaboration blocks describe the dynamics of the hydraulic capacitor $C_3$ and of the hydraulic resistor $R_3$. 
\label{pompa_elettro_idraulica_POG}
     }

Figure~\ref{pompa_elettro_idraulica} and Figure~\ref{pompa_elettro_idraulica_POG} also highlight the one-to-one correspondence between every single physical element $L_1$, $R_1$, $K_{12}$, $J_2$, $b_2$, $K_{23}$, $C_3$, $R_3$ in the physical system and the blocks between the corresponding pair of power sections \Circled{\scriptsize 1},$\ldots$,\Circled{\scriptsize 9} 
in the POG block scheme. The POG block scheme of Figure~\ref{pompa_elettro_idraulica_POG} is in a one-to-one correspondence with the following state-space model:
\begin{sequation}\label{stateSPompa}
  \begin{array}{@{}r@{}c@{}l@{}}
 \underbrace{
 \!\mat{@{}c@{}c@{}c@{}}{
 L_{1} & 0 & 0\\
 0 & 
 J_{2} & 0\\
 0 & 0 & 
 C_{3}}\!
 }_{\ds\L}
 \underbrace{\!
 \mat{@{}c@{}}{\dot{I}_{1}\\ \dot{\omega}_{2}\\ \dot{P}_{3}}\!
 }_{\ds\dot{\x}}
 &=&
 \underbrace{\!
 \mat{@{}c@{\;}c@{\;}c@{}}{
 -R_{1} & -K_{12} & 0\\
 K_{12} & -b_{2} & -K_{23}\\
 0 & K_{23} &
 -R_3\!
 }}_{\ds\A}
 \underbrace{\!
 \mat{@{}c@{}}{
 {I_{1}}\\ 
 \omega_{2}\\ 
 P_{3}}\!
 }_{\ds\x} \!+\!
 \underbrace{\!
 \mat{@{}c@{\;\,}c@{}}{
 1 & 0\\
 0 & 0\\
 0 & -1}\!
 }_{\ds\B}
 \underbrace{\!
 \mat{@{}c@{}}{V_{a}\\Q_{b}}\!
 }_{\ds\u},
\\[14mm]
 \underbrace{\!
 \mat{@{}c@{}}{I_{a}\\P_{b}}\!
 }_{\ds\y}
&=&
\underbrace{
\mat{@{}ccc@{}}{
1 & 0 & 0\\
0 & 0 & 1}
}_{\ds\C}
\x+
 \underbrace{\!
 \mat{@{}c@{\;\;}c@{}}{
 0 & 0\\
 0 & 0}\!
 }_{\ds\D}\,
 \underbrace{\!
 \mat{@{}c@{}}{V_{a}\\Q_{b}}\!
 }_{\ds\u},
  \end{array} \!\!\!\!\!\!\!\!\!\!
\end{sequation}
obtained by applying Property~\ref{To_StSp}. System~\eqref{stateSPompa} is in the POG state-space form $\S$ in~\eqref{POG_stsp}.

\end{sidebar}
As an example, the sidebar ``First Case Study: A DC Motor Driving an Hydraulic Pump'' involves physical elements in three different energetic domains, that are electrical, mechanical rotational and hydraulic. 
 Each energetic domain is also characterized by four different types of variables: two {\it energy variables} $\mygreen q_{e}$ and $\myred q_{f}$, which are used to define the amount of energy stored within the
dynamic elements, and two {\it power variables} $\mygreen v_{e}$ and $\myred v_{f}$, which are used to define the energy movement within the system. 
%
%
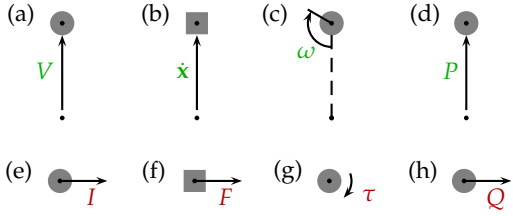
\begin{figure}[t!]
   \setlength{\unitlength}{2.1mm}
  \psset{unit=1.0\unitlength}
\begin{center}
  \hspace{-5mm}
   \begin{pspicture}(-4,0)(3,7.5)
 \pscircle[fillstyle=solid,fillcolor=gray,linecolor=gray](0,6){0.7}
 \psline{->}(0,0.5)(0,5.25)
 \pscircle*(0,6){0.15}
 \pscircle*(0,0){0.15}
 \rput[r](-0.5,3){$\mygreen V$}
\rput(-2.6,6.5){(a)}
 \end{pspicture}
 \hspace{1.0mm}
 \begin{pspicture}(-4,0)(3,7.5)
 \rput(0,6){\psframe[fillstyle=solid,fillcolor=gray,linecolor=gray](-0.7,-0.7)(0.7,0.7)}
 \pscircle*(0,0){0.15}
 \pscircle*(0,6){0.15}
 \rput[r](-0.5,3){$\mygreen \dot\x$}
 \psline{->}(0,0.5)(0,5.25)
\rput(-2.6,6.5){(b)}
 \end{pspicture}
 \hspace{1.0mm}
 \begin{pspicture}(-4,0)(3,7.5)
 \pscircle*(0,0){0.15}
 \rput(0,3.0){
 \pscircle[fillstyle=solid,fillcolor=gray,linecolor=gray](0,3){0.7}
 \pscircle*(0,3){0.15}
 \pscircle*(0,-3){0.15}
 \psarc{<-}(0,3){1.5}{150}{-90}
 \psline[linewidth=0.8pt,linestyle=dashed]{-}(0,-3)(0,2.25)
 \rput(0,3){\psline[linewidth=0.8pt]{-}(0,0)(1.75;150)}
 \rput[rt](-1,1.5){$\mygreen \omega$}
 }
 \rput(-3.6,6.5){(c)}
 \end{pspicture}
 \hspace{1.0mm}
 \begin{pspicture}(-4,0)(3,7.5)
 \pscircle[fillstyle=solid,fillcolor=gray,linecolor=gray](0,6){0.7}
 \pscircle*(0,0){0.15}
 \pscircle*(0,6){0.15}
 \rput[r](-0.5,3){$\mygreen P$}
 \psline{->}(0,0.5)(0,5.25)
 \rput(-2.6,6.5){(d)}
 \end{pspicture} 
 \\
 \hspace{7mm}
 \begin{picture}(7,5)(-1,-1)
 \pscircle[fillstyle=solid,fillcolor=gray,linecolor=gray](0,0){0.7}
 \pscircle*(0,0){0.15}
 \psline{->}(0,0)(3,0)
 \rput[t](2.0,-0.5){$\myred I$}
 \rput(-2.6,0.6){(e)}
 \end{picture}
 \hspace{1mm}
 \begin{picture}(7,5)(-1,-1)
 \psframe[fillstyle=solid,fillcolor=gray,linecolor=gray](-0.7,-0.7)(0.7,0.7)
 \pscircle*(0,0){0.15}
 \psline{->}(0,0)(3,0)
 \rput[t](2.0,-0.5){$\myred F$}
 \rput(-2.6,0.6){(f)}
 \end{picture}
 \hspace{1mm}
 \begin{picture}(7,5)(-1,-1)
 \pscircle[fillstyle=solid,fillcolor=gray,linecolor=gray](0,0){0.7}
 \pscircle*(0,0){0.15}
 \psarc{<-}(0,0){1.4}{-50}{20}
 \rput[t](2.5,-0.5){$\myred \tau$}
 \rput(-2.6,0.6){(g)}
 \end{picture}
 \hspace{1mm}
 \begin{picture}(7,5)(-1,-1)
 \pscircle[fillstyle=solid,fillcolor=gray,linecolor=gray](0,0){0.7}
 \pscircle*(0,0){0.15}
 \psline{->}(0,0)(3,0)
 \rput[t](2.0,-0.5){$\myred Q$}
 \rput(-2.6,0.6){(h)}
 \end{picture}
 \end{center}
 \caption{From (a) to (d): the voltage $\mygreen V$, the velocity $\mygreen \dot\x$, the angular velocity $\mygreen \omega$, and the pressure $\mygreen P$ are defined between two points of the space. From (e) to (h): the current $\myred I$, the force $\myred F$, the torque $\myred \tau$ and the volume flow rate $\myred Q$ are defined at each point of the space.}\label{variables_def_in_space}
\end{figure}
 The power variables $\mygreen v_{e}$ in Table~\ref{Ambiti_Energetici} are called {\it across} power variables because they are defined between two points of the space, as shown in Figure~\ref{variables_def_in_space}: these variables coincide with the {\it effort} power variables defined in the BG technique as far as the electrical and hydraulic domains are concerned, and with the {\it flow} power variables defined in the BG technique as far as the mechanical translational and rotational domains are concerned~\cite{Gawthrop2007}. The power variables $\myred v_{f}$ in Table~\ref{Ambiti_Energetici} are called {\it through} power variables
because they are defined at each point of the space, as shown in Figure~\ref{variables_def_in_space}: these variables coincide with the {\it flow} power variables defined in the BG technique as far as the electrical and hydraulic domains are concerned, and with the {\it effort} power variables defined in the BG technique as far as the mechanical translational and rotational domains are concerned \cite{Gawthrop2007}. 

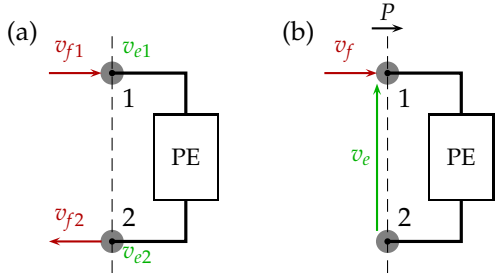
\begin{figure}[t!]
   \setlength{\unitlength}{2.8mm}
  \psset{unit=1.0\unitlength}
 \hspace{10mm}
 \begin{pspicture}(-3,-9.5)(5,2.5)
 \pscircle[fillstyle=solid,fillcolor=gray,linecolor=gray](0,0){0.5}
 \pscircle[fillstyle=solid,fillcolor=gray,linecolor=gray](0,-8){0.5}
 \pscircle*(0,0){0.15}
 \psline[linewidth=1.2pt]{-}(0,0)(3.5,0)(3.5,-2)
 \rput[t](3.5,-2){\framebox(3,4){PE}}
 \psline[linewidth=1.2pt]{-}(3.5,-6)(3.5,-8)(0,-8)
 \pscircle*(0,-8){0.15}
 \psline[linecolor=myred]{->}(-3,0)(-0.5,0)
 \rput[b](-2,0.5){$\myred v_{f1}$} %
 \psline[linecolor=myred]{<-}(-3,-8)(-0.5,-8)
 \rput[b](-2,-7.5){$\myred v_{f2}$} %
 %
 \rput[lb](0.5, 0.75){$\mygreen v_{e1}$}
 \rput[lt](0.5,-0.75){\large $1$}
 \rput[lb](0.5,-7.25){\large $2$}
 \rput[lt](0.5,-8.25){$\mygreen v_{e2}$}
 \psline[linewidth=0.4pt,linestyle=dashed]{-}(0,-9.5)(0,1.75)
 \rput[lt](-5,2.5){\large  (a)}
 \end{pspicture}
 \hspace{12mm}
 %
 %
 \begin{pspicture}(-3,-9.5)(5,2.5)
 \pscircle[fillstyle=solid,fillcolor=gray,linecolor=gray](0,0){0.5}
 \pscircle[fillstyle=solid,fillcolor=gray,linecolor=gray](0,-8){0.5}
 \pscircle*(0,0){0.15}
 \psline[linewidth=1.2pt]{-}(0,0)(3.5,0)(3.5,-2)
 \rput[t](3.5,-2){\framebox(3,4){PE}}
 \psline[linewidth=1.2pt]{-}(3.5,-6)(3.5,-8)(0,-8)
 \pscircle*(0,-8){0.15}
 \psline[linewidth=0.8pt,linecolor=mygreen]{->}(-0.5,-7.5)(-0.5,-0.5)
 \rput[r](-0.75,-4){$\mygreen v_e$}
 \psline[linecolor=myred]{->}(-3,0)(-0.5,0)
 \rput[b](-2,0.5){$\myred v_f$} %
 \rput[b](0,2.5){$P$}
 \psline{->}(-0.75,2.125)(0.75,2.125)
 \rput[lt](0.5,-0.75){\large $1$}
 \rput[lb](0.5,-7.25){\large $2$}
 \psline[linewidth=0.4pt,linestyle=dashed]{-}(0,-9.5)(0,1.75)
 \rput[lt](-5,2.5){\large  (b)}
 \end{pspicture}
 \caption{Physical element (PE) representations. (a) The two terminals of the PE are characterized by the pairs of power variables ($\mygreen v_{e1}$, $\myred v_{f1}$) and ($\mygreen v_{e2}$, $\myred v_{f2}$). (b) An analogous representation of the physical element PE, where its interaction with other physical elements is described through the power section $P$ characterized by the power variables ($\mygreen v_e$, $\myred v_f$). The horizontal arrow located at the top of the power section indicates that the power $P$ is positive when entering the system.}\label{PE_represent}
\end{figure}

\begin{figure}[t!]
   \begin{center}
  \begin{tabular}{|@{}c@{}|@{\hspace{0.1cm}}c@{\hspace{0.1cm}}|@{\hspace{0.1cm}}c@{\hspace{0.1cm}}|@{\hspace{0.1cm}}c@{\hspace{0.1cm}}|@{\hspace{0.1cm}}c@{\hspace{1mm}}|}
  \hline
   &
  \bf  \small \multirow{2}{1.6cm}{\centering Electrical} &
  \bf  \small Mechanical  &
  \bf  \small Mechanical  &
  \bf \small \multirow{2}{1.6cm}{\centering Hydraulic} \\  
  &&&&\\[-4mm]
   &
  \bf  \small &
  \bf  \small Translational  &
  \bf  \small Rotational  &
  \bf \small \\   \hline
  &&&&\\[1mm]
      \setlength{\unitlength}{1.3mm}
  \psset{unit=1.0\unitlength}
   \begin{picture}(2,12)(-1,-1.5)
   \rput[c]{90}(-0.2,4){\scriptsize Physical Element $\mygreen \cD_{e}$}
  \end{picture}
  &
  \psfrag{V1}[r][r]{$\scr \mygreen V$}
  \psfrag{V3}[r][r]{$\scr \mygreen $}
  \psfrag{I1}[b][b]{$\scr \myred   I$}
  \psfrag{I3}[r][r]{$\scr \myred   $}
  \psfrag{C}[l][l]{$\scr C$}
   \includegraphics[clip,width=1.5cm]{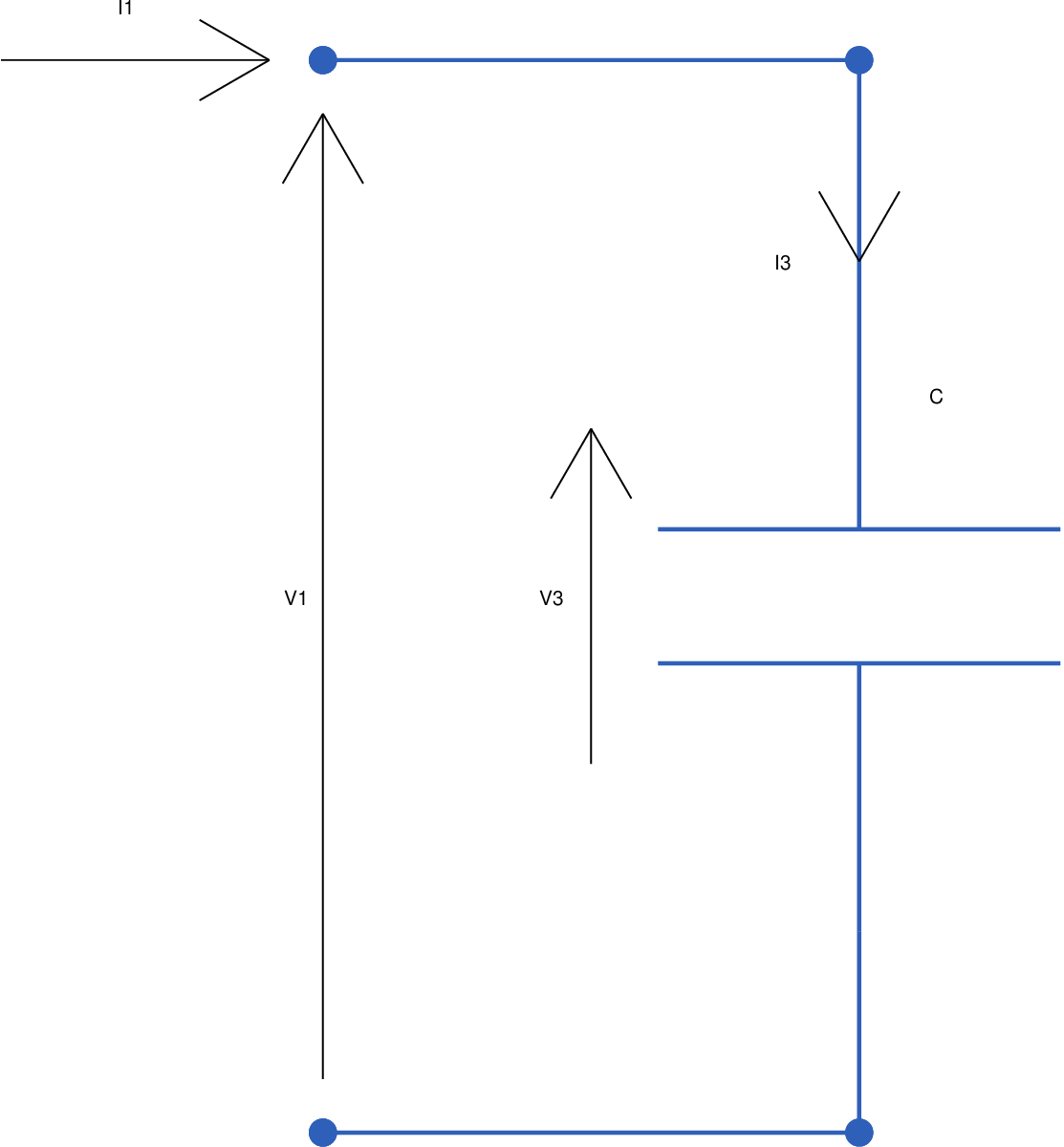}
  &
  \psfrag{v1}[r][r]{$\scr \mygreen v$}
  \psfrag{v3}[r][r]{$\scr \mygreen $}
  \psfrag{F1}[b][b]{$\scr \myred   F$}
  \psfrag{F3}[r][r]{$\scr \myred   $}
  \psfrag{M}[l][l]{$\scr M$}
  \includegraphics[clip,width=1.5cm]{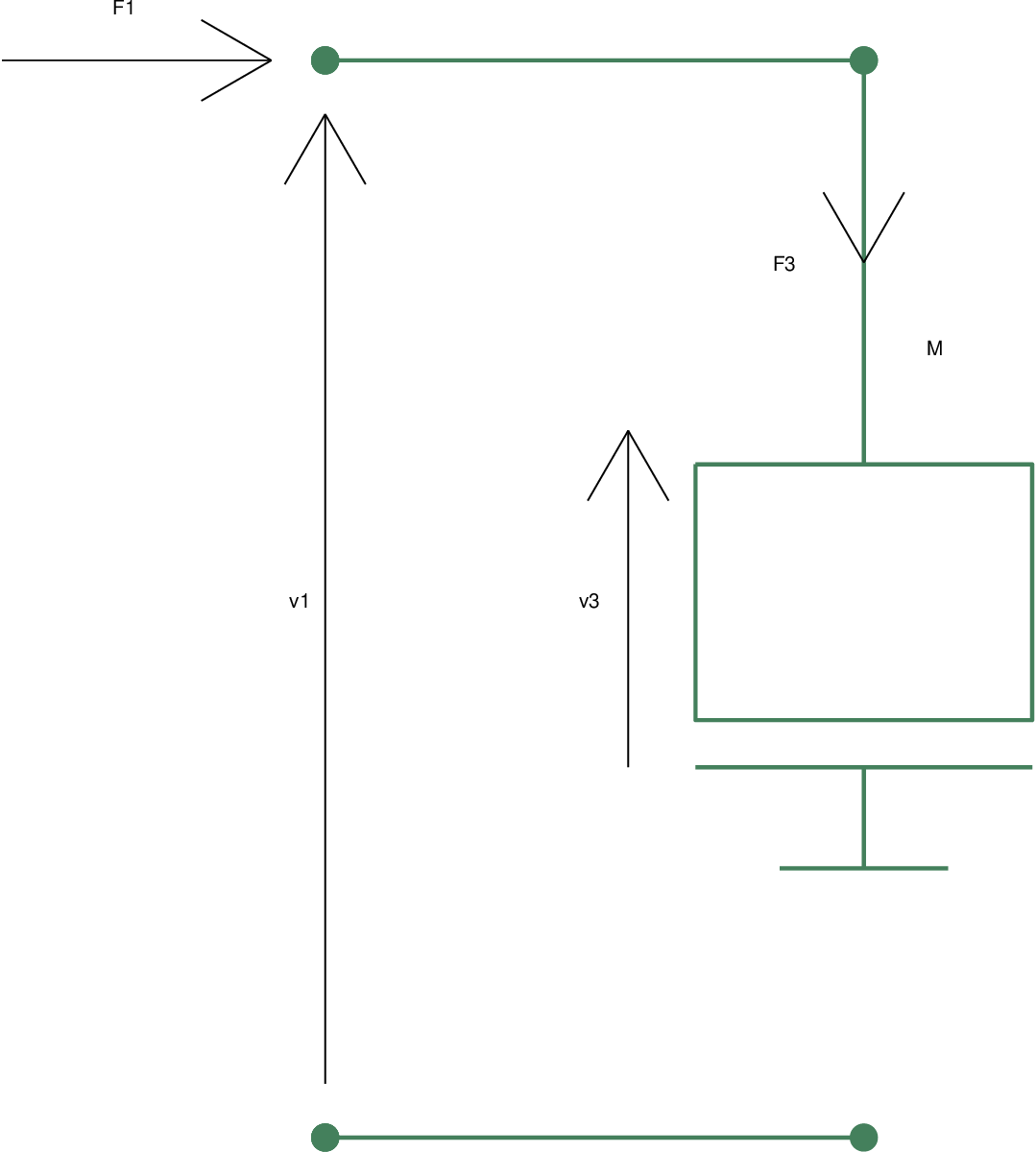}
  &
  \psfrag{w1}[r][r]{$\scr \mygreen \omega$}
  \psfrag{w3}[r][r]{$\scr \mygreen $}
  \psfrag{T1}[b][b]{$\scr \myred   \tau$}
  \psfrag{T3}[r][r]{$\scr \myred   $}
  \psfrag{J}[l][l]{$\scr J$}
  \includegraphics[clip,width=1.5cm]{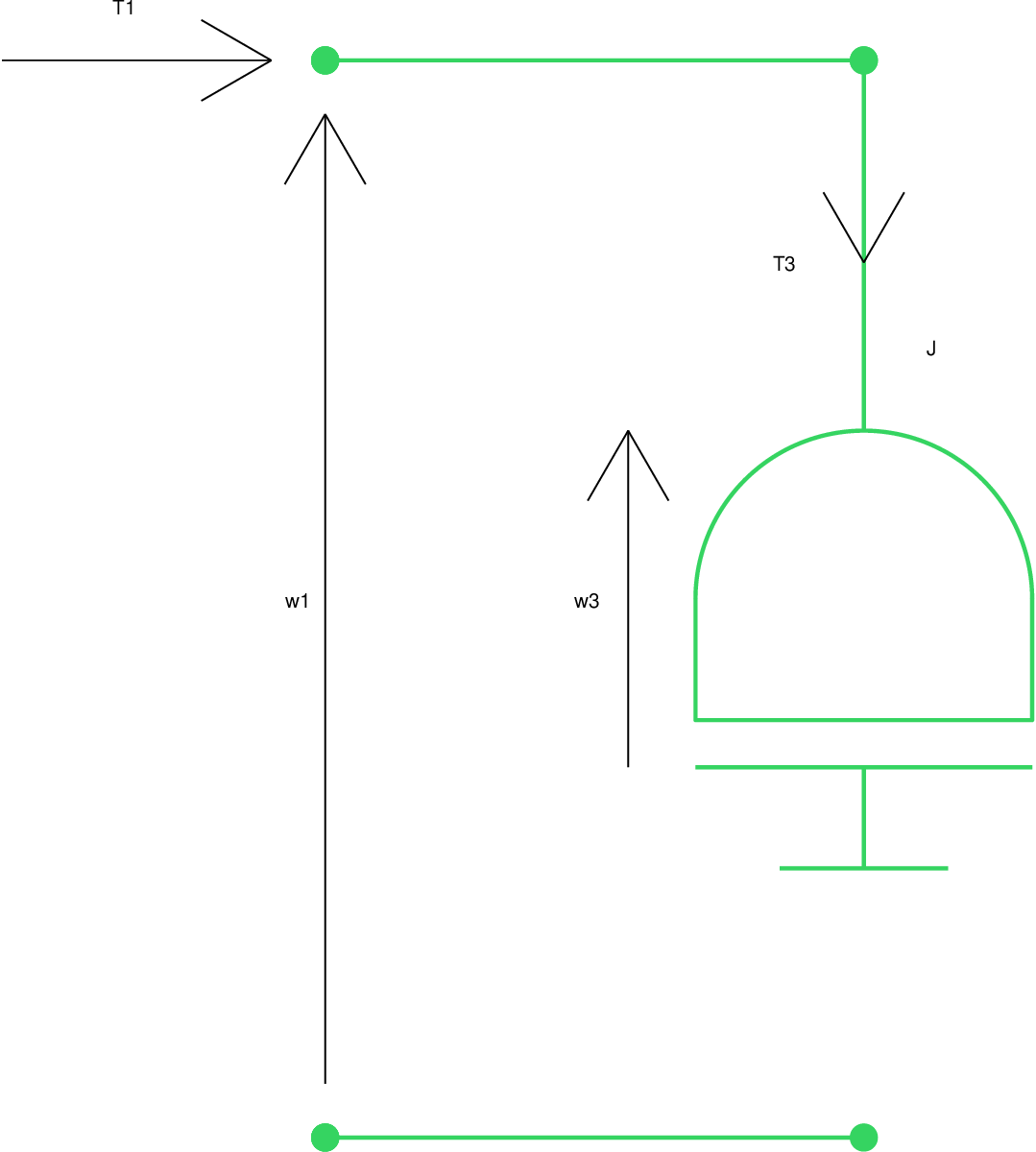}
   &
  \psfrag{P1}[r][r]{$\scr \mygreen P$}
  \psfrag{P3}[r][r]{$\scr \mygreen $}
  \psfrag{Q1}[b][b]{$\scr \myred   Q$}
  \psfrag{Q3}[r][r]{$\scr \myred   $}
  \psfrag{Ci}[l][l]{$\scr C_I$}
  \includegraphics[clip,width=1.5cm]{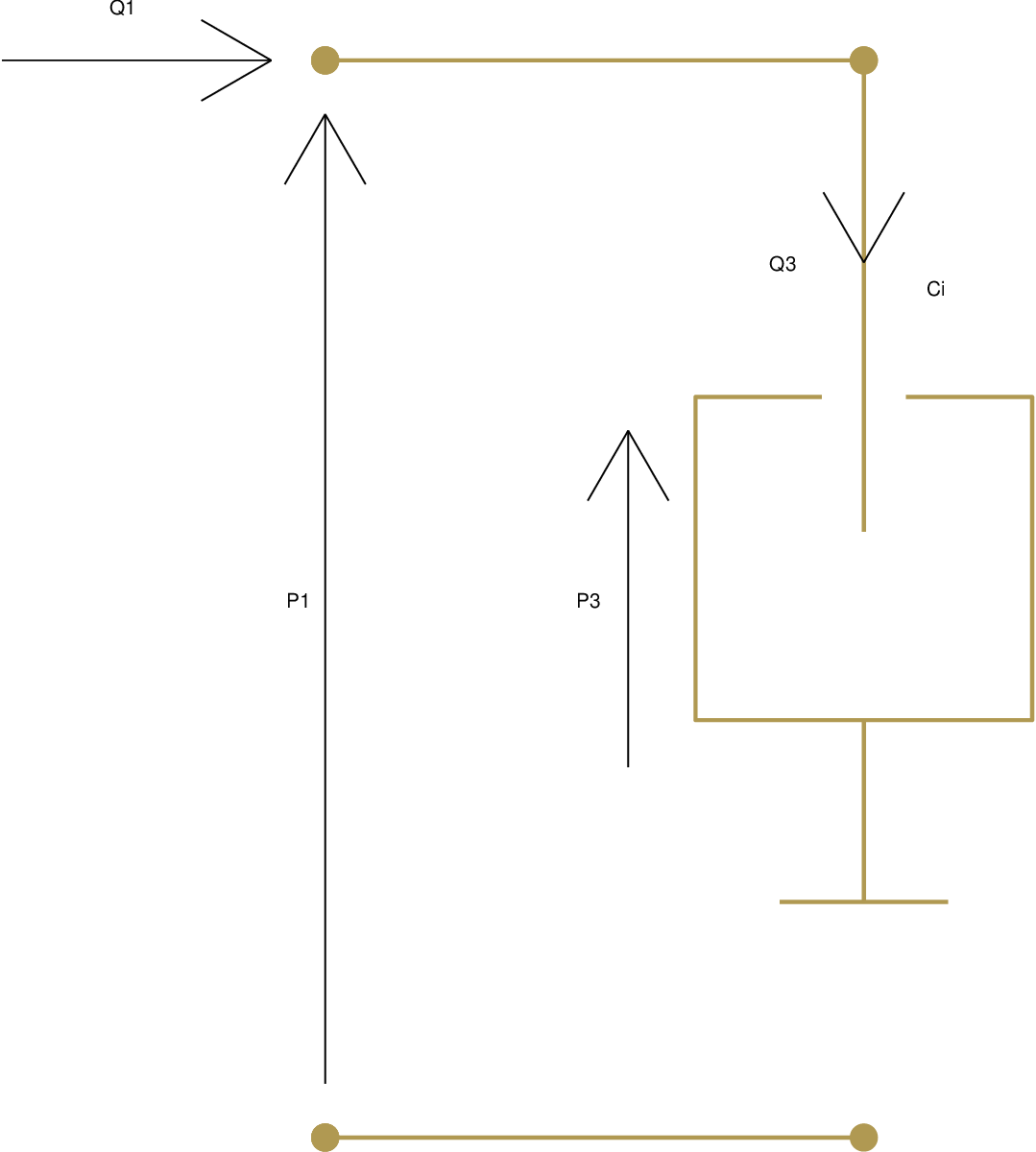}
    \\[1mm]
    &   \small   \multirow{2}{1.6cm}{\centering Capacitor}
    &   \small  \multirow{2}{1.6cm}{\centering Mass}
    &  \small   \multirow{2}{1.6cm}{\centering Inertia}
    & \small   Hydraulic 
    \\
    &   \small   
    &   \small  
    &  \small   
    & \small   Capacitor
  \\[2mm]
       \hline 
       \hline 
  &&&&\\[0mm]
  \setlength{\unitlength}{1.3mm}
  \psset{unit=1.0\unitlength}
   \begin{picture}(2,13)(-1,-1.5)
   \rput[c]{90}(-0.2,5){\scriptsize Physical Element $\myred \cD_{f}$}
  \end{picture}
  &
  \psfrag{V1}[r][r]{$\scr \mygreen V$}
  \psfrag{V3}[r][r]{$\scr \mygreen $}
  \psfrag{I1}[b][b]{$\scr \myred   I$}
  \psfrag{I3}[r][r]{$\scr \myred   $}
  \psfrag{L}[l][l]{$\scr L$}
  \includegraphics[clip,width=1.5cm]{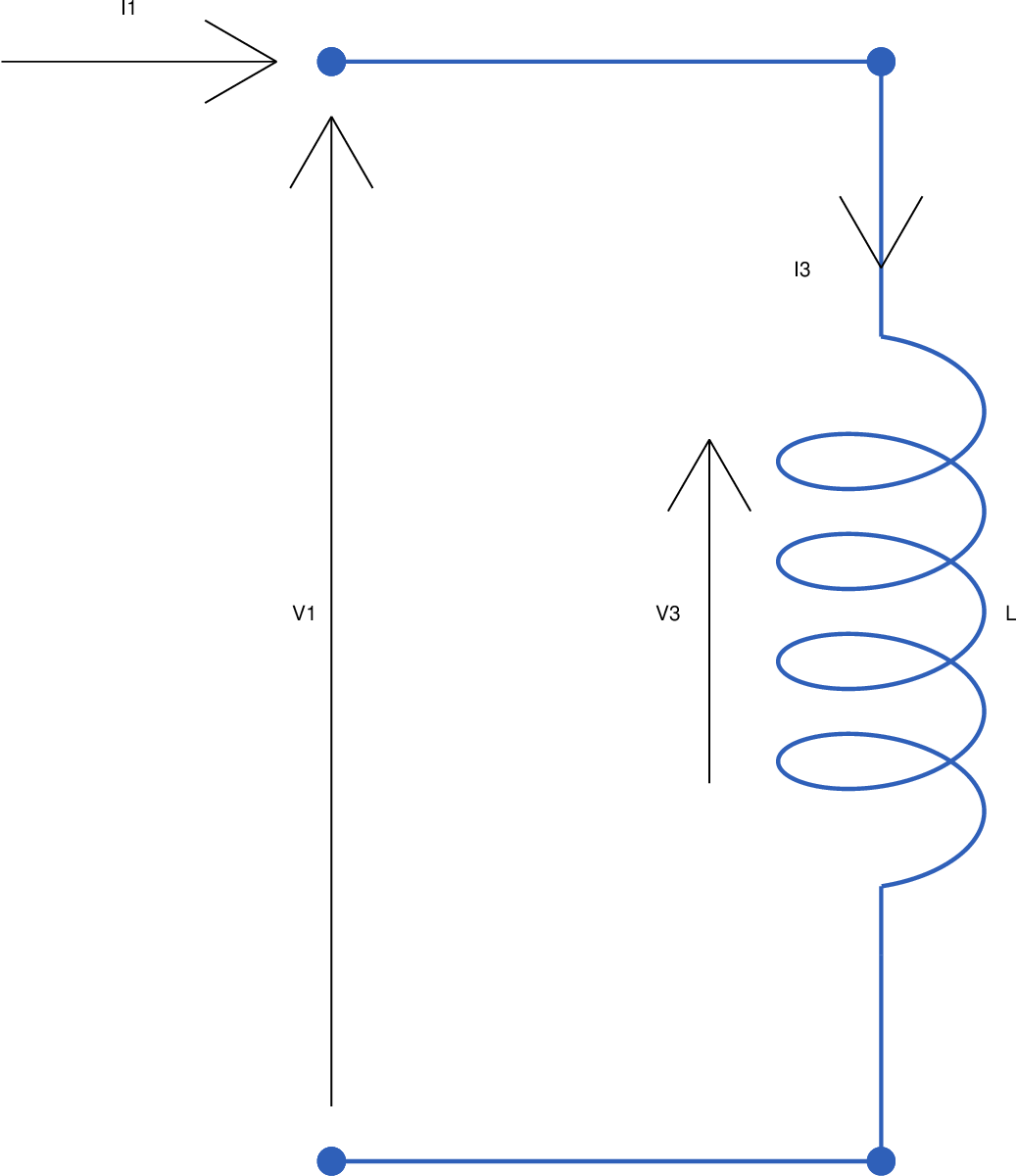}
   &
  \psfrag{v1}[r][r]{$\scr \mygreen v$}
  \psfrag{v3}[r][r]{$\scr \mygreen $}
  \psfrag{F1}[b][b]{$\scr \myred   F$}
  \psfrag{F3}[r][r]{$\scr \myred   $}
  \psfrag{K}[l][l]{$\scr E$}
  \psfrag{E}[l][l]{$\scr E$}
  \includegraphics[clip,width=1.5cm]{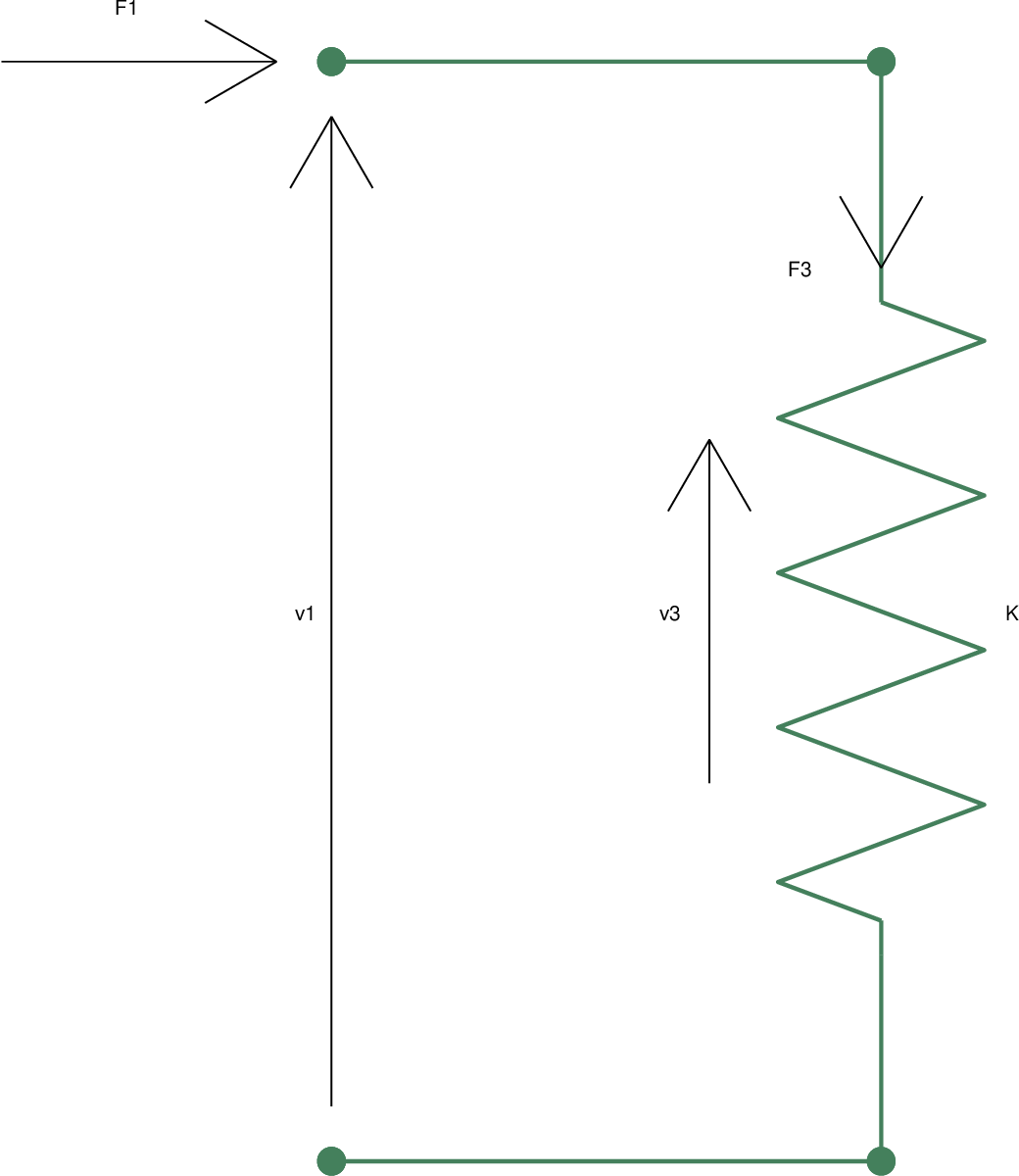}
   &
  \psfrag{w1}[r][r]{$\scr \mygreen \omega$}
  \psfrag{w3}[r][r]{$\scr \mygreen $}
  \psfrag{T1}[b][b]{$\scr \myred   \tau$}
  \psfrag{T3}[r][r]{$\scr \myred   $}
  \psfrag{K}[l][l]{$\!\scr  E_r$}
  \psfrag{Et}[l][l]{$\scr E_r$}
  \includegraphics[clip,width=1.5cm]{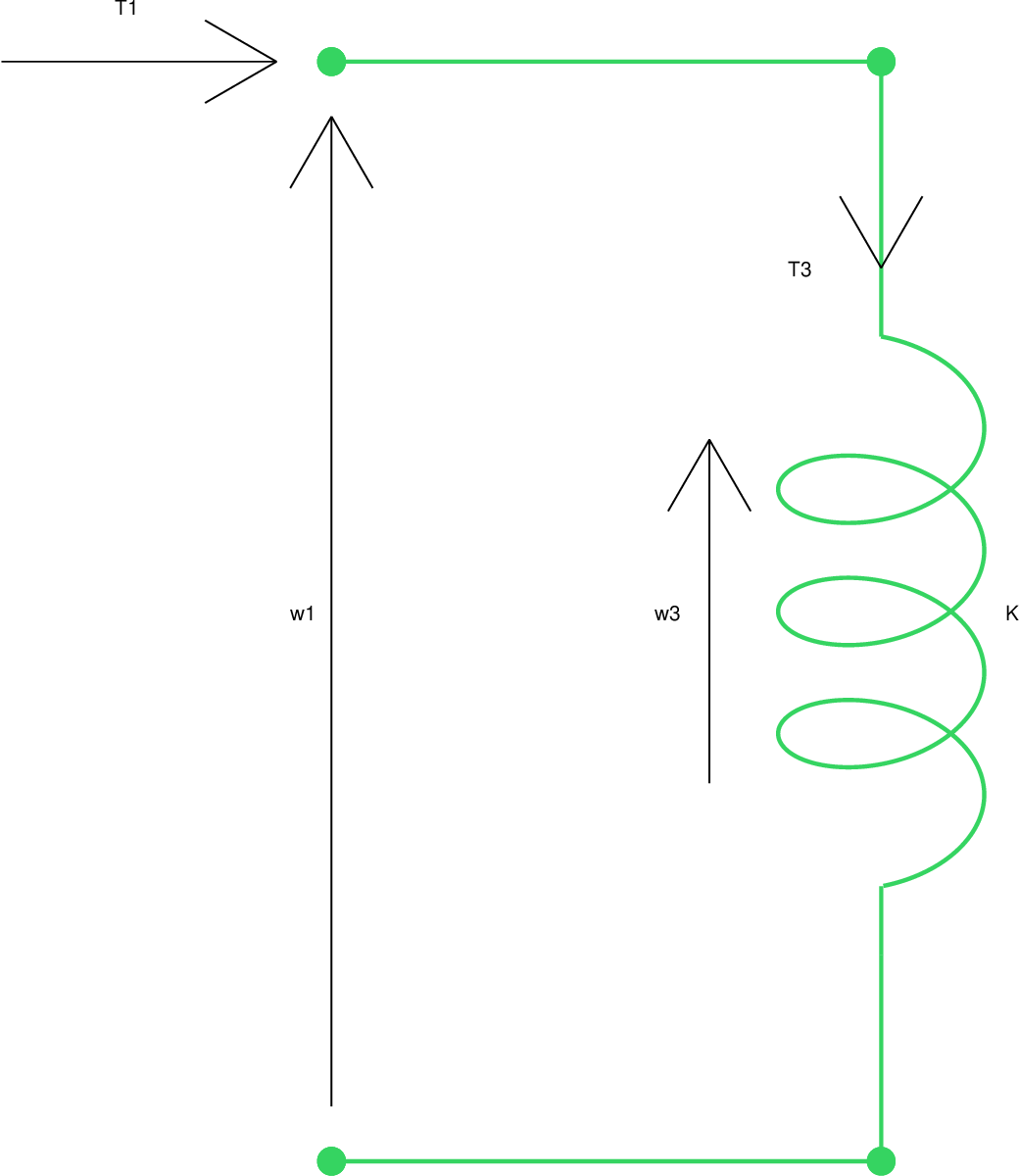}
   &
  \psfrag{P1}[r][r]{$\scr \mygreen P$}
  \psfrag{P3}[r][r]{$\scr \mygreen $}
  \psfrag{Q1}[b][b]{$\scr \myred   Q$}
  \psfrag{Q3}[r][r]{$\scr \myred   $}
  \psfrag{Li}[l][l]{$\!\scr L_{I}$}
  \includegraphics[clip,width=1.5cm]{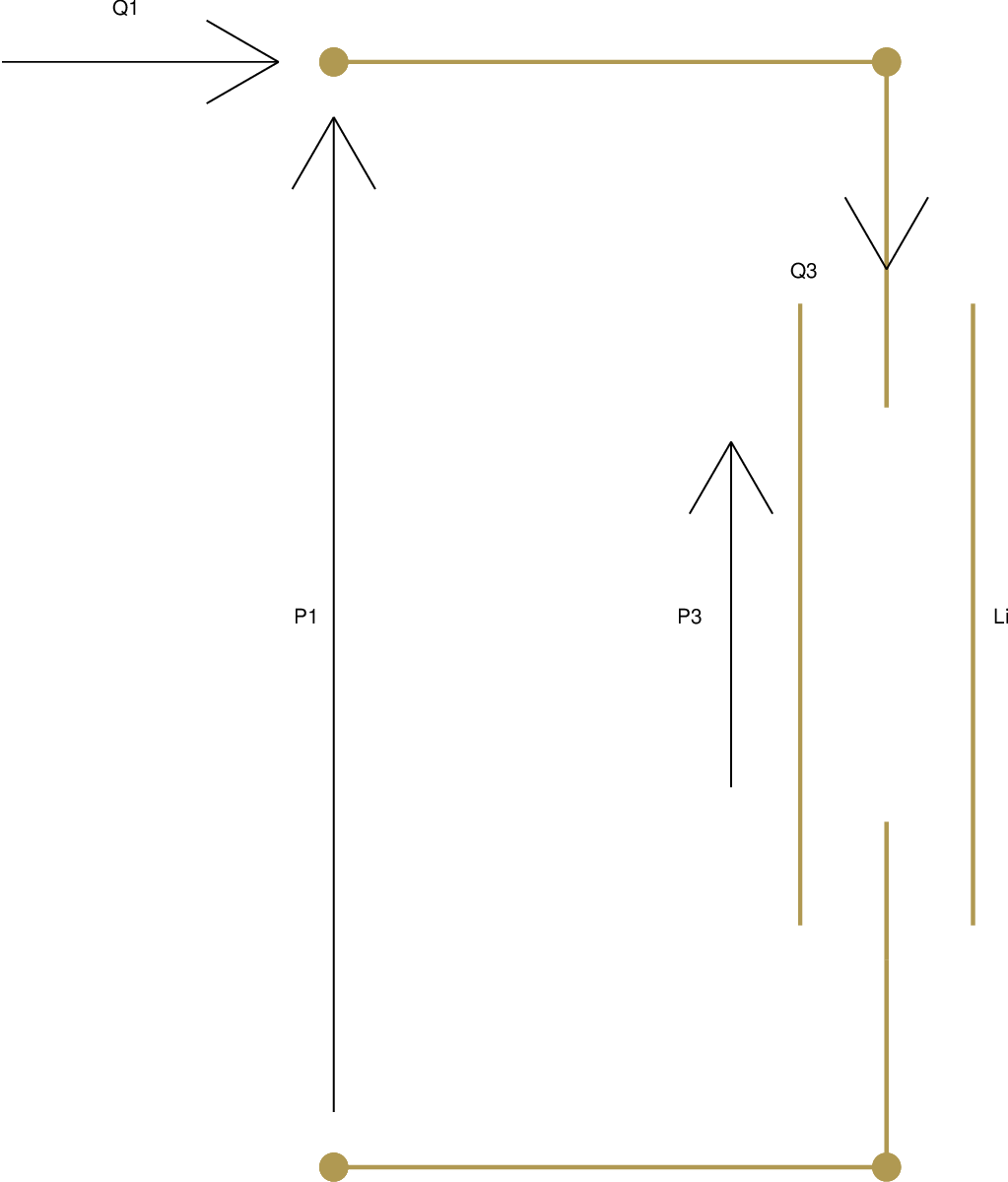}
    \\[1mm]
    &   \small   \multirow{2}{1.6cm}{\centering Inductor}
    &   \small  \multirow{2}{1.6cm}{\centering Spring}
    &  \small   Rotational
    & \small   Hydraulic 
    \\
    &   \small   
    &   \small  
    &  \small   Spring
    & \small   Inductor
  \\[2mm]
     \hline 
     \hline 
  &&&&\\[0mm]
  \setlength{\unitlength}{1.3mm}
  \psset{unit=1.0\unitlength}
   \begin{picture}(2,13)(-1,-1.5)
   \rput[c]{90}(-0.2,5){\scriptsize Physical Element $R$}
  \end{picture}
  &
  \psfrag{V1}[r][r]{$\scr \mygreen V$}
  \psfrag{I1}[b][b]{$\scr \myred   I$}
  \psfrag{R}[l][l]{$\scr R$}
  \includegraphics[clip,width=1.5cm]{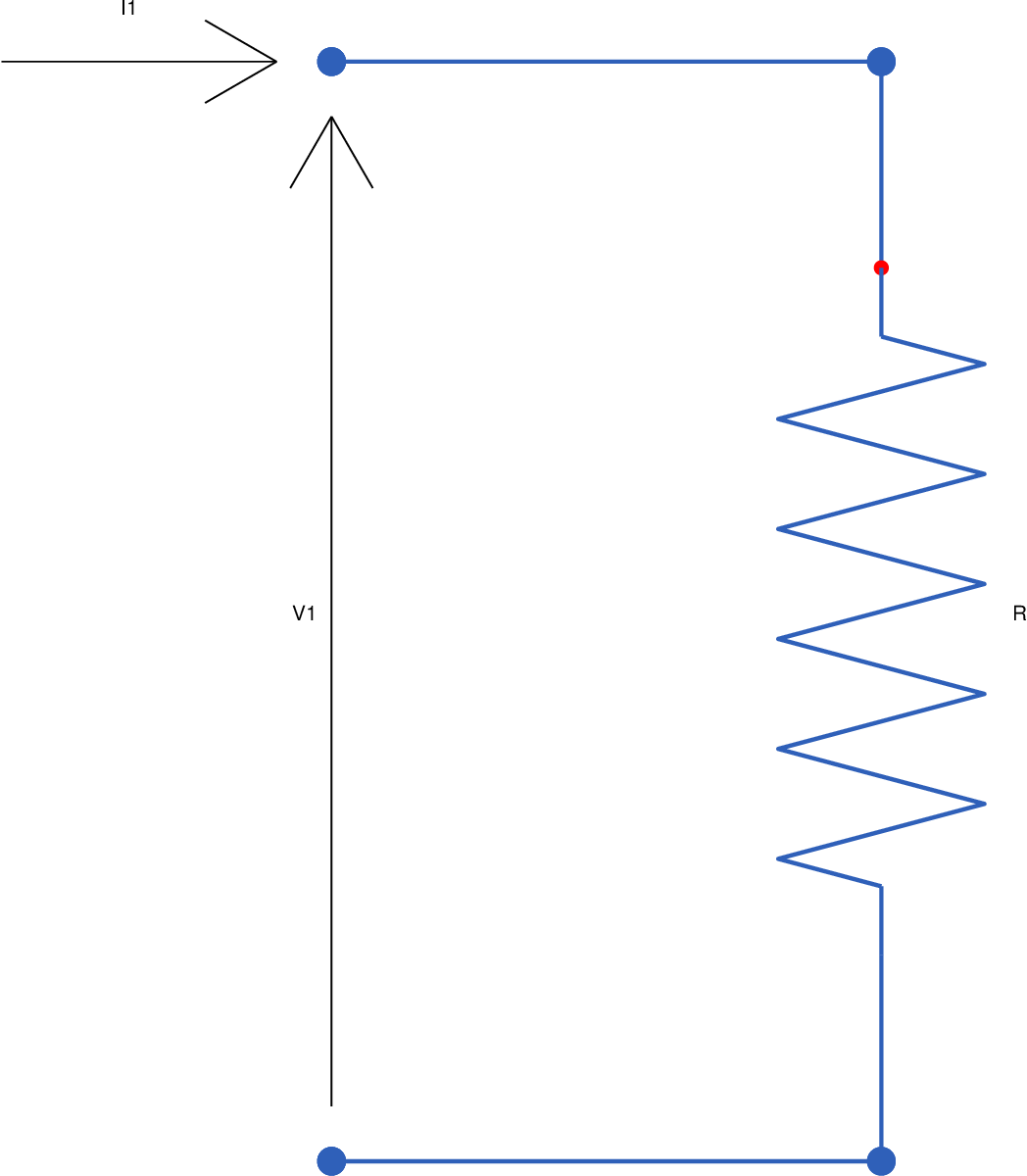}
 &
  \psfrag{v1}[r][r]{$\scr \mygreen v$}
  \psfrag{F1}[b][b]{$\scr \myred   F$}
  \psfrag{b}[l][l]{$\scr b$}
  \includegraphics[clip,width=1.5cm]{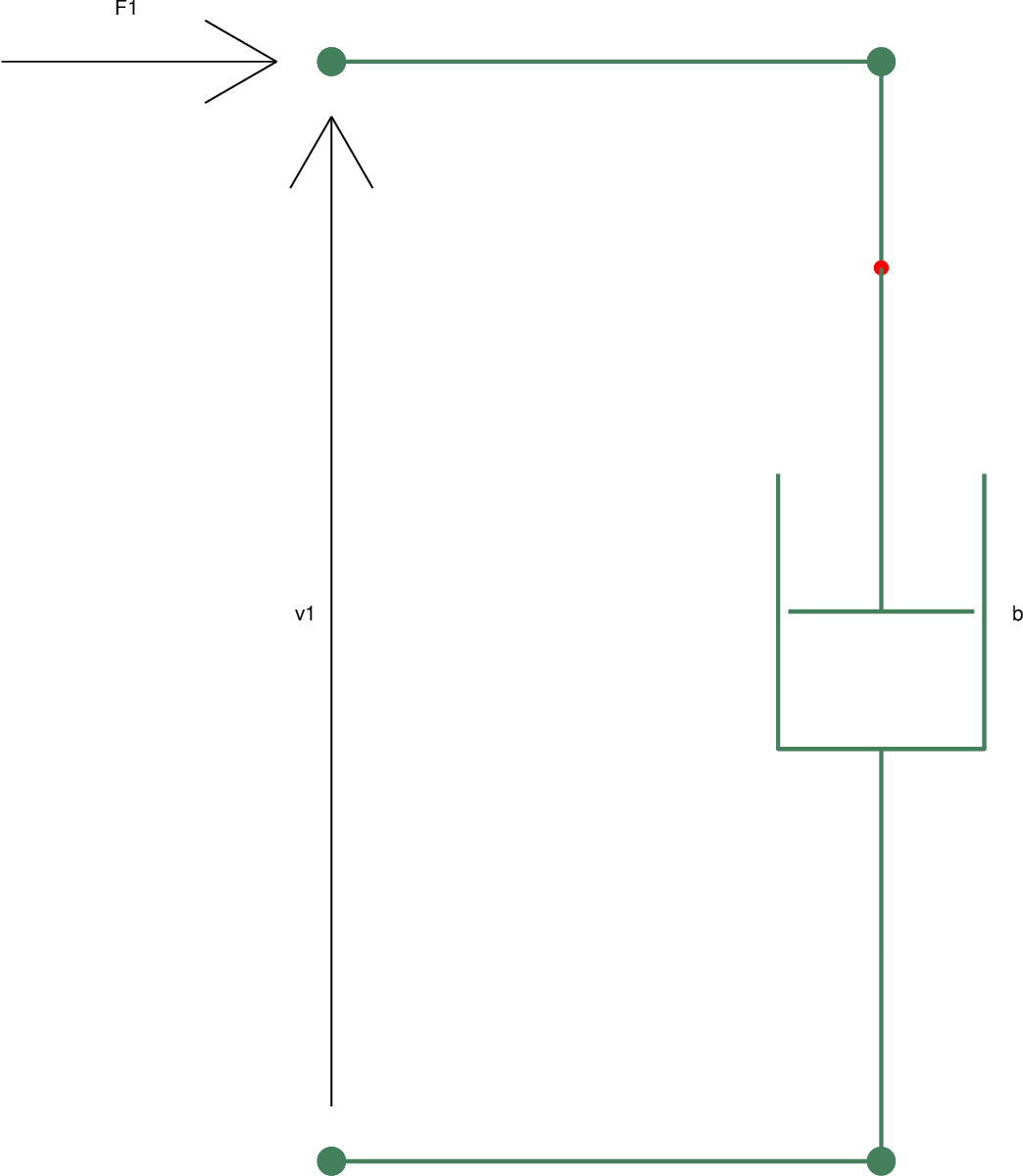}
 &
  \psfrag{w1}[r][r]{$\scr \mygreen \omega$}
  \psfrag{T1}[b][b]{$\scr \myred   \tau$}
  \psfrag{b}[l][l]{$\scr d$}
  \includegraphics[clip,width=1.5cm]{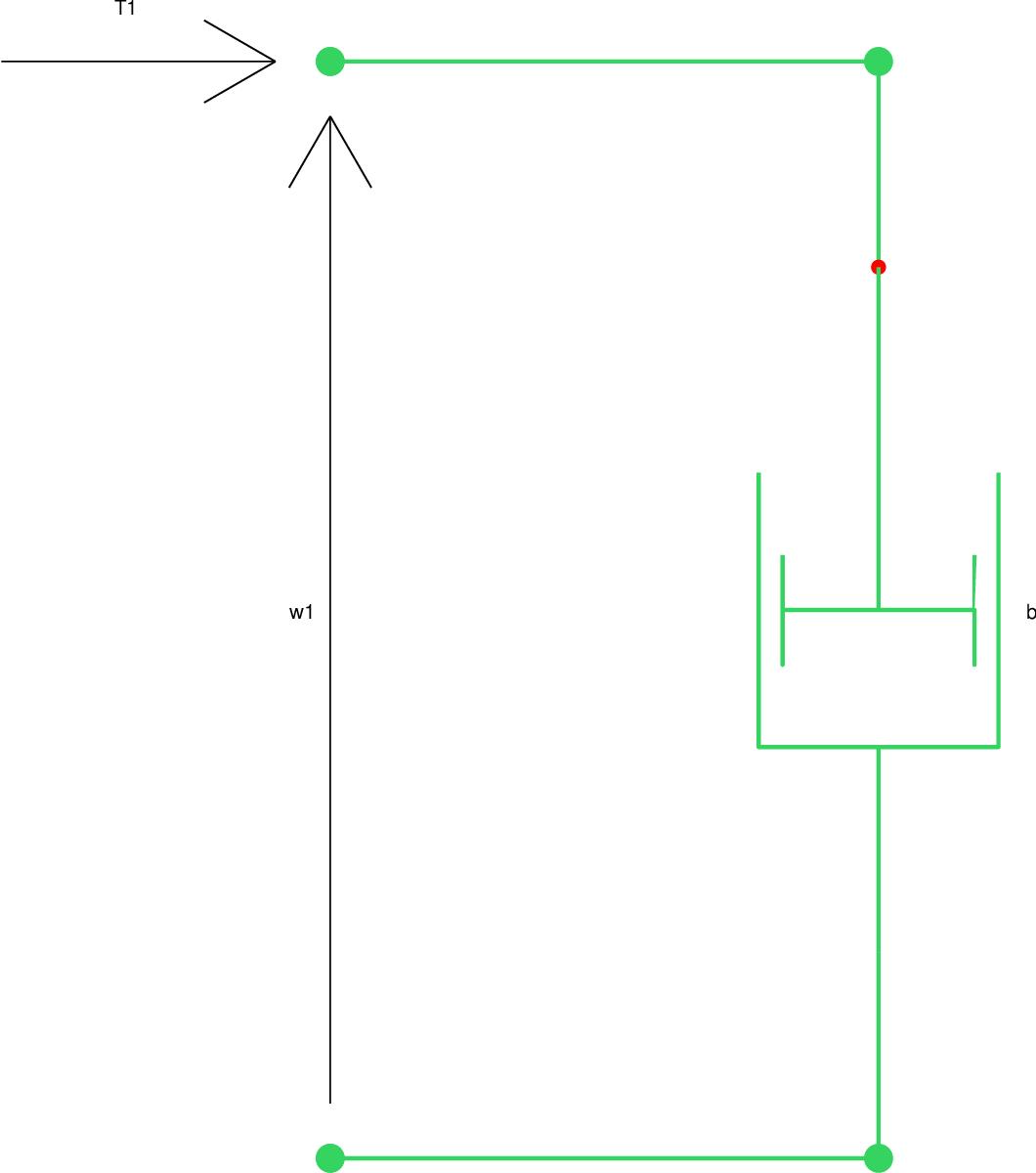}
 &
  \psfrag{P1}[r][r]{$\scr \mygreen P$}
  \psfrag{Q1}[b][b]{$\scr \myred   Q$}
  \psfrag{Ri}[l][l]{$\!\scr R_{I}$}
  \includegraphics[clip,width=1.5cm]{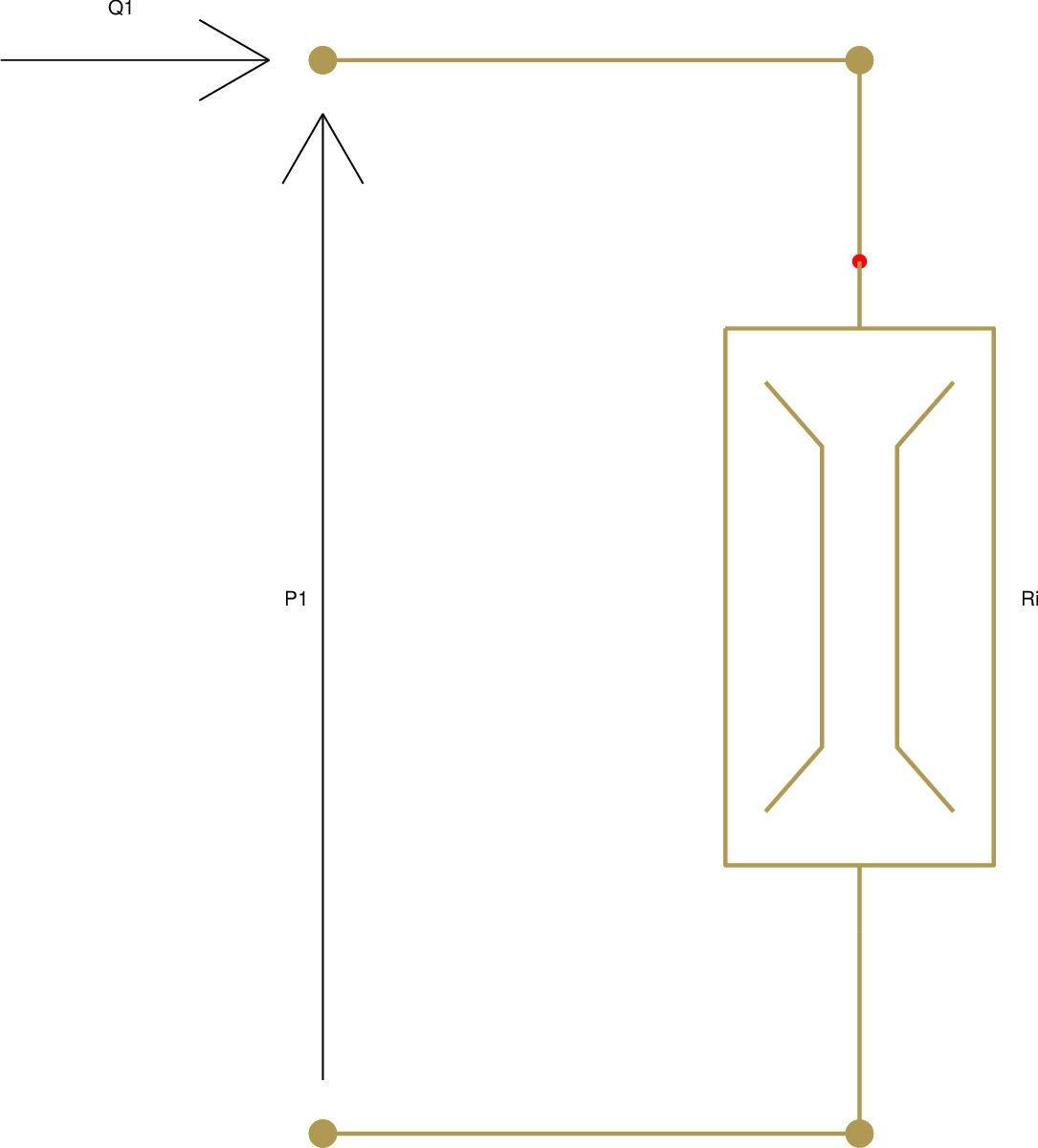}
    \\[1mm]
    &   \small   \multirow{2}{1.6cm}{\centering Resistor}
    &   \small  \multirow{2}{1.6cm}{\centering Friction}
    &  \small   Angular
    & \small   Hydraulic 
    \\
    &   \small   
    &   \small  
    &  \small   Friction
    & \small   Resistor
  \\[2mm]
 \hline 
  \end{tabular}
  \end{center}
  \caption{Simscape graphical representation of the dynamic Physical Elements (PEs) $\mygreen D_e$, $\myred D_f$ and of the static PEs $R$ in the electrical, mechanical translational, mechanical rotational and hydrauli energetic domains. The PEs are shown using the representation proposed in Figure~\ref{PE_represent}(b). 
}\label{Internal_Physical_Elements_Simscape}
\end{figure}

\begin{figure}[t!]
   \begin{center}
  \begin{tabular}{|@{}c@{\,}|@{\hspace{0.02cm}}c@{\hspace{0.02cm}}|@{\hspace{0.02cm}}c@{\hspace{0.02cm}}|@{\hspace{0.02cm}}c@{\hspace{0.02cm}}|@{\hspace{0.01cm}}c@{\hspace{0.2mm}}|}
  \hline
   &
  \bf  \small \multirow{2}{1.6cm}{\centering Electrical} &
  \bf  \small Mechanical  &
  \bf  \small Mechanical  &
  \bf \small \multirow{2}{1.6cm}{\centering Hydraulic} \\  
  &&&&\\[-4mm]
   &
  \bf  \small &
  \bf  \small Translational  &
  \bf  \small Rotational  &
  \bf \small \\   \hline
  &&&&\\[0mm]
  \setlength{\unitlength}{1.1mm}
  \psset{unit=1.0\unitlength}
   \begin{picture}(2,13)(-1,-1.5)
   \rput[c]{90}(0,6){\bf \scriptsize  Across Generators \mygreen  $G_{e}$}
  \end{picture}
  &
  \psfrag{V1}[r][r]{$\scr \mygreen V$}
  \psfrag{V3}[r][r]{$\scr \mygreen $}
  \psfrag{I1}[b][b]{$\scr \myred   I$}
  \psfrag{I3}[r][r]{$\scr \myred   $}
  \psfrag{L}[l][l]{$\scr L$}
  \includegraphics[clip,width=1.125cm]{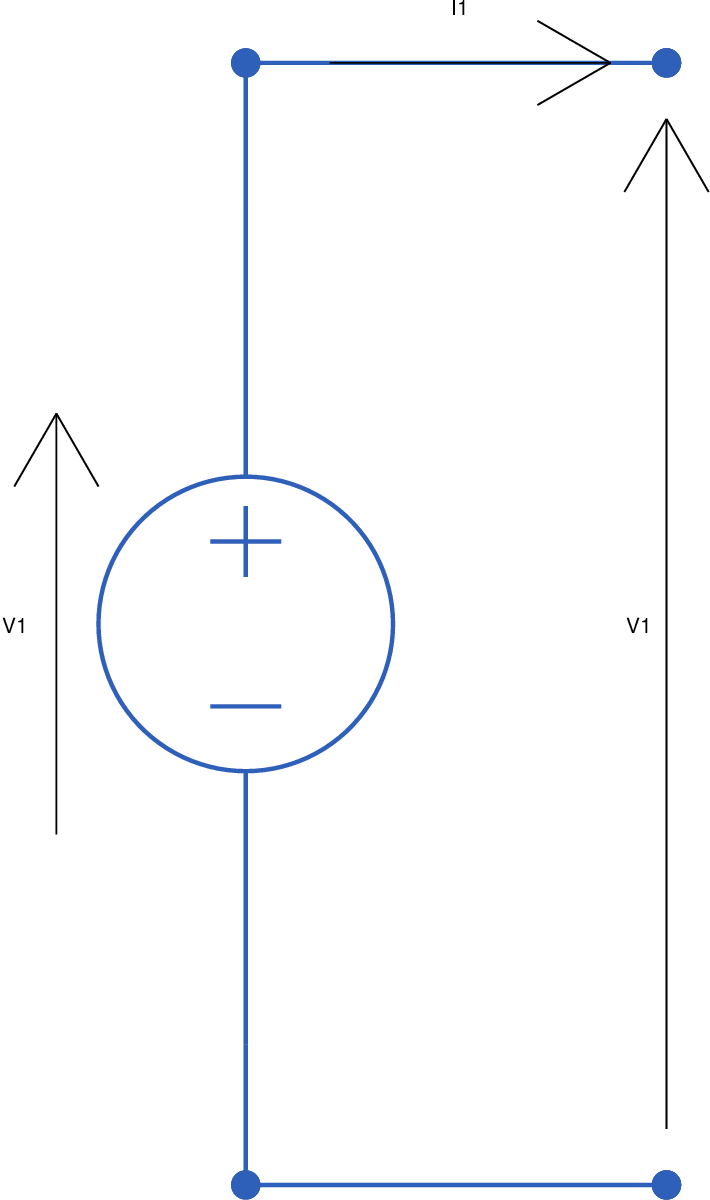}
   &
  \psfrag{v1}[r][r]{$\scr \mygreen v$}
  \psfrag{v3}[r][r]{$\scr \mygreen $}
  \psfrag{F1}[b][b]{$\scr \myred   F$}
  \psfrag{F3}[r][r]{$\scr \myred   $}
  \psfrag{K}[l][l]{$\scr E$}
  \psfrag{E}[l][l]{$\scr E$}
 \includegraphics[clip,width=1.125cm]{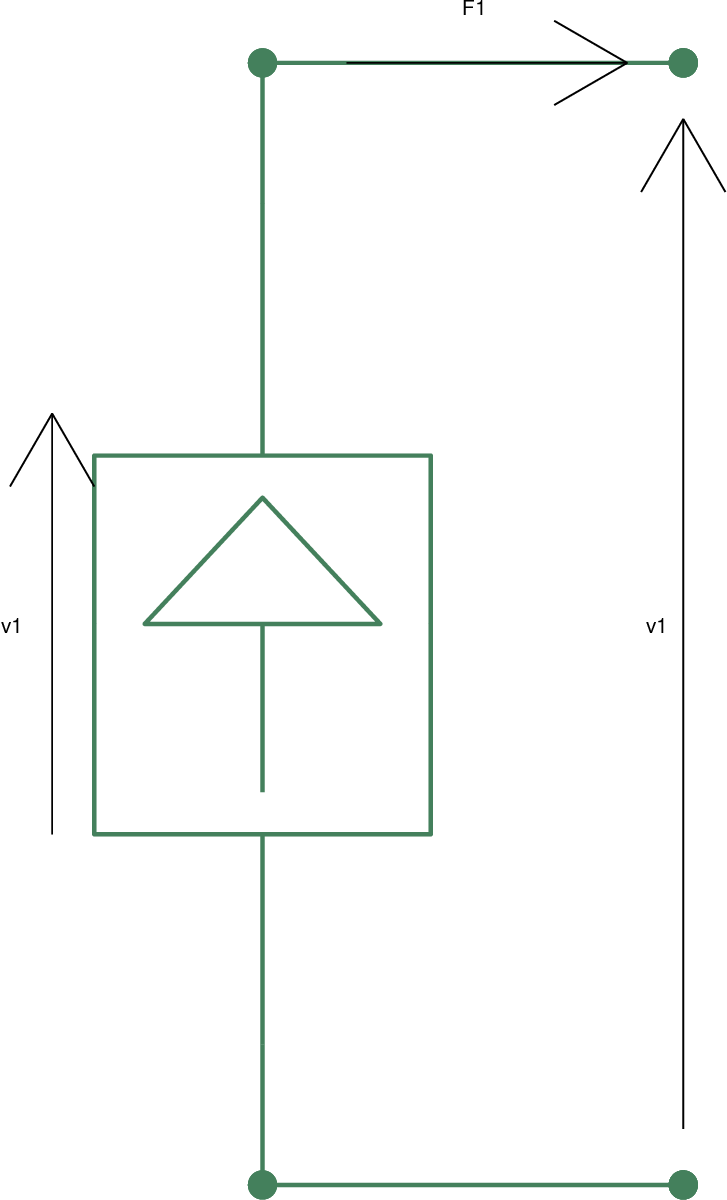}
   &
  \psfrag{w1}[r][r]{$\scr \mygreen \omega$}
  \psfrag{w3}[r][r]{$\scr \mygreen $}
  \psfrag{T1}[b][b]{$\scr \myred   \tau$}
  \psfrag{T3}[r][r]{$\scr \myred   $}
  \psfrag{K}[l][l]{$\scr E_r$}
  \psfrag{Et}[l][l]{$\scr E_r$}
  \includegraphics[clip,width=1.125cm]{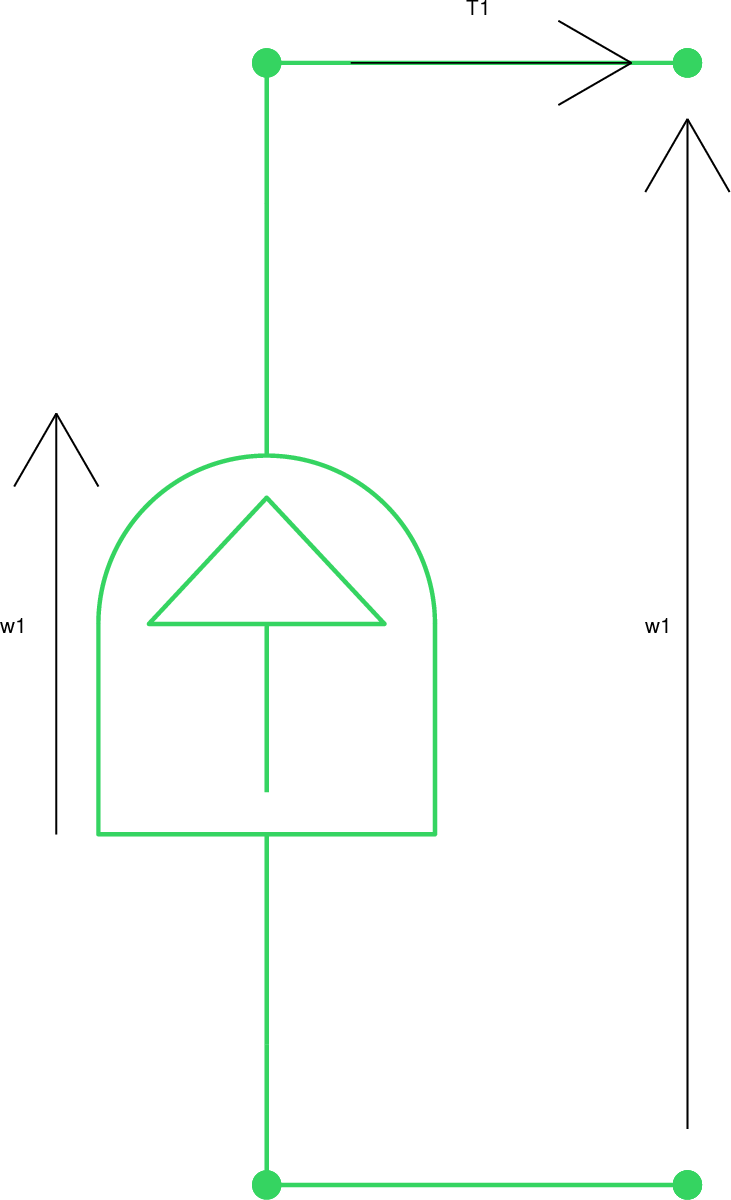}
   &
  \psfrag{P1}[r][r]{$\scr \mygreen P$}
  \psfrag{P3}[r][r]{$\scr \mygreen $}
  \psfrag{Q1}[b][b]{$\scr \myred   Q$}
  \psfrag{Q3}[r][r]{$\scr \myred   $}
  \psfrag{Li}[l][l]{$\scr L_{I}$}
  \includegraphics[clip,width=1.125cm]{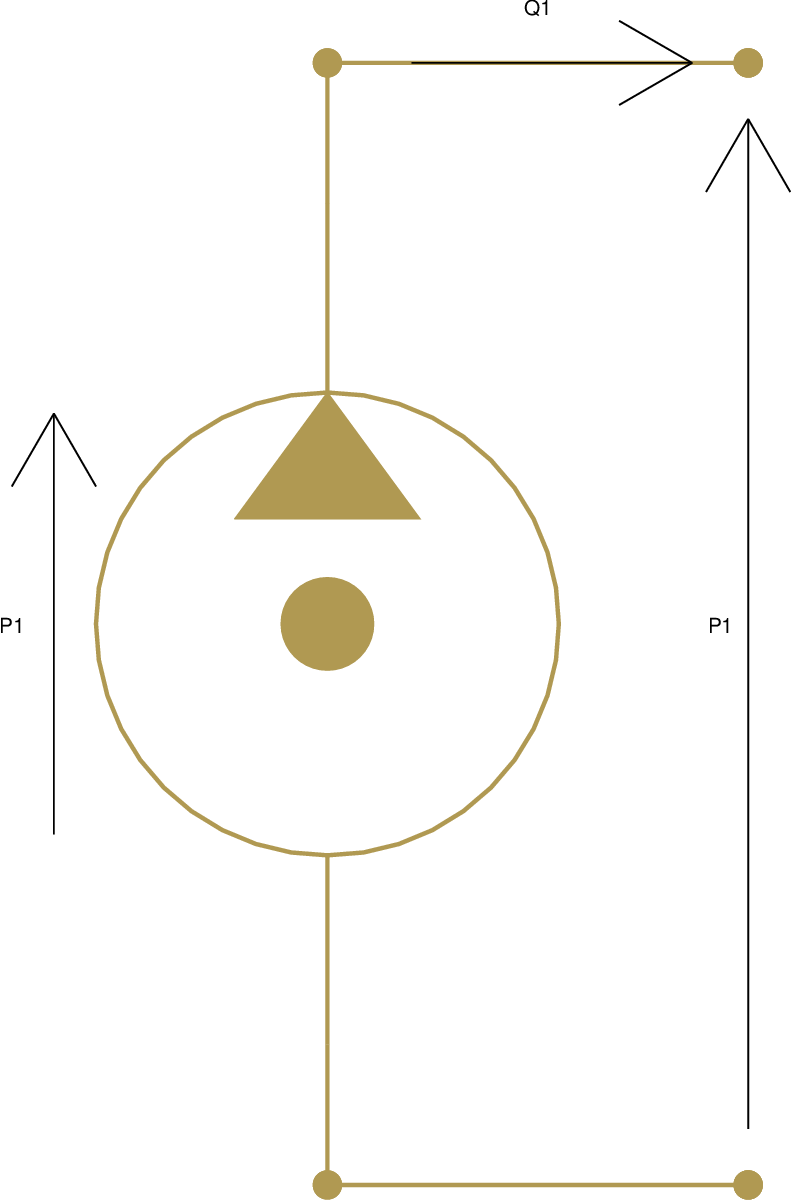}
    \\[1mm]
    &   \small    Voltage
    &   \small  Velocity
    &  \small   Angular Velocity
    & \small   Pressure
    \\
    &   \small Generator
    &   \small  Generator
    &  \small   Generator
    & \small   Generator
  \\[1mm]\hline 
    \hline 
   &&&&\\[0mm]
  \setlength{\unitlength}{1.1mm}
  \psset{unit=1.0\unitlength}
   \begin{picture}(2,13)(-1,-1.5)
   \rput[c]{90}(0,5){\bf\scriptsize   Through Generators  \myred   $G_{f}$}
  \end{picture}
  &
  \psfrag{V1}[r][r]{$\scr \mygreen V$}
  \psfrag{V3}[r][r]{$\scr \mygreen $}
  \psfrag{I1}[b][b]{$\scr \myred   I$}
  \psfrag{I3}[r][r]{$\scr \myred   $}
  \psfrag{L}[l][l]{$\scr L$}
   \includegraphics[clip,width=0.9cm]{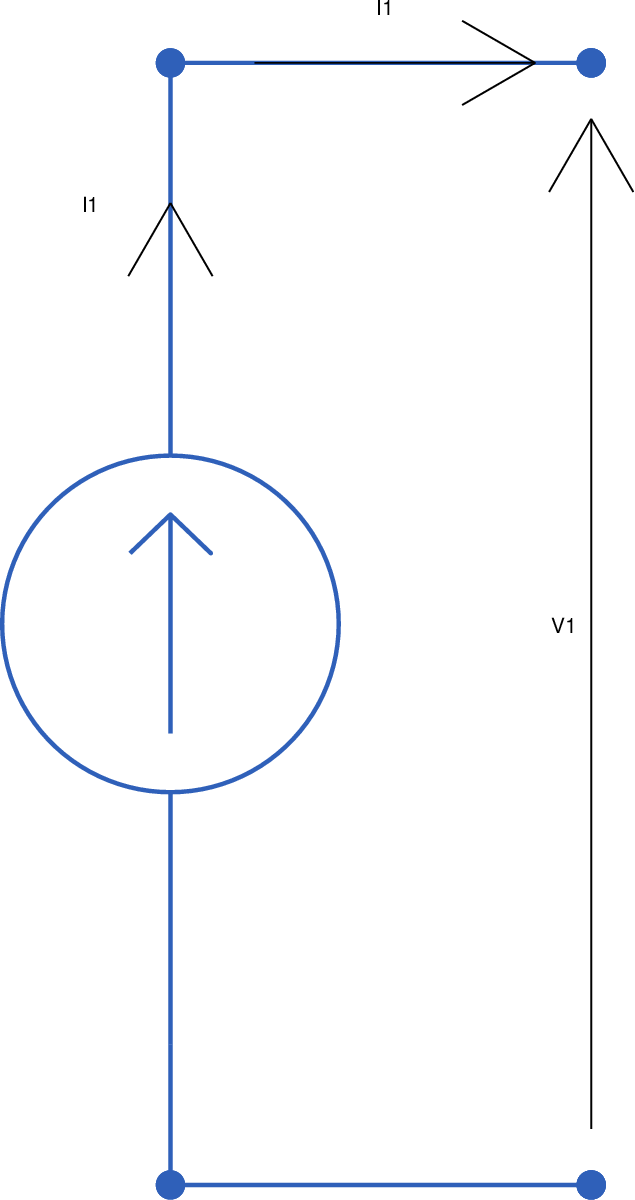}
  &
  \psfrag{v1}[r][r]{$\scr \mygreen v$}
  \psfrag{v3}[r][r]{$\scr \mygreen $}
  \psfrag{F1}[b][b]{$\scr \myred   F$}
  \psfrag{F3}[r][r]{$\scr \myred   $}
  \psfrag{K}[l][l]{$\scr E$}
  \psfrag{E}[l][l]{$\scr E$}
  \includegraphics[clip,width=0.9cm]{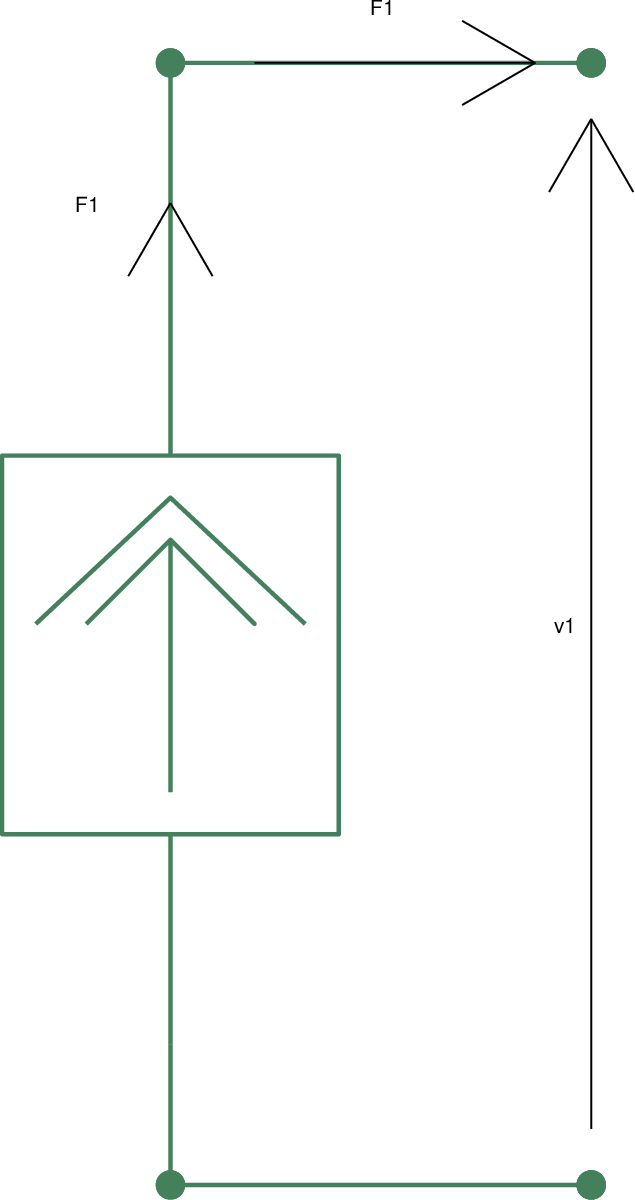}
   &
  \psfrag{w1}[r][r]{$\scr \mygreen \omega$}
  \psfrag{w3}[r][r]{$\scr \mygreen $}
  \psfrag{T1}[b][b]{$\scr \myred   \tau$}
  \psfrag{T3}[r][r]{$\scr \myred   $}
  \psfrag{K}[l][l]{$\scr E_r$}
  \psfrag{Et}[l][l]{$\scr E_r$}
  \includegraphics[clip,width=0.9cm]{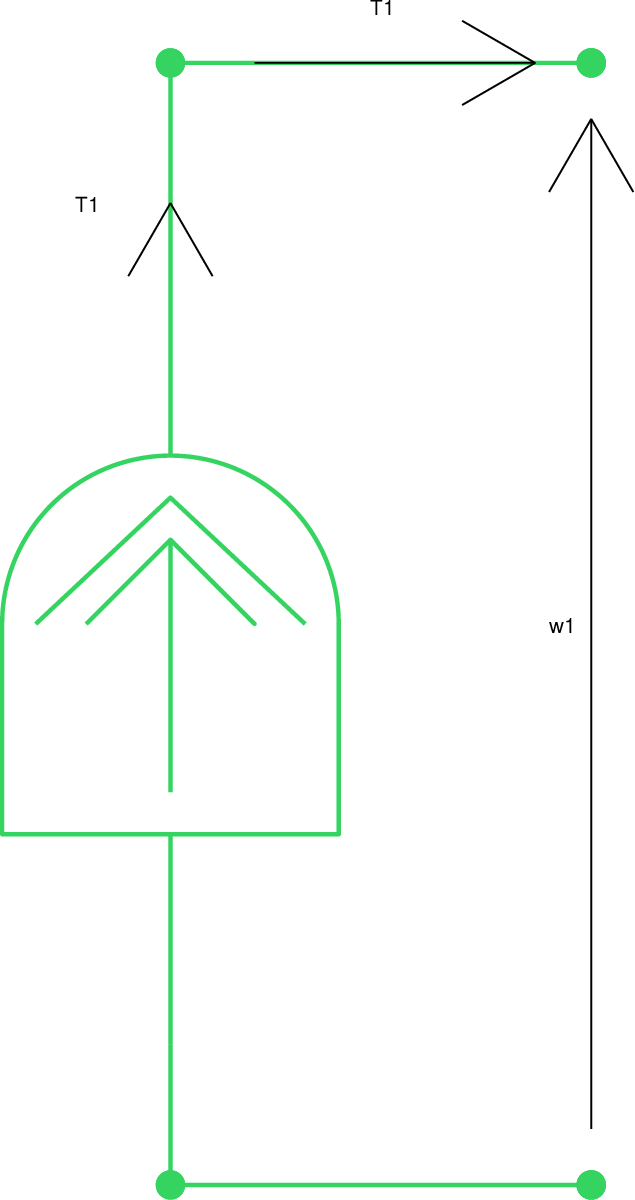}
   &
  \psfrag{P1}[r][r]{$\scr \mygreen P$}
  \psfrag{P3}[r][r]{$\scr \mygreen $}
  \psfrag{Q1}[b][b]{$\scr \myred   Q$}
  \psfrag{Q3}[r][r]{$\scr \myred   $}
  \psfrag{Li}[l][l]{$\scr L_{I}$}
  \includegraphics[clip,width=0.9cm]{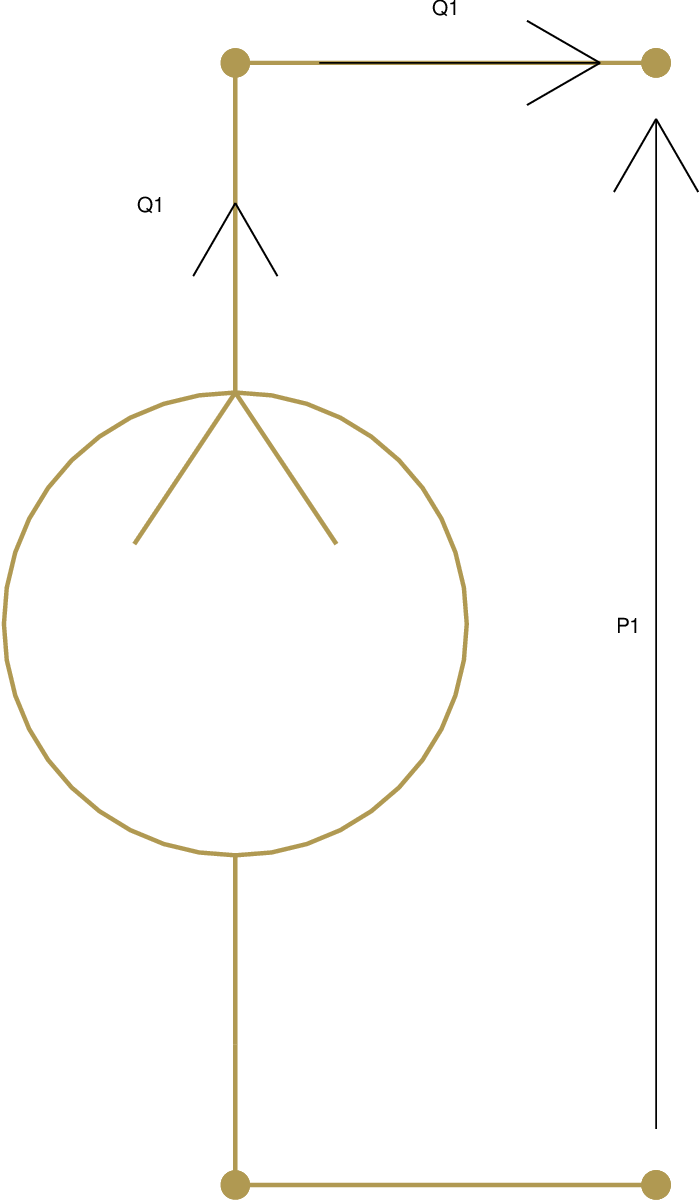}
    \\[1mm]
    &   \small    Current
    &   \small  Force
    &  \small   Torque
    & \small   Flow Rate
    \\
    &   \small Generator
    &   \small  Generator
    &  \small   Generator
    & \small   Generator
  \\[1mm]\hline 
  \end{tabular}
  \end{center}
    \caption{Simscape graphical representation of the across and through generators $\mygreen D_e$ and $\myred D_f$ in the electrical, mechanical translational, mechanical rotational and hydrauli energetic domains. The generators are shown using the representation proposed in Figure~\ref{PE_represent}(b).
}\label{External_Physical_Elements_Simscape}
\end{figure}

The dynamic elements $\mygreen D_{e}$ are called {\it across elements} as they produce an across variable $\mygreen v_{e}$ as output power variable. At the same time, the dynamic elements $\myred D_{f}$ are {\it through elements} as they produce a through variable $\myred v_{f}$ as output power variable. 
The dynamic elements $\mygreen D_{e}$ and $\myred D_{f}$, and the static elements $\cR$, 
are characterized by two terminals, each of which characterized by a pair of power variables ($\mygreen v_{e1}$, $\myred v_{f1}$) and ($\mygreen v_{e2}$, $\myred v_{f2}$), as shown in Figure~\ref{PE_represent}(a). By defining $\mygreen v_{e}=\mygreen v_{e1}-\mygreen v_{e2}$ and $\myred v_{f}=\myred v_{f1}=\myred v_{f2}$ as new power variables,
the power interaction of the Physical Element PE with the external world can be described using the {\it power section} $P$ shown in Figure~\ref{PE_represent}(b): in the POG block schemes, the power sections are denoted by a dashed line, and the product of the two power variables $\mygreen v_{e}$ and $\myred v_{f}$  has the physical meaning of a power flowing through the section.
Figure~\ref{Internal_Physical_Elements_Simscape} shows, for each energetic domain, the Simscape graphical representation of the dynamic elements $\mygreen D_e$, $\myred D_f$ and of the static elements $R$ reported in Table~\ref{Ambiti_Energetici}. 
%
%
%
Each energetic domain is also characterized by two power variable generators: an {\it across generator} $ \mygreen G_{e}$ producing as output an across power variable $ \mygreen v_{e}$  and a {\it through generator} $ \myred G_{f}$ producing as output a through power variable $ \myred v_{f}$. 
Figure~\ref{External_Physical_Elements_Simscape} shows the Simscape graphical representation of the across and through generators $\mygreen D_e$ and $\myred D_f$. 


\subsection{Modular Structure of the POG}\label{Block_Types_sect}

\begin{figure}[t!]
  \centering
   \setlength{\unitlength}{2.8mm}
  \psset{unit=1.0\unitlength}
   \schemaPOG{(-2.625,-3)(8.25,13.0)}{2.9mm}{
 \thicklines %
 \bloin{\mygreen \v_{e1}}{\myred \v_{f1}}
 \rput[t](3,13.5){\mbox{\footnotesize (a)}}
%
\blovisuintm{\mygreen \boldsymbol{\Phi_{e}}\muno({\mygreen \q_{e}})}{\mygreen \v_e}{\dst}
 \rput(-2,4.5){\mygreen \q_e}
 \rput(-2,0.75){\myred \v_f}
 \bloout{\mygreen \v_{e2}}{\myred \v_{f2}}
 }
    \schemaPOG{(-2.625,-3)(8.25,13)}{2.9mm}{
 \thicklines %
 \bloin{\mygreen \v_{e1}}{\myred \v_{f1}}
 \rput[t](3,13.5){\mbox{\footnotesize (b)}}
\blovigiuintm{\myred \boldsymbol{\Phi_{f}}\muno({\myred \q_{f}})}{\myred \v_f}{\dst}
 \rput(-2,5.5){\myred \q_f}
 \rput(-2,9.25){\mygreen \v_e}
 \bloout{\mygreen \v_{e2}}{\myred \v_{f2}}
 }
 
    \schemaPOG{(-3.625,-1)(8.25,13)}{2.9mm}{
 \thicklines %
 \bloin{\x_{1}}{\y_{1}}
 \rput[t](3,13.5){\mbox{\footnotesize (c)}}
 \rput(3.5,8.5){\x}
 \POGvigiu{\ds \boldsymbol{\R}}{}{}{\dst}
 %
 \rput(-2.5,1.5){\y}
 \bloout{\x_{2}}{\y_{1}}
 }
    \schemaPOG{(-2.625,-1)(8.25,13)}{2.9mm}{
 \bloin{\x_{1}}{\y_{1}}
 \rput[t](2.5,13.5){\mbox{\footnotesize (d)}}
 \blokxyor{3.0}{3.0}{$\K$}{$\K$\tras}{}{}
 \bloout{\x_{2}}{\y_{2}}
 %
 }
 \vspace{4mm}
 \caption{Elaboration Blocks (EBs) and Connection Blocks (CBs). (a) Basic EB describing the dynamic elements $\mygreen \cD_{e}$. (b) Basic EB describing the dynamic elements $\myred \cD_{f}$. (c) Basic EB describing the static elements $\R$. (d) CB describing a power conversion occurring in the system. The power variables $\mygreen \v_{e1}$, $\mygreen \v_{e2}$, $\mygreen \v_{e}$, $\myred \v_{f1}$, $\myred \v_{f2}$, $\myred \v_{f}$ in (a) and (b), as well as the inverted constitutive relations $\mygreen \boldsymbol{\Phi_{e}}\muno({\mygreen \q_{e}})$ and $\myred \boldsymbol{\Phi_{f}}\muno({\myred \q_{f}})$, are boldface to denote the fact that they can be vectors and matrices, respectively, for the case of multidimensional dynamic elements $\mygreen \boldsymbol{\cD_{e}}$ and $\myred \boldsymbol{\cD_{f}}$. The same consideration applies to the power variables $\x_1$, $\x_2$, $\x$, $\y_1$, $\y_2$, $\y$, and to the matrices $\boldsymbol{\R}$, $\K$ in (c) and (d).
 }
 \label{basic_EBs_and_CBs}
\end{figure}
The POG graphical rapresentations of the  dynamic elements $\mygreen D_e$ and $\myred D_f$ are shown in Figure~\ref{basic_EBs_and_CBs}(a) and Figure~\ref{basic_EBs_and_CBs}(b), respectively.
These graphical representation respect the integral causality~\cite{Gawthrop2007}, since the the output power variables $\mygreen \v_{e}$ (for $\mygreen D_{e}$) and $\myred \v_{f}$ (for $\myred D_{f}$) exhibit an integral relationship with respect to the input power variables.
The dynamic element $\mygreen D_{e}$ is characterized by the internal energy variable $\mygreen \q_{e}$, the input power variable $\myred \v_{f}$, the output power variable $\mygreen \v_{e}$, the constitutive relation $\mygreen \q_{e}\black =\mygreen \Phi_{e}( \v_{e})$, and the differential equation ${\mygreen \dot{\q}_{e}}={\myred \v_{f}}$. Conversely, the dynamic element $\myred D_{f}$ is characterised by the internal energy variable $\myred \q_{f}$, the input power variable $\mygreen \v_{e}$, the output power variable $\myred \v_{f}$, the constitutive relation $\myred \q_{f}\black =\myred \Phi_{f}( \v_{f})$, and the differential equation ${\myred \dot{\q}_{f}}={\mygreen \v_{e}}$. The integral causality is adopted in the POG technique since the objective is to have block schemes that are suitable for simulation.
%
%
The EBs in Figure~\ref{basic_EBs_and_CBs}(a) and Figure~\ref{basic_EBs_and_CBs}(b),
describing the PEs $\mygreen D_{e}$ and $\myred D_{f}$, respectively, are characterized by the following nonlinear integral relations:
\[
\mygreen \v_{e}\black=\mygreen\Phi_{e}\muno\black\left(\int_{0}^t \myred \v_{f}\black dt\right), \hspace{6mm}
\myred \v_{f}\black=\myred\Phi_{f}\muno\black\left(\int_{0}^t \mygreen \v_{e}\black dt\right), \hspace{6mm}
\]
which are graphically represented using the Laplace transform. The power variables and the constitutive relations in Figure~\ref{basic_EBs_and_CBs} are boldface to denote the fact that they can be vectors and matrices, respectively, for the case of multidimensional dynamic elements $\mygreen D_{e}$ and $\myred D_{f}$. 

\begin{sidebar}{Second Case Study: An Hydraulic Continuous Variable Transmission}

 \setcounter{sequation}{1}
\renewcommand{\thesequation}{S\arabic{sequation}}
\setcounter{stable}{1}
\renewcommand{\thestable}{S\arabic{stable}}
\setcounter{sfigure}{2}
\renewcommand{\thesfigure}{S\arabic{sfigure}}

Figure~\ref{POG_CVT_Grafico} shows a schematic representation of the second case study, which is an hydraulic Continuous Variable Transmission (CVT)~\cite{TebaldiZanasiCDC23}.

\sdbarfig{  
\centering
 \setlength{\unitlength}{2.46mm}
 \psset{unit=\unitlength}
 \SpecialCoor
 \begin{pspicture}(-3,-25)(29,18.5)
 \newrgbcolor{color_0}{0 0 0}
  \newrgbcolor{color_1}{1         0.2         0.2}
  \newrgbcolor{amethyst}{0.6, 0.4, 0.8}
\psline[fillstyle=solid,fillcolor=color_1,linewidth=0.9pt](19,0)(19,4.5)(17,4.5)(17,5)(19.5,5)(19.5,0.5)(26,0.5)(26,0)
 \psline[fillstyle=solid,fillcolor=color_1,linewidth=0.9pt](19,0)(19,-4.5)(17,-4.5)(17,-5)(19.5,-5)(19.5,-0.5)(26,-0.5)(26,0)
 \newrgbcolor{color_2}{0         0.8           0}
\psline[fillstyle=solid,fillcolor=color_2,linewidth=0.9pt](17,5)(17,6.5)(18,6.5)(18,5)(17,5)
 \psline[fillstyle=solid,fillcolor=color_2,linewidth=0.9pt](17,4.5)(17,3)(18,3)(18,4.5)(17,4.5)
 \psline[fillstyle=solid,fillcolor=color_2,linewidth=0.9pt](17,-4.5)(17,-3)(18,-3)(18,-4.5)(17,-4.5)
 \psline[fillstyle=solid,fillcolor=color_2,linewidth=0.9pt](17,-5)(17,-6.5)(18,-6.5)(18,-5)(17,-5)
 \newrgbcolor{color_3}{0.9         0.9           0}
\psline[fillstyle=solid,fillcolor=color_3,linewidth=0.9pt](1,0)(1,0.5)(4,0.5)(4,3.5)(5,3.5)(5,0.5)(17,0.5)(17,3)(18,3)(18,0)
 \psline[fillstyle=solid,fillcolor=color_3,linewidth=0.9pt](1,0)(1,-0.5)(4,-0.5)(4,-3.5)(5,-3.5)(5,-0.5)(17,-0.5)(17,-3)(18,-3)(18,0)
\newrgbcolor{color_4}{0         0         1}   \psline[fillstyle=solid,fillcolor=color_4,linewidth=0.9pt](17,6.5)(17,8.5)(18,8.5)(18,6.5)(17,6.5)
 \psline[fillstyle=solid,fillcolor=color_4,linewidth=0.9pt](17,-6.5)(17,-8.5)(18,-8.5)(18,-6.5)(17,-6.5)
 \newrgbcolor{color_5}{0.1         0.8         0.8}
 \psline[fillstyle=solid,fillcolor=color_5,linewidth=0.9pt](4,6)(4,8.5)(5,8.5)(5,6.5)(10,6.5)(10,9.5)(11,9.5)(11,6.5)(11,6.5)(11,9.5)(11,9.5)(11,9)(10,9)(10,8.5)(11,8.5)(11,8)(10,8)(10,7.5)(11,7.5)(11,7)(10,7)(10,6.5)(11,6.5)(11,6)
 \psline[fillstyle=solid,fillcolor=color_5,linewidth=0.9pt](4,6)(4,3.5)(5,3.5)(5,5.5)(10,5.5)(10,2.5)(11,2.5)(11,5.5)(11,5.5)(11,2.5)(11,2.5)(11,3)(10,3)(10,3.5)(11,3.5)(11,4)(10,4)(10,4.5)(11,4.5)(11,5)(10,5)(10,5.5)(11,5.5)(11,6)
 \newrgbcolor{color_6}{1         0.47         1}
 \psline[fillstyle=solid,fillcolor=color_6,linewidth=0.9pt](18,12)(18,15.5)(17,15.5)(17,12.5)(11,12.5)(11,14.5)(10,14.5)(10,12.5)(10,12.5)(10,14.5)(10,14.5)(10,14)(11,14)(11,13.5)(10,13.5)(10,13)(11,13)(11,12.5)(10,12.5)(10,12)
 \psline[fillstyle=solid,fillcolor=color_6,linewidth=0.9pt](18,12)(18,8.5)(17,8.5)(17,11.5)(11,11.5)(11,9.5)(10,9.5)(10,11.5)(10,11.5)(10,9.5)(10,9.5)(10,10)(11,10)(11,10.5)(10,10.5)(10,11)(11,11)(11,11.5)(10,11.5)(10,12)
 \pcline[linecolor=color_1,linewidth=1.25pt]{->}(26.25,0)(27.75,0)
  \aput[1pt]{0}(0.5){$\omega_c$} \bput[1pt]{0}(0.5){$\tau_c$}
 \pcline[linecolor=color_2,linewidth=1.25pt]{->}(20.25,4.75)(21.75,4.75)
  \aput[1pt]{0}(0.5){$\omega_p$} \bput[1pt]{0}(0.5){$\tau_p$}
 \pcline[linecolor=color_3,linewidth=1.25pt]{->}(-0.75,0)(0.75,0)
  \aput[1pt]{0}(0.5){$\omega_s$} \bput[1pt]{0}(0.5){$\tau_s$}
 \pcline[linecolor=color_4,linewidth=1.25pt]{->}(16.75,0)(18.25,0)
  \aput[1pt]{0}(0.5){$\omega_r$} \bput[1pt]{0}(0.5){$\tau_r$}
 \pcline[linecolor=color_5,linewidth=1.25pt]{->}(2.25,6)(3.75,6)
  \aput[1pt]{0}(0.5){$\omega_d$} \bput[1pt]{0}(0.5){$\tau_d$}
 \pcline[linecolor=color_6,linewidth=1.25pt]{->}(18.25,12)(19.75,12)
  \aput[1pt]{0}(0.5){$\omega_e$} \bput[1pt]{0}(0.5){$\tau_e$}
 \rput{0}(0,0){\psline[linewidth=0.8pt,linestyle=dashed,linecolor=amethyst](7.5,1.5)(7.5,17.81)(13.5,17.81)(13.5,1.5)(7.5,1.5)}
 \pcline[linewidth=0.4pt]{|->}(21,0)(21,3) \bput[1pt]{0}(0.5){$r_{s}$}
 \psline[linewidth=0.4pt,linestyle=dashed]{-}(21,3)(17,3)
 \pcline[linewidth=0.4pt]{|->}(15,0)(15,4.75) \rput(14.5,2.5){$r_{c}$}
 \psline[linewidth=0.4pt,linestyle=dashed]{-}(15,4.75)(17,4.75)
 \pcline[linewidth=0.4pt]{|->}(23,0)(23,6.5) \bput[1pt]{0}(0.5){$r_{r}$}
 \psline[linewidth=0.4pt,linestyle=dashed]{-}(23,6.5)(18,6.5)
 \pcline[linewidth=0.4pt]{|->}(15,-4.75)(15,-3) \aput[1pt]{0}(0.5){$r_{p}$}
 \psline[linewidth=0.4pt,linestyle=dashed]{-}(15,-3)(17,-3)
 \psline[linewidth=0.4pt,linestyle=dashed]{-}(15,-4.75)(17,-4.75)
 \pcline[linewidth=0.4pt]{|->}(25,0)(25,8.5) \bput[1pt]{0}(0.5){$r_{re}$}
 \psline[linewidth=0.4pt,linestyle=dashed]{-}(25,8.5)(18,8.5)
 \pcline[linewidth=0.4pt]{|->}(21,12)(21,15.5) \bput[1pt]{0}(0.5){$r_{e}$}
 \psline[linewidth=0.4pt,linestyle=dashed]{-}(21,15.5)(18,15.5)
 \pcline[linewidth=0.4pt]{|->}(6,0)(6,3.5) \rput(6.5,1.82){$r_{a}$}
 \psline[linewidth=0.4pt,linestyle=dashed]{-}(6,3.5)(5,3.5)
 \pcline[linewidth=0.4pt]{|->}(1.5,6)(1.5,8.5) \aput[1pt]{0}(0.5){$r_{d}$}
 \psline[linewidth=0.4pt,linestyle=dashed]{-}(1.5,8.5)(4,8.5)
 \pcline[linewidth=0.4pt]{|->}(9,6)(9,9.5) \rput(7.16,8.2){$h_{p}(\theta)$}
 \psline[linewidth=0.4pt,linestyle=dashed]{-}(9,9.5)(9,9.5)
 \pcline[linewidth=0.4pt]{|->}(9,12)(9,14.5) \aput[1pt]{0}(0.6){$h_{q}$}
 \psline[linewidth=0.4pt,linestyle=dashed]{-}(9,14.5)(10,14.5)
 \pcline[linecolor=red,linewidth=1.25pt]{->}(16.75,2.25)(16.75,3.75) \aput[1pt]{0}(0.5){$F_{sp}$}
 \pcline[linecolor=red,linewidth=1.25pt]{-}(18,3)(17,3)
 \pcline[linecolor=red,linewidth=1.25pt]{->}(16.75,5.75)(16.75,7.25) \aput[1pt]{0}(0.5){$F_{pr}$}
 \pcline[linecolor=red,linewidth=1.25pt]{-}(18,6.5)(17,6.5)
 \pcline[linecolor=red,linewidth=1.25pt]{->}(3.75,2.75)(3.75,4.25) \aput[1pt]{0}(0.5){$F_{sd}$}
 \pcline[linecolor=red,linewidth=1.25pt]{-}(5,3.5)(4,3.5)
 \pcline[linecolor=red,linewidth=1.25pt]{->}(16.75,7.75)(16.75,9.25) \aput[1pt]{0}(0.5){$F_{re}$}
 \pcline[linecolor=red,linewidth=1.25pt]{-}(18,8.5)(17,8.5)
 \pcline[linecolor=red,linewidth=1.25pt]{->}(11.25,8.75)(11.25,10.25) \bput[1pt]{0}(0.5){$P_{de}$}
 \pcline[linecolor=red,linewidth=1.25pt]{-}(11,9.5)(10,9.5)
\rput(10.5,17){\textcolor{amethyst}{ \scriptsize Hydro-}} \rput(10.5,16.1){\textcolor{amethyst}{
\scriptsize Mechanical}} \rput(10.5,15.2){\textcolor{amethyst}{\scriptsize Part}}
 \newrgbcolor{myorange}{0, 0.68, 0}
 \psline[linewidth=0.8pt,linecolor=myorange,linestyle=dashed]{-}(-1,-9)(-1,9)(7,9)(7,1)(14,1)(14,16)(28,16)(28,-9)(-1,-9)
 \rput(5,-7.5){\textcolor{myorange}{\scriptsize Planetary Gear Set}}
%
 \newrgbcolor{brown}{0.59, 0.29, 0.0}
 \newrgbcolor{grey}{0.66, 0.66, 0.66}
  \newrgbcolor{myblue}{0.16, 0.32, 0.75}
 \newrgbcolor{color_5}{0.1         0.8         0.8}
 \newrgbcolor{color_6}{1         0.47         1}
\rput(0,3.2){
\rput(0,14){(a)}
\rput(0,-14.68){(b)}
 \psline[linewidth=0.8pt,linecolor=black,linestyle=dashed](-1,-13)(28,-13)
 \setlength{\unitlength}{2.68mm}
 \psset{unit=\unitlength}
\rput(-3,-18.5){
 \psline[fillstyle=solid,fillcolor=color_5,linewidth=0.9pt](4,-2)(4,-1)(6.5,-1)(5.5,1)(6.5,1.5)(9,-3.5)(8,-4)(7,-2)(4,-2)
 %
\rput(-0.125,0.25){
 \psline[fillstyle=solid,fillcolor=brown,linewidth=0.9pt](6.85,0.8)(7.1,0.3)(11,0.3)(11,0.8)(6.85,0.8)
 \psline[fillstyle=solid,fillcolor=brown,linewidth=0.9pt](7.45,-0.4)(7.7,-0.9)(11,-0.9)(11,-0.4)(7.45,-0.4)
 \psline[fillstyle=solid,fillcolor=brown,linewidth=0.9pt](8.05,-1.6)(8.3,-2.1)(11,-2.1)(11,-1.6)(8.05,-1.6)
 \psline[fillstyle=solid,fillcolor=brown,linewidth=0.9pt](8.65,-2.8)(8.9,-3.3)(11,-3.3)(11,-2.8)(8.65,-2.8)
}
%
 \psline[fillstyle=solid,fillcolor=grey,linewidth=0.9pt](10.875,-4.5)(10.875,2)(11.875,2)(11.875,-0.75)(14.875,-0.75)(14.875,0.25)(15.875,0.25)(15.875,-0.75)
 (18.875,-0.75)(18.875,2)(19.875,2)(19.875,-4.5)(18.875,-4.5)(18.875,-1.75)(11.875,-1.75)(11.875,-4.5)(10.875,-4.5)
 \psarc[fillstyle=solid,fillcolor=grey,linewidth=0.9pt](15.375,1.5){1.5}{-73}{253}
 %
 \psarc[fillstyle=solid,fillcolor=myblue,linecolor=myblue,linewidth=0.9pt](15.375,1.5){1.106}{160}{20}
 %
\rput(8.875,0.25){
 \psline[fillstyle=solid,fillcolor=brown,linewidth=0.9pt](15.15,0.8)(14.9,0.3)(11,0.3)(11,0.8)(15.15,0.8)
 \psline[fillstyle=solid,fillcolor=brown,linewidth=0.9pt](14.55,-0.4)(14.3,-0.9)(11,-0.9)(11,-0.4)(14.55,-0.4)
 \psline[fillstyle=solid,fillcolor=brown,linewidth=0.9pt](13.95,-1.6)(13.7,-2.1)(11,-2.1)(11,-1.6)(13.95,-1.6)
 \psline[fillstyle=solid,fillcolor=brown,linewidth=0.9pt](13.35,-2.8)(13.1,-3.3)(11,-3.3)(11,-2.8)(13.35,-2.8)
}
%
 \psline[fillstyle=solid,fillcolor=color_6,linewidth=0.9pt](24,1)(24.25,1.5)(25.25,1)(24.25,-1)(26.75,-1)(26.75,-2)(23.75,-2)(22.75,-4)(21.75,-3.5)(24,1)
%
%
\rput(7.3,2){$\overbrace{\hspace{17.1mm}}_{\mbox{}}$}
\rput(7.3,3.5){$\scriptsize\mbox{Hydraulic Pump}$}
\rput(15.375,3.5){$\overbrace{\hspace{24mm}}_{\mbox{}}$}
\rput(15.375,5){$\scriptsize\mbox{Hydraulic Capacitor}$}
\rput(23.3,2){$\overbrace{\hspace{17.1mm}}_{\mbox{}}$}
\rput(23.3,3.5){$\scriptsize\mbox{Hydraulic Motor}$}
 \psline[linecolor=black,linestyle=dashed,linewidth=0.6pt](7.25,-1.25)(9,-4.75)(5,-4.75)
  \psarc[linecolor=black,linewidth=1pt]{<-}(9,-4.75){2.25}{116}{180}
\rput(6.75,-3.25){$\theta$}
 \pcline[linecolor=color_5,linewidth=1.25pt]{->}(2,-1.5)(3.5,-1.5)
  \aput[1pt]{0}(0.5){$\omega_d$} \bput[1pt]{0}(0.5){$\tau_d$}
   \pcline[linecolor=color_6,linewidth=1.25pt]{->}(27.25,-1.5)(28.75,-1.5)
  \aput[1pt]{0}(0.5){$\omega_e$} \bput[1pt]{0}(0.5){$\tau_e$}
  \pcline[linecolor=black,linewidth=1.25pt]{->}(12.75,-1.262)(14.25,-1.262)
  \rput(13.5,-2.6){\small $Q_p(\theta)$}
  \pcline[linecolor=black,linewidth=1.25pt]{->}(16.6,-1.262)(18.1,-1.262)
    \rput(17.35,-2.6){\small $Q_m$}
    \pcline[linecolor=black,linewidth=0.635pt]{->}(15.5,2.25)(17.25,2.75)
    \rput(18.1,2.82){$P_{de}$}
    \rput(13.4,0.2){$C_{de}$}
    }}
 \end{pspicture}
 \vspace{-0.68cm}
   }{
    Second Case Study: An hydraulic Continuous Variable Transmission (CVT). (a) The structure of the hydraulic CVT system, which is composed of a planetary gear set and of a hydro-mechanical part. (b) The structure of the CVT hydro-mechanical part, which is composed of an hydraulic pump and an hydraulic motor charging and discharging an hydraulic capacitor $C_{de}$. The pressure $P_{de}$ within the hydraulic capacitor $C_{de}$ is function of the hydraulic pump volume flow rate $Q_p(\theta)$ and of the hydraulic motor volume flow rate $Q_m$: $C_{de} \dot{P}_{de}= Q_p(\theta) - Q_m $.
    \label{POG_CVT_Grafico}
     }

The planetary gear set interacts with the other physical elements through the ring (blue element in Figure~\ref{POG_CVT_Grafico}(a)), sun (yellow element), planetary (green elements), and carrier (red element) power sections, characterized by the power variables $(\omega_r,\,\tau_r)$, $(\omega_s,\,\tau_s)$, $(\omega_p,\,\tau_p)$, and $(\omega_c,\,\tau_c)$, respectively. The sun is connected to the hydraulic pump (light blue element), while the ring is connected to the hydraulic motor (pink element). The gears 
exchange the tangential forces $F_{sp}$, $F_{pr}$ and $F_{re}$ while elastically interacting with each other through the springs characterized by the stiffness coefficients $K_{sp}$, $K_{pr}$ and $K_{re}$, and the sun exchanges the tangential force $F_{sd}$ with the hydraulic pump
while elastically interacting with it through the stiffness coefficients $K_{sd}$. 
As shown in Figure~\ref{POG_CVT_Grafico}(b), the pressure $P_{de}$ within the hydraulic capacitor $C_{de}$ is a function of the hydraulic pump volume flow rate $Q_p(\theta)$ and of the hydraulic motor volume flow rate $Q_m$ as described by the following differential equation: $C_{de} \dot{P}_{de}= Q_p(\theta) - Q_m $. By controlling the tilt angle $\theta$ of the hydraulic pump plates, as highlighted in Figure~\ref{POG_CVT_Grafico}(b), the hydraulic pressure $P_{de}$ can be properly controlled. 

 The hydraulic CVT of Figure~\ref{POG_CVT_Grafico}(a) can be modeled using the vectorial POG block scheme of Figure~\ref{Ravigneaux_POG_scheme}, which is said to be multidimensional since the matrices composing it contain more than one dynamic or static PE. 
 \sdbarfig{  
\begin{pspicture}(-2,8)(5,42)
\centering
 \schemaPOG{(-1.5,-3.82)(26,19.02)}{2.6mm}{
 \thicklines
 \bloin{\boldsymbol{\tau}}{\boldsymbol{\omega}}
 \blovigiumxm{\ds \J\muno}{}{\dst}{\frac{1}{s}}{}
 \bloivsuaor{\ds \B_{J}}{}{}
 \blokxyaor{3.5}{2.5}{\R\tras(t)}{\R(t)}{}{}
 \blovisumxmaor{\ds \K}{\scr\F}{}{\frac{1}{s}}{}
 \bloivsuaor{\ds \B_{K}}{}{}   
 \bloout{\boldsymbol{0}}{\F}
 }
\pscircle[linewidth=0.4pt](-26,2){1.25mm}
\rput(-26,2){\scriptsize $1$}
\pscircle[linewidth=0.4pt](-20,2){1.25mm}
\rput(-20,2){\scriptsize $2$}
\pscircle[linewidth=0.4pt](-16,2){1.25mm}
\rput(-16,2){\scriptsize $3$}
\pscircle[linewidth=0.4pt](-10.5,2){1.25mm}
\rput(-10.5,2){\scriptsize $4$}
\pscircle[linewidth=0.4pt](-4.5,2){1.25mm}
\rput(-4.5,2){\scriptsize $5$}
\pscircle[linewidth=0.4pt](-0.5,2){1.25mm}
\rput(-0.5,2){\scriptsize $6$}
%
%
\end{pspicture}
 }{
 POG block scheme of the hydraulic CVT in Figure~\ref{POG_CVT_Grafico}(a). Starting from the left-hand side, the first two parallel-connected elaboration blocks  describe the dynamics of the rotational elements in matrix $\J$, affected by the friction coefficients in matrix $\B_J$.
 The connection block characterized by matrix $\R(t)$ describes the energy conversions in the system, between the mechanical rotational/translational domains and between the mechanical rotational/hydraulic domains. The last two parallel-connected elaboration blocks describe the dynamics of the elastic and hydraulic elements in matrix $\K$, affected by the dissipative terms in matrix $\B_K$.
\label{Ravigneaux_POG_scheme}
     }
 The parallel-connected elaboration blocks between power sections 
 \Circled{\scriptsize 1} - \Circled{\scriptsize 2} and  \Circled{\scriptsize 2} - \Circled{\scriptsize 3},
 describing the mechanical rotational part of the system, are composed of the inertia and friction matrices $\J$ and $\B_J$ shown in \eqref{Ravigneaux_POG_scheme_vect_matr}. The parallel-connected elaboration blocks between power sections 
  \Circled{\scriptsize 4} - \Circled{\scriptsize 5} and  \Circled{\scriptsize 5} - \Circled{\scriptsize 6},
 describing the mechanical translational and hydraulic parts of the system, are composed of the coupling and dissipative matrices $\K$ and $\B_K$ shown in \eqref{Ravigneaux_POG_scheme_vect_matr}. The vectors $\boldsymbol{\tau}$,  
 $\boldsymbol{\omega}$ and $\F$ are also given in \eqref{Ravigneaux_POG_scheme_vect_matr}, while the energy conversion matrix $\R(t)$ 
 between power sections  \Circled{\scriptsize 3} - \Circled{\scriptsize 4}
 is given in \eqref{POG_CVT_Def_R}.

 \begin{sequation}\label{Ravigneaux_POG_scheme_vect_matr}
      \begin{array}{c}
 \boldsymbol\omega\!=\!\!
\left[\begin{array}{@{\!}c@{\!}}
\omega_{c} \\
\omega_{p} \\
\omega_{s} \\
\omega_{r} \\
\omega_{d} \\
\omega_{e}
\end{array}\right]
 \!\!, \boldsymbol\tau\!=\!\!
\left[\begin{array}{@{\!}c@{\!}}
\tau_{c} \\
\tau_{p} \\
\tau_{s} \\
\tau_{r} \\
\tau_{d} \\
\tau_{e} \end{array} \right]
 \!\!, \boldsymbol J\!\!=\!\!
\left[\begin{array}{@{\!}c@{}c@{}c@{}c@{}c@{}c@{\!}}
J_{c} & 0 & 0 & 0 & 0 & 0 \\
0 & J_{p} & 0 & 0 & 0 & 0 \\
0 & 0 & J_{s} & 0 & 0 & 0 \\
0 & 0 & 0 & J_{r} & 0 & 0 \\
0 & 0 & 0 & 0 & J_{d} & 0 \\
0 & 0 & 0 & 0 & 0 & J_{e}
\end{array} \right]
 \!\!, \B_{J}\!\!=\!\!
\left[\begin{array}{@{\!}c@{}c@{}c@{}c@{}c@{}c@{\!}}
b_{c} & 0 & 0 & 0 & 0 & 0 \\
0 & b_{p} & 0 & 0 & 0 & 0 \\
0 & 0 & b_{s} & 0 & 0 & 0 \\
0 & 0 & 0 & b_{r} & 0 & 0 \\
0 & 0 & 0 & 0 & b_{d} & 0 \\
0 & 0 & 0 & 0 & 0 & b_{e}
\end{array} \right]
 \!\!,\\[12.5mm]
 \F\!=\!\! \left[\begin{array}{@{\!}c@{\!}}
F_{sp} \\
F_{pr} \\
F_{sd} \\
F_{re} \\
P_{de} \\
\end{array}\right]
 \!\!, \boldsymbol K\!\!=\!\!
\left[\begin{array}{@{\!}c@{\!\!}c@{\!\!}c@{\!\!}c@{\!\!}c@{\!}}
K_{sp} & 0 & 0 & 0 & 0 \\
0 & K_{pr} & 0 & 0 & 0 \\
0 & 0 & K_{sd} & 0 & 0 \\
0 & 0 & 0 & K_{re} & 0 \\
0 & 0 & 0 & 0 & {C_{de}}^{\!\!-\!1}
\end{array} \right]
 \!\!, \boldsymbol B_K\!\!=\!\!
\left[\begin{array}{@{\!}c@{\!}c@{\!}c@{\!}c@{\!}c@{\!}}
d_{sp} & 0 & 0 & 0 & 0 \\
0 & d_{pr} & 0 & 0 & 0 \\
0 & 0 & d_{sd} & 0 & 0 \\
0 & 0 & 0 & d_{re} & 0 \\
0 & 0 & 0 & 0 & R_{de}
\end{array} \right]
 \!\!.
\end{array} 
 \end{sequation}

\end{sidebar}

 \begin{sidebar}{\continuesidebar}

 \begin{sequation}\label{POG_CVT_Def_R}
 \R(t)=\mat{cccccc}{
 -r_{c} & r_{p} & r_{s} & 0 & 0 & 0 \\[1mm] r_{c} & r_{p} & 0 & -r_{r} & 0 & 0 \\[1mm] 0 & 0 & r_{a} & 0 & r_{d} & 0 \\[1mm] 0 & 0 & 0 & r_{ {re}} & 0 & r_{e} \\[1mm] 0 & 0 & 0 & 0 & h_{p}(\theta) & -h_{q}
}.\end{sequation}

The procedure to obtain the matrix $\R(t)$ from the schematics of the hydraulic CVT system shown in Figure~\ref{POG_CVT_Grafico} is described in \cite{Nostro_1}.
The POG block scheme of Figure~\ref{Ravigneaux_POG_scheme} is in a one-to-one correspondence with the following POG state-space model:

 \begin{sequation}\label{Ravigneaux_Model_0}
 \begin{array}{@{\!}r@{}c@{}l@{\!}}
 \underbrace{\mat{@{}c@{\;\;}c@{}}{\J&\mathbf{0}\\\mathbf{0}&\K\muno}}_{\L}\!\dot{\x}
 &=&
 \underbrace{\mat{@{\!}c@{\;\;}c@{\!}}{-\!\B_J\!-\!\R\tras(t)\B_K\R(t) & -\!\R\tras(t)\\ \R(t) &\mathbf{0}}}_{\A(t)}\!\x
 \!+\!
 \underbrace{\mat{@{\;}c@{\;}}{\I\\
 \boldsymbol{0}}}_{\B}\,
 \underbrace{\u}_{\boldsymbol{\tau}}, \\[10.5mm]
 \y & = &\x=\mat{@{\!}c@{\!}}{\boldsymbol{\omega} \\[2mm] \F},
 \end{array}
 \end{sequation}
 where the definition of the different matrices and vectors is given in \eqref{Ravigneaux_POG_scheme_vect_matr} and \eqref{POG_CVT_Def_R}.
Model~\eqref{Ravigneaux_Model_0} is obtained by applying Property~\ref{To_StSp}, and is in the POG state-space form $\S$ in~\eqref{POG_stsp}. 
 From the definition of the energy conversion matrix $\R(t)$ in \eqref{POG_CVT_Def_R}, it can be noticed that the coefficients handling the mechanical rotational/translational energy conversions are the radii of all the gears present in the hydraulic CVT system depicted in Figure~\ref{POG_CVT_Grafico}(a), while the coefficients responsible for the mechanical rotational/hydraulic energy conversions are the coefficients $h_{p}(\theta)$ and $h_q$ in Figure~\ref{POG_CVT_Grafico}(a). Since the coefficient $h_{p}(\theta)$ is function of the time-variant tilt angle $\theta=\theta(t)$ of the hydraulic pump plates, the energy conversion matrix $\R(t)$ is time-variant as well and, consequently, the considered hydraulic CVT system exhibits a time-variant nature.

 By disregarding the gears elastic interaction and the hydraulic capacitor, namely by letting $\K \rightarrow \infty$ and $C_{de} \rightarrow 0$, from
the first equation in \eqref{Ravigneaux_Model_0} the set
$\R(t)\boldsymbol \omega=\boldsymbol 0$ of vectorial constraints can be obtained. The latter constraint implies that the speed vector $\boldsymbol
\omega$ can be expressed as a function of a new reduced speed vector vector $\hx$ of the reduced-order system: $
\boldsymbol \omega=\Q_1(t) \hx$. Choosing $\hx=\omega_s$, the following can be written:
\begin{sequation}\label{CVT_hydr_red}
 \underbrace{\mat{@{\!}c@{\!}}{\boldsymbol{\omega}\\ \F}}_{\ds \x}
 \!\!\!= \!\!\underbrace{\mat{@{\!}c@{\!}}{\Q(t)\\[1mm] \mathbf{0}} }_{\ds
 \T(t)}\!\underbrace{\hx}_{\omega_s}
 \!,
 \;\; \mbox{where}\;\;
 \Q(t)\!\!=\!\!\left[\begin{array}{@{\!}c@{\!}}
 \frac{h_q r_d r_{re} r_s+K_p r_a r_e r_r \theta  }{2 h_q r_d r_{re} r_c} \\[1mm]
 \frac{K_p r_a r_e r_r \theta  -h_q r_d r_{re} r_s }{2 h_q r_d r_p r_{re}} \\[1mm]
 1 \\[1mm]
 \frac{K_p\, r_a\,r_e\,\theta }{h_q\,r_d\,r_{re}} \\[1mm]
 -\frac{r_a}{r_d} \\[1mm]
 -\frac{K_p\, r_a\,\theta }{h_q\,r_d}
 \end{array}\right]\!\!.
\end{sequation}
The congruent transformation $ \x=\T(t) \hx$ in \eqref{CVT_hydr_red}
is a particular case,  when $\T_u=0$, of the congruent transformation defined in \eqref{POG_stsp_transf}. This
congruent time-variant transformation 
relates the state vector $\x$ of the original hydraulic CVT model \eqref{Ravigneaux_Model_0} to the chosen new state vector
$\hx=\omega_s$ of the reduced-order hydraulic CVT model. By applying the transformation $ \x=\T(t)
\hx$ in \eqref{CVT_hydr_red} to system \eqref{Ravigneaux_Model_0}, the
following reduced state-space time-variant model of the hydraulic CVT can be obtained:
 \begin{sequation}\label{Ravigneaux_reduced_bis}
 \begin{array}{c}  \hL(t)\,\dot\hx=\hA(t)\,\hx+\hB(t)\, \boldsymbol
 \tau,\;
 \end{array}
 \end{sequation}
where the new system matrices $\hL(t)$, $\hA(t)$, and $\hB(t)$ are computed as described in  \eqref{POG_stsp_transf_matr}. 





  \end{sidebar}
Two examples of this type is shown in the sidebars ``Second Case Study: An Hydraulic Continuous Variable Transmission'' and ``Third Case Study: A Permanent Magnet Synchronous Motor''. 
The static element $R$ can be graphically represented using the EB of  Figure~\ref{basic_EBs_and_CBs}(c). In this case there is no integral causality issue, meaning that the static block $R$ can be freely input-output inverted without any problem.
%
%
%
\begin{sidebar}{Third Case Study: A Permanent Magnet Synchronous Motor}

\setcounter{sequation}{6}
\renewcommand{\thesequation}{S\arabic{sequation}}
\setcounter{stable}{1}
\renewcommand{\thestable}{S\arabic{stable}}
\setcounter{sfigure}{4}
\renewcommand{\thesfigure}{S\arabic{sfigure}}

Figure~\ref{POG_PMSM_Grafico} shows a schematic representation of the third case study, which is a Permanent Magnet Synchronous Motor (PMSM)~\cite{Nostro_4}.

\sdbarfig{
 \centering
 \setlength{\unitlength}{3.5mm}
 \psset{unit=\unitlength}
 \psset{linewidth=.4pt}
 \psset{arrowlength=.8}
 \psset{arrowinset=.1}
 \begin{pspicture}(0,5)(22,18.5)
 \put( 2.5, 11.5){\vector(0,-1){0}}
 \put( 5.5, 11.5){\vector(0,-1){0}}
 \put( 8.5, 11.5){\vector(0,-1){0}}
 \put( 4.25, 12){\makebox(0,0)[r]{$I_{d}$}}
 \put( 7.25, 12){\makebox(0,0)[r]{$I_{q}$}}
 \put( 2.5,6){\line(1,0){6}}
 \put( 2.5, 6){\indver{L_{s}}{}}
 \put( 2.5,12){\molver{R_{s}}{}}
 \put( 2.5,18){\punto}
  \put( 4,18.5){\makebox(0,0)[b]{$V_{d}$}}
 \put( 5.5, 6){\indver{}{}}
 \put( 5.5,12){\molver{}{}}
 \put( 5.5,18){\punto}
  \put( 7,18.5){\makebox(0,0)[b]{$V_{q}$}}
 %
 \put( 8.5, 6){\indver{}{}}
 \put( 8.5,12){\molver{}{}}
 \put( 8.5,18){\punto}
 %
 %
 \put( 5.5, 6){\punto}
 \put(7, 7){\makebox(0,0)[b]{$\scr L_{s_0}$}}
 \put(4, 7){\makebox(0,0)[b]{$\scr L_{s_0}$}}
 \put(12, 8.5){\begin{picture}(4,8)   
  \put(0,3.5){\line(1,0){1}}
  \put(0,3.5){\line(0,1){1}}
  \put(8,3.5){\line(0,1){1}}
  \put(0,4.5){\line(1,0){1}}
  \put(5.5,3.5){\line(1,0){2.5}}
  \put(5.5,4.5){\line(1,0){2.5}}
  \put(1,2){\framebox(4.5,4){$J_{m}$}}
  \multiput(1,2)(0.5,0){7}{\line(1,1){1}}
  \multiput(1,5)(0.5,0){7}{\line(1,1){1}}
  \put(6,4.75){\framebox(1.5,0.5){}}
  \multiput(6,4.85)(0,0.1){4}{\line(1,0){1.5}}
  \put(6,2.75){\framebox(1.5,0.5){}}
  \multiput(6,2.85)(0,0.1){4}{\line(1,0){1.5}}
  \put(6,5.5){\makebox(1.5,2){$b_{m}$}}
  \put(   1,6){\makebox(4.5,2){$\omega$}}
  \put(8.68,2){\makebox(0,2)[r]{${\tau}$}}
\end{picture}}
 \end{pspicture}
  }{
    Third Case Study: A Permanent Magnet Synchronous Motor (PMSM). On the left: the electrical part of the PMSM, composed of the stator phase inductors $L_s$, which are mutually coupled through the mutual inductors coefficient having maximum value $L_{s_0}$, and of the stator phases resistors $R_s$. On the right: the mechanical part of the PMSM, composed of the rotor inertia $J_m$ rotating at angular speed $\omega$, subject to the friction coefficient $b_m$ and to the load torque ${\tau}$.
    \label{POG_PMSM_Grafico}
     }
The electrial part of the PMSM is composed of the stator phase inductors and resistors: $L_s$ and $R_s$, respectively. The inductors $L_s$ are mutually coupled through the mutual inductance coefficient having maximum value  $L_{s_0}$. 
The input electrical power in the d-q reference frame is given by ${\V}\tras\,\I$, where the voltage and current vectors ${\V}$ and $\I$ are given by:

\begin{sequation}\label{PMSM_POG_scheme_VI}
 \V\!=\!\!
\left[\begin{array}{@{\!}c@{\!}}
V_{d} \\
V_{q}
\end{array}\right]
 \!\!, \hspace{11mm} \mbox{and} \hspace{11mm} \I\!=\!\!
\left[\begin{array}{@{\!}c@{\!}}
I_{d} \\
I_{q}  \end{array} \right],
 \end{sequation}
%
 and is then converted to the mechanical rotational domain, after the impact of the stator dynamics and some
 dissipation associated with the electrical resistors $R_s$. The resulting output power flux is given by $\omega\,{\tau}$, being $\omega$ and $\tau$ the rotor angular speed and load torque, respectively, after the impact of the rotor dynamics and some
 dissipation associated with the rotor friction coefficient $b_m$. 
 
 The physical system in Figure~\ref{POG_PMSM_Grafico} can be modeled using the POG block scheme shown in Figure~\ref{POG_scheme_PMSM}. 
As observed in the sidebar ``Second Case Study: An Hydraulic Continuous Variable Transmission'', even in this case it is remarkable to notice that, if the input and output variables of a POG block scheme are multidimensional, as for the voltage and current vectors ${\V}$ and $\I$ in \eqref{PMSM_POG_scheme_VI}, the involved elaboration and connection blocks become multidimensional as well. In fact, the series-connected elaboration blocks between power sections  \Circled{\scriptsize 1} - \Circled{\scriptsize 2} and  \Circled{\scriptsize 2} - \Circled{\scriptsize 3},
describing the electrical part of the system, are composed of the matrices $\L_e$ and $\L_{en}$ defined in \eqref{PMSM_POG_scheme_vect_matr}. The elaboration block between power sections  \Circled{\scriptsize 3} - \Circled{\scriptsize 4},
describing the dissipation taking place in the electrical part of the PMSM, is composed of the resistance matrix $\R_e$ defined in \eqref{PMSM_POG_scheme_vect_matr}. 
\sdbarfig{
\begin{pspicture}(-20,0.86)(40,34)
\centering
 \schemaPOG{(2,-2)(20,46)}{2.325mm}{
 \thicklines
 \bloin{{\V}}{\I}
 \blovigiumxm{\L_e\muno}{}{\dst}{\frac{1}{s}}{}
 \bloivsuaor{\mbox{\footnotesize $ \L_{en}$}}{}{}
 \bloivsuaor{\R_e}{}{}
 \blokxyaor{3.7}{2.5}{\mbox{$\K_{\tau}$}}{\mbox{\small $\K_{\tau}\tras$}}{}{\overline{\tau}}
 \blovisumxm{\ds\frac{1}{J_m}}{}{\dst}{\frac{1}{s}}{}
 \bloivgiuor{b_m}{}{}
 \bloout{\omega}{{\tau}}
 }
\pscircle[linewidth=0.4pt](-20,0){1.25mm}
\rput(-20,0){\scriptsize $1$}
\pscircle[linewidth=0.4pt](-14,0){1.25mm}
\rput(-14,0){\scriptsize $2$}
\pscircle[linewidth=0.4pt](-10,0){1.25mm}
\rput(-10,0){\scriptsize $3$}
\pscircle[linewidth=0.4pt](-6,0){1.25mm}
\rput(-6,0){\scriptsize $4$}
\pscircle[linewidth=0.4pt](-0.25,0){1.25mm}
\rput(-0.25,0){\scriptsize $5$}
\pscircle[linewidth=0.4pt](5.75,0){1.25mm}
\rput(5.75,0){\scriptsize $6$}
\pscircle[linewidth=0.4pt](9.75,0){1.25mm}
\rput(9.75,0){\scriptsize $7$}
%
%
\end{pspicture}
 }{POG block scheme of the Permanent Magnet Synchronous Motor schematized in Figure~\ref{POG_PMSM_Grafico}. Starting from the left-hand side, the first three series-connected elaboration blocks  describe the dynamics of the electric part, which is characterized by the dynamic elements in matrices $\L_e$ and $\L_{en}$ and by the static elements in matrix $\R_e$.
 The connection block characterized by vector 
  $\K_{\tau}\!=\! \left[\begin{array}{@{\!}c@{\;\;}c@{\!}}
 K_{d} &
 K_{q}
 \end{array}\right]\tras
 $ 
 describes the energy conversion between the electrical and the mechanical rotational energetic domains. The last two parallel-connected elaboration blocks describe the dynamics of the mechanical part, characterized by the rotor inertia $J_m$ and by the friction coefficient $b_m$.
\label{POG_scheme_PMSM}
     }
     The matrices $\L_e$, $\L_{en}$ and $\R_e$ are defined as:
\begin{sequation}\label{PMSM_POG_scheme_vect_matr}
      \begin{array}{@{\!\!}c}
\L_e\!\!=\!\!
\left[\begin{array}{@{\!}c@{\!}c@{\!}}
p L_{e} & 0 \\[2mm]
0 & p L_{e}
\end{array} \right]
 \!\!, \L_{en}\!\!=\!\!
\left[\begin{array}{@{\!}c@{\!\!}c@{\!}}
0 & -p^2 \omega L_{e} \\[2mm]
p^2 \omega L_{e} & 0
\end{array} \right]
 \!\!, \R_e\!\!=\!\!
\left[\begin{array}{@{\!}c@{\!\!}c@{\!}}
-p R_s & 0 \\[2mm]
0 & -p R_s
\end{array} \right]
 \!\!,
\end{array}
 \end{sequation}
     where $L_{e}=L_s+\frac{L_{s0}}{2}$, $L_s$ is the auto-inductance coefficient, and $p$ is the number of polar pairs.
The connection block between power sections \Circled{\scriptsize 4} - \Circled{\scriptsize 5},
describing the energy conversion between the electrical and mechanical rotational domains, is composed of the torque vector $\K_{\tau}\!=\! \left[\begin{array}{@{\!}c@{\;\;}c@{\!}}
 K_{d} &
 K_{q}
 \end{array}\right]\tras
 $.
It is in correspondence of this connection block that the energy conversion between the two-dimensional electrical part and the one-dimensional mechanical part of the PMSM occurs: the parallel-connected elaboration blocks between power sections  
\Circled{\scriptsize 5} - \Circled{\scriptsize 6} and \Circled{\scriptsize 6} - \Circled{\scriptsize 7}
are no longer two-dimensional, but one-dimensional instead. These two elaboration blocks describe the mechanical rotational part of the system, and are composed of the rotor inertia $J_m$ and of the rotor friction coefficient $b_m$.

The POG block scheme of Figure~\ref{POG_scheme_PMSM} is in a one-to-one correspondence with the following state-space model:

\begin{sequation}\label{PMSM_POG_scheme_stsp}
 \begin{array}{@{}c}
 \underbrace{
 \mat{@{\!}c@{\;}c@{\!}}{
 \L_{e}  & 0  \\
 0 &  J_{m}}}_{\L}
  \underbrace{
 \mat{@{\!}c@{\!}}{
  \dot \I \\
  \dot \omega}}_{\dot \x}
 \!=\!
  \underbrace{
 \mat{@{\!}c@{\;}c@{\!}}{
 \L_{en}+\R_{e} & -\K_\tau  \\
 \tras\K_\tau & -b_{m}}}_{\A}
 
   \underbrace{
\mat{@{\!}c@{\!}}{
  \I \\
 \omega}}_{\x}
  \!+
     \underbrace{
 \mat{@{\!}c@{\;}c@{\!}}{
 1 & 0  \\
 0 & -1} }_{\B}
    \underbrace{
 \mat{@{\!}c@{\!}}{
  {\V} \\
  {\tau}}}_{\u}\!, 
%
 \end{array}
 \end{sequation}
and $\y=\x$ 
and $\D=\boldsymbol{0}$. System~\eqref{PMSM_POG_scheme_stsp} is obtained by applying Property~\ref{To_StSp}, and is in the POG state-space form $\S$ defined in~\eqref{POG_stsp}. 
\end{sidebar}

 \begin{sidebar}{\continuesidebar}
The dynamics of the PMSM can also be expressed in the static reference frame starting from the model of Figure~\ref{POG_scheme_PMSM} by using a proper transformation~\cite{Nostro_4}, thus allowing effective simulations of the PMSM in the static frame as well. A simulation example is shown in Figure~\ref{PMSM_Simulation_Figure}, where the PMSM is simulated using the parameters available from the dataset in~\cite{Nostro_4}. In this case, a speed control using a proportional-integral-derivative regulator is applied to the PMSM, in order to make the rotor speed $\omega$ track the desired speed profile $\omega_{d}$ shown in Figure~\ref{PMSM_Simulation_Figure}(a):
\[
\overline{\tau}_{d} \! = \! K_p 
\left(1 + T_d s + \dfrac{1}{T_i s}\right)\left(\omega_{d}-\omega\right),
\]
where $K_p$, $T_d$ and $T_i$ are the proportional action and the time constants of the derivative and integral actions, respectively, and are set as the dataset in~\cite{Nostro_4}. In the simulation of Figure~\ref{PMSM_Simulation_Figure}, the PMSM is supposed to be subject to a load torque $\tau=200$ Nm.
Figure~\ref{PMSM_Simulation_Figure}(b) shows that the actual motive torque $\overline{\tau}$ properly tracks the desired profile $\overline{\tau}_{d}$, resulting in 
the rotor speed $\omega$ properly tracking the desired profile $\omega_{d}$, as shown in Figure~\ref{PMSM_Simulation_Figure}(a). Figure~\ref{PMSM_Simulation_Figure}(c) and Figure~\ref{PMSM_Simulation_Figure}(d) report the three-phase voltages and currents in the static frame in a zoomed-in time interval, showing the detailed simulations that can be performed starting from the PMSM model of Figure~\ref{POG_scheme_PMSM}.

\setcounter{sfigure}{6}
\renewcommand{\thesfigure}{S\arabic{sfigure}}

\sdbarfig{
 \psfrag{[rpm]}[][][0.85]{[rpm]}
 \psfrag{[Nm]}[][][0.85]{[Nm]}
 \psfrag{[V]}[][][0.85]{[V]}
 \psfrag{[A]}[][][0.85]{[A]}
 \psfrag{Wmd}[][][0.68]{$\omega_d$}
 \psfrag{Wm}[][][0.68]{$\omega$}
 \psfrag{Tmd}[][][0.68]{$\overline{\tau}_{d}$}
 \psfrag{Tm}[][][0.68]{$\overline{\tau}$}
 \psfrag{Is1}[][][0.68]{$I_1$}
 \psfrag{Is2}[][][0.68]{$I_2$}
 \psfrag{Is3}[][][0.68]{$I_3$}
 \psfrag{Vs1}[][][0.68]{$V_1$}
 \psfrag{Vs2}[][][0.68]{$V_2$}
 \psfrag{Vs3}[][][0.68]{$V_3$}
 \psfrag{(a)}[b][b][0.68]{(a)}
 \psfrag{(b)}[b][b][0.68]{(b)}
 \psfrag{(c)}[b][b][0.68]{(c)}
 \psfrag{(d)}[b][b][0.68]{(d)}
  \psfrag{Time [s]}[t][t][0.85]{Time [s]}
\includegraphics[clip,width=1\columnwidth]{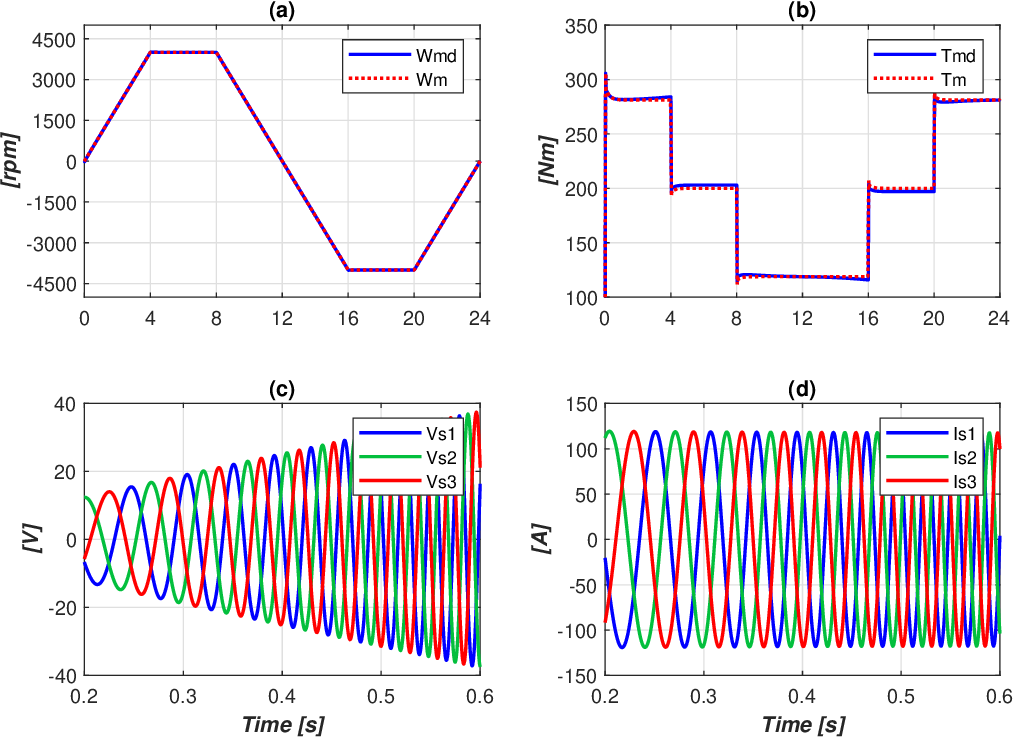}
  }{Simulation of the Permanent Magnet Synchronous Motor (PMSM). A speed control is applied to the PMSM in order to make the rotor speed track the desired profile while subject to a load torque. (a) The desired and actual rotor speeds $\omega_d$ and $\omega$. (b) The desired and actual motive torques $\overline{\tau}_d$ and $\overline{\tau}$. (c) The three-phase voltages $V_1$, $V_2$, and $V_3$. (d) The three-phase currents $I_1$, $I_2$, and $I_3$.
\label{PMSM_Simulation_Figure}
     } 
     
     \end{sidebar}
Whenever an energy conversion needs to be graphically represented, use can be made of the connection block of  Figure~\ref{basic_EBs_and_CBs}(d). If the power variables $\x_1$ and $\x_2$ are of the same type, meaning that the power variables $\y_1$ and $\y_2$ are also of the same type, the connection block assumes the physical meaning of a {\it transformer}. Conversely, the connection block is said to be a {\it gyrator}. 





\subsection{Series and Parallel Connections}\label{series_paral_sect}

A complex physical system can be composed of a large number of PEs, and 
%
%
\begin{figure}[t!]
 \begin{center}
 \setlength{\unitlength}{2.7mm}
 \psset{unit=1.0\unitlength}
 \begin{picture}(14,12.5)(-3,-8)
 \pscircle[fillstyle=solid,fillcolor=gray,linecolor=gray](0,0){0.5}
 \pscircle[fillstyle=solid,fillcolor=gray,linecolor=gray](8,0){0.5}
 \rput(0,0){\punto}
 \rput(8,0){\punto}
 \psline[linewidth=1.2pt]{-}(0,0)(2,0)
 \psline[linewidth=1.2pt]{-}(6,0)(8,0)
 \psline[linewidth=0.4pt]{-}(0,-6)(8,-6)
 \rput[l](2,0){\framebox(4,3){PE}}
 \rput(0,-6){\punto}
 \psline[linecolor=mygreen]{->}(-0.5,-5.5)(-0.5,-0.5)
 \rput[r](-0.75,-3){$\mygreen v_{e1}$}
 \rput[r](-0.5,-6){\footnotesize $0$}
 \psline[linecolor=myred]{->}(-3,0)(-0.5,0)
 \rput[b](-2,0.5){$\myred v_{f1}$} 
 \rput(8,-6){\punto}
 \psline[linecolor=mygreen]{->}(8.5,-5.5)(8.5,-0.5)
 \rput[l](8.75,-3){$\mygreen v_{e2}$}
 \rput[l](8.5,-6){\footnotesize $0$}
 \psline[linecolor=myred]{->}(8.5,0)(11,0)
 \rput[b](9.5,0.5){$\myred v_{f2}$}
 \rput[b](0,3.25){$P_1$}
 \psline{->}(-1,2.75)(1,2.75)
 \rput[b](8,3.25){$P_2$}
 \psline{->}(7,2.75)(9,2.75)
 \rput[bl](0.5,0.75){\footnotesize $1$}
 \rput[br](7.5,0.75){\footnotesize $2$}
 \psline[linewidth=0.4pt,linestyle=dashed]{-}(0,-7.5)(0,1.75)
 \psline[linewidth=0.4pt,linestyle=dashed]{-}(8,-7.5)(8,1.75)
 \rput(4,5){(a)}
   \psellipticarc[linestyle=dashed,linecolor=mygreen,linewidth=0.3pt]{->}(4,-3.75)(2.1,2.1){-70}{250}
   \rput(4,-4.1){\tiny \mbox{\shortstack{Voltage\\[-.5mm] Kirchhoff's \\[-.5mm] Law\\{\mygreen VKL}}}}
 \end{picture}
 \hspace{5mm}
 \begin{picture}(13,11)(-3,-8)
 \pscircle[fillstyle=solid,fillcolor=gray,linecolor=gray](4.0,0){0.5}
 \pscircle[fillstyle=solid,fillcolor=gray,linecolor=gray](4.0,-6){0.5}
 \rput(0,0){\punto}
 \psline[linewidth=1.2pt]{-}(0,0)(8,0)
 \rput(8,0){\punto}
 \psline[linewidth=1.2pt]{-}(4.0,0)(4.0,-1.5)
 \rput(4.0,0){\punto}
 \rput[t](4.0,-1.5){\framebox(3,3){PE}}
 \psline[linewidth=1.2pt]{-}(8,-6)(0,-6)
 \rput(8,-6){\punto}
 \rput(4.0,-6){\punto}
 \psline[linewidth=1.2pt]{-}(4.0,-4.5)(4.0,-6)
 \rput(0,-6){\punto}
 \psline[linecolor=mygreen]{->}(-0.5,-5.5)(-0.5,-0.5)
 \rput[r](-0.75,-3){$\mygreen v_{e1}$}
 \psline[linecolor=mygreen]{->}(8.5,-5.5)(8.5,-0.5)
 \rput[l](8.75,-3){$\mygreen v_{e2}$}
 \psline[linecolor=myred]{->}(-3,0)(-0.5,0)
 \rput[b](-2,0.5){$\myred v_{f1}$}
 \psline[linecolor=myred]{->}(8.5,0)(10,0)
 \rput[b](10,0.5){$\myred v_{f2}$} %
 \rput[b](0,3.25){$P_1$}
 \psline{->}(-1,2.75)(1,2.75)
 \rput[b](8,3.25){$P_2$}
 \psline{->}(7,2.75)(9,2.75)
 \rput[b](3.25,0.25){\footnotesize $1$}
 \rput[t](4.0,-6.75){\footnotesize $2$}
 \psline[linewidth=0.4pt,linestyle=dashed]{-}(0,-7.5)(0,1.75)
 \psline[linewidth=0.4pt,linestyle=dashed]{-}(8,-7.5)(8,1.75)
 \rput(4,5){(b)}
  \psellipse[linestyle=dashed,linecolor=myred,linewidth=0.3pt](4,0.1)(2.2,1.3)
  \rput(4,3){\tiny \mbox{\shortstack{Current\\ Kirchhoff's Law\\\myred CKL}}}
 \end{picture}
 \end{center}
\caption{Connections of the Physical Elements (PEs) with each other or with the external world. (a) For what concerns a PE connected in series, $\mygreen{v_e}\black=\mygreen{v_{e1}}\black-\mygreen{v_{e2}}$ holds, where such relation is named generalized Voltage Kirchhoff's Law ({\mygreen VKL}). (b) For what concerns a PE connected in parallel, $\myred{v_f }\black=\myred{v_{f1}}\black-\myred{v_{f2}}$ holds, where such relation is named generalized Current Kirchhoff's Law ({\myred CKL}).}\label{terminali}
\end{figure}
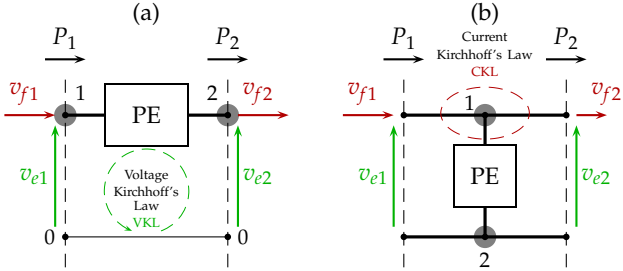
can be modeled using a right combination of EBs and CBs. The PEs interact with each other and with the external world through series connections or parallel connections, as highlighted in Figure~\ref{terminali}(a) and in Figure~\ref{terminali}(b), respectively. As shown in Figure~\ref{PE_represent}(a), each of the two PE terminals is characterized by a pair of power variables: ($\mygreen v_{e1}$, $\myred v_{f1}$) and ($\mygreen v_{e2}$, $\myred v_{f2}$), respectively.

When the PE is connected in series, the effort power variables are related through the following relation $\mygreen{v_e}\black=\mygreen{v_{e1}}\black-\mygreen{v_{e2}}$, which is named generalized Voltage Kirchhoff's Law ({\mygreen VKL}). In this case, the three possible elaboration block configurations that can be obtained from Figure~\ref{basic_EBs_and_CBs}(b)
are shown in Figure~\ref{terminali_POG_series}.
\begin{figure}[t!]
 \begin{center}
 \small
  \schemaPOG{(-0.625,0)(6.25,13.0)}{2.7mm}{
 \thicklines %
 \rput(3,11.5){\tiny \mygreen \mbox{\mygreen VKL}}
 \psellipse[linestyle=dashed,linecolor=mygreen,linewidth=0.3pt](3,10)(2.0,1.0)
 \bloin{\mygreen v_{e1}}{\myred v_{f1}}
 \flowrightsu{P_1}
 \rput[t](3,13.5){\mbox{\footnotesize (a)}}
\blovigiuintm{\myred \Phi_{f}\muno({\myred q_{f}})}{}{\dst}
 \rput(-2,5.5){\myred q_f}
 \rput(-2,0.5){\myred v_f}
 \rput(-1.75,8.75){\mygreen v_e}
 %
 \bloout{\mygreen v_{e2}}{\myred v_{f2}}
 \flowrightsu{P_2}
 }
 \schemaPOG{(-2.625,0)(6.25,13.0)}{2.7mm}{
 \thicklines %
 \rput(3,11.5){\tiny \mygreen \mbox{\mygreen VKL}}
 \psellipse[linestyle=dashed,linecolor=mygreen,linewidth=0.3pt](3,10)(2.0,1.0)
 \bloin{\mygreen v_{e1}}{\myred v_{f1}}
 \flowrightsu{P_1}
 \loopvidstaor{linewidth=1.2pt,linecolor=blue,linestyle=dashed}
%
 \rput[t](3,13.5){\mbox{\footnotesize (b)}}
\blovisuintmaor{\mygreen \Phi_{e}\muno({\mygreen q_{e}})}{}{}
 \rput(-2,4){\mygreen q_e}
 \rput(-2,0.5){\myred v_f}
  \rput(-1.68,8.85){\mygreen v_e}
 \bloout{\mygreen v_{e2}}{\myred v_{f2}}
 \flowrightsu{P_2}
 }
  \schemaPOG{(-2.625,0)(6.25,13.0)}{2.7mm}{
 \thicklines %
 \rput(3,11.5){\tiny \mygreen \mbox{\mygreen VKL}}
 \psellipse[linestyle=dashed,linecolor=mygreen,linewidth=0.3pt](3,10)(2.0,1.0)
 \bloin{\mygreen v_{e1}}{\myred v_{f1}}
 \flowrightsu{P_1}
%
 \rput[t](3,13.5){\mbox{\footnotesize (c)}}
\blovisuintmor{\mygreen \Phi_{e}\muno({\mygreen q_{e}})}{}{\giu}
 \rput(-3.75,4){\mygreen q_e}
 \rput(-3.75,0.5){\myred v_f}
  \rput(-4.28,8.85){\mygreen v_e}
 \loopvisntor{linewidth=1.2pt,linecolor=blue,linestyle=dashed}
 \bloout{\mygreen v_{e2}}{\myred v_{f2}}
 \flowrightsu{P_2}
 }
 \end{center}
\caption{Elaboration Block configurations of physical elements connected in series. (a) Basic configuration, where 
{\mygreen $v_{e1}$} and {\mygreen $v_{e2}$} are the input power variables while {\myred $v_{f1}$} and {\myred $v_{f2}$} are the output power variables. (b) This configuration can be obtained from (a) by inverting the path highlighted by the blue arrow. In this case, {\myred $v_{f1}$} and {\mygreen $v_{e2}$} are the input power variables while {\mygreen $v_{e1}$} and {\myred $v_{f2}$} are the output power variables. (c) This configuration can be obtained from (a) by inverting the path highlighted by the blue arrow. In this case, {\mygreen $v_{e1}$} and {\myred $v_{f2}$} are the input power variables while {\myred $v_{f1}$} and {\mygreen $v_{e2}$} are the output power variables.}\label{terminali_POG_series}
\vspace{-3mm}
\end{figure}
Figure~\ref{terminali_POG_series}(a) shows the basic configuration for the series connection, 
characterized by the input power variables {\mygreen $v_{e1}$} and {\mygreen $v_{e2}$} and by the output power variables {\myred $v_{f1}\black = \myred v_{f2}$}. 
By inverting either the left or the right power paths, the configurations of Figure~\ref{terminali_POG_series}(b) or Figure~\ref{terminali_POG_series}(c) are obtained, respectively. In the first case, the input power variables are {\myred $v_{f1}$} and {\mygreen $v_{e2}$}, while the output power variables are {\mygreen $v_{e1}$} and {\myred $v_{f2}$}. In the second case, the input power variables are {\mygreen $v_{e1}$} and {\myred $v_{f2}$}, while the output power variables are {\myred $v_{f1}$} and {\mygreen $v_{e2}$}. 
In order to respect the integral causality, from Figure~\ref{terminali_POG_series} it is evident that: 1) the through dynamic elements $\myred D_f$ connected in series can only be graphically represented using the POG graphical representation shown in Figure~\ref{terminali_POG_series}(a); 2) the across dynamic elements $\mygreen D_e$ connected in series can be graphically represented using the two POG graphical representations shown in Figure~\ref{terminali_POG_series}(b) and Figure~\ref{terminali_POG_series}(c).

When the physical element is connected in parallel, the through power variables are related through the following relation $\myred{v_f }\black=\myred{v_{f1}}\black-\myred{v_{f2}}$, which is named generalized Current Kirchhoff's Law ({\myred CKL}). In this case, the three possible elaboration block configurations that can be obtained from Figure~\ref{basic_EBs_and_CBs}(a) are shown in Figure~\ref{terminali_POG_parallel}. Figure~\ref{terminali_POG_parallel}(a) shows the basic configuration, 
%
\begin{figure}[t!]
 \begin{center}
 \small
   \schemaPOG{(-0.625,-1)(6.25,11.5)}{2.7mm}{
 \thicklines %
 \rput(3,-1.5){\tiny \myred \mbox{\myred CKL}}
 \psellipse[linestyle=dashed,linecolor=myred,linewidth=0.3pt](3,0)(2.0,1.0)
 \bloin{\mygreen v_{e1}}{\myred v_{f1}}
 \flowrightsu{P_1}
 \rput[t](3,13.5){\mbox{\footnotesize (a)}}
\blovisuintm{\mygreen \Phi_{e}\muno({\mygreen q_{e}})}{}{\dst}
 \rput(-2,4.5){\mygreen q_e}
 \rput(-2,9.45){\mygreen v_e}
 \rput(-1.4,1){\myred v_f}
 \bloout{\mygreen v_{e2}}{\myred v_{f2}}
 \flowrightsu{P_2}
 }
  \schemaPOG{(-2.625,-1)(6.25,11.5)}{2.7mm}{
 \thicklines %
 \rput(3,-1.5){\tiny \myred \mbox{\myred CKL}}
 \psellipse[linestyle=dashed,linecolor=myred,linewidth=0.3pt](3,0)(2.0,1.0)
 \bloin{\mygreen v_{e1}}{\myred v_{f1}}
 \flowrightsu{P_1}
 \loopvidstor{linewidth=1.2pt,linecolor=blue,linestyle=dashed}
%
 \rput[t](3,13.5){\mbox{\footnotesize (b)}}
\blovigiuintmor{\myred \Phi_{f}\muno({\myred q_{f}})}{}{}
 \rput(-2.15,6){\myred q_f}
 \rput(-1.3,1.0){\myred v_f}
  \rput(-2.15,9.45){\mygreen v_e}
 \bloout{\mygreen v_{e2}}{\myred v_{f2}}
 \flowrightsu{P_2}
 }
 \schemaPOG{(-2.625,-1)(6.25,11.5)}{2.7mm}{
 \thicklines %
 \rput(3,-1.5){\tiny \myred \mbox{\myred CKL}}
 \psellipse[linestyle=dashed,linecolor=myred,linewidth=0.3pt](3,0)(2.0,1.0)
 \bloin{\mygreen v_{e1}}{\myred v_{f1}}
 \flowrightsu{P_1}
%
 \rput[t](3,13.5){\mbox{\footnotesize (c)}}
\blovigiuintmaor{\myred \Phi_{f}\muno({\myred q_{f}})}{}{\su}
 \rput(-3.75,6){\myred q_f}
 \rput(-4.85,1.0){\myred v_f}
  \rput(-3.75,9.45){\mygreen v_e}
 \loopvisntaor{linewidth=1.2pt,linecolor=blue,linestyle=dashed}
 \bloout{\mygreen v_{e2}}{\myred v_{f2}}
 \flowrightsu{P_2}
 }
 \end{center}
\caption{Elaboration Block configurations of physical elements connected in parallel. (a) Basic configuration, where 
{\myred $v_{f1}$} and {\myred $v_{f2}$} are the input power variables while {\mygreen $v_{e1}$} and {\mygreen $v_{e2}$} are the output power variables. (b) This configuration can be obtained from (a) by inverting the path highlighted by the blue arrow. In this case, {\mygreen $v_{e1}$} and {\myred $v_{f2}$} are the input power variables while {\myred $v_{f1}$} and {\mygreen $v_{e2}$} are the output power variables. (c) This configuration can be obtained from (a) by inverting the path highlighted by the blue arrow. In this case, {\myred $v_{f1}$} and {\mygreen $v_{e2}$} are the input power variables while {\mygreen $v_{e1}$} and {\myred $v_{f2}$} are the output power variables.}\label{terminali_POG_parallel}
\end{figure}
%
characterized by the input power variables {\myred $v_{f1}$} and {\myred $v_{f2}$} and by the output power variables {$\mygreen v_{e1} \black = \mygreen v_{e2}$}. 
By inverting either the left or the right power paths, the configurations of Figure~\ref{terminali_POG_parallel}(b) or Figure~\ref{terminali_POG_parallel}(c) are obtained, respectively. In the first case, the input power variables are {\mygreen $v_{e1}$} and {\myred $v_{f2}$}, while the output power variables are {\mygreen $v_{e2}$} and {\myred $v_{f1}$}. In the second case, the input power variables are {\mygreen $v_{e2}$} and {\myred $v_{f1}$}, while the output power variables are {\mygreen $v_{e1}$} and {\myred $v_{f2}$}.
In order to respect the integral causality, from Figure~\ref{terminali_POG_parallel} it is evident that: 1) the across dynamic elements $\mygreen D_e$  connected in parallel can only be graphically represented using the POG graphical representation shown in Figure~\ref{terminali_POG_parallel}(a); 2)  the through dynamic elements $\myred D_f$ connected in parallel can be graphically represented using the two POG graphical representations shown in Figure~\ref{terminali_POG_parallel}(b) and Figure~\ref{terminali_POG_parallel}(c).

Based on the previous considerations, the following three properties hold.


\begin{figure}[t!]
    \centering \footnotesize
 \setlength{\unitlength}{1.5mm}
 \psset{unit=1.0\unitlength}
 \begin{center}
 \begin{tabular}{|@{\;}c@{\;}|@{}c@{\;\;}|@{\;}c@{\;\;}|}
   \hline  
   & &\\[18mm]
 \multirow{2}{*}{
 \schemaPOG{(-1.1,0)(10,8)}{2.0mm}{
 \thicklines %
 \rput[r](-0.5,5){\mbox{(A)}}
 \bloin{\Effi}{\Floi}
 \rput[l](4.25,8){\Effiii}
 \rput[l](4.25,1.5){\Floiii}
 \blogiu{\ds\frac{1}{\Kn\,s}}{}{\dst}
 \bloout{\Effii}{\Floii}
  }}
   &
  \psfrag{V1}[r][r]{$\scr {\mygreen V_{1}}$}
  \psfrag{V2}[l][l]{$\scr {\mygreen V_{2}}$}
  \psfrag{V3}[t][t]{$\scr {\mygreen V}$}
  \psfrag{I1}[b][b]{$\scr {\myred I_{1}}$}
  \psfrag{I2}[b][b]{$\scr {\myred I_{2}}$}
  \psfrag{I3}[t][t]{$\scr {\myred I}$}
  \psfrag{L}[b][b]{$\scr L$}
  \rput(0.5,6.25){
  \rput[b](7.7,12.5){\mbox{(a)}}
  \rput[b](7.7,-2.5){$\scr\Kn= L\;$}
  \includegraphics[clip,width=2.4cm]{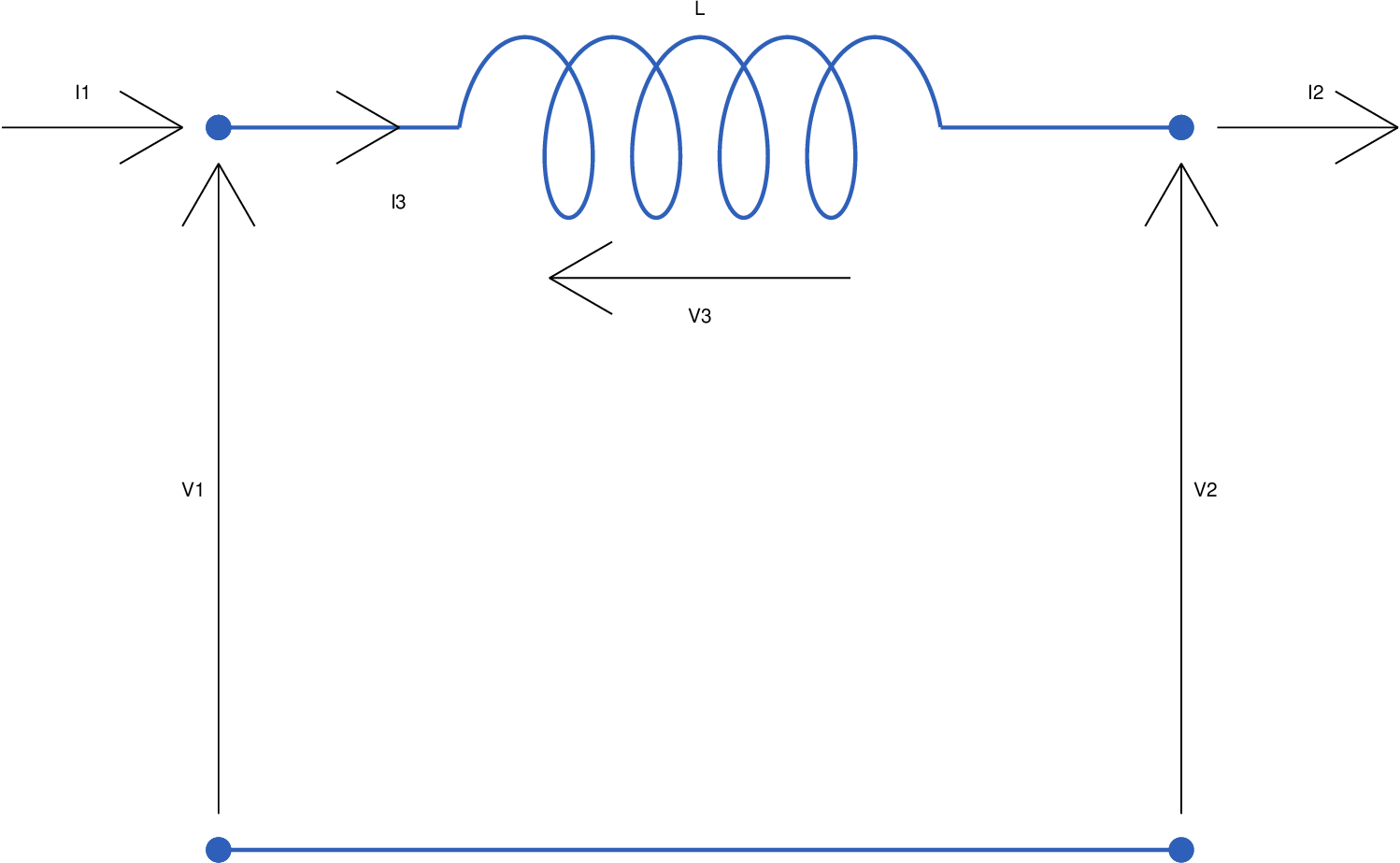}
  }
   &
  \psfrag{v1}[r][r]{$\scr {\mygreen v_{1}}$}
  \psfrag{v2}[l][l]{$\scr {\mygreen v_{2}}$}
  \psfrag{v3}[t][t]{$\scr {\mygreen v}$}
  \psfrag{F1}[b][b]{$\scr {\myred F_{1}}$}
  \psfrag{F2}[b][b]{$\scr {\myred F_{2}}$}
  \psfrag{F3}[t][t]{$\scr {\myred F}$}
  \psfrag{K}[b][b]{$\scr E$}
  \psfrag{E}[b][b]{$\scr E$}
  \rput(0.5,6.25){
  \rput[b](7.7,12.5){\mbox{(b)}}
  \rput[b](7.7,-2.5){$\scr\Kn= E\;$}
  \includegraphics[clip,width=2.4cm]{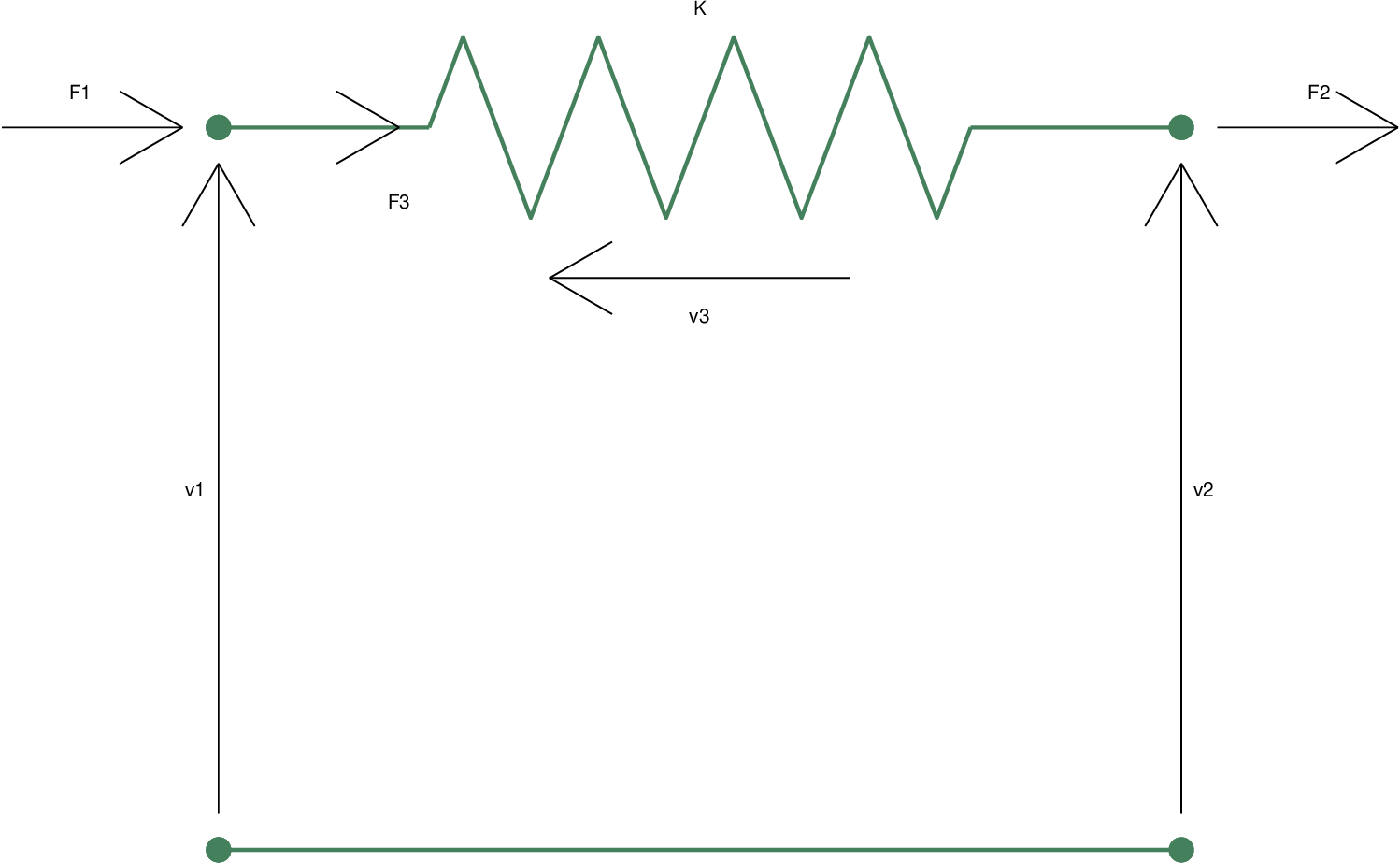}
  }
   \\[0mm]
   & &\\[-1mm]
   \cline{2-3}
   & &\\[-1mm] 
    \begin{minipage}[b]{2.0cm}
    $ $  \\         
    $ $  \\         
    $ $  \\         
    $ $  \\[1mm]         
    \end{minipage}
   &
  \psfrag{w1}[r][r]{$\scr {\mygreen \omega_{1}}$}
  \psfrag{w2}[l][l]{$\scr {\mygreen \omega_{2}}$}
  \psfrag{w3}[t][t]{$\scr {\mygreen \omega}$}
  \psfrag{T1}[b][b]{$\scr {\myred \tau_{1}}$}
  \psfrag{T2}[b][b]{$\scr {\myred \tau_{2}}$}
  \psfrag{T3}[t][t]{$\scr {\myred \tau}$}
  \psfrag{K}[b][b]{$\scr E_{r}$}
  \psfrag{Et}[b][b]{$\scr E_{r}$}
  \rput(0.5,2){
  \rput[b](7.7,11.95){\mbox{(c)}}
  \rput[b](7.7,-2.5){$\scr\Kn= E_r\;$}
  \includegraphics[clip,width=2.4cm]{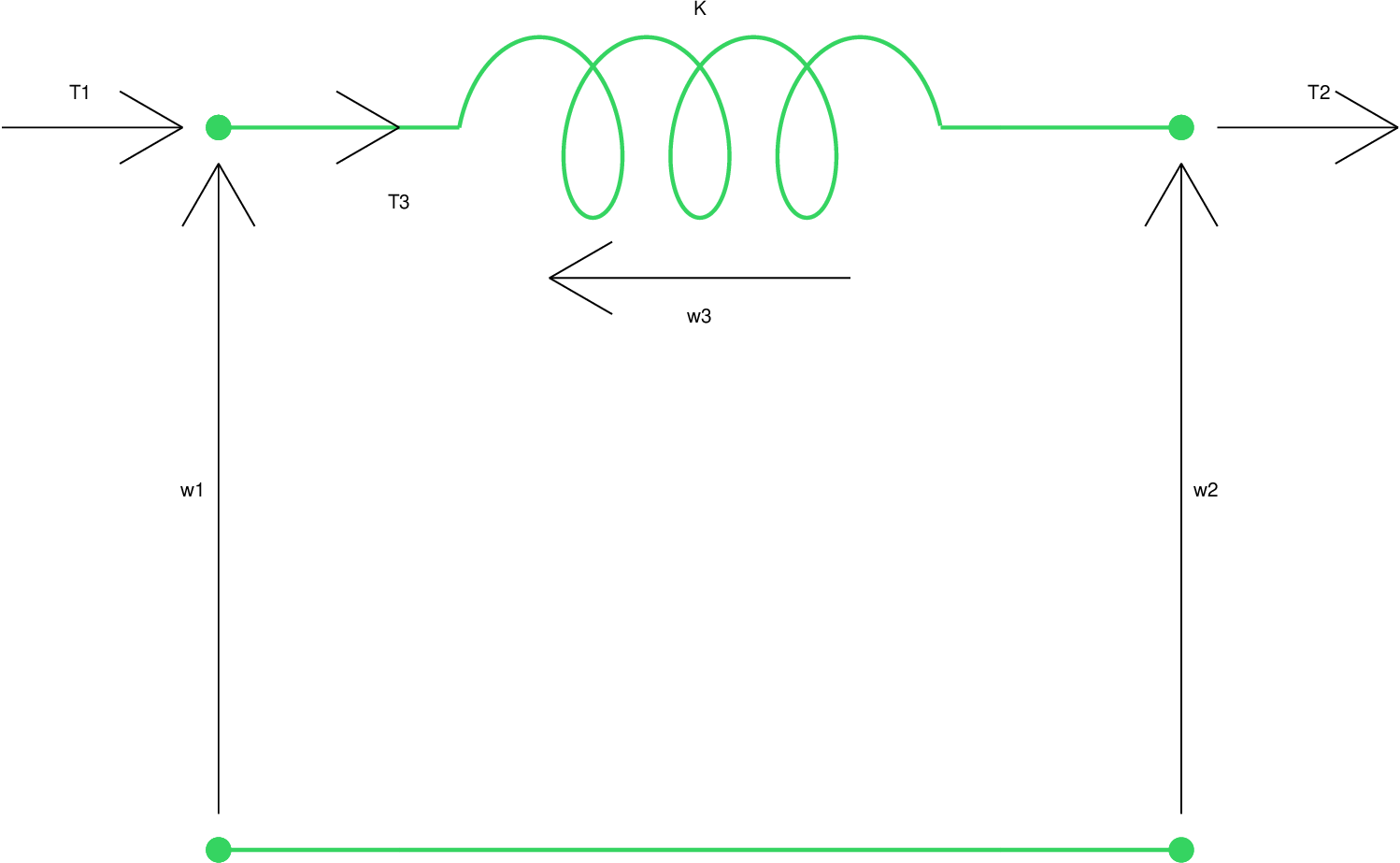}
  }
   &
  \psfrag{P1}[r][r]{$\scr {\mygreen P_{1}}$}
  \psfrag{P2}[l][l]{$\scr {\mygreen P_{2}}$}
  \psfrag{P3}[t][t]{$\scr {\mygreen P}$}
  \psfrag{Q1}[b][b]{$\scr {\myred Q_{1}}$}
  \psfrag{Q2}[b][b]{$\scr {\myred Q_{2}}$}
  \psfrag{Q3}[t][t]{$\scr {\myred Q}$}
  \psfrag{Li}[b][b]{$\scr L_{I}$}
  \rput(0.5,2){
  \rput[b](7.7,11.75){\mbox{(d)}}
  \rput[b](7.7,-2.5){$\scr\Kn= L_I\;$}
  \includegraphics[clip,width=2.4cm]{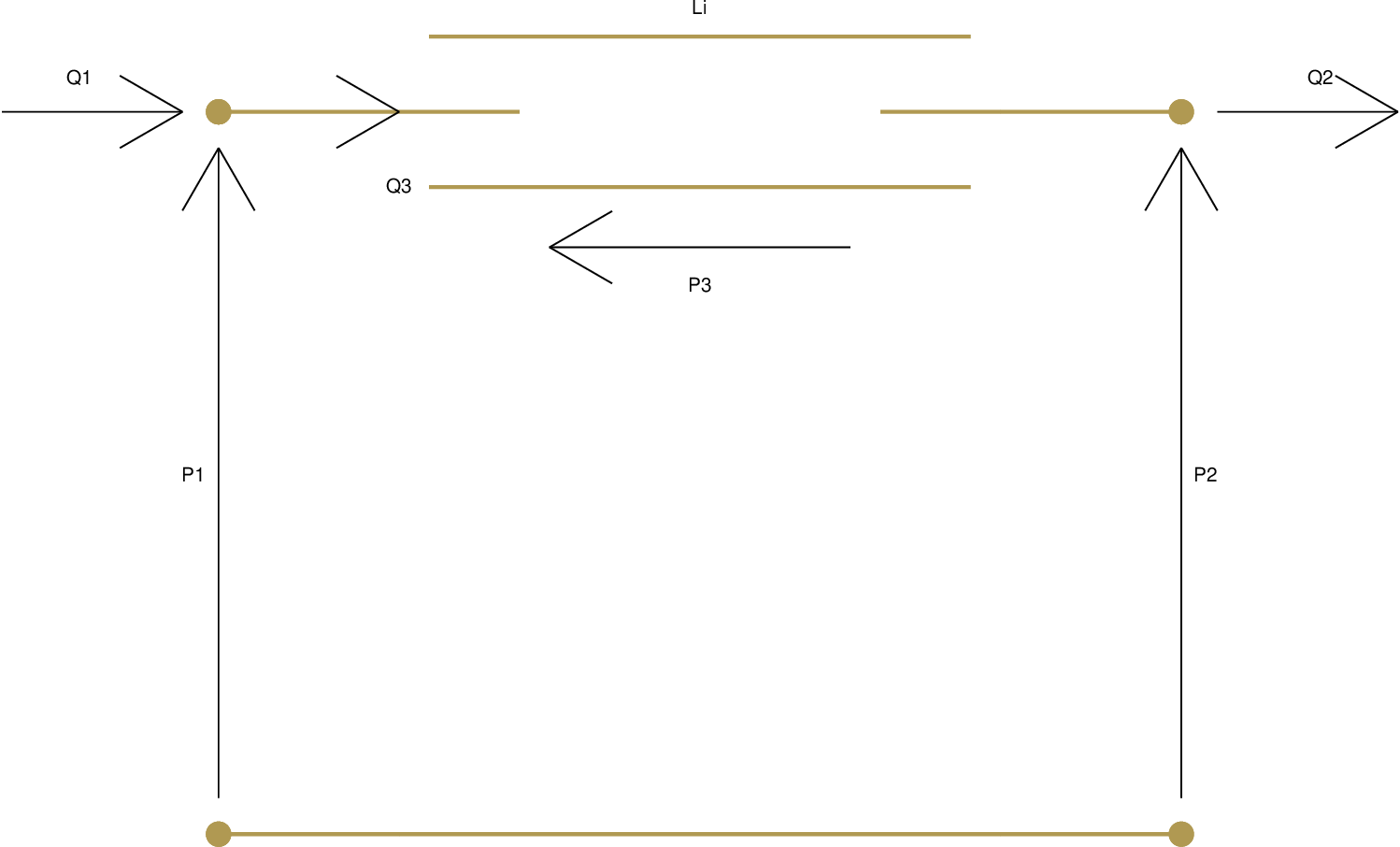}
  }
   \\[8mm]
   \hline
   \hline  
   & &\\[19mm]
 \multirow{2}{*}{
 \schemaPOG{(-1.1,0)(10,8)}{2.0mm}{
 \thicklines %
 \bloin{\Effi}{\Floi}
 \rput[r](-0.5,5){\mbox{(B)}}
 \rput[l](4.25,8){\Effiii}
 \rput[l](4.25,1.5){\Floiii}
 \blosu{\ds\frac{1}{\Kn\,s}}{}{\dst}
 \bloout{\Effii}{\Floii}
  }}
  &
  \psfrag{V1}[r][r]{$\scr {\mygreen V_{1}}$}
  \psfrag{V2}[l][l]{$\scr {\mygreen V_{2}}$}
  \psfrag{V3}[r][r]{$\scr {\mygreen V}$}
  \psfrag{I1}[b][b]{$\scr {\myred I_{1}}$}
  \psfrag{I2}[b][b]{$\scr {\myred I_{2}}$}
  \psfrag{I3}[r][r]{$\scr {\myred I}$}
  \psfrag{C}[l][l]{$\scr C$}
  \rput(0.5,6.5){
  \rput[b](7.7,13.0){\mbox{(e)}}
  \rput[b](7.7,-2.5){$\scr\Kn= C\;$}
  \includegraphics[clip,width=2.4cm]{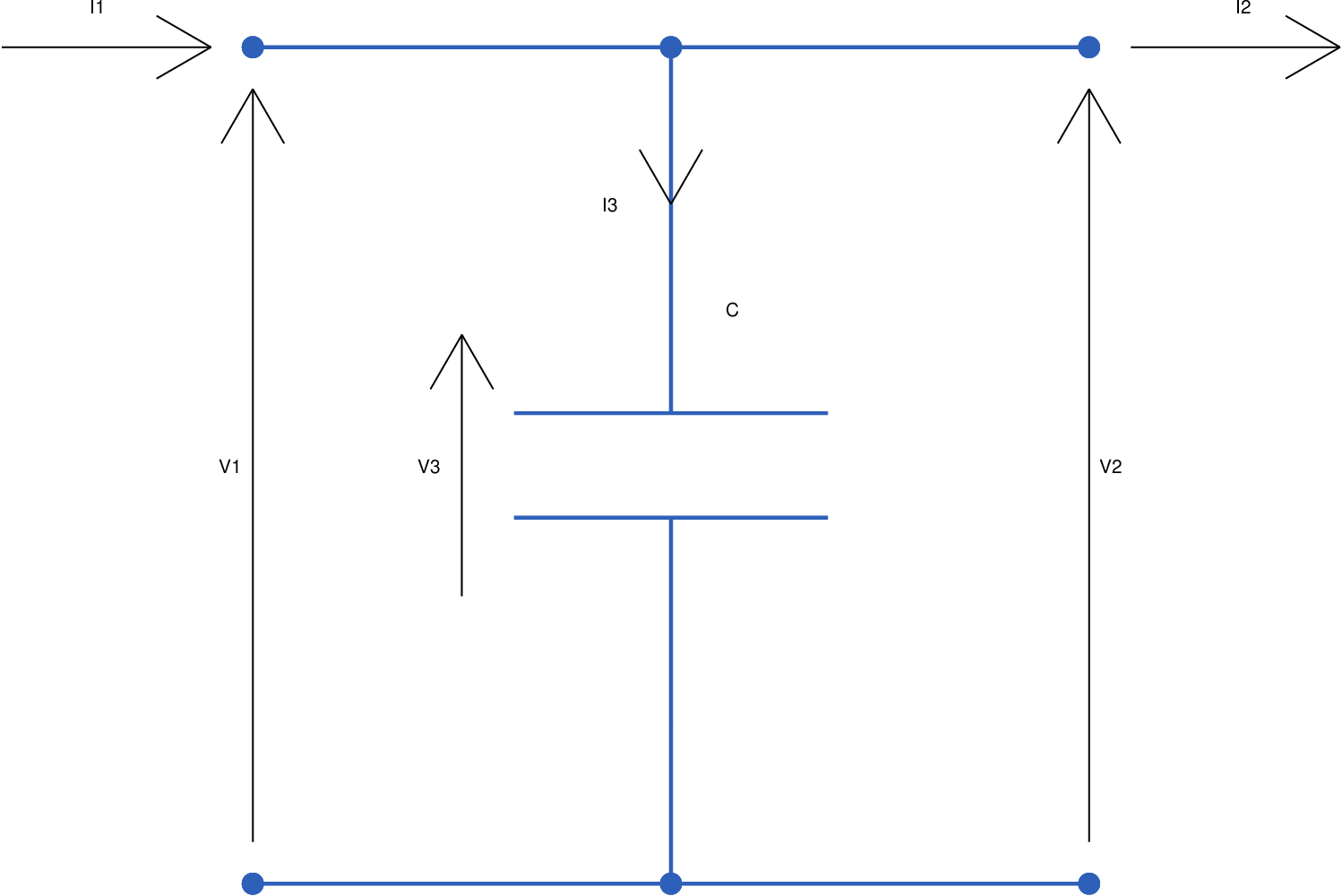}
  }
   &
  \psfrag{v1}[r][r]{$\scr {\mygreen v_{1}}$}
  \psfrag{v2}[l][l]{$\scr {\mygreen v_{2}}$}
  \psfrag{v3}[r][r]{$\scr {\mygreen v}$}
  \psfrag{F1}[b][b]{$\scr {\myred F_{1}}$}
  \psfrag{F2}[b][b]{$\scr {\myred F_{2}}$}
  \psfrag{F3}[r][r]{$\scr {\myred F}$}
  \psfrag{M}[l][l]{$\scr M$}
  \rput(0.5,6.5){
  \rput[b](7.7,13.0){\mbox{(f)}}
  \rput[b](7.7,-2.5){$\scr\Kn= M\;$}
  \includegraphics[clip,width=2.4cm]{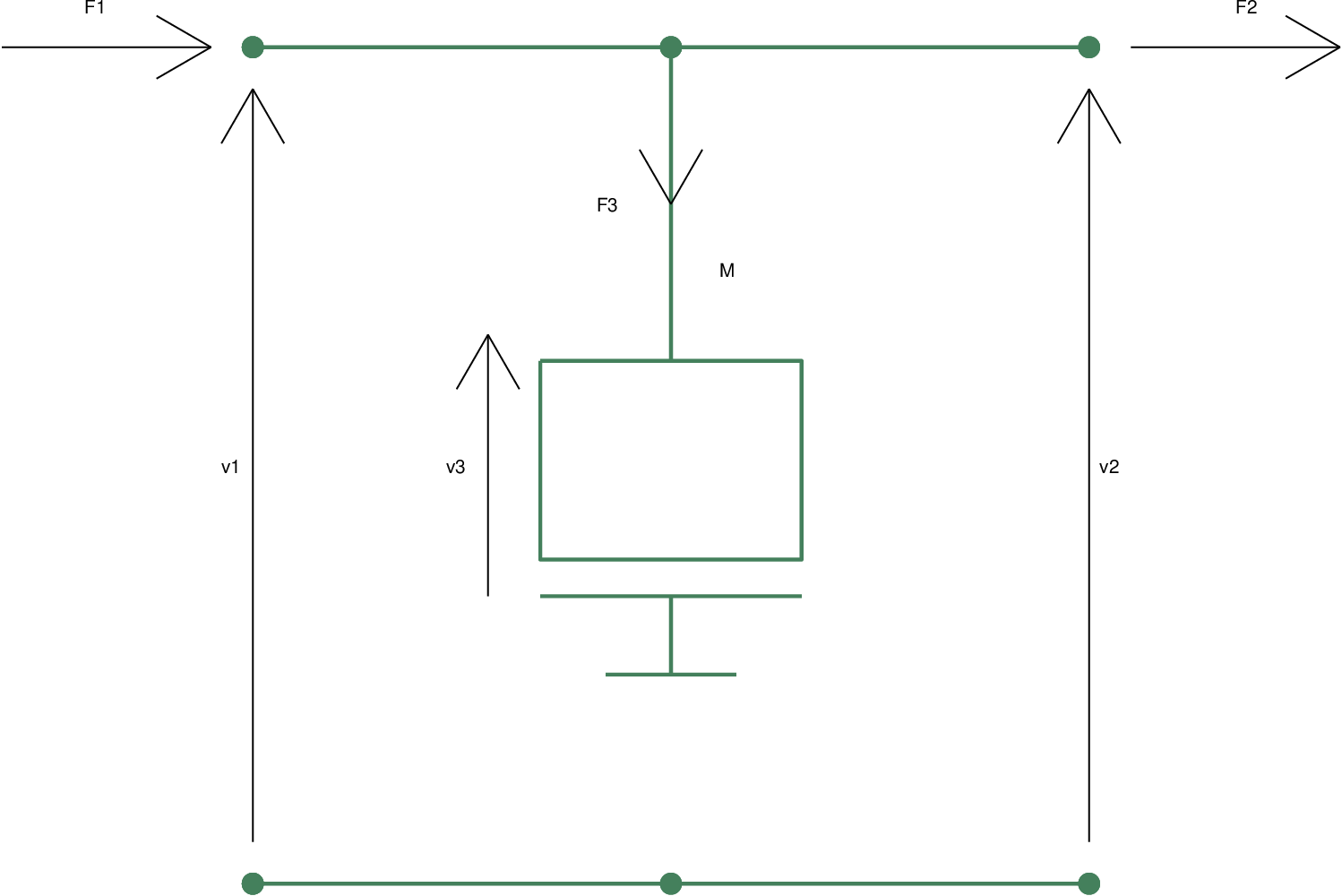}
  }
  \\ 
  & &\\[-1mm]
 \cline{2-3}
   & &\\[1mm] 
   \begin{minipage}[b]{2.1cm}
    $ $  \\         
    $ $  \\         
    $ $  \\         
    $ $  \\[5mm]    
   \end{minipage}
 &
  \psfrag{w1}[r][r]{$\scr {\mygreen \omega_{1}}$}
  \psfrag{w2}[l][l]{$\scr {\mygreen \omega_{2}}$}
  \psfrag{w3}[r][r]{$\scr {\mygreen \omega}$}
  \psfrag{T1}[b][b]{$\scr {\myred \tau_{1}}$}
  \psfrag{T2}[b][b]{$\scr {\myred \tau_{2}}$}
  \psfrag{T3}[r][r]{$\scr {\myred \tau}$}
  \psfrag{J}[l][l]{$\scr J$}
  \rput[b](7.7,13){\mbox{(g)}}
  \rput[b](7.7,-2.75){$\scr\Kn= J\;$}
  \;\includegraphics[clip,width=2.4cm]{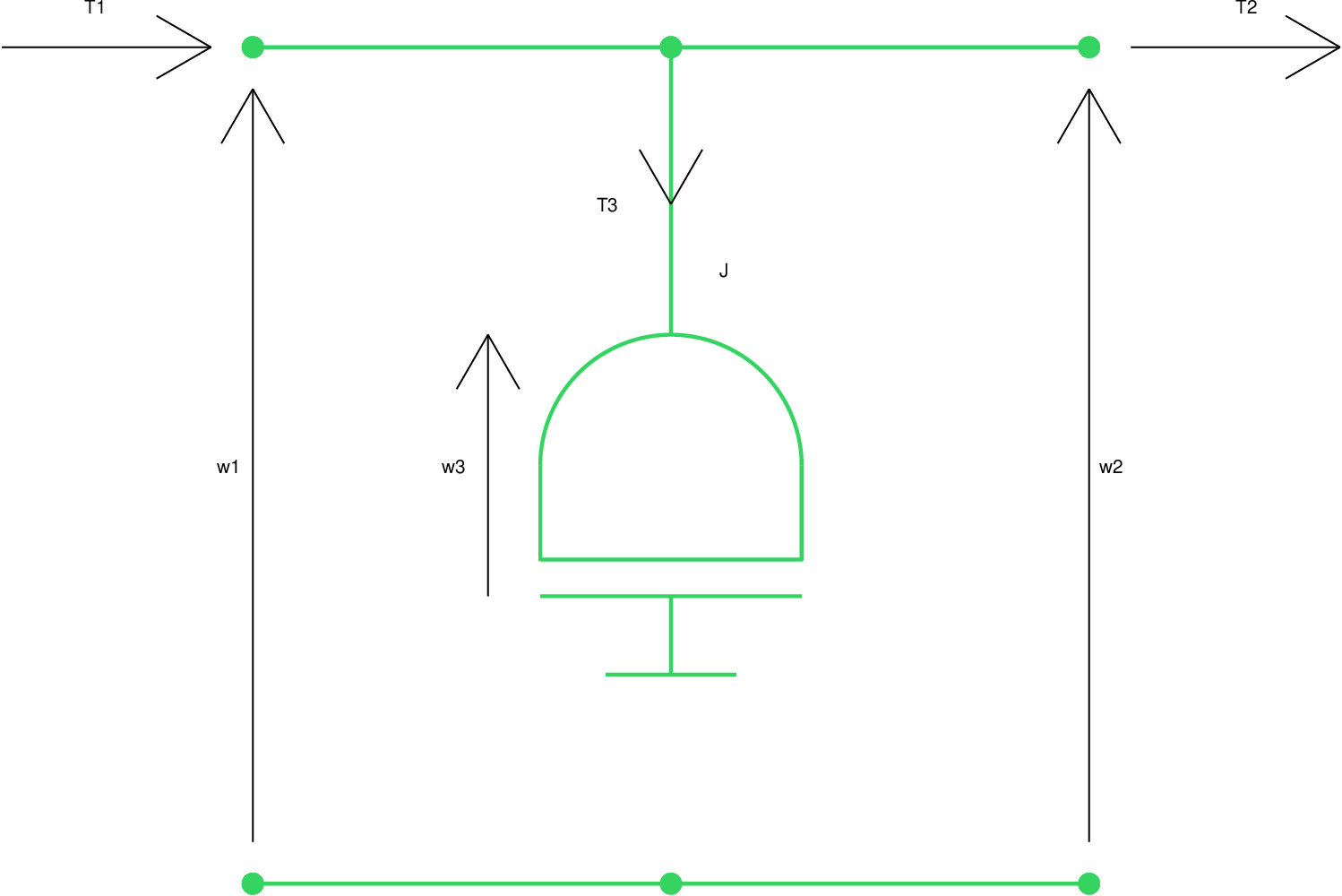}
   &
  \psfrag{P1}[r][r]{$\scr {\mygreen P_{1}}$}
  \psfrag{P2}[l][l]{$\scr {\mygreen P_{2}}$}
  \psfrag{P3}[r][r]{$\scr {\mygreen P}$}
  \psfrag{Q1}[b][b]{$\scr {\myred Q_{1}}$}
  \psfrag{Q2}[b][b]{$\scr {\myred Q_{2}}$}
  \psfrag{Q3}[r][r]{$\scr {\myred Q}$}
  \psfrag{Ci}[l][l]{$\scr C_{I}$}
  \rput[b](7.7,13){\mbox{(h)}}
  \rput[b](7.7,-2.75){$\scr\Kn= C_I\;$}
  \includegraphics[clip,width=2.4cm]{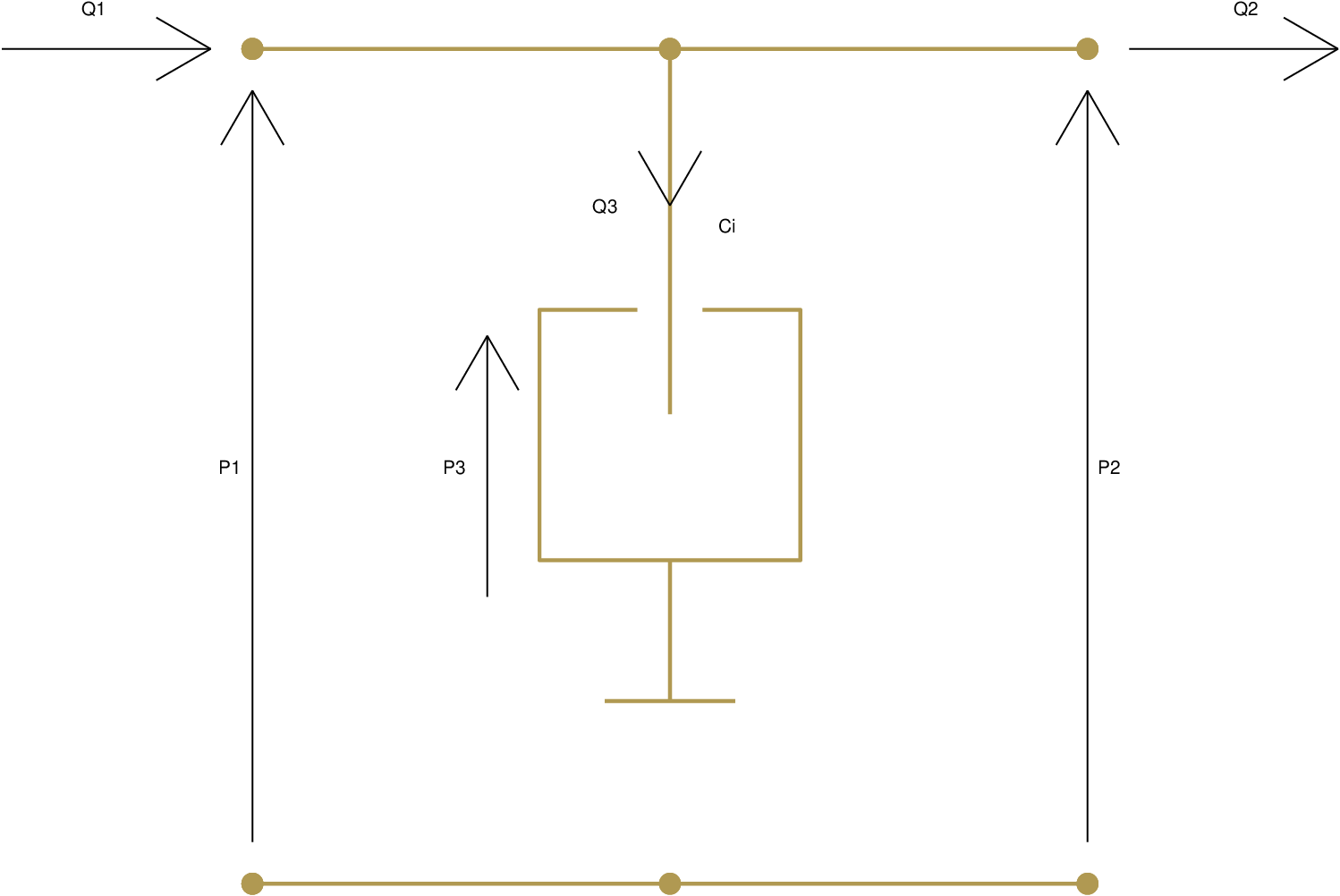}
   \\[4mm]
   \hline  
\end{tabular}
\end{center}
    \caption{Dynamic Physical Elements (PEs) admitting one Elaboration Block (EB) configuration only. (A) EB configuration for the through dynamic elements {\myred $D_f$}
    connected 
    in series: inductor $L$ (a), spring $E$ (b), rotational spring $E_r$ (c), and hydraulic inductor $L_I$ (d). (B) EB configuration for the across dynamic elements {\mygreen $D_e$}
    connected in parallel: capacitor $C$ (e), mass $M$ (f), inertia $J$ (g), and hydraulic capacitor $C_I$ (h).}
    \label{fig:un_grafico}
\end{figure}

\begin{Prop}\label{through_s_across_p_Prop}
Through dynamic elements $\myred D_f$ connected in series and across dynamic elements $\mygreen D_e$ connected in parallel only admit the EB configurations of Figure~\ref{terminali_POG_series}(a) and Figure~\ref{terminali_POG_parallel}(a), respectively.
In the linear case, the EB configuration of Figure~\ref{terminali_POG_series}(a) translates into the configuration of Figure~\ref{fig:un_grafico}(A). 
The coefficient $ K$ characterizing the EB transfer function
\[
\myred v_f \black = \dfrac{1}{ K \black s} \mygreen v_e
\]
depends on the considered through dynamic element $\myred D_f$ connected in series: $ K=L$ for the inductor,  $ K=E$ for the spring,  $ K=E_r$ for the rotational spring,  $ K=L_I$ for the hydraulic inductor.  Similarly, 
the EB configuration of Figure~\ref{terminali_POG_parallel}(a) translates into the configuration of Figure~\ref{fig:un_grafico}(B). The coefficient $ K$ characterizing the EB transfer function
\[
\mygreen v_e \black = \dfrac{1}{ K \,\black s} \myred v_f
\]
depends on the considered through dynamic element $\mygreen D_e$ connected in parallel: $ K=C$ for the capacitor,  $ K=M$ for the mass,  $ K=J$ for the inertia,  $ K=C_I$ for the hydraulic capacitor. 
\end{Prop}

\begin{figure}[t!]
    \centering\footnotesize
 \setlength{\unitlength}{1.5mm}
 \psset{unit=1.0\unitlength}
 \begin{center}
 \begin{tabular}{|@{\;}c@{\;}|@{\;}c@{\;\;}|@{\;}c@{\;\;}|}
   \hline  
   & &\\[0mm]
 \schemaPOG{(-2,0)(10,10)}{2.0mm}{
 \thicklines %
 \rput[r](-0.5,5){\mbox{(A)}}
 \bloin{\Effi}{\Floi}
 \rput[l](4.25,8){\Effiii}
 \rput[l](4.25,1.5){\Floiii}
 \blogiuor{\ds\frac{1}{\Kn\,s}}{}{}
 \bloout{\Effii}{\Floii}
  }
   &
  \psfrag{V1}[r][r]{$\scr {\mygreen V_{1}}$}
  \psfrag{V2}[l][l]{$\scr {\mygreen V_{2}}$}
  \psfrag{V3}[r][r]{$\scr {\mygreen V}$}
  \psfrag{I1}[b][b]{$\scr {\myred I_{1}}$}
  \psfrag{I2}[b][b]{$\scr {\myred I_{2}}$}
  \psfrag{I3}[r][r]{$\scr {\myred I}$}
  \psfrag{L}[l][l]{$\scr L$}
  \rput(0,6){
  \rput[b](7.7,13){\mbox{(a)}}
  \rput[b](7.7,-2){$\scr\Kn= L\;$}
  \includegraphics[clip,width=2.4cm]{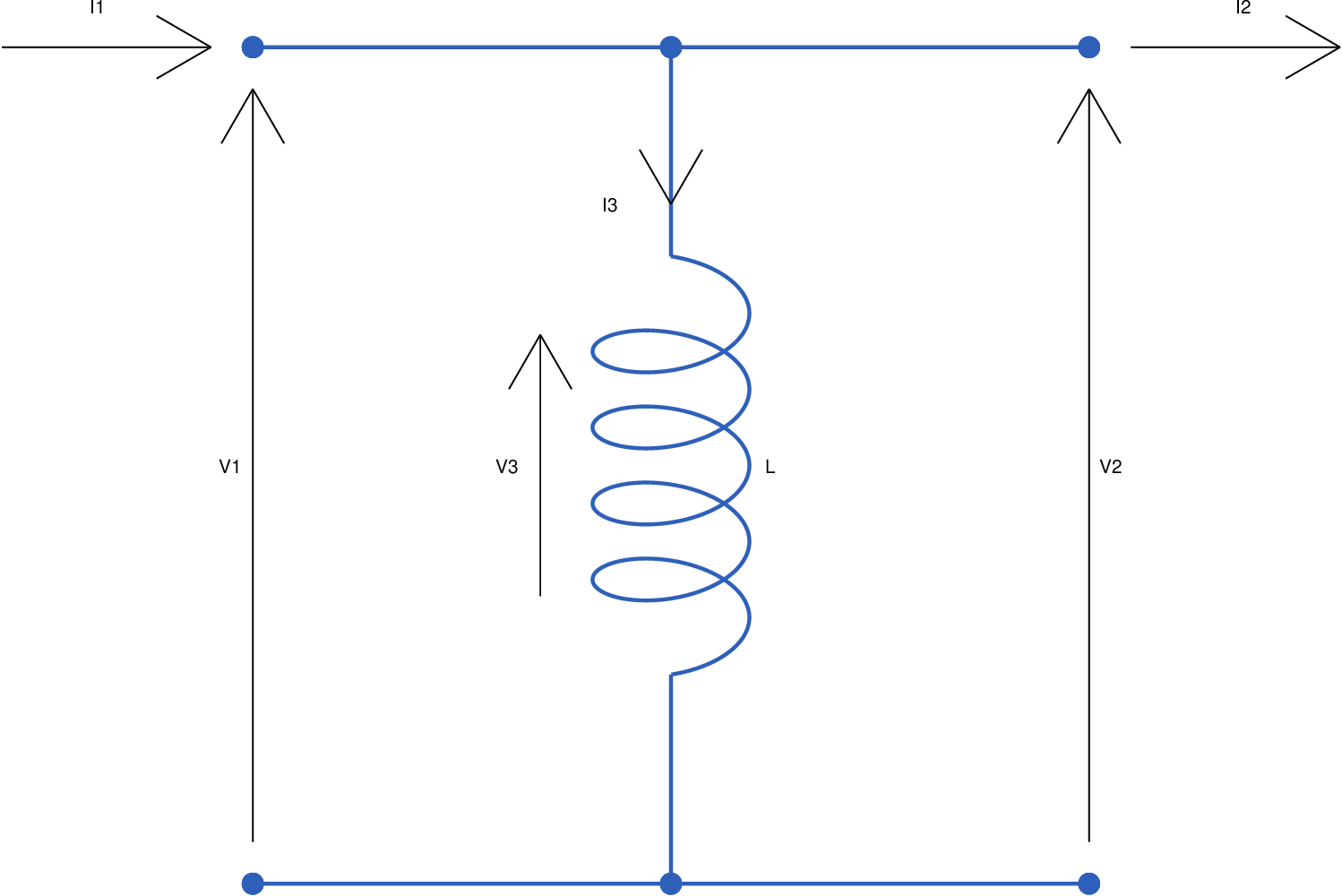}
  }
   &
  \psfrag{v1}[r][r]{$\scr {\mygreen v_{1}}$}
  \psfrag{v2}[l][l]{$\scr {\mygreen v_{2}}$}
  \psfrag{v3}[r][r]{$\scr {\mygreen v}$}
  \psfrag{F1}[b][b]{$\scr {\myred F_{1}}$}
  \psfrag{F2}[b][b]{$\scr {\myred F_{2}}$}
  \psfrag{F3}[r][r]{$\scr {\myred F}$}
  \psfrag{K}[l][l]{$\scr E$}
  \psfrag{E}[l][l]{$\scr E$}
  \rput(0,6){
  \rput[b](7.7,13){\mbox{(b)}}
  \rput[b](7.7,-2){$\scr\Kn= E\;$}
  \includegraphics[clip,width=2.4cm]{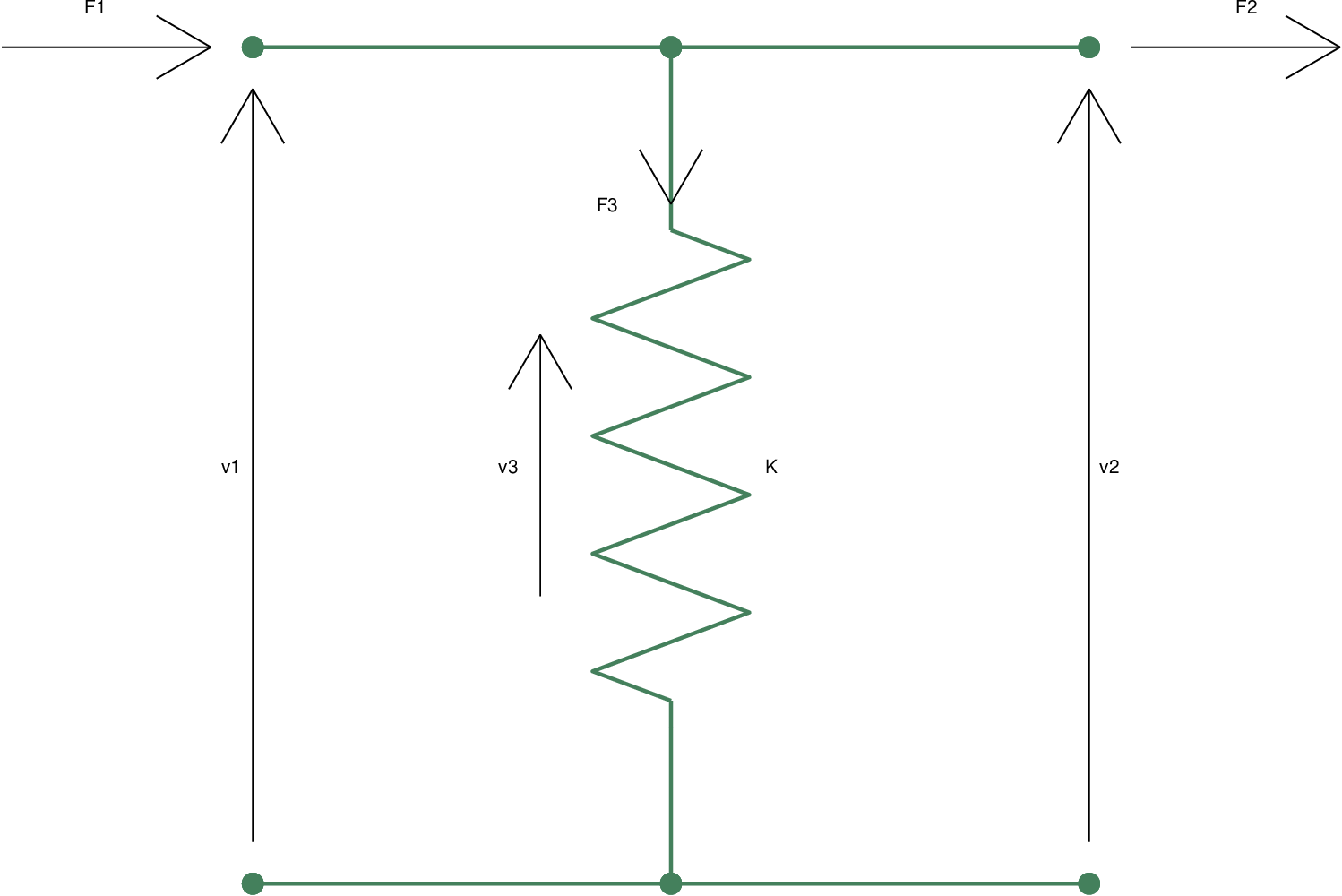}
  }
   \\
   & &\\[-2mm]
   \hline  
   & &\\[0mm] 
 \schemaPOG{(-2,0)(10,10)}{2.0mm}{
 \thicklines %
 \rput[r](-0.5,5){\mbox{(B)}}
 \bloin{\Effi}{\Floi}
 \rput[l](4.25,8){\Effiii}
 \rput[l](4.25,1.5){\Floiii}
 \blogiuaor{\ds\frac{1}{\Kn\,s}}{}{\su}
 \bloout{\Effii}{\Floii}
  }
   &
  \psfrag{w1}[r][r]{$\scr {\mygreen \omega_{1}}$}
  \psfrag{w2}[l][l]{$\scr {\mygreen \omega_{2}}$}
  \psfrag{w3}[r][r]{$\scr {\mygreen \omega}$}
  \psfrag{T1}[b][b]{$\scr {\myred \tau_{1}}$}
  \psfrag{T2}[b][b]{$\scr {\myred \tau_{2}}$}
  \psfrag{T3}[r][r]{$\scr {\myred \tau}$}
  \psfrag{K}[l][l]{$\scr E_{r}$}
  \psfrag{Et}[l][l]{$\scr E_{r}$}
  \rput(0,6){
  \rput[b](7.7,13){\mbox{(c)}}
  \rput[b](7.7,-2.3){$\scr\Kn= E_r\;$}
  \includegraphics[clip,width=2.4cm]{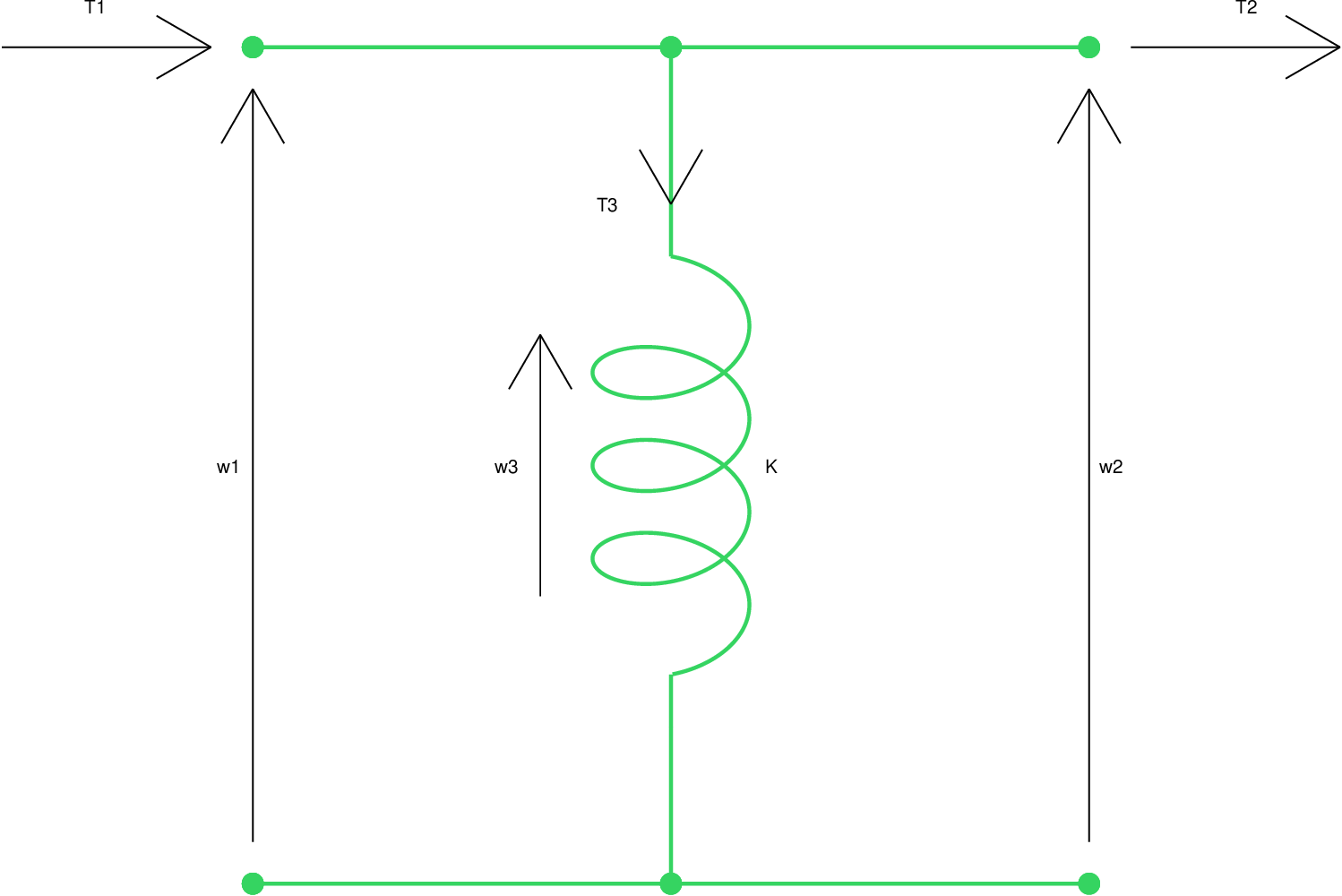}
  }
   &
  \psfrag{P1}[r][r]{$\scr {\mygreen P_{1}}$}
  \psfrag{P2}[l][l]{$\scr {\mygreen P_{2}}$}
  \psfrag{P3}[r][r]{$\scr {\mygreen P}$}
  \psfrag{Q1}[b][b]{$\scr {\myred Q_{1}}$}
  \psfrag{Q2}[b][b]{$\scr {\myred Q_{2}}$}
  \psfrag{Q3}[r][r]{$\scr {\myred Q}$}
  \psfrag{Li}[l][l]{$\scr L_{I}$}
  \rput(0,6){
  \rput[b](7.7,13){\mbox{(d)}}
  \rput[b](7.7,-2.3){$\scr\Kn= L_I\;$}
  \includegraphics[clip,width=2.4cm]{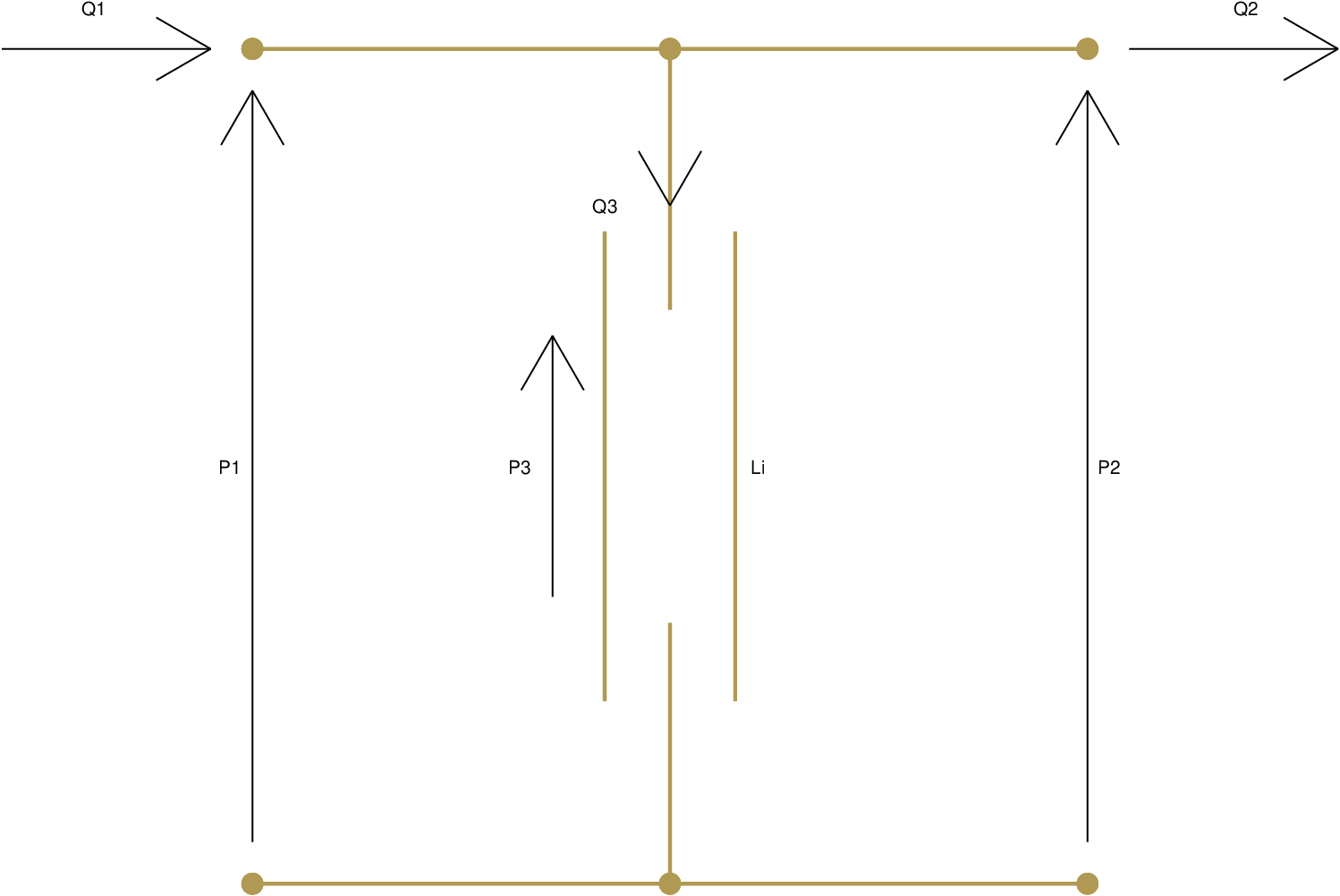}
  }
   \\[2mm]
   \hline
   \hline  
  & &\\[0mm]
 \schemaPOG{(-2,0)(10,10)}{2.0mm}{
 \thicklines %
 \bloin{\Effi}{\Floi}
 \rput[r](-0.5,5){\mbox{(C)}}
 \rput[l](4.25,8){\Effiii}
 \rput[l](4.25,1.5){\Floiii}
 \blosuaor{\ds\frac{1}{\Kn\,s}}{}{}
 \bloout{\Effii}{\Floii}
  }
  &
  \psfrag{V1}[r][r]{$\scr {\mygreen V_{1}}$}
  \psfrag{V2}[l][l]{$\scr {\mygreen V_{2}}$}
  \psfrag{V3}[t][t]{$\scr {\mygreen V}$}
  \psfrag{I1}[b][b]{$\scr {\myred I_{1}}$}
  \psfrag{I2}[b][b]{$\scr {\myred I_{2}}$}
  \psfrag{I3}[t][t]{$\scr {\myred I}$}
  \psfrag{C}[rb][rb]{$\scr C$}
  \rput(0,6){
  \rput[b](7.7,13){\mbox{(e)}}
  \rput[b](7.7,-2){$\scr\Kn= C\;$}
  \includegraphics[clip,width=2.4cm]{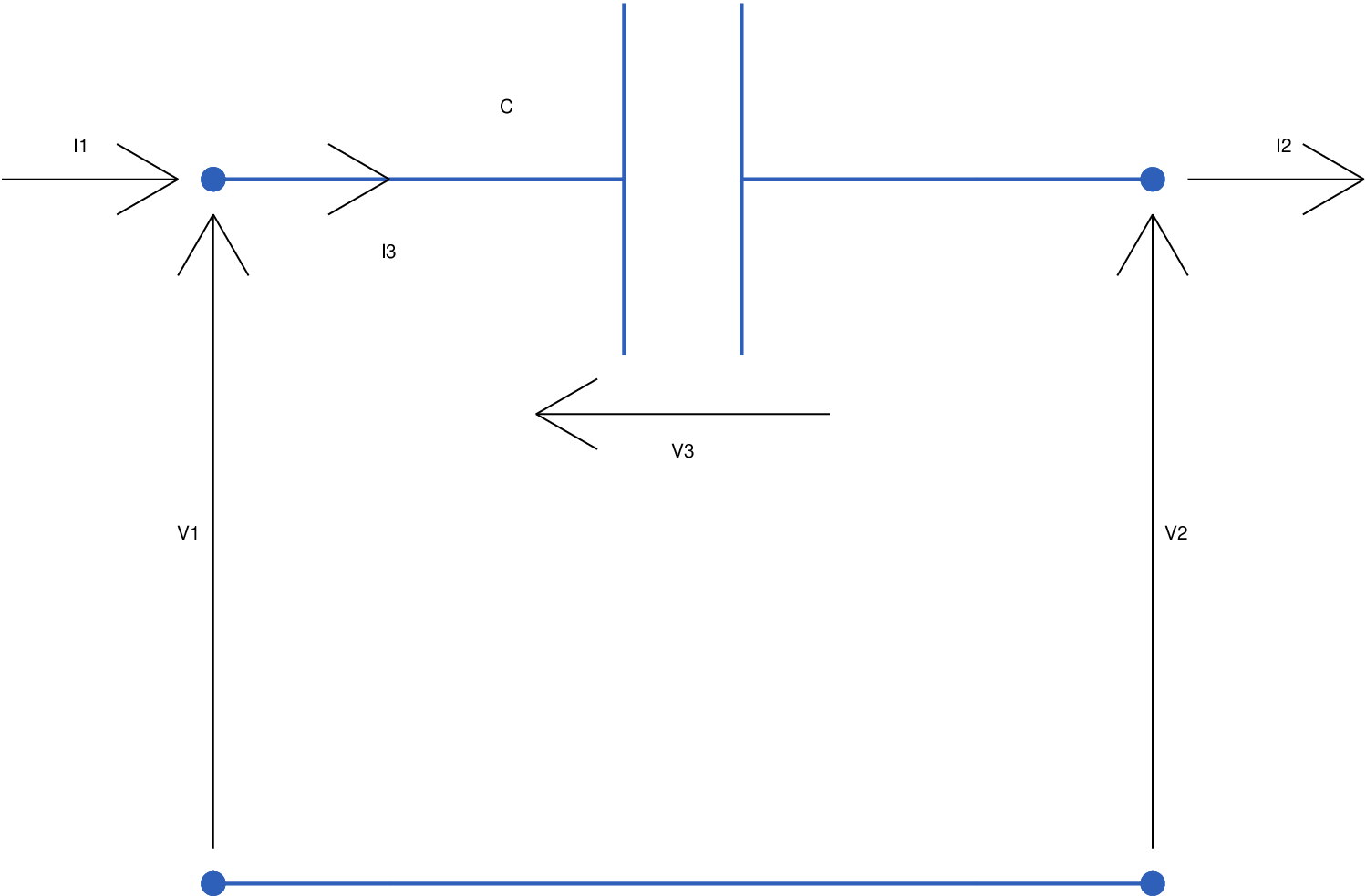}
  }
   &
  \psfrag{v1}[r][r]{$\scr {\mygreen v_{1}}$}
  \psfrag{v2}[l][l]{$\scr {\mygreen v_{2}}$}
  \psfrag{v3}[t][t]{$\scr {\mygreen v}$}
  \psfrag{F1}[b][b]{$\scr {\myred F_{1}}$}
  \psfrag{F2}[b][b]{$\scr {\myred F_{2}}$}
  \psfrag{F3}[t][t]{$\scr {\myred F}$}
  \psfrag{M}[rb][rb]{$\scr M$}
  \rput(0,6){
  \rput[b](7.7,13){\mbox{(f)}}
  \rput[b](7.7,-2){$\scr\Kn= M$}
  \includegraphics[clip,width=2.4cm]{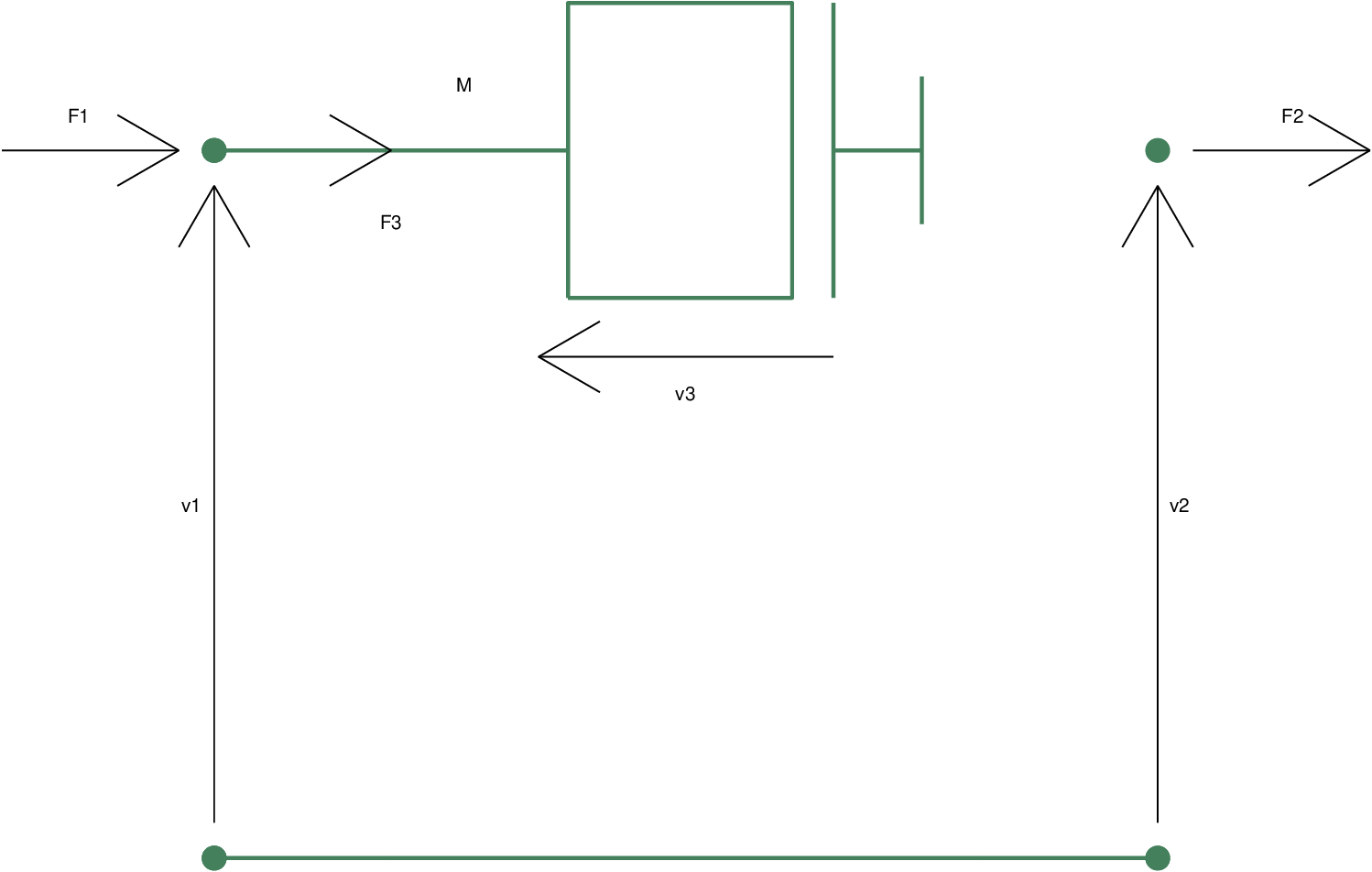}
  }
  \\ 
  & &\\[-2mm]
   \hline  
   & &\\[0mm] 
 \schemaPOG{(-2,0)(10,9.5)}{2.0mm}{
 \thicklines %
 \rput(4,4){
 \bloin{\Effi}{\Floi}
 \rput[r](-0.5,5){\mbox{(D)}}
 \rput[l](4.25,8){\Effiii}
 \rput[l](4.25,1.5){\Floiii}
 \blosuor{\ds\frac{1}{\scr\Kn\,s}}{}{\giu}
 \bloout{\Effii}{\Floii}
  }  }
  &
  \psfrag{w1}[r][r]{$\scr {\mygreen \omega_{1}}$}
  \psfrag{w2}[l][l]{$\scr {\mygreen \omega_{2}}$}
  \psfrag{w3}[t][t]{$\scr {\mygreen \omega}$}
  \psfrag{T1}[b][b]{$\scr {\myred \tau_{1}}$}
  \psfrag{T2}[b][b]{$\scr {\myred \tau_{2}}$}
  \psfrag{T3}[t][t]{$\scr {\myred \tau}$}
  \psfrag{J}[rb][rb]{$\scr J$}
  \rput[b](7.7,13){\mbox{(g)}}
  \rput[b](7.7,-3){$\scr\Kn= J\;$}
  \includegraphics[clip,width=2.4cm]{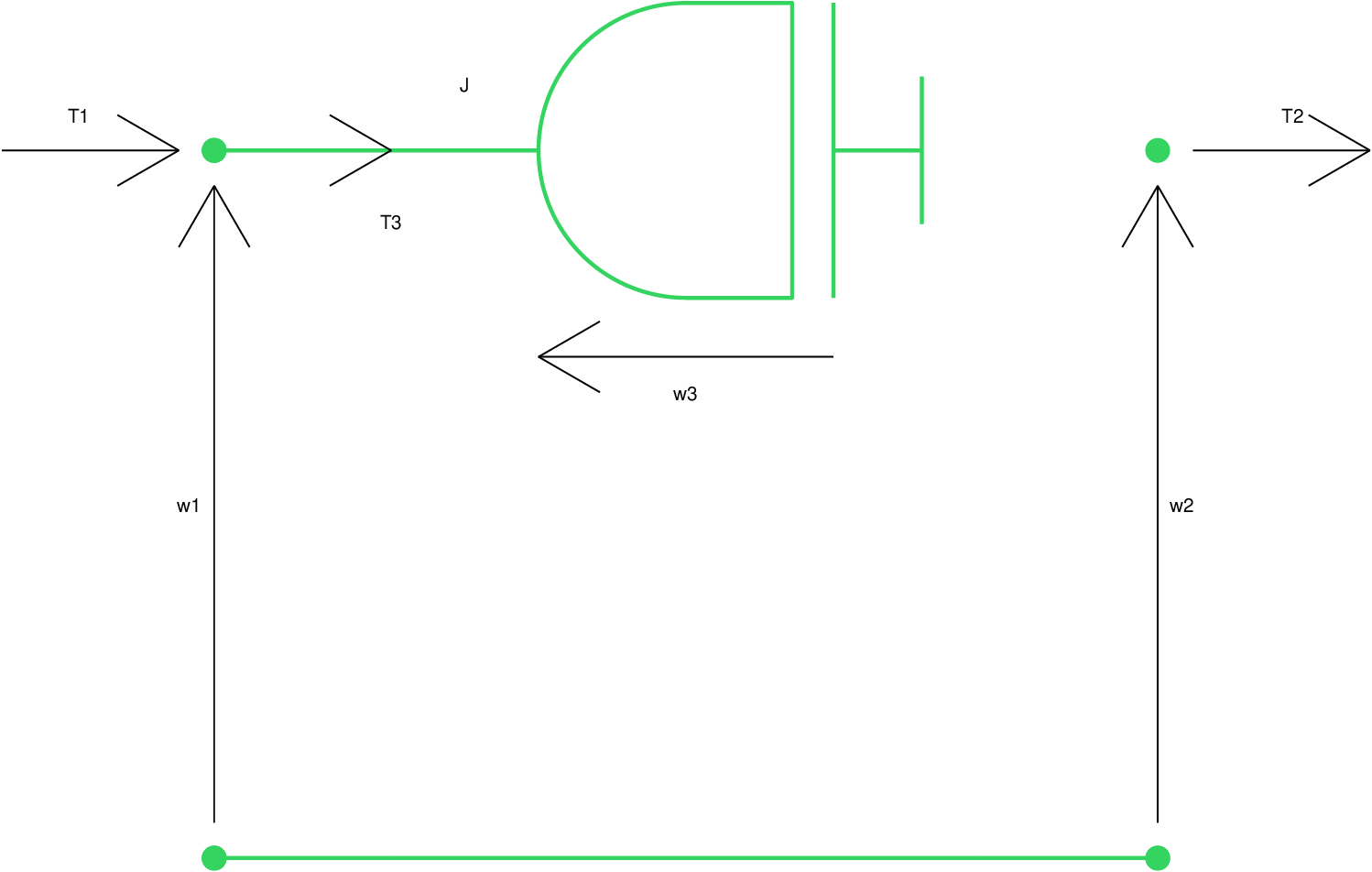}
   &
  \psfrag{P1}[r][r]{$\scr {\mygreen P_{1}}$}
  \psfrag{P2}[l][l]{$\scr {\mygreen P_{2}}$}
  \psfrag{P3}[t][t]{$\scr {\mygreen P}$}
  \psfrag{Q1}[b][b]{$\scr {\myred Q_{1}}$}
  \psfrag{Q2}[b][b]{$\scr {\myred Q_{2}}$}
  \psfrag{Q3}[t][t]{$\scr {\myred Q}$}
  \psfrag{Ci}[rb][rb]{$\scr C_{I}$}
  \rput[b](7.7,13){\mbox{(h)}}
  \rput[b](7.7,-3){$\scr\Kn= C_I$}
  \includegraphics[clip,width=2.4cm]{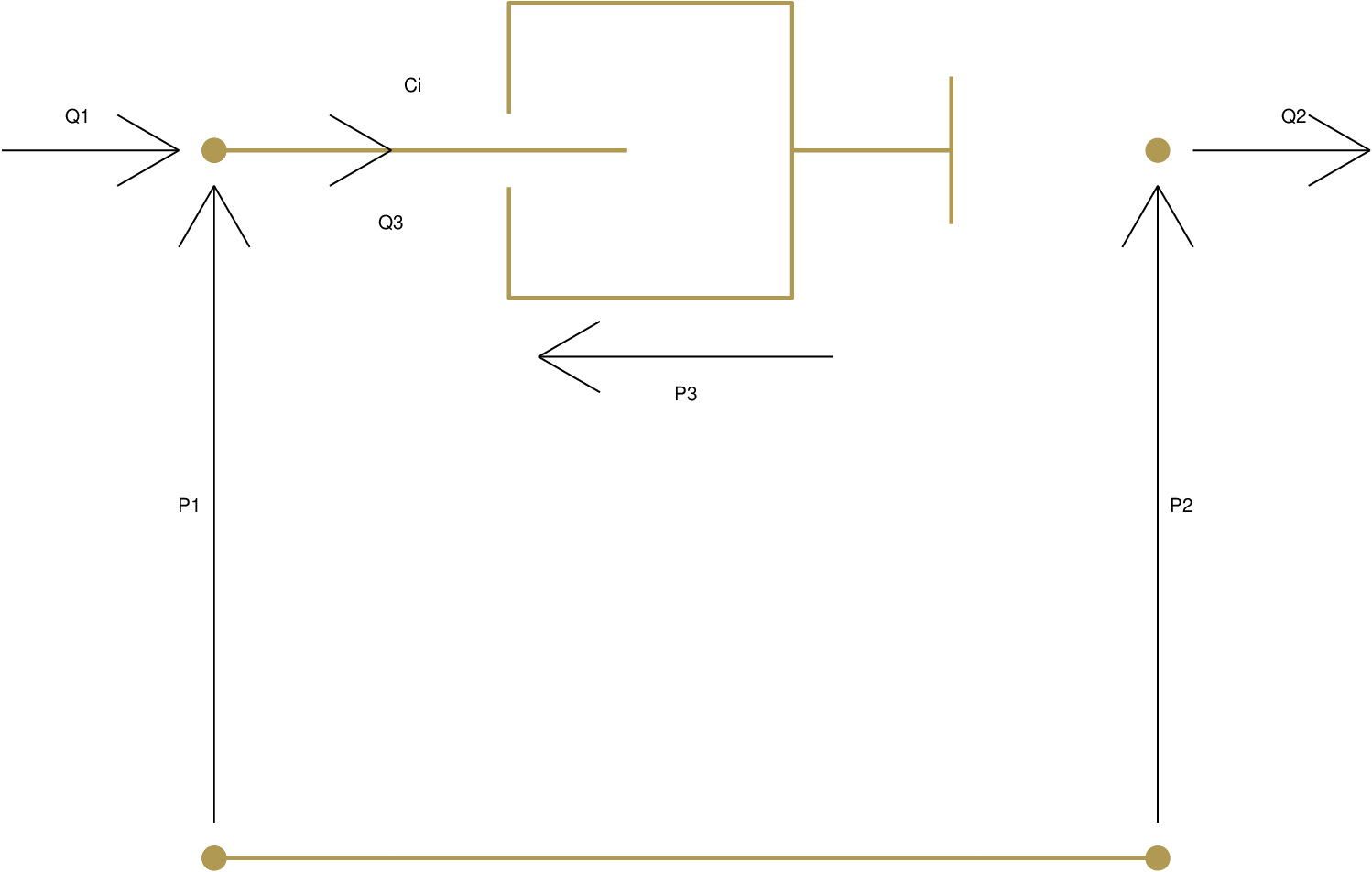}
  \\[4mm] \hline 
\end{tabular}
\end{center}
    \caption{Dynamic Physical Elements (PEs) admitting two Elaboration Block (EB) configurations. (A) and (B) EB configurations for the through dynamic elements {\myred $D_f$} connected in parallel: inductor $L$ (a), spring $E$ (b), rotational spring $E_r$ (c), and hydraulic inductor $L_I$ (d). (C) and (D) EB configuration for the across dynamic elements {\mygreen $D_e$} connected in series: capacitor $C$ (e), mass $M$ (f), inertia $J$ (g), and hydraulic capacitor $C_I$ (h).}    
    \label{fig:due_grafici}
\end{figure}

\begin{Prop}\label{across_s_through_p_Prop}
Through dynamic elements $\myred D_f$ connected in parallel and across dynamic elements $\mygreen D_e$ connected in series admit two EB configurations: those of Figure~\ref{terminali_POG_parallel}(b)-(c) and Figure~\ref{terminali_POG_series}(b)-(c), respectively.
In the linear case, the EB configurations of Figure~\ref{terminali_POG_parallel}(b)-(c) translate into the configurations of Figure~\ref{fig:due_grafici}(A)-(B). 
%
%
Similarly, the EB configurations of Figure~\ref{terminali_POG_series}(b)-(c) translate into the configurations of Figure~\ref{fig:due_grafici}(C)-(D). 
\end{Prop}


\begin{Prop}\label{static_el_Prop}
 Static elements $R$ admit all the possible EB configurations shown in Figure~\ref{terminali_POG_series} and in Figure~\ref{terminali_POG_parallel} for the series and parallel connections, respectively. This follows from the fact that static elements are not characterized by differential equations, and therefore do not exhibit the integral causality issue.
\end{Prop}
%

From Figure~\ref{terminali_POG_series}, Figure~\ref{terminali_POG_parallel}, Figure~\ref{fig:un_grafico}, and Figure~\ref{fig:due_grafici}, it can be concluded that the symbolism of the POG technique is composed of standard blocks that can be found in standard Simulink libraries, which makes this technique particularly suitable and intuitive for non-expert users.
\begin{pullquote}
The symbolism of the POG technique is composed of standard blocks available in standard Simulink libraries, which makes this technique particularly suitable and intuitive for non-expert users.
\end{pullquote}





\subsection{POG State-Space Model}\label{POG_to_stsp_sect}

A system  $\S$ in a POG state-space representation has the following structure:
\begin{equation}\label{POG_stsp}
  \S\!=\!\left\{
  \begin{array}{@{}r@{\,}c@{\,}l}
  \L\,\dot{\x} &=& \A\x+\B\u,\\
          \y &=& \C\x+\D\u,
  \end{array}
  \right.
\end{equation} 
where $\L$ is the energy matrix, $\A$ is the power matrix, $\B$ is the input power matrix, $\C$ is the output matrix, and $\D$ is the input/output matrix. For a system $\S$ in a POG state-space form as in \eqref{POG_stsp}, matrix $\L=\L\tras\geq 0$ is always a symmetric semidefinite matrix. Furthermore,  the names \emph{energy} and \emph{power} matrices, for matrices  $\L$ and $\A$, come from the fact that they allow to compute the energy $E_s$ stored within the system and the power $P_d$  dissipated within the system, respectively, as follows:  

\begin{equation}\label{POG_stsp_EsPd}
    E_s=\frac{1}{2}\x\tras\L\x\geq 0,
     \hspace{5mm}
     \mbox{and}
     \hspace{5mm}
     P_d=\x\tras \A \x=\x\tras \A_s \x,
\end{equation}
where $\A_s=\ts\frac{\A+\A\tras}{2}$ is the symmetric part of matrix $\A$. In the POG state-space representation \eqref{POG_stsp}, the matrix $\L$ is always a symmetric and positive semidefinite 
matrix, 
%
%
as opposed to the so-called descriptor state-space model~\cite{DescriptorStSp2024}.
A system  $\S$ in the POG state-space form 
\eqref{POG_stsp} can always be converted into a system $\ooS$ in a classical state-space form by performing the following transformations: 
\begin{equation}\label{POG_to_classical_stsp}
  \S\!=\!\left\{
  \begin{array}{@{}r@{\;}c@{\;}l}
  \dot{\x} &=& \underbrace{\L\muno\A}_{\ooA}\x+\underbrace{\L\muno\B}_{\ooB}\u,\\
          \y &=& \C\x+\D\u,
  \end{array}
  \right.
  \hspace{.5mm}
  \Leftrightarrow
  \hspace{.5mm}
  \ooS\!=\!\left\{
  \begin{array}{@{}r@{\;}c@{\;}l}
  \dot{\x} &=& \ooA\x+\ooB\u,\\[3mm]
          \y &=& \C\x+\D\u.
  \end{array}
  \right.
\end{equation}
Additionally, it can be proven that the input/output transfer matrix $\H(s)$ can always be obtained from $\S$ in \eqref{POG_stsp} and from $\ooS$ in \eqref{POG_to_classical_stsp}
%
%
as follows: 
%
\[ 
\H(s) = \C(\L\, s -\A )\muno \B +\D=\C(\I\, s -\ooA )\muno \ooB +\D.
\]
Furthermore, the following property holds.

 \begin{Prop}\label{To_StSp}
From a linear POG block scheme, the matrices $\L$, $\A$, $\B$, $\C$ and $\D$ of the POG state-space model $\S$ in \eqref{POG_stsp} can always be read by direct inspection of the POG block scheme itself, as described in the following.
 \end{Prop}


Property~\ref{To_StSp} is described by referring to the system in
 Figure~\ref{pompa_elettro_idraulica} 
 of the sidebar ``First case study: A DC Motor Driving an Hydraulic Pump''. Let $n$ be the number of dynamic elements $\mygreen D_{e}$ and $\myred D_{f}$ in the system, and let $m$ be the number of inputs and
outputs, defining the dimensions of the state, input and output vectors $\x\in \mathds{R}^{n \times 1}$, $\u,\y\in \mathds{R}^{m \times 1}$ of system $\S$ in \eqref{POG_stsp}, respectively.
The components $x_i$ of the state vector $\x$ must be chosen equal to the output power variables of the dynamic elements that are present in the system. For the system under consideration, $n=3$ and $m=2$ hold, and the state, input, and output vectors are $\x \!=\! \mat{@{}c@{\;\;}c@{\;\;}c@{}}{I_1 &\omega_2 &P_3}^T$, $\u \!= \!\mat{@{}c@{\;\;}c@{}}{V_a & Q_b}^T$, and  $\y \!= \!\mat{@{}c@{\;\;}c@{}}{I_a & P_b}^T$. 
The coefficients of the system matrices $\L$, $\A$, $\B$, $\C$ and $\D$ in \eqref{POG_stsp} can be directly read by direct inspection of the POG block scheme as follows.
\noindent 1) the $l_{ii}$ element of the diagonal matrix $\L \in \mathds{R}^{n \times n}$, for $i \in {1,\,\ldots,\,n}$, is given by the coefficient characterizing the $i$-th PE.
For the system in Figure~\ref{pompa_elettro_idraulica}, the elements $l_{ii}$ of the
diagonal matrix $\L$ are
%
%
$L_1$, $J_2$ and $C_3$, as shown by the resulting matrix $\L$ in \eqref{stateSPompa}.
\noindent 2) The $a_{ij}$ element of matrix $\A\in \mathds{R}^{n \times n}$ is the {\it global gain} of the static paths linking the $j$-th state variables $x_j$ to the input of the integrator associated with the $i$-th PE, for $i,j \in {1,\,\ldots,\,n}$. Note that the {\it global gain} is the sum of the gains of all the static paths connecting the two considered points.
For example, the static path linking the first state variable $I_1$ in Figure~\ref{pompa_elettro_idraulica_POG} to the integrator of $L_1$ is $-R_1$, forming the $a_{11}$ element of the resulting matrix $\A$ in \eqref{stateSPompa}.
\noindent 3) The $b_{ij}$ element of matrix $\B\in \mathds{R}^{n \times m}$ is the global gain of the static paths linking the $j$-th input
variable $u_j$ to the integrator of the $i$-th PE, for $i \in {1,\,\ldots,\,n}$ and $j \in {1,\,\ldots,\,m}$. For example, the path linking the first input variable $V_a$ in Figure~\ref{pompa_elettro_idraulica_POG} to the integrator of $L_1$ is $1$, forming the $b_{11}$ element of the resulting matrix $\B$ in \eqref{stateSPompa}.
\noindent 4) The $c_{ij}$ element of matrix $\C \in \mathds{R}^{m \times n}$ is the global gain of the static paths linking the $j$-th state
variable $x_j$ to the $i$-th output variable $y_i$, for $i \in {1,\,\ldots,\,m}$ and $j \in {1,\,\ldots,\,n}$. For example, the path linking the first state variable $I_1$ in Figure~\ref{pompa_elettro_idraulica_POG} to first output variable $I_a$ is $1$, as $I_1=I_a$. This forms the $c_{11}$ element of the resulting matrix $\C$ in \eqref{stateSPompa}.
\noindent 5) The $d_{ij}$ element of matrix $\D \in \mathds{R}^{m \times m}$ is the global gain of the static paths linking the $j$-th input
variable to the $i$-th output variable, for $i,j \in {1,\,\ldots,\,m}$. 
For example, the path linking the first input variable $V_a$ in Figure~\ref{pompa_elettro_idraulica_POG} to first output variable $I_a$ is $0$, as no such path exists. This forms the $d_{11}$ element of the resulting matrix $\D$ in \eqref{stateSPompa}.


Other three examples of application of Property~\ref{To_StSp} can be found in the sidebars ``Second Case Study: An Hydraulic Continuously Variable Transmission'', ``Third Case Study: A Permanent Magnet Synchronous Motor'', and ``Fourth Case Study: An Hydraulic Clutch''.

\begin{sidebar}{Fourth Case Study: An Hydraulic Clutch}

\setcounter{sequation}{9}
\renewcommand{\thesequation}{S\arabic{sequation}}
\setcounter{stable}{1}
\renewcommand{\thestable}{S\arabic{stable}}
\setcounter{sfigure}{7}
\renewcommand{\thesfigure}{S\arabic{sfigure}}

Figure~\ref{POG_TransmSyst_Grafico} shows a schematic representation of the fourth case study, which is an hydraulic clutch.

\sdbarfig{
\centering
\setlength{\unitlength}{2.5mm}
 \psset{unit=1.0\unitlength} \thicklines
 \begin{pspicture}(2,-12.75)(27,5)
 \psline(2,2)(5,2)(5,3)(2,3)
 \multirput(5,0)(1,0){4}{\psframe[fillstyle=solid,fillcolor=lightgray](0.1,0)(1,5)}
 \pscurve[linewidth=1.4pt](9,0)(10,1.5)(11,2)
 \pscurve[linewidth=1.4pt](9,5)(10,3.5)(11,3)
 \rput(11,1.25){\psframe[fillstyle=solid,fillcolor=lightgray](4,2.5)}
 \psline(11,1.15)(17,1.15)(17,-5)(17.5,-5)(17.5,1.15)
  (25,1.15)(25,3.85)(11,3.85)
  \rput(14,-9){\psframe(6.5,4)}
  \rput(17.25,-7){\shortstack{Valve\\[2mm] ($R_v$)}}
  \rput(14,2.5){$m_{p}$}
  \rput[t](12,1){$b_P$}
  \rput[bl](9.05,2){\large $K_{m}$}
  \psline{->}(15.25,-13)(15.25,-11)
  \rput[t](14.0,-12.25){$P_a$}
  \psline[linestyle=dashed,linewidth=0.6pt]{-}(13.5,-10)(17,-10)
   \rput(12.5,-10){\scriptsize $1$}
  \pscircle[linewidth=0.4pt](12.5,-10){1.36mm} 
 \psline[linestyle=dashed,linewidth=0.6pt]{-}(15.5,-4)(19,-4)
  \rput(14.5,-4){\scriptsize $2$}
  \pscircle[linewidth=0.4pt](14.5,-4){1.36mm} 
  \psline[linestyle=dashed,linewidth=0.6pt]{-}(15,-0.5)(15,5)
  \rput(15,-1.25){\scriptsize $3$}
  \pscircle[linewidth=0.4pt](15,-1.25){1.36mm} 
  \rput(15,5.75){\scriptsize $4$}
  \pscircle[linewidth=0.4pt](15,5.75){1.36mm} 
  \psline[linestyle=dashed,linewidth=0.6pt]{-}(13,-0.5)(13,5)
  \rput(13,-1.25){\scriptsize $5$}
  \pscircle[linewidth=0.4pt](13,-1.25){1.36mm} 
  \psline[linestyle=dashed,linewidth=0.6pt]{-}(11,-0.5)(11,5)
  \rput(11,-1.25){\scriptsize $6$}
  \pscircle[linewidth=0.4pt](11,-1.25){1.36mm} 
  \psline[linestyle=dashed,linewidth=0.6pt]{-}(9,-1.5)(9,6)
  \rput(9,-2.25){\scriptsize $7$}
  \pscircle[linewidth=0.4pt](9,-2.25){1.36mm} 
  \psline{->}(19.25,-11)(19.25,-13)
  \multirput(15,-11)(4,0){2}{\psline(0,-1)(0,2)(0.5,2)(0.5,-1)}
  \rput(20,2.5){$P_m$}
  \pscurve(18,1)(20.25,0.9)(22.5,1)
  \pscurve(18,4)(20.25,4.1)(22.5,4)
  \rput[b](20.25,4.5){$C_{m}$}
  \psline{->}(17,2.5)(15,2.5)
  \psline{->}(17.25,-2)(17.25,0)
  \rput[l](17.75,-1){$Q_v$}
  \end{pspicture}
   }{Fourth Case Study: An hydraulic clutch. The supply pressure $P_a$ generates the volume flow rate $Q_v$ entering the hydraulic capacitor $C_m$ through an hydraulic resistor (valve) $R_v$. The pressure $P_m$ originating within the hydraulic capacitor $C_m$ pushes the piston having mass $m_p$ subject to friction $b_p$. The piston movement in turn compresses the mechanical spring characterized by stiffness $K_m$, generating the compression of the clutch plates locking the clutch.  \label{POG_TransmSyst_Grafico}
     }
By controlling the supply pressure $P_a$, or by modulating the hydraulic resistor $R_v$, the volume of fluid within 
\sdbarfig{
 \begin{pspicture}(-26.5,1.5)(40,32.02)
\centering
 \schemaPOG{(6,-2)(20,46)}{2.125mm}{
 \thicklines
 \bloin{P_a}{Q_a}
 \bloivgiu{R_v}{}{\dst}
 \blovisumxm{\dfrac{1}{C_m}}{P_1}{\dst}{\frac{1}{s}}{}
 \blokxyor{3.7}{2.5}{A}{A}{}{}
 \blovigiumxm{\dfrac{1}{m_p}}{v_p}{\dst}{\frac{1}{s}}{}
 \bloivsuaor{b_p}{F_p}{}
 \blovisumxm{K_m}{}{\dst}{\frac{1}{s}}{}
 \bloout{F_d}{v_d}
 }
\pscircle[linewidth=0.4pt](-20,0){1.25mm}
\rput(-20,0){\scriptsize $1$}
\pscircle[linewidth=0.4pt](-16,0){1.25mm}
\rput(-16,0){\scriptsize $2$}
\pscircle[linewidth=0.4pt](-10,0){1.25mm}
\rput(-10,0){\scriptsize $3$}
\pscircle[linewidth=0.4pt](-4.25,0){1.25mm}
\rput(-4.25,0){\scriptsize $4$}
\pscircle[linewidth=0.4pt](1.75,0){1.25mm}
\rput(1.75,0){\scriptsize $5$}
\pscircle[linewidth=0.4pt](5.75,0){1.25mm}
\rput(5.75,0){\scriptsize $6$}
\pscircle[linewidth=0.4pt](11.75,0){1.25mm}
\rput(11.75,0){\scriptsize $7$}
%
%
\end{pspicture} 
 }{POG block scheme of the Hydraulic clutch in Figure~\ref{POG_TransmSyst_Grafico}. Starting from the left-hand side, the first two elaboration blocks describe the hydraulic resistance (valve) $R_v$ and the hydraulic capacitor $C_m$. The following connection block handles the energy conversion from the hydraulic to the mechanical translational mechanical domain and viceversa. The next three elaboration blocks describe the piston mass $m_p$, the piston friction coefficient $b_p$ and the spring stiffness $K_m$.
\label{POG_TransmSyst}
     }
the hydraulic capacitor $C_m$ can be properly controlled, thus controlling the movement of the piston $m_p$ and, consequently, the compression of the clutch plates by means of the mechanical spring $K_m$.

 The physical system in Figure~\ref{POG_TransmSyst_Grafico} can be modeled using the POG block scheme shown in Figure~\ref{POG_TransmSyst}, which is in a one-to-one correspondence with the following POG state-space model: 
%
\begin{sequation}\label{TransmSyst_POG_scheme_stsp}
\begin{array}{@{}r@{\;}c@{\;}l@{}}
\underbrace{
\left[\begin{array}{@{}c@{\;}c@{\;}c@{}} C_{m} & 0 & 0\\ 0 & m_{p} & 0\\ 0 & 0 & \frac{1}{K_{m}} \end{array}\right]
}_{\L}\!\dot{\x} &\!=\!& \underbrace{
\left[\begin{array}{@{}c@{\;}c@{\;}c@{}} - R_v & - A & 0\\ A & - b_p & -1\\ 0 & 1 & 0 \end{array}\right]
}_{\A}\underbrace{
\left[\begin{array}{@{}c@{}} P_m\\ v_{p}\\ F_{m} \end{array}\right]
}_{\x}\!+\!\!\underbrace{
\left[\begin{array}{@{}c@{\;}c@{}} R_v & 0\\ 0 & 0\\ 0 & -1 \end{array}\right]
}_{\B}\underbrace{
\left[\begin{array}{@{}c@{}} P_{a}\\ v_{d} \end{array}\right]
}_{\u}\!,\\[10.8mm]\underbrace{
\left[\begin{array}{@{}c@{}} Q_{a}\\ F_{d} \end{array}\right]
}_{\y}&=&\underbrace{
\left[\begin{array}{@{}c@{\;}c@{\;\;\;}c@{}} - R_v & 0 & 0\\ 0 & 0 & 1 \end{array}\right]
}_{\C}\x+\underbrace{
\left[\begin{array}{@{}c@{\;\;}c@{}} R_v & 0\\ 0 & 0 \end{array}\right]
}_{\D}\u. \end{array}
 \end{sequation}
%
The hydraulic capacitance $C_m$ is usually very small, meaning that it can be neglected in order to reduce the system dimension. For this purpose, the following state-space congruent transformation 
\begin{sequation}\label{TransmSyst_POG_scheme_stsp_TR1}
\underbrace{
\left[\begin{array}{@{}c@{}} P_m\\ v_p\\ F_{m} \end{array}\right]
}_{\x}
 =
 \underbrace{
 \mat{@{}c@{\;\;}c@{}}{-\frac{A}{R_v}  & 0 \\ 1 & 0 \\0 & 1  }
 }_{\T}
 \underbrace{
\left[\begin{array}{@{}c@{}} v_p\\ F_{m} \end{array}\right]
}_{\hx}
 +
 \underbrace{
  \mat{@{}c@{\;\;\;}c@{}}{1  & 0 \\ 0 & 0 \\0 & 0  }
 }_{\T_u}
 \underbrace{
\left[\begin{array}{@{}c@{}} P_{a}\\ v_d \end{array}\right]
}_{\u}
\end{sequation}
can be applied to system \eqref{TransmSyst_POG_scheme_stsp}, leading to the following reduced-order system: 
\begin{sequation}\label{TransmSyst_POG_scheme_stsp_red}
\begin{array}{@{}r@{\;}c@{\;}l@{}}
\underbrace{
\left[\begin{array}{@{}c@{\;\;}c@{}} m_{p} & 0\\ 0 & \frac{1}{K_{m}} \end{array}\right]
}_{\hL}\dot{\hx} &=& \underbrace{
\left[\begin{array}{@{}c@{\;\;}c@{}}  - b_p - \frac{A^2}{R_v} & -1\\ 1 & 0 \end{array}\right]
}_{\hA}\underbrace{
\left[\begin{array}{@{}c@{}} v_p\\ F_{m} \end{array}\right]
}_{\hx}+\underbrace{
\left[\begin{array}{@{}c@{\;\;}c@{}} A & 0\\ 0 & -1 \end{array}\right]
}_{\hB}\u,\\[10.8mm]\underbrace{
\left[\begin{array}{@{}c@{}} Q_{a}\\ F_{d} \end{array}\right]
}_{\y}&=&\underbrace{
\left[\begin{array}{@{}c@{\;\;}c@{}} A & 0\\ 0 & 1 \end{array}\right]
}_{\hC}\x+\underbrace{
\left[\begin{array}{@{}c@{\;\;}c@{}} 0 & 0\\ 0 & 0 \end{array}\right]
}_{\hD}\underbrace{
\left[\begin{array}{@{}c@{}} P_{a}\\ v_d \end{array}\right]
}_{\u}, \end{array}
\end{sequation}
where the new system matrices $\hL$, $\hA$, and $\hB$ are computed as described in  \eqref{POG_stsp_transf_matr}.

\end{sidebar}

\subsection{Congruent State-Space Transformations and Model Reduction}\label{POG_model_reduct_sect}

The typical type of state-space transformation that can be applied to a system $\ooS$ in the classical state-space form \eqref{POG_to_classical_stsp} is a {\it similitude} transformation $\x=\T\,\tx$, which transforms system $\ooS$ into the transformed system $\tS$ as follows:
%
%
\begin{equation}\label{classical_stsp_transf}
  \ooS\!=\! \left\{
 \begin{array}{@{}r@{\;}c@{\;}l@{}}
 \dot{\x}&=&\ooA\, \x + \ooB\, \u \\[1mm]
 \y &=& \C\,\x+ \D\, \u
 \end{array}
 \right.
 \hspace{2mm}
 \stackrel{\x=\T\,\tx}{\Longrightarrow}
 \hspace{2mm}
 \tS\!=\!\left\{
 \begin{array}{@{\,}r@{\;}c@{\;}l@{\,}}
 \dot{\tx}&=&\tA \,\tx\! +\! \tB\, \u \\[1mm]
 \y &=& \tC\,\tx+\! \D\, \u
 \end{array}
 \right.,
\end{equation}

where the matrices of the transformed system $\tS$ 
are:
\begin{equation}\label{classical_stsp_transf_matr}
 \tA=\T\muno[\ooA\T-\!\dot{\T}],
 \hspace{8mm}
 \tB=\T\muno\ooB,
 \hspace{8mm}
 \tC=\C\T.
\end{equation} 
In this case, the transformation matrix $\T$ must be square, not singular and can be, in general, time-variant, even though its explicit dependence ``(t)'' on time
is not shown in \eqref{classical_stsp_transf} for simplicity of notation.
On the other hand, a system $\S$ in the POG state-space form as in \eqref{POG_stsp} can be transformed into a new POG system $\hat{S}$ using the
{\it congruent} transformation $\x=\T\,\z+\T_u\,\u$ as follows:
\begin{equation}\label{POG_stsp_transf}
\S\!=\!\!\left\{
\begin{array}{@{}r@{}c@{}l@{}}
\L\dot{\x} &=& \A\x\!+\!\B\u\\[1mm]
\y&=&\C\x+\D\u
 \end{array}
 \right.
 \hspace{1mm}
 \stackrel{\x\!=\!\T\hx\!+\!\T_u\u}{\Longrightarrow}
 \hspace{1mm}
\hS\!=\!\!\left\{
\begin{array}{@{}r@{}c@{}l@{}}
\hL\dot{\hx} 
&=& \hA\hx\!+\!\hB\u\\[1mm]
\y&=&\hC\hx\!+\!\hD\u
 \end{array}
 \right.
\end{equation}
\begin{pullquote}
The utility of congruent state-space transformations 
within the POG technique becomes evident when considering model reduction.
\end{pullquote}
where the new matrices of 
system $\hS$ 
 are defined as follows:
\begin{equation}\label{POG_stsp_transf_matr}
\begin{array}{c}
 \hL\!=\!\T\tras\L\T,
 \hspace{5mm}
 ,
 \hspace{5mm}
 \hA\!=\!\T\tras\left[\A\T\!-\!\L\dot{\T}
 \right], \\[4mm]
\hB\!=\!\T\tras\left[\A\T_u\!+\!\B\right],
 \hspace{4.6mm}
 \hC\!=\!\C\T,
 \hspace{4.6mm}
 \hD\!=\!\C\T_u\!+\!\D,
\end{array}
\end{equation}
supposing $\T_u$ constant and $\T\tras\L\T_u=0$. Note that, in the case of congruent transformations, matrix $\T$ can also be rectangular and  not full rank. 
The utility of congruent state-space transformations 
within the POG technique becomes evident when considering model reduction.
When an eigenvalue of the energy matrix $\L$ tends to zero or to
infinite, the system model degenerates to a lower dimension, and the matrices of the reduced-order model can be obtained using \eqref{POG_stsp_transf_matr} by properly defining the matrices $\T$ and $\T_u$ of the state-space congruent transformation $\x=\T\,\hx+\T_u\,\u$.
Two examples of model reduction using this technique can be found in the sidebars ``Second Case Study: An Hydraulic Continuously Variable Transmission'' and ``Fourth Case Study: An Hydraulic Clutch''.




\begin{sidebar}{Comparison With Other Graphical Modeling Techniques}

Figure~\ref{POG_BG_EMR_Comparison_Figure} shows the dynamic modeling of the DC motor driving an hydraulic pump of Figure~\ref{pompa_elettro_idraulica} using the three main graphical modeling techniques available in the literature: POG in Figure~\ref{POG_BG_EMR_Comparison_Figure}(a), BG in Figure~\ref{POG_BG_EMR_Comparison_Figure}(b), and EMR in Figure~\ref{POG_BG_EMR_Comparison_Figure}(c). 

\setcounter{sequation}{12}
\renewcommand{\thesequation}{S\arabic{sequation}}
\setcounter{stable}{0}
\renewcommand{\thestable}{S\arabic{stable}}
\setcounter{sfigure}{9}
\renewcommand{\thesfigure}{S\arabic{sfigure}}

\sdbarfig{
\input{Figure_Comparison_POG_BG_EMR}
   }{Different graphical modeling techniques available in the literature applied to model the DC motor driving an hydraulic pump of Figure~\ref{pompa_elettro_idraulica}. (a) Power-Oriented Graphs (POG) block scheme. 
   (b) Bond Graph (BG) of the system.
   (c) Energetic Macroscopic Representation (EMR) of the system.
   \label{POG_BG_EMR_Comparison_Figure}
     }

The main characteristics of POG, BG and EMR are compared in Table~\ref{POG_BG_EMR_Comparison_Table}. The three graphical modeling techniques have different properties and characteristics that make them suitable for different objectives, as indicated in the last row of Table~\ref{POG_BG_EMR_Comparison_Table}. The main advantages provided by POG over BG and EMR include: a) the symbolism, which is composed of standard blocks that can be found in standard Simulink libraries. b) the capability to derive the POG state-space model
directly from the graphical description. 
The POG state-space model has a well defined energetic meaning, as described in the Section~``POG State-Space Model'', and enables the application of congruent state-space transformations which are particularly suitable for model reduction, as described in the Section ``Congruent State-Space Transformations and Model Reduction''. c) the ease of use for non-expert users, thanks to 
%
%
a more intuitive graphical description which does not require 
the users to learn new
specific graphical symbols. 
Furthermore, the POG schemes can be directly implemented in the Matlab/Simulink environment for direct simulation of the systems.
%

\sdbartable{A point-by-point comparison between the properties of the three main graphical modeling techniques available in the literature for modeling physical systems: Power-Oriented Graphs (POG), Bond Graph (BG), and Energetic Macroscopic Representation (EMR).\label{POG_BG_EMR_Comparison_Table}}
{\begin{tabular*}{18.5pc}{@{}|p{35.25pt}<{\centering}|p{45.6pt}<{\centering}|p{45.6pt}<{\centering}|p{45.6pt}<{\centering}|@{}}
\hline
\textbf{Property} & \textbf{POG} & \textbf{BG} & \textbf{EMR} \\
\hline
\scriptsize 
Author~-~Year
& \centering \scriptsize 
R.~Zanasi~-~1991~\cite{zanasi1991}
& \scriptsize H.~M.~Paynter~-~1959~\cite{paynter1961analysis}
& \scriptsize $\!\!$A.~Bouscayrol~-~2000~\cite{Bouscayrol_2000}
\\
\hline
\multirow{3}{33.6pt}{
\centering 
\scriptsize  Extended Name
} 
& 
\multirow{3}{45.6pt}{\centering 
\scriptsize Power-Oriented Graphs
}
& 
\multirow{3}{45.6pt}{\centering 
\scriptsize Bond Graph
}
& 
\scriptsize Energetic Macroscopic Representation
\\
\hline
\multirow{5}{33.6pt}{\centering \scriptsize Symbolism }
& \scriptsize Gains, Integrators and Summation Nodes
& \multirow{4}{45.6pt}{\centering \scriptsize Arrows and Junctions}
& \multirow{4}{45.6pt}{\centering \scriptsize Energy domain-dependent Pictograms}
\\
\hline
\scriptsize \multirow{2}{35.25pt}{\centering \scriptsize Causality}
& \scriptsize Preferably integral 
& \scriptsize Preferably integral 
& \scriptsize Exclusively integral 
\\
\hline \scriptsize 
Power Variables Direction Visibility  &
\multirow{3}{45.6pt}{\centering \scriptsize Yes} 
& 
\multirow{3}{45.6pt}{\centering \scriptsize No} 
& \multirow{3}{45.6pt}{\centering \scriptsize Yes} 
\\
\hline
\multirow{4}{35.25pt}{\centering \scriptsize  $\!\!\!\!$ POG State-Space Model} & 
\scriptsize Directly Obtainable from Graphical Description
& 
\scriptsize Not Directly Obtainable from Graphical Description
& 
\scriptsize Not Directly Obtainable from Graphical Description\\
\hline
 \vspace{-1.6mm}\scriptsize   Simulink Implementation
& \scriptsize Use of Basic Blocks from Standard Libraries
& \multirow{3}{45.6pt}{\centering \scriptsize Need of Dedicated Blocks} 
& \multirow{3}{45.6pt}{\centering \scriptsize  Need of Dedicated Blocks}
\\
\hline \scriptsize \centering 
$\!\!\!\!$ Ease of Use for Non-Expert Users
 & \multirow{3}{45.6pt}{\centering \scriptsize More Intuitive }
& \multirow{3}{45.6pt}{\centering \scriptsize Less Intuitive }
& \multirow{3}{45.6pt}{\centering \scriptsize Average} 
\\
\hline
\scriptsize Main Scope
 & \scriptsize Simulation and Analysis
& \scriptsize Simulation and Design
& \scriptsize Simulation and Control
\\
\hline
\end{tabular*}}

\end{sidebar}

\section{The POG Technique: General Rules}\label{general_rules_sect}
\begin{figure}[t!]
  \begin{center}
 \centering \scriptsize
 \setlength{\unitlength}{3.6mm}
 \psset{unit=\unitlength}
 \SpecialCoor
  \begin{pspicture}(-11,-4)(11,4.5)
  \psfrag{Va}[rc][rc][0.88]{$\mygreen V_a$}
   \psfrag{C1}[ct][ct][0.88]{$\black C_1$}
   \psfrag{V1}[cb][cb][0.88]{$\mygreen V_1$}
   \psfrag{I1}[ct][ct][0.88]{$\!\!\!\! \myred I_1$}
   \psfrag{L2}[lc][lc][0.88]{$\black L_2$}
   \psfrag{I2}[rc][rc][0.88]{$\myred I_2$}
   \psfrag{L3}[cb][cb][0.88]{$\black L_3$}
   \psfrag{I3}[ct][ct][0.88]{$\myred I_3$}
   \psfrag{V3}[ct][ct][0.88]{$\mygreen V_{L3}$}
   \psfrag{R3}[cb][cb][0.88]{$\black R_3$}
   \psfrag{Vr3}[ct][ct][0.88]{$\mygreen V_{R3}$}
   \psfrag{Ir3}[ct][ct][0.88]{$\myred I_{R3}$}
   \psfrag{C4}[lc][lc][0.88]{$\black C_4$}
   \psfrag{V4}[rc][rc][0.88]{$\mygreen V_4$}
   \psfrag{V2}[rc][rc][0.88]{$\mygreen V_2$}
   \psfrag{I4}[rc][rc][0.88]{$\myred I_4$}
   \psfrag{R4}[cb][cb][0.88]{$\black R_4$}
   \psfrag{Ir4}[ct][ct][0.88]{$\myred I_{R4}$}
   \psfrag{Vr4}[ct][ct][0.88]{$\mygreen V_{R4}$}
   \psfrag{Vb}[lc][lc][0.88]{$\mygreen V_b$}
   \hspace*{0cm}
   \rput(0,0){\includegraphics[clip,width=8cm]{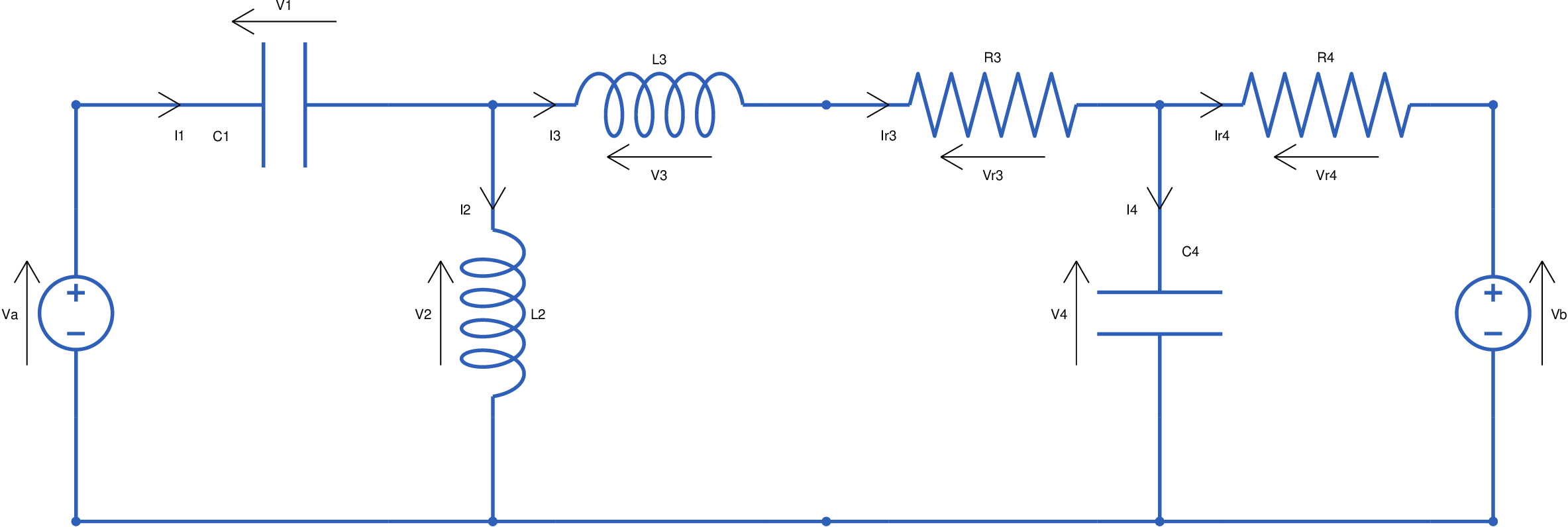}}
    \rput(-10.0,4){(a)}
   \psellipticarc[linestyle=dashed,linecolor=mygreen,linewidth=0.3pt]{<-}(-7,-0.8)(1,1){-250}{70}
   \rput(-7,-0.8){\tiny \mbox{\shortstack{\mygreen $\mbox{VKL}_1$}}}
   \psellipticarc[linestyle=dashed,linecolor=myred,linewidth=0.3pt]{<-}(-4.1,2.16)(0.92,0.48){-250}{70}
   \rput(-4.1,3.06){\tiny \mbox{\shortstack{\myred $\mbox{CKL}_1$}}}
   \psellipticarc[linestyle=dashed,linecolor=mygreen,linewidth=0.3pt]{<-}(-1.1,-0.8)(1,1){-250}{70}
   \rput(-1.1,-0.8){\tiny \mbox{\shortstack{\mygreen $\mbox{VKL}_2$}}}
   \psellipticarc[linestyle=dashed,linecolor=mygreen,linewidth=0.3pt]{<-}(2.1,-0.8)(1,1){-250}{70}
   \rput(2.1,-0.8){\tiny \mbox{\shortstack{\mygreen $\mbox{VKL}_3$}}}
   \psellipticarc[linestyle=dashed,linecolor=myred,linewidth=0.3pt]{<-}(5.25,2.16)(0.92,0.48){-250}{70}
   \rput(5.25,3.06){\tiny \mbox{\shortstack{\myred $\mbox{CKL}_2$}}}
   \psellipticarc[linestyle=dashed,linecolor=mygreen,linewidth=0.3pt]{<-}(7.75,-0.8)(1,1){-250}{70}
   \rput(7.75,-0.8){\tiny \mbox{\shortstack{\mygreen $\mbox{VKL}_4$}}}
   \psline[linecolor=black,linewidth=0.3pt](0.55,-3)(0.55,1.5)(0.375,1.25)
   \psline[linecolor=black,linewidth=0.3pt](0.55,1.52)(0.725,1.25)
   \rput(1,-2){\tiny $V_x$}
\end{pspicture}
  \end{center}
\vspace{.1cm}
  \begin{center}
 \centering \scriptsize
 \setlength{\unitlength}{3.6mm}
 \psset{unit=\unitlength}
 \SpecialCoor
  \begin{pspicture}(-11,-4)(11,4.7)
   \psfrag{Ia}[rc][rc]{$I_a$}
   \psfrag{Va}[rc][rc]{$V_a$}
   \psfrag{Ib}[lc][lc]{$I_b$}
   \psfrag{Vb}[lc][lc]{$V_b$}
   \hspace*{0cm}
   \rput(0,0){\includegraphics[clip,width=7.9cm]{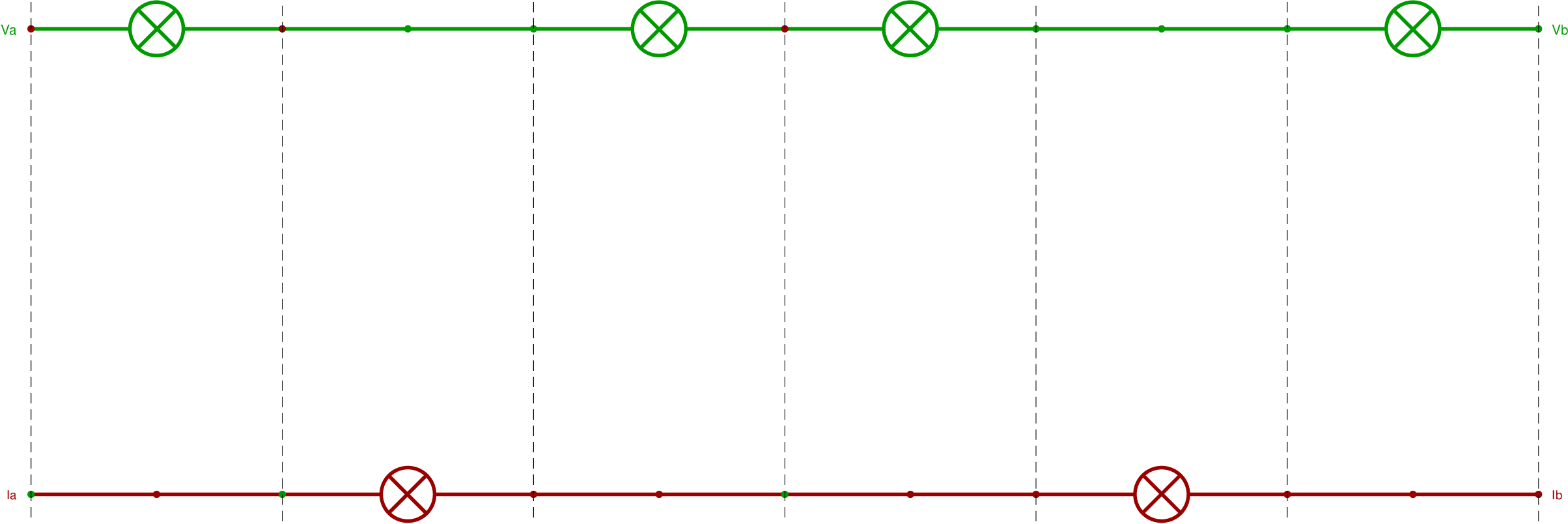}}
   \rput(-10.5,4.25){$1^\circ$}
   \rput(-6.88,4.25){$2^\circ$}
   \rput(-3.4,4.25){$3^\circ$}
   \rput(0.16,4.25){$4^\circ$}
   \rput(3.68,4.25){$5^\circ$}
   \rput(7.3,4.25){$6^\circ$}
   \rput(10.82,4.25){$7^\circ$}
       \rput(-10,4.88){(b)}
   \psellipticarc[linestyle=dashed,linecolor=mygreen,linewidth=0.3pt]{<-}(-8.75,3.22)(1.25,0.75){-250}{70}
   \rput(-8.75,2.0){\tiny \mbox{\shortstack{\mygreen $\mbox{VKL}_1$}}}
   \rput(-8.35,4.25){\tiny \mbox{\shortstack{\mygreen $1^\circ \mbox{SN}$}}}
   \psellipticarc[linestyle=dashed,linecolor=myred,linewidth=0.3pt]{<-}(-5.25,-3.22)(1.25,0.75){-250}{70}
   \rput(-5.25,-2.1){\tiny \mbox{\shortstack{\myred $\mbox{CKL}_1$}}}
   \rput(-4.85,-4.35){\tiny \mbox{\shortstack{\myred $2^\circ \mbox{SN}$}}}
   \psellipticarc[linestyle=dashed,linecolor=mygreen,linewidth=0.3pt]{<-}(-1.75,3.22)(1.25,0.75){-250}{70}
   \rput(-1.75,2.0){\tiny \mbox{\shortstack{\mygreen $\mbox{VKL}_2$}}}
   \rput(-1.35,4.25){\tiny \mbox{\shortstack{\mygreen $3^\circ \mbox{SN}$}}}
   \psellipticarc[linestyle=dashed,linecolor=mygreen,linewidth=0.3pt]{<-}(1.75,3.22)(1.25,0.75){-250}{70}
   \rput(1.75,2.0){\tiny \mbox{\shortstack{\mygreen $\mbox{VKL}_3$}}}
   \rput(2.15,4.25){\tiny \mbox{\shortstack{\mygreen $4^\circ \mbox{SN}$}}}
   \psellipticarc[linestyle=dashed,linecolor=myred,linewidth=0.3pt]{<-}(5.25,-3.22)(1.25,0.75){-250}{70}
   \rput(5.25,-2.1){\tiny \mbox{\shortstack{\myred $\mbox{CKL}_2$}}}
   \rput(5.65,-4.35){\tiny \mbox{\shortstack{\myred $5^\circ \mbox{SN}$}}}
   \psellipticarc[linestyle=dashed,linecolor=mygreen,linewidth=0.3pt]{<-}(8.75,3.22)(1.25,0.75){-250}{70}
   \rput(8.75,2.0){\tiny \mbox{\shortstack{\mygreen $\mbox{VKL}_4$}}}    \rput(9.15,4.25){\tiny \mbox{\shortstack{\mygreen $6^\circ \mbox{SN}$}}}
\rput(0,1.1){$\mygreen \underbrace{\hspace{7.5cm}}_{\mbox{across power line}}$}
\rput(0,-1.1){$\myred \overbrace{\hspace{7.5cm}}^{\mbox{through power line}}$}
\end{pspicture}
  \end{center}
\vspace{.1cm}
  \begin{center}
   \centering \scriptsize
 \setlength{\unitlength}{3.6mm}
 \psset{unit=\unitlength}
 \SpecialCoor
  \begin{pspicture}(-11,-4)(11,4.7)
   \psfrag{Ia}[rc][rc]{$I_a$}
   \psfrag{Va}[rc][rc]{$V_a$}
   \psfrag{1/s}[cc][cc]{$\black \frac{1}{s}$}
   \psfrag{1/C1}[cc][cc]{$\black \frac{1}{C_1}$}
   \psfrag{V1}[lb][lb]{$V_1$}
   \psfrag{1/s}[cc][cc]{$\black \frac{1}{s}$}
   \psfrag{1/L2}[cc][cc]{$\black \frac{1}{L_2}$}
   \psfrag{I2}[lt][lt]{$I_2$}
   \psfrag{1/s}[cc][cc]{$\black \frac{1}{s}$}
   \psfrag{1/L3}[cc][cc]{$\black \frac{1}{L_3}$}
   \psfrag{I3}[ct][ct]{$I_3$}
   \psfrag{1/s}[cc][cc]{$\black \frac{1}{s}$}
   \psfrag{1/C4}[cc][cc]{$\black \frac{1}{C_4}$}
   \psfrag{V4}[cb][cb]{$V_4$}
   \psfrag{Ib}[lc][lc]{$I_b$}
   \psfrag{Vb}[lc][lc]{$V_b$}
   \hspace*{0cm}
   \rput(0,0){\includegraphics[clip,width=7.9cm]{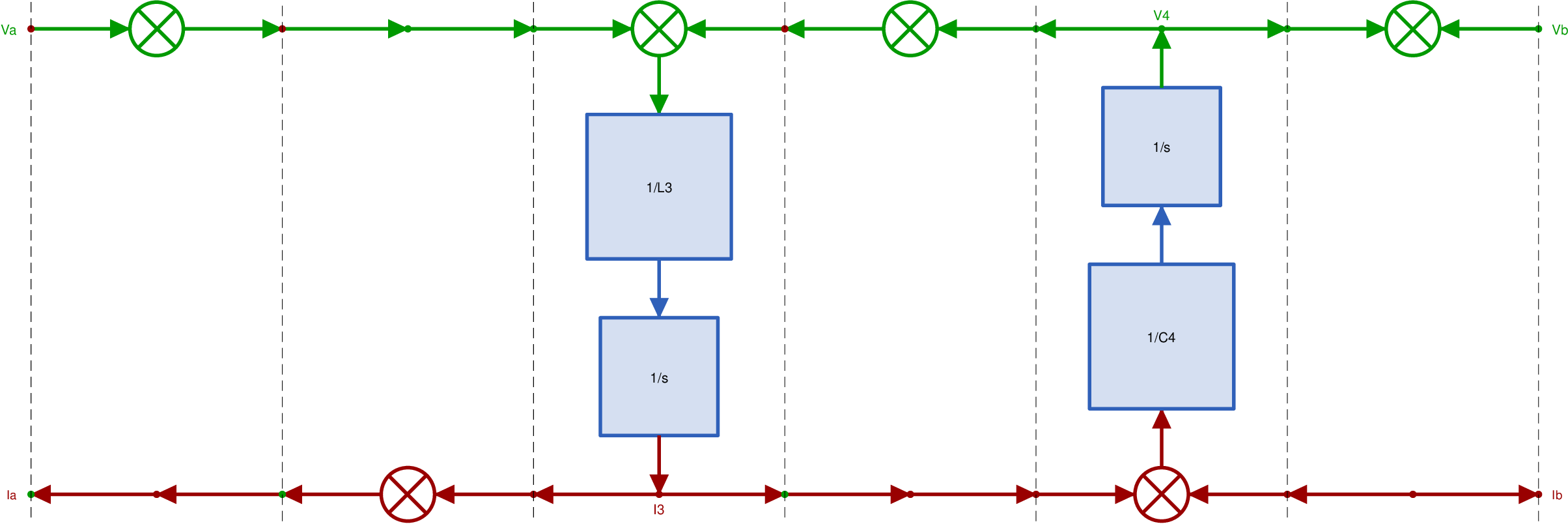}}
      \rput(-10.5,4.25){$1^\circ$}
   \rput(-6.88,4.25){$2^\circ$}
   \rput(-3.4,4.25){$3^\circ$}
   \rput(0.16,4.25){$4^\circ$}
   \rput(3.68,4.25){$5^\circ$}
   \rput(7.3,4.25){$6^\circ$}
   \rput(10.82,4.25){$7^\circ$}
     \rput(-5.32,3.9){$V_{2}$}
     \rput(5.88,-2.56){$I_{4}$}
     \rput(-0.9,2.52){$V_{L3}$}
         \rput(-10,4.88){(c)}
%
\psellipse[linestyle=dashed,linecolor=myorange](-1.8,0)(1.6,4.4)
\psellipse[linestyle=dashed,linecolor=myorange](5.28,0)(1.6,4.4)
\rput(-8.8,-3.82){$I_1$}
\rput(-0.55,3.82){$V_x$}
%
\end{pspicture}
  \end{center}
\vspace{.1cm}
  \begin{center}
   \centering \scriptsize
 \setlength{\unitlength}{3.6mm}
 \psset{unit=\unitlength}
 \SpecialCoor
  \begin{pspicture}(-11,-4)(11,4.7)
   \psfrag{Ia}[rc][rc]{$I_a$}
   \psfrag{Va}[rc][rc]{$V_a$}
   \psfrag{1/s}[cc][cc]{$\black \frac{1}{s}$}
   \psfrag{1/C1}[cc][cc]{$\black \frac{1}{C_1}$}
   \psfrag{V1}[lb][lb]{$V_1$}
   \psfrag{1/s}[cc][cc]{$\black \frac{1}{s}$}
   \psfrag{1/L2}[cc][cc]{$\black \frac{1}{L_2}$}
   \psfrag{I2}[lt][lt]{$I_2$}
   \psfrag{1/s}[cc][cc]{$\black \frac{1}{s}$}
   \psfrag{1/L3}[cc][cc]{$\black \frac{1}{L_3}$}
   \psfrag{I3}[ct][ct]{$I_3$}
   \psfrag{1/s}[cc][cc]{$\black \frac{1}{s}$}
   \psfrag{1/C4}[cc][cc]{$\black \frac{1}{C_4}$}
   \psfrag{V4}[cb][cb]{$V_4$}
   \psfrag{Ib}[lc][lc]{$I_b$}
   \psfrag{Vb}[lc][lc]{$V_b$}
   \hspace*{0cm}
   \rput(0,0){\includegraphics[clip,width=7.9cm]{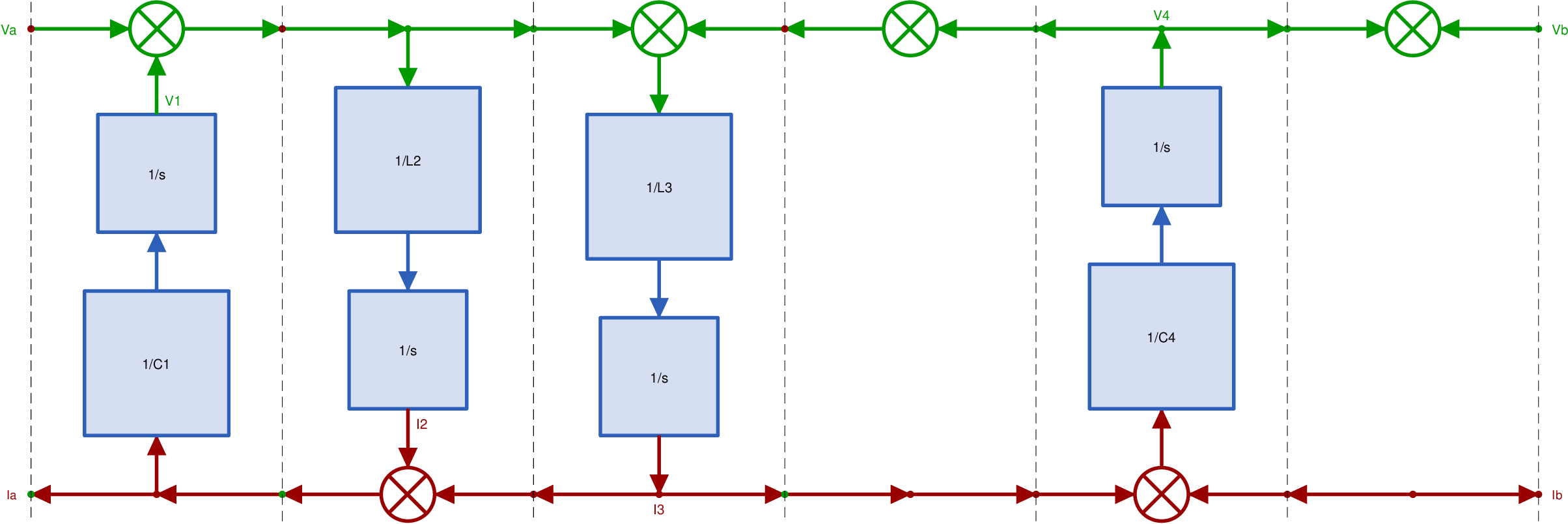}}
      \rput(-10.5,4.25){$1^\circ$}
   \rput(-6.88,4.25){$2^\circ$}
   \rput(-3.4,4.25){$3^\circ$}
   \rput(0.16,4.25){$4^\circ$}
   \rput(3.68,4.25){$5^\circ$}
   \rput(7.3,4.25){$6^\circ$}
   \rput(10.82,4.25){$7^\circ$}
     \rput(-5.32,3.9){$V_{2}$}
     \rput(5.88,-2.56){$I_{4}$}
     \rput(-0.9,2.52){$V_{L3}$}
         \rput(-10,4.88){(d)}
%
%
\psellipse[linestyle=dashed,linecolor=blue](-5.28,0)(1.6,4.4)
\psellipse[linestyle=dashed,linecolor=blue](-8.8,0)(1.6,4.4)
\rput(-8.8,-3.82){$I_1$}
\rput(-0.55,3.82){$V_x$}
%
\end{pspicture}
  \end{center}
\vspace{.1cm}
    \begin{center}
   \centering \scriptsize
 \setlength{\unitlength}{3.6mm}
 \psset{unit=\unitlength}
 \SpecialCoor
  \begin{pspicture}(-11,-4)(11,4.7)
   \psfrag{Ia}[rc][rc]{$I_a$}
   \psfrag{Va}[rc][rc]{$V_a$}
   \psfrag{1/s}[cc][cc]{$\black \frac{1}{s}$}
   \psfrag{1/C1}[cc][cc]{$\black \frac{1}{C_1}$}
   \psfrag{V1}[lb][lb]{$V_1$}
   \psfrag{1/s}[cc][cc]{$\black \frac{1}{s}$}
   \psfrag{1/L2}[cc][cc]{$\black \frac{1}{L_2}$}
   \psfrag{I2}[lt][lt]{$I_2$}
   \psfrag{1/s}[cc][cc]{$\black \frac{1}{s}$}
   \psfrag{1/L3}[cc][cc]{$\black \frac{1}{L_3}$}
   \psfrag{I3}[ct][ct]{$I_3$}
   \psfrag{1/s}[cc][cc]{$\black \frac{1}{s}$}
   \psfrag{1/C4}[cc][cc]{$\black \frac{1}{C_4}$}
   \psfrag{V4}[cb][cb]{$V_4$}
   \psfrag{Ib}[lc][lc]{$I_b$}
   \psfrag{Vb}[lc][lc]{$V_b$}
   \psfrag{Vr3}[lc][lc]{$V_{R3}$}
   \psfrag{R3}[cc][cc]{$R_3$}
   \psfrag{Ir4}[lc][lc]{$I_{R4}$}
   \psfrag{1/R4}[cc][cc]{$\frac{1}{R_4}$}
   \hspace*{0cm}
   \rput(0,0){\includegraphics[clip,width=7.9cm]{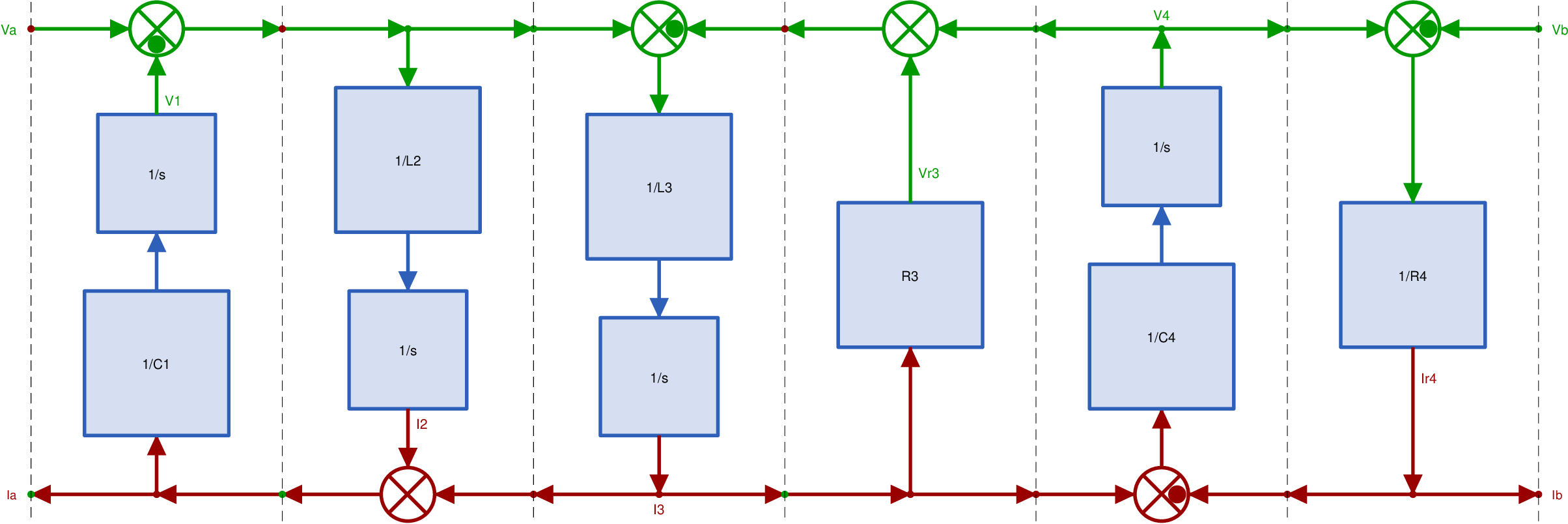}}
      \rput(-10.5,4.25){$1^\circ$}
   \rput(-6.88,4.25){$2^\circ$}
   \rput(-3.4,4.25){$3^\circ$}
   \rput(0.16,4.25){$4^\circ$}
   \rput(3.68,4.25){$5^\circ$}
   \rput(7.3,4.25){$6^\circ$}
   \rput(10.82,4.25){$7^\circ$}
     \rput(-5.32,3.9){$V_{2}$}
     \rput(5.88,-2.56){$I_{4}$}
     \rput(-0.9,2.52){$V_{L3}$}
     \rput(9.62,1.68){$V_{R4}$}
         \rput(-10,4.88){e)}
\psellipse[linestyle=dashed,linecolor=red](1.8,0)(1.6,4.4)
\psellipse[linestyle=dashed,linecolor=red](8.8,0)(1.6,4.4)
\rput(-8.8,-3.82){$I_1$}
\rput(-0.55,3.82){$V_x$}
%
\end{pspicture}
  \end{center}
  \caption{Electrical circuit taken as a case study. a) Simscape schematic, including the definition of the positive direction of the input and state variables representing {\bf Step 1} of the modeling procedure. b) {\bf Step 2} of the modeling procedure. c) {\bf Step 3} of the modeling procedure. d) {\bf Step 4} of the modeling procedure. e) {\bf Step 5} and {\bf Step 6} of the modeling procedure, resulting in the final POG block scheme of the system.}\label{Electric_Circuit_1}
  \vspace{-.15cm}
  \end{figure}

In this section, a few  general POG modeling rules
are formally given with reference to an electrical circuit case study, whose Simscape schematic is shown in Figure~\ref{Electric_Circuit_1}(a).

\noindent {\bf Rule 1}: {\em The positive directions of the
power variables $\mygreen v_e$ and $\myred v_f$ of all the PEs in the considered system 
must be chosen such that the
power $P=\mygreen v_e \myred v_f$ is positive when entering the considered PE, see Figure~\ref{PE_represent}(a)}. 

\noindent {\bf Rule 2}: {\em The power variables $\mygreen v_e$ of all the PEs connected in parallel and the power variables $\myred v_f$ of all the PEs connected in series
must share the same positive direction}.
In the example of Figure~\ref{Electric_Circuit_1}(a), this implies that the positive direction of current $I_3$ must be the same for both inductor $L_3$ and for resistor $R_3$.

\noindent {\bf Rule 3}: {\em The dynamic PEs {\mygreen $D_e$} and {\myred $D_f$} in a POG block scheme must always be modeled guaranteeing integral causality~\cite{zanasi1991,zanasi2010}}. This implies that across elements {\mygreen $D_e$} connected in parallel, such as capacitor $C_4$ in Figure~\ref{Electric_Circuit_1}(a), only admit the Elaboration Block (EB) configuration shown in Figure~\ref{terminali_POG_parallel}(a), as detailed in Figure~\ref{fig:un_grafico}(B) and Figure~\ref{fig:un_grafico}(e) for the capacitor $C_4$.
At the same time, through elements {\myred $D_f$} connected in series, such as inductor $L_3$ in Figure~\ref{Electric_Circuit_1}(a), only admit the EB configuration shown in Figure~\ref{terminali_POG_series}(a), as detailed in Figure~\ref{fig:un_grafico}(A) and Figure~\ref{fig:un_grafico}(a) for the inductor $L_3$. 
This rule comes from the application of Properties~\ref{through_s_across_p_Prop} and \ref{across_s_through_p_Prop}. 

\noindent {\bf Rule 4}: {\em Static elements $R$ admit all possible EB configurations in Figure~\ref{terminali_POG_series} and in Figure~\ref{terminali_POG_parallel}.} This rule comes from the application of Property~\ref{static_el_Prop}. 

\noindent {\bf Rule 5}: 
{\em If the considered physical system is stable, then all the loops of the corresponding POG block scheme must contain an odd number of minus signs.} This rule comes from the fact that the series and parallel connections between physical elements can only be made using  the EBs of the type reported in Figure~\ref{terminali_POG_series} and in Figure~\ref{terminali_POG_parallel}.

\section{Step-By-Step Modeling of Physical Systems}\label{fast_modeling_section}
In this section, a step-by-step modeling procedure called Fast Modeling Power Oriented Graphs (FMPOG) is presented. 
The Fast Modeling POG is a guided, step-by-step procedure, based on the POG technique. It allows the transformation of physical system schematics into the corresponding POG block schemes and, ultimately, into the state-space models of the considered physical systems.
%
%
In order to present the procedure, reference is 
made to the electrical circuit shown in Figure~\ref{Electric_Circuit_1}(a). 

\noindent {\bf Step 1}: {\em Choose the positive directions of the state and input variables of the considered physical system}. For the system in Figure~\ref{Electric_Circuit_1}(a), the input variables are the voltages $V_a$ and $V_b$, while the state variables
are the output power variables of all the system dynamic elements:
the currents $I_2$, $I_3$ and the voltages $V_1$, $V_4$. The chosen positive direction for each variable is highlighted by the corresponding arrow placed next to each variable, as denoted by
 \mbox{ \hspace{-1mm}
  \psline[linewidth=0.4pt,linecolor=black](0,1)(3,1)
  \rput(3,1){\psline[linewidth=0.4pt,linecolor=black](-1,0.75)(0,0)(-1,-0.75)} \hspace{4.1mm}}
and  \mbox{ \hspace{-1mm}
  \psline[linewidth=0.4pt,linecolor=black](0,0)(0,3)
  \rput{90}(0,3){\psline[linewidth=0.4pt,linecolor=black](-1,0.75)(0,0)(-1,-0.75)} \hspace{2.1mm}}
in Figure~\ref{Electric_Circuit_1}(a). Since the power variable { $I_3 \black = I_{R3}$} is in common to both elements $L_3$ and $R_3$, it has the same positive direction for both the elements according to {\bf Rule 2}.

 \noindent {\bf Step 2}: {\em Choose the
``{\mygreen across power line}'' and the ``{\myred through power line}'' and draw the series/parallel structure of the POG block scheme: draw a summation node on the {\mygreen across power line} if the considered PE is connected in series and draw a summation node on the {\myred through power line} if the considered PE is connected in parallel}. In this case, the {\mygreen across power line} and the {\myred through power line} have been chosen to be the upper and the lower ones, respectively, as shown in
Figure~\ref{Electric_Circuit_1}(b).
By comparing the connections in Figure~\ref{terminali} with those of the PEs in the system of Figure~\ref{Electric_Circuit_1}(a), it can be noticed that: a) PEs $C_1$, $L_3$, $R_3$ and $R_4$ are connected in series, therefore the corresponding summation nodes  {\mygreen $1^\circ \mbox{SN}$}, {\mygreen $3^\circ \mbox{SN}$}, {\mygreen$4^\circ \mbox{SN}$} and {\mygreen$6^\circ \mbox{SN}$} have been added in the {\mygreen across power line} as shown in Figure~\ref{Electric_Circuit_1}.b, implementing the generalized Voltage Kirchhoff's Laws {\mygreen $\mbox{VKL}_1$}, {\mygreen $\mbox{VKL}_2$}, {\mygreen $\mbox{VKL}_3$} and {\mygreen $\mbox{VKL}_4$},
respectively; b) PEs $L_2$ and $C_4$ are connected in parallel, therefore the corresponding summation nodes {\myred $2^\circ \mbox{SN}$} and {\myred $5^\circ \mbox{SN}$} have been added in the {\myred through power line}, see Figure~\ref{Electric_Circuit_1}.b, implementing the generalized Current Kirchhoff's Laws {\myred $\mbox{CKL}_1$} and {\myred $\mbox{CKL}_2$}, respectively.

\noindent {\bf Step 3}: {\em Add all the EBs of the dynamic elements {\myred $D_f$} and {\mygreen $D_e$} that only admit one possible configuration to the POG block scheme}. This can be directly done by making use of {\bf Rule 3}, which derives directly from Property~\ref{through_s_across_p_Prop}. Since $L_3$ is a through element {\myred $D_f$} connected in series, the only EB configuration guaranteeing integral causality is the one in Figure~\ref{fig:un_grafico}(A), that is the one encircled in green between power sections $3^\circ$ and $4^\circ$ in Figure~\ref{Electric_Circuit_1}(c) introducing the transfer function $\frac{I_3(s)}{V_{L3}(s)}=\frac{1}{L_3 s}$.
At the same time, since $C_4$ is an across PE {\mygreen $D_e$} connected in parallel, the only EB configuration guaranteeing integral causality is the one in Figure~\ref{fig:un_grafico}(B), that is the one encircled in green between power sections $5^\circ$ and $6^\circ$ in Figure~\ref{Electric_Circuit_1}(c)  introducing the transfer function $\frac{V_4(s)}{I_4(s)}=\frac{1}{C_4 s}$.

\noindent {\bf Step 4}: {\em Add all the EBs of the dynamic elements {\myred $D_f$} and {\mygreen $D_e$} that admit two possible 
configurations to the POG block scheme}.
Once again, this can be directly done by recalling {\bf Rule 3} and making the following observations, which derives directly from Property~\ref{across_s_through_p_Prop}. Since $L_2$ is a through PE {\myred $D_f$} connected in parallel, both the EB configurations in Figure~\ref{fig:due_grafici}(A) and Figure~\ref{fig:due_grafici}(B) guarantee integral causality. However, since its EB must be connected to the EB of $L_3$, the only matching EB for $L_2$ is the one in Figure~\ref{fig:due_grafici}(A), as shown by the EB encircled in blue between power sections $2^\circ$ and $3^\circ$ in Figure~\ref{Electric_Circuit_1}(d) introducing the transfer function $\frac{I_2(s)}{V_2(s)}=\frac{1}{L_2 s}$.  
Since $C_1$ is an across PE {\mygreen $D_e$} connected in series, both the EB configurations in Figure~\ref{fig:due_grafici}(C) and Figure~\ref{fig:due_grafici}(D)
%
%
guarantee integral causaslity. However, since its EB must be connected to the EB of $L_2$ and with voltage generator $V_a$, the only matching EB for PE $C_1$ is the one in Figure~\ref{fig:due_grafici}(D), as shown by the EB encircled in blue between power sections $1^\circ$ and $2^\circ$ in Figure~\ref{Electric_Circuit_1}(d) introducing the transfer function $\frac{V_1(s)}{I_a(s)}=\frac{1}{C_1 s}$.

\noindent {\bf Step 5}: {\em Add all the EBs of the static elements $R$ to the POG block scheme}. According to {\bf Rule 4}, resulting from the application of Property~\ref{static_el_Prop}, static elements $R$ admit all the three EB configurations in Figure~\ref{terminali_POG_series} when connected in series and all the three EB configurations in Figure~\ref{terminali_POG_parallel} when connected in parallel. However, since the element $R_3$ in Figure~\ref{Electric_Circuit_1}(a) is connected in series and its EB must be connected to the EBs of elements $L_3$ and $C_4$, the only possible configuration is the one shown in Figure~\ref{fig:due_grafici}(C), as shown by the EB encircled in red between power sections $4^\circ$ and $5^\circ$ in Figure~\ref{Electric_Circuit_1}(e) introducing the static relation $V_{R3}=R_3\,I_3$.
At the same time, since the element $R_4$ in Figure~\ref{Electric_Circuit_1}(a) is connected in series and its EB must be connected to that of the PE $C_4$ and with voltage generator $V_b$, the only possible configuration is the one shown in Figure~\ref{terminali_POG_series}(a), as shown by the EB encircled in red between power sections $6^\circ$ and $7^\circ$ in Figure~\ref{Electric_Circuit_1}(d) introducing the static relation $I_{R4}=\frac{1}{R_4}V_{R4}$.

\noindent {\bf Step 6}: {\em Add the correct signs to the summation nodes of the POG block scheme according to the positive directions defined at {\bf Step 1}}.
From the positive directions
chosen at {\bf Step 1} and recalling {\bf Rule 1}, it is possible to correctly determine the signs $p_1$ and $p_2$ of the input power variables $u_1$ and $u_2$ of each summation node in order to have its output power variable $y=\sum_{i=1}^{2} p_i u_i$ correctly defined.
For the considered system, as shown in Figure~\ref{Electric_Circuit_1}(e), the output variables $y$ of the summation nodes are given by:
\begin{equation}\label{SNs_first_esempio}
    \begin{array}{c@{\hspace{.68mm}\rightarrow\hspace{.68mm}}c@{}c@{}c@{\hspace{6.8mm}}c@{\hspace{.68mm}\rightarrow\hspace{.68mm}}c@{}c@{}c}
{\mygreen 1^\circ \mbox{SN}} & V_2 & = & V_a-V_1, &
{\myred 2^\circ \mbox{SN}} & I_1 & = & I_2+I_3, \\[1mm]
{\mygreen 3^\circ \mbox{SN}} & V_{L3} & = & V_2-V_x, &
{\mygreen 4^\circ \mbox{SN}} & V_x & = & V_{R3}+V_4, \\[1mm]
{\myred 5^\circ \mbox{SN}} & I_4 & = & I_3-I_{R4}, &
{\mygreen 6^\circ \mbox{SN}} & V_{R4} & = & V_4-V_b.
\end{array}
\end{equation}
As an example, for what concerns the {\mygreen $1^\circ \mbox{SN}$}, the output power variable $y$ 
is voltage $V_2$, whereas the input power variables $u_1$ and $u_2$ 
are voltages $V_a$ and $V_1$, as shown in Figure~\ref{Electric_Circuit_1}(d). Applying {\mygreen $\mbox{VKL}_1$} in Figure~\ref{Electric_Circuit_1}(a), with the signs defined at {\bf Step 1}, yields $y=V_2=u_1-u_2=V_a-V_1$ as in \eqref{SNs_first_esempio}. By making similar considerations, the remaining relations in \eqref{SNs_first_esempio} can be obtained, characterizing all the summation nodes in the
resulting final POG block scheme illustrated in Figure~\ref{Electric_Circuit_1}(e).
\begin{pullquote}
The Fast Modeling POG is a guided, step-by-step procedure, based on the POG technique. It allows the transformation of physical system schematics into the corresponding POG block schemes and, ultimately, into the state-space models of the considered physical systems.
\end{pullquote}

It can be proven that, by applying the described FMPOG procedure, the obtained POG block scheme is always unique if  no algebraic loops affect the considered physical system.

 A program implemented in the Matlab/Simulink environment has been created for the automatic conversion of Simscape schematics into POG block schemes and, ultimately, into the corresponding state-space equations of the system. The program is called ``POG\_Modeler'' and is freely available in the repository \cite{Suppl_Mat_Ref}, together with some additional examples and instructions. The latter also contains a video tutorial describing the content of the given repository and showing how to use the POG\_Modeler program. Specifically, the video also describes where to find the two main outputs of the program: the POG block scheme and the state-space model.


\subsection{Modeling Example in the Hydraulic Domain}\label{Modeling_example_in_the_hydraulic_domain}

  \begin{figure}[t!]
  \begin{center}
 \centering \footnotesize
 \setlength{\unitlength}{5.52mm}
 \psset{unit=\unitlength}
 \SpecialCoor
  \begin{pspicture}(-7,-3.5)(8,3.5)
   \psfrag{Pa}[rc][rc][0.68]{$\mygreen \footnotesize P_a$}
   \psfrag{L1}[cb][cb][0.68]{$\black \footnotesize L_1$}
   \psfrag{Q1}[rb][rb][0.68]{$\myred \footnotesize \!\!Q_1$}
   \psfrag{P1}[ct][ct][0.68]{$\mygreen \footnotesize P_1$}
   \psfrag{R1}[cb][cb][0.68]{$\black \footnotesize R_1$}
   \psfrag{Pr1}[ct][ct][0.68]{$\mygreen \footnotesize P_{R1}$}
   \psfrag{Qr1}[rb][rb][0.68]{$\myred \footnotesize \!\! Q_{R1}$}
   \psfrag{C2}[lc][lc][0.68]{$\black \footnotesize C_2$}
   \psfrag{P2}[rc][rc][0.68]{$\mygreen \footnotesize P_2$}
   \psfrag{Q2}[rc][rc][0.68]{$\myred \footnotesize Q_2$}
   \psfrag{R2}[t][t]{}
   \psfrag{Qr2}[lc][lc][0.68]{$\myred \footnotesize Q_{R2}$}
   \psfrag{Pr2}[lc][lc][0.68]{$\mygreen \scriptsize P_{R2}$}
   \psfrag{L3}[cb][cb][0.68]{$\black \footnotesize L_3$}
   \psfrag{Q3}[cb][cb][0.68]{$\myred \footnotesize Q_3$}
   \psfrag{P3}[ct][ct][0.68]{$\mygreen \footnotesize P_3$}
   \psfrag{C4}[lc][lc][0.68]{$\black \footnotesize C_4$}
   \psfrag{P4}[rc][rc][0.68]{$\mygreen \scriptsize P_4$}
   \psfrag{Q4}[rc][rc][0.68]{$\myred \footnotesize Q_4$}
   \psfrag{R4}[lc][lc][0.68]{$\black \footnotesize R_4$}
   \psfrag{Qr4}[lc][lc][0.68]{$\myred \footnotesize Q_{R4}$}
   \psfrag{Pr4}[rc][rc]{}
   \psfrag{R5}[cb][cb][0.68]{$\black \footnotesize R_5$}
   \psfrag{Qr5}[rb][rb][0.68]{$\myred \footnotesize \!\!Q_{R5}$}
   \psfrag{Pr5}[ct][ct][0.68]{$\mygreen \footnotesize P_{R5}$}
   \psfrag{Qb}[lc][lc][0.68]{$\myred \footnotesize Q_b$}
   \hspace*{0cm}
   \rput(0,0){\includegraphics[clip,width=8cm]{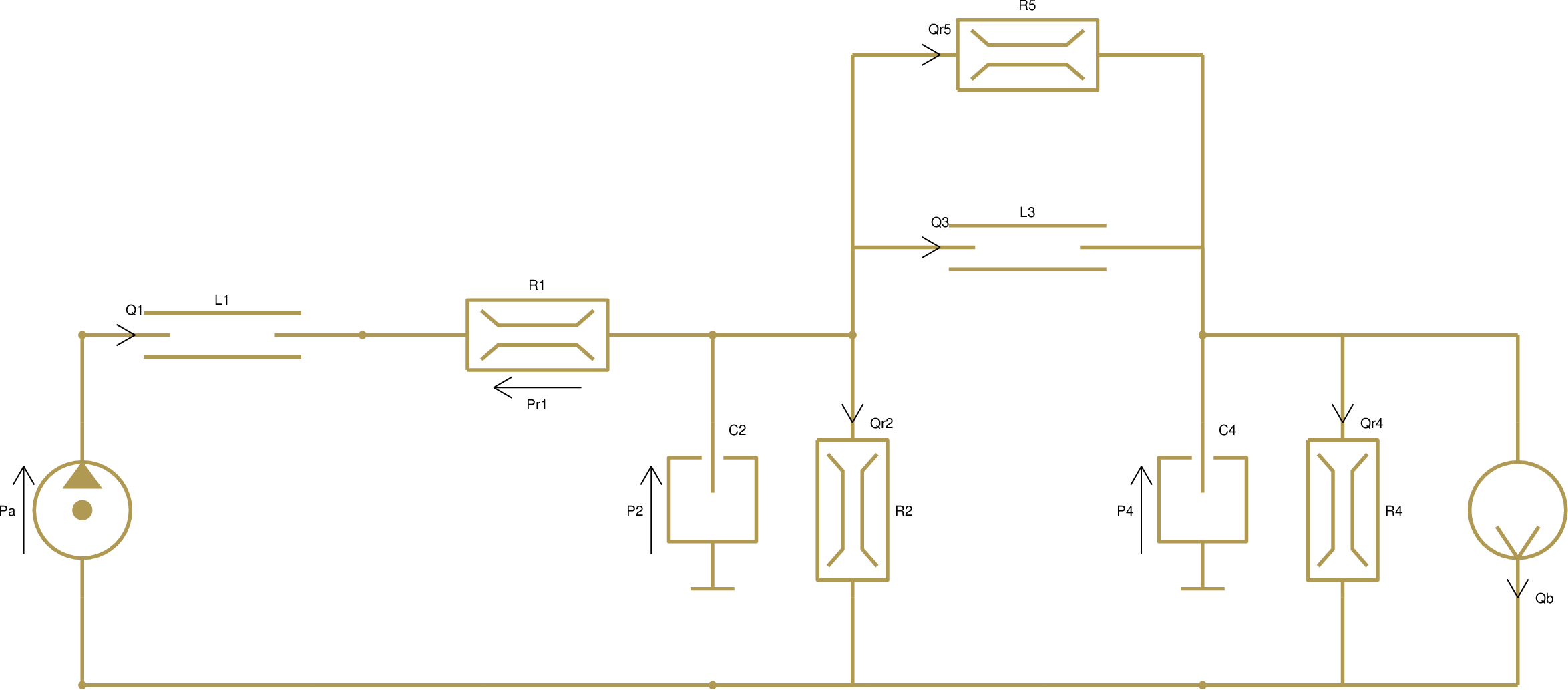}}
     \rput[rt](0.5,-2.5){\tiny $\black R_{2}$}
     \rput(5.3,-2.5){\tiny $\mygreen P_{R4}$}
     \rput(-6.5,2.5){(a)}
    \psellipticarc[linestyle=dashed,linecolor=mygreen,linewidth=0.3pt]{<-}(2.27,-1.5)(0.5,0.5){-250}{70}
   \rput(2.27,-1.5){\tiny \mbox{\shortstack{\mygreen $\mbox{VKL}$}}}
   \psellipticarc[linestyle=dashed,linecolor=myred,linewidth=0.3pt]{<-}(0.7,0.9)(0.5,0.3){-250}{70}
   \rput[r](0.1,0.9){\tiny \mbox{\shortstack{\myred $\mbox{CKL}$}}}
   \rput(-0.02,0){\psline[linecolor=black,linewidth=0.6pt](0.8,0.35)(0.7,0.5)(0.6,0.35)}
   \rput[lt](0.8,0.5){\tiny $\black Q_{2}'$}
\end{pspicture}
  \end{center}
  \vspace{.1cm}
    \begin{center}
   \centering \scriptsize
 \setlength{\unitlength}{5.52mm}
 \psset{unit=\unitlength}
 \SpecialCoor
  \begin{pspicture}(-7,-2.5)(7.5,2.5)
   \psfrag{Qa}[rc][rc]{$Q_a$}
   \psfrag{Pa}[rc][rc]{$P_a$}
   \psfrag{Pb}[lc][lc]{$P_b$}
   \psfrag{Qb}[lc][lc]{$Q_b$}
   \hspace*{0cm} 
   \rput(0,0){\includegraphics[clip,width=7.8cm]{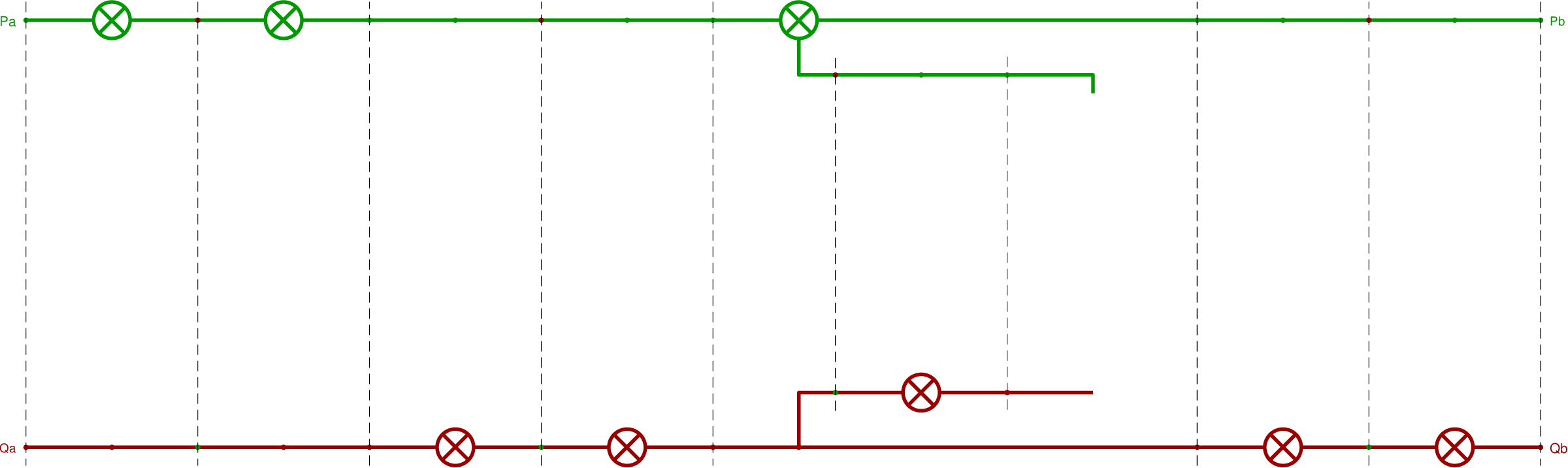}}
   \rput(-6.9,2.6){(b)}
\rput(0,0.75){$\mygreen \underbrace{\hspace{7.5cm}}_{\mbox{across power line}}$}
\rput(0,-0.65){$\myred \overbrace{\hspace{7.5cm}}^{\mbox{through power line}}$}
\end{pspicture}
  \end{center}
  \vspace{.1cm}
    \begin{center}
   \centering \scriptsize
 \setlength{\unitlength}{5.52mm}
 \psset{unit=\unitlength}
 \SpecialCoor
  \begin{pspicture}(-7,-2.5)(7.5,2.5)
   \psfrag{Qa}[rc][rc]{$Q_a$}
   \psfrag{Pa}[rc][rc]{$P_a$}
   \psfrag{1/L1}[cc][cc]{$\black \frac{1}{L_1}$}
   \psfrag{1/s}[cc][cc]{$\black \frac{1}{s}$}
   \psfrag{Q1}[ct][ct]{$Q_1$}
   \psfrag{1/C2}[cc][cc]{$\black \frac{1}{C_2}$}
   \psfrag{1/s}[cc][cc]{$\black \frac{1}{s}$}
   \psfrag{P2}[cb][cb]{$P_2$}
   \psfrag{1/C4}[cc][cc]{$\black \frac{1}{C_4}$}
   \psfrag{1/s}[cc][cc]{$\black \frac{1}{s}$}
   \psfrag{P4}[cb][cb]{$P_4$}
   \psfrag{Pb}[lc][lc]{$P_b$}
   \psfrag{Qb}[lc][lc]{$Q_b$}
   \hspace*{0cm} 
   \rput(0,0){\includegraphics[clip,width=7.8cm]{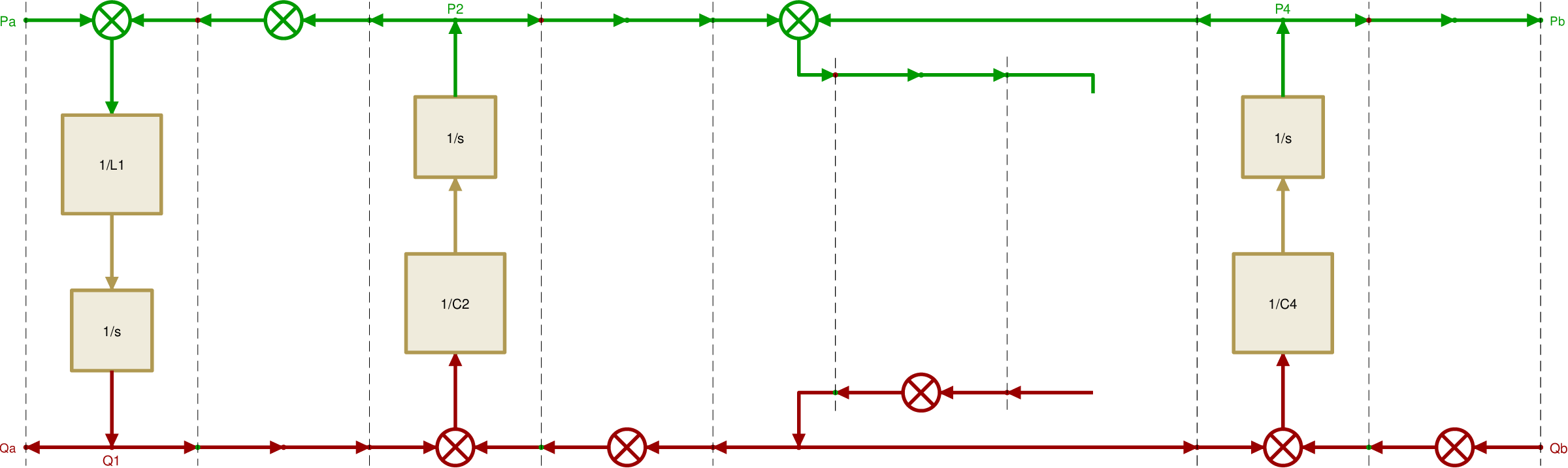}}
   \rput(-6.9,2.6){(c)}
   \psellipticarc[linestyle=dashed,linecolor=mygreen,linewidth=0.3pt]{<-}(0.15,1.95)(0.5,0.3){-250}{70}
   \rput[rb](-0.25,2.2){\tiny \mbox{\shortstack{\mygreen $\mbox{VKL}$}}}
   \psellipticarc[linestyle=dashed,linecolor=myred,linewidth=0.3pt]{<-}(1.22,-1.4)(0.5,0.3){-250}{70}
   \rput[lb](1.7,-1.2){\tiny \mbox{\shortstack{\myred $\mbox{CKL}$}}}
\psline[linestyle=dashed,linecolor=myorange](0.3,-1.8)(3.4,-1.8)(3.4,1.7)(0.3,1.7)(0.3,-1.8)
\psline[linecolor=myorange]{->}(0.8,1.7)(1.5,2.4)
\rput[lb](1.5,2.4){\textcolor{myorange}{\scriptsize \shortstack{Nested\\Structure}}}
  \psellipse[linestyle=dashed,linecolor=orange](-6.05,0)(0.75,2.5)
  \psellipse[linestyle=dashed,linecolor=orange](-2.95,0)(0.75,2.5)
  \psellipse[linestyle=dashed,linecolor=orange]( 4.5,0)(0.75,2.5)
   %
\end{pspicture}
  \end{center}
  \vspace{.1cm}
    \begin{center}
   \centering \scriptsize
 \setlength{\unitlength}{5.52mm}
 \psset{unit=\unitlength}
 \SpecialCoor
  \begin{pspicture}(-7,-2.5)(7.5,2.5)
   \psfrag{Qa}[rc][rc]{$Q_a$}
   \psfrag{Pa}[rc][rc]{$P_a$}
   \psfrag{1/L1}[cc][cc]{$\black \frac{1}{L_1}$}
   \psfrag{1/s}[cc][cc]{$\black \frac{1}{s}$}
   \psfrag{Q1}[ct][ct]{$Q_1$}
   \psfrag{1/C2}[cc][cc]{$\black \frac{1}{C_2}$}
   \psfrag{1/s}[cc][cc]{$\black \frac{1}{s}$}
   \psfrag{P2}[cb][cb]{$P_2$}
   \psfrag{1/L3}[cc][cc]{$\black \frac{1}{L_3}$}
   \psfrag{1/s}[cc][cc]{$\black \frac{1}{s}$}
   \psfrag{Q3}[lt][lt]{$Q_3$}
   \psfrag{1/C4}[cc][cc]{$\black \frac{1}{C_4}$}
   \psfrag{1/s}[cc][cc]{$\black \frac{1}{s}$}
   \psfrag{P4}[cb][cb]{$P_4$}
   \psfrag{Pb}[lc][lc]{$P_b$}
   \psfrag{Qb}[lc][lc]{$Q_b$}
   \hspace*{0cm} 
   \rput(0,0){\includegraphics[clip,width=7.8cm]{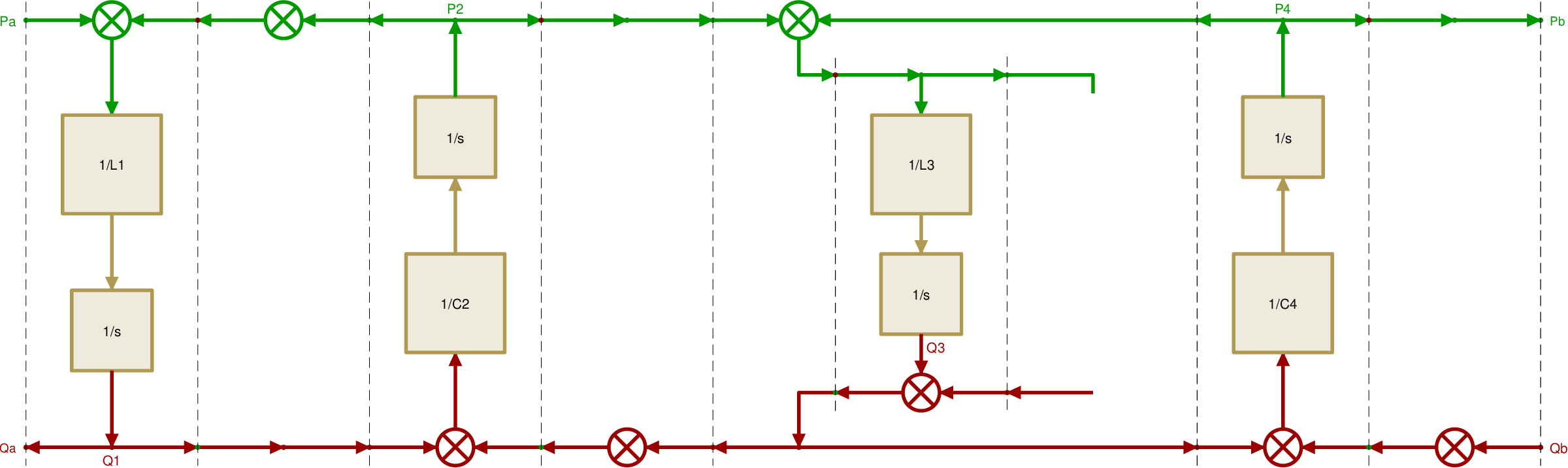}}
   \rput(-6.9,2.6){(d)}
  \psellipse[linestyle=dashed,linecolor=blue](1.2,0)(0.75,1.85)
\end{pspicture}
  \end{center}
  \vspace{.1cm}
    \begin{center}
   \centering \scriptsize
 \setlength{\unitlength}{5.52mm}
 \psset{unit=\unitlength}
 \SpecialCoor
  \begin{pspicture}(-7,-2.5)(7.5,2.5)
  \psfrag{Qa}[rc][rc]{$Q_a$}
   \psfrag{Pa}[rc][rc]{$P_a$}
   \psfrag{1/L1}[cc][cc]{$\black \frac{1}{L_1}$}
   \psfrag{1/s}[cc][cc]{$\black \frac{1}{s}$}
   \psfrag{Q1}[ct][ct]{$Q_1$}
   \psfrag{R1}[cc][cc]{$\black R_1$}
   \psfrag{Pr1}[lb][lb]{$P_{r1}$}
   \psfrag{1/C2}[cc][cc]{$\black \frac{1}{C_2}$}
   \psfrag{1/s}[cc][cc]{$\black \frac{1}{s}$}
   \psfrag{P2}[cb][cb]{$P_2$}
   \psfrag{1/R2}[cc][cc]{$\black \frac{1}{R_2}$}
   \psfrag{Qr2}[lt][lt]{$Q_{R2}$}
   \psfrag{1/L3}[cc][cc]{$\black \frac{1}{L_3}$}
   \psfrag{1/s}[cc][cc]{$\black \frac{1}{s}$}
   \psfrag{Q3}[lt][lt]{}
   \psfrag{1/R5}[cc][cc]{$\black \frac{1}{R_5}$}
   \psfrag{Qr5}[lt][lt]{}
   \psfrag{1/C4}[cc][cc]{$\black \frac{1}{C_4}$}
   \psfrag{1/s}[cc][cc]{$\black \frac{1}{s}$}
   \psfrag{P4}[cb][cb]{$P_4$}
   \psfrag{1/R4}[cc][cc]{$\black \frac{1}{R_4}$}
   \psfrag{Qr4}[lt][lt]{$Q_{R4}$}
   \psfrag{Pb}[lc][lc]{$P_b$}
   \psfrag{Qb}[lc][lc]{$Q_b$}   
   \hspace*{0cm}
   \rput(0,0){\includegraphics[clip,width=7.8cm]{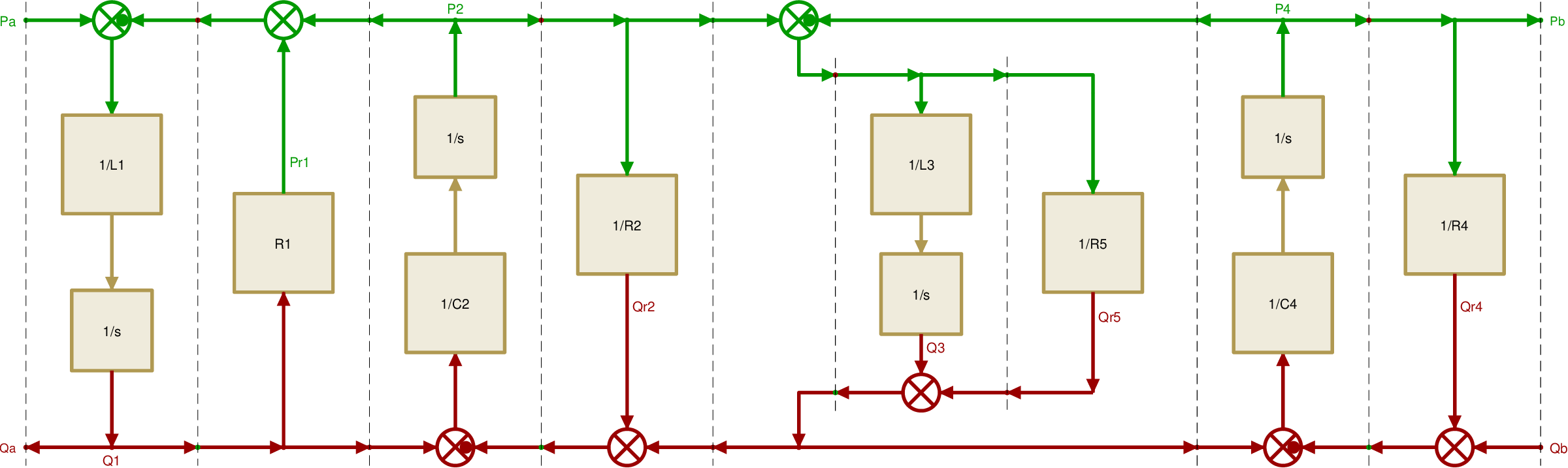}}
   \rput(-6.9,2.6){e)}
   \rput[rt](0,1.5){\tiny $\black P_{3}$}
   \rput[rt](1,-0.9){\tiny $\black Q_{3}$}
   \rput[l](2.8,-1){\tiny $\black Q_{R5}$}
   \rput[rb](0,-1.7){\tiny $\black Q_{2}'$}
%
\psellipse[linestyle=dashed,linecolor=red](-4.5,0)(0.75,2.5)
\psellipse[linestyle=dashed,linecolor=red](-1.35,0)(0.75,2.5)
\psellipse[linestyle=dashed,linecolor=red](6.05,0)(0.75,2.5)
\psellipse[linestyle=dashed,linecolor=red](2.8,0)(0.75,1.85)
\end{pspicture}
\end{center}
 \vspace{-3mm}
  \caption{
  Hydraulic system taken as a second case study. a) Simscape schematic, including the definition of the positive direction of the input and state variables representing {\bf Step 1} of the modeling procedure. b) {\bf Step 2} of the modeling procedure. c) {\bf Step 3} of the modeling procedure. d) {\bf Step 4} of the modeling procedure. e) {\bf Step 5} and {\bf Step 6} of the modeling procedure, resulting in the final POG block scheme of the system.
  }\label{Hydraulic_Circuit_1}
  \end{figure}

Reference is made to the hydraulic system in Figure~\ref{Hydraulic_Circuit_1}(a), composed of the PEs $L_1$, $R_1$, $C_2$, $R_2$, $L_3$, $R_5$, $C_4$, $R_4$ in the hydraulic domain, see Table~\ref{Ambiti_Energetici}. This system can be straightforwardly modeled using the FMPOG step-by-step procedure
described in Section~``Step-By-Step Modeling Of Physical Systems'', and making similar observations as in the previous case study. The system input variables are the pressure $P_a$ and the volume flow rate $Q_b$, while the system state variables are the volume flow rates $Q_1$ and $Q_{3}$, and the pressures $P_2$ and $P_4$.
The positive directions of the state and input variables are determined following {\bf Step 1}, as shown in
Figure~\ref{Hydraulic_Circuit_1}(a).
As per {\bf Step 2}, the {\mygreen across power line} and the {\myred through power line} have been chosen to be the upper and the lower ones, respectively, as shown in
Figure~\ref{Hydraulic_Circuit_1}(b).
In this example, it is possible to note the presence of a series PE element which is given by the parallel two more PEs, that are resistor $R_5$ and inductor $L_3$ in Figure~\ref{Hydraulic_Circuit_1}(a), which leads to the introduction of the Nested Structure present in Figure~\ref{Hydraulic_Circuit_1}(c). 
As per {\bf Step 3}, the EBs of the dynamic elements admitting one configuration only are added to the POG block scheme, as shown by the EBs highlighted in orange in
Figure~\ref{Hydraulic_Circuit_1}(c). Subsequently, the EBs of the dynamic elements admitting two configurations are added to the POG block scheme according to {\bf Step 4}, as shown by the EBs highlighted in blue in
Figure~\ref{Hydraulic_Circuit_1}(d). The next  {\bf Step 5} consists in adding the EBs of the static elements, as shown by the EBs highlighted in red in
Figure~\ref{Hydraulic_Circuit_1}(e). The correct signs are finally added to the summation nodes as described in {\bf Step 6} resulting in the final POG block scheme of Figure~\ref{Hydraulic_Circuit_1}(e). 
The Nested Structure highlighted in Figure~\ref{Hydraulic_Circuit_1}(c)
is characterized by the combination of the following {\mygreen VKL} and {\myred CKL}: $P_3=P_{2}-P_{4}$
and $Q_2'=Q_3+Q_{R5}$.

\subsection{DC Motor Supplying an Hydraulic Pump}\label{DC_mot_and_pump_sect}
  \begin{figure}[t!]
  \begin{center}
 \centering \footnotesize
 \setlength{\unitlength}{5.52mm}
 \psset{unit=\unitlength}
 \SpecialCoor
  \begin{pspicture}(-8.0,-1.9)(7.5,1.7)
   \psfrag{Va}[rc][rc][0.8]{$\mygreen  \footnotesize V_a$}
   \psfrag{L1}[cb][cb][0.8]{$\black  \footnotesize L_1$}
   \psfrag{I1}[ct][ct][0.8]{$\myred  \footnotesize I_1$}
   \psfrag{V1}[ct][ct][0.8]{$\mygreen \footnotesize V_1$}
   \psfrag{R1}[cb][cb][0.8]{$\footnotesize R_1$}
   \psfrag{Ir1}[ct][ct][0.8]{$\myred \footnotesize I_{R1}$}
   \psfrag{Vr1}[lt][lt][0.8]{$\mygreen \footnotesize V_{R1}$}
   \psfrag{T=Km}[cc][cc][0.8]{$\black  \footnotesize K_{12}$}
   \psfrag{J2}[lc][lc][0.8]{$\black \footnotesize  J_2$}
   \psfrag{Em}[r][r][0.8]{$\black \tiny  E_{12}$}
   \psfrag{Tm}[r][r][0.8]{$\myred \footnotesize  \tau_{12}$}
   \psfrag{vb2}[r][r][0.8]{$\mygreen \footnotesize \omega_{b2}$}
   \psfrag{w2}[rc][rc][0.8]{$\mygreen  \footnotesize \omega_2$}
   \psfrag{T2}[rc][rc][0.8]{$\myred  \footnotesize \tau_2$}
   \psfrag{b2}[lc][lc][0.8]{$\footnotesize b_2$}
   \psfrag{Tb2}[lr][lr][0.8]{$\myred  \footnotesize \tau_{b2}$}
   \psfrag{Vb2}[lc][lc][0.8]{$\mygreen \footnotesize \omega_{b2} $}
   \psfrag{G=Kp}[cc][cc][0.8]{$\black K_{23}$}
   \psfrag{Tp}[rc][rc][0.8]{$\myred\footnotesize  \tau_{23}$}
   \psfrag{Qp}[rc][rc][0.8]{$\myred \footnotesize Q_{23}$}
   \psfrag{C3}[lc][lc][0.8]{$\black\footnotesize  C_3$}
   \psfrag{P3}[rc][rc][0.8]{$\mygreen \footnotesize P_3$}
   \psfrag{Q3}[rc][rc][0.8]{$\myred \footnotesize Q_3$}
   \psfrag{QR4}[rc][rc][0.68]{$\myred \footnotesize Q_{R3}$}
   \psfrag{PR4}[rb][rb][0.68]{$\mygreen \footnotesize P_{R3}$}
   \psfrag{R4}[lc][lc][0.8]{$\black \footnotesize R_3$}
   \psfrag{Qb}[lc][lc][0.8]{$\myred \footnotesize Q_b$}
   \psfrag{Pb}[lc][lc][0.8]{$\mygreen\footnotesize  P_b$}
   \hspace*{-2mm}
   \rput(0,0){\includegraphics[clip,width=8.3cm,height=2.0cm]{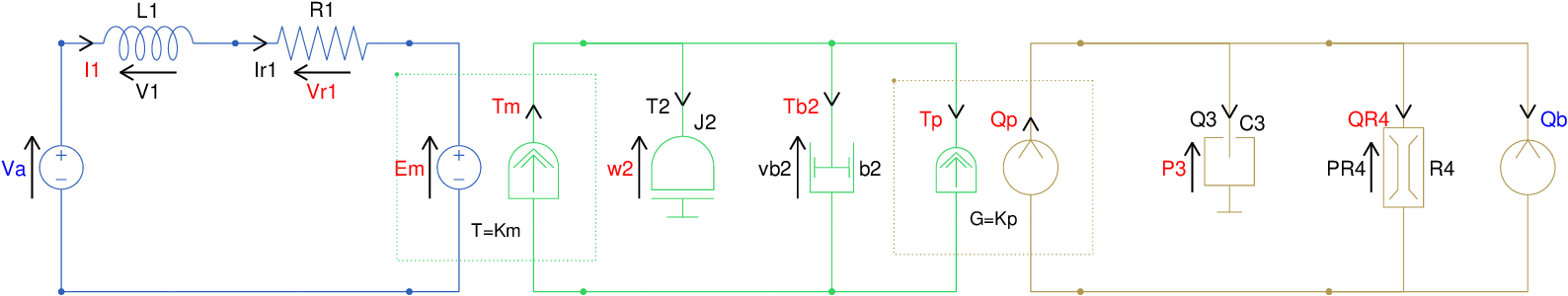}}
    \rput[lb](-7.5,1.5){(a)}
\end{pspicture}
  \end{center}
\vspace{-0.3cm}
 \begin{center}
 \centering \footnotesize
 \setlength{\unitlength}{5.52mm}
 \psset{unit=\unitlength}
 \SpecialCoor
  \begin{pspicture}(-7.3,-1.5)(7.7,3)
   \psfrag{I}[rc][rc]{$I_a$}
   \psfrag{V}[rc][rc]{$V_a$}
   \psfrag{P}[lc][lc]{$P_b$}
   \psfrag{Q}[lc][lc]{$Q_b$}
   \hspace*{0cm}
   \rput(1.22,3){\includegraphics[clip,width=9.3cm]{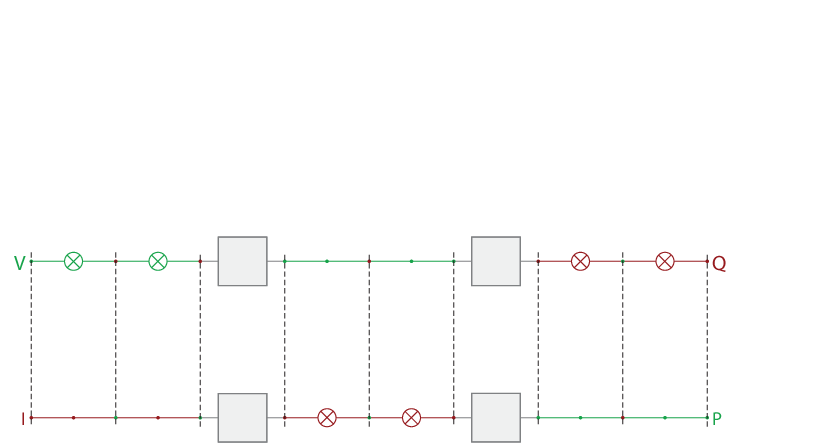}}
    \rput[lb](-7.3,2.9){(b)}
\rput(-4.8,1.5){$\mygreen \underbrace{\hspace{1.9cm}}_{\mbox{\tiny across power line}}$}
\rput(-4.8,-0.4){$\myred \overbrace{\hspace{1.9cm}}^{\mbox{\tiny through power line}}$}
\rput(0.35,1.5){$\mygreen \underbrace{\hspace{1.9cm}}_{\mbox{\tiny across power line}}$}
\rput(0.35,-0.4){$\myred \overbrace{\hspace{1.9cm}}^{\mbox{\tiny through power line}}$}
\rput(5.55,1.5){$\myred \underbrace{\hspace{1.9cm}}_{\mbox{\tiny through power line}}$}
\rput(5.55,-0.4){$\mygreen \overbrace{\hspace{1.9cm}}^{\mbox{\tiny across power line}}$} 
 \rput{90}(-2.25,0.5){\tiny Transformer}
 \rput{90}(3,0.5){\tiny Gyrator}
\end{pspicture}
  \end{center}
%
  \begin{center}
   \centering \footnotesize
 \setlength{\unitlength}{5.52mm}
 \psset{unit=\unitlength}
 \SpecialCoor
  \begin{pspicture}(-7.6,-2.0)(7.6,2.5)
   \psfrag{Ia}[rc][rc]{$I_a$}
   \psfrag{Va}[rc][rc]{$V_a$}
   \psfrag{1/s}[cc][cc]{$\black \frac{1}{s}$}
   \psfrag{1/L1}[cc][cc]{$\black \frac{1}{L_1}$}
   \psfrag{I1}[ct][ct]{$I_1$}
   \psfrag{R1}[cc][cc]{$\black R_1$}
   \psfrag{Vr1}[lb][lb]{}
   \psfrag{Km}[cc][cc]{$\black K_{12}$}
   \psfrag{Km'}[cc][cc]{$\black K_{12}\tras$}
   \psfrag{1/s}[cc][cc]{$\black \frac{1}{s}$}
   \psfrag{1/J2}[cc][cc]{$\black \frac{1}{J_2}$}
   \psfrag{w2}[cb][cb]{$\omega_2$}
   \psfrag{b2}[cc][cc]{$\black b_2$}
   \psfrag{Tb2}[lt][lt]{}
   \psfrag{Kp}[cc][cc]{$\black K_{23}$}
   \psfrag{Kp'}[cc][cc]{$\black K_{23}\tras$}
   \psfrag{1/s}[cc][cc]{$\black \frac{1}{s}$}
   \psfrag{C3}[cc][cc]{$\black \frac{1}{C_3}$}
   \psfrag{R4}[cc][cc]{$R_3$}
   \psfrag{QR4}[lb][lb]{}
   \psfrag{P3}[ct][ct]{$P_3$}
   \psfrag{Pb}[lc][lc]{$P_b$}
   \psfrag{Qb}[lc][lc]{$Q_b$}
   \hspace*{0cm}
   \rput(0,0){\includegraphics[clip,width=8.0cm]{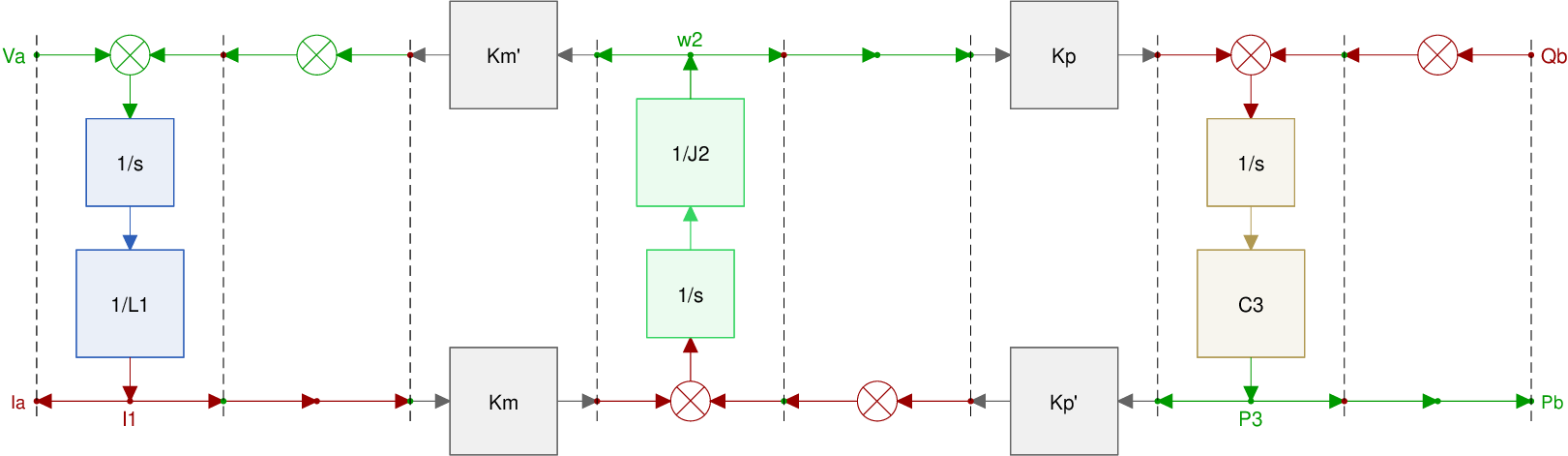}}
    \rput(-7.3,2.5){(c)}
     \rput[t](-1.5,-1.7){\tiny $\tau_2$}
     \rput[b](3.7,1.7){\tiny $Q_3$}
    \psellipse[linestyle=dashed,linecolor=myorange](-6.05,0)(0.75,2.5)
    \psellipse[linestyle=dashed,linecolor=myorange](-0.85,0)(0.75,2.5)
    \psellipse[linestyle=dashed,linecolor=myorange]( 4.35,0)(0.75,2.5)
\end{pspicture}
  \end{center}
  \begin{center}
   \centering \footnotesize
 \setlength{\unitlength}{5.52mm}
 \psset{unit=\unitlength}
 \SpecialCoor
  \begin{pspicture}(-7.6,-2.0)(7.6,2.5)
   \psfrag{Ia}[rc][rc]{$I_a$}
   \psfrag{Va}[rc][rc]{$V_a$}
   \psfrag{1/s}[cc][cc]{$\black \frac{1}{s}$}
   \psfrag{1/L1}[cc][cc]{$\black \frac{1}{L_1}$}
   \psfrag{I1}[ct][ct]{$I_1$}
   \psfrag{R1}[cc][cc]{$\black R_1$}
   \psfrag{Vr1}[lb][lb]{}
   \psfrag{Km}[cc][cc]{$\black K_{12}$}
   \psfrag{Km'}[cc][cc]{$\black K_{12}\tras$}
   \psfrag{1/s}[cc][cc]{$\black \frac{1}{s}$}
   \psfrag{1/J2}[cc][cc]{$\black \frac{1}{J_2}$}
   \psfrag{w2}[cb][cb]{$\omega_2$}
   \psfrag{b2}[cc][cc]{$\black b_2$}
   \psfrag{Tb2}[lt][lt]{}
   \psfrag{Kp}[cc][cc]{$\black K_{23}$}
   \psfrag{Kp'}[cc][cc]{$\black K_{23}\tras$}
   \psfrag{1/s}[cc][cc]{$\black \frac{1}{s}$}
   \psfrag{C3}[cc][cc]{$\black \frac{1}{C_3}$}
   \psfrag{R4}[cc][cc]{$R_3$}
   \psfrag{QR4}[lb][lb]{}
   \psfrag{P3}[ct][ct]{$P_3$}
   \psfrag{Pb}[lc][lc]{$P_b$}
   \psfrag{Qb}[lc][lc]{$Q_b$}
   \hspace*{0cm}
   \rput(0,0){\includegraphics[clip,width=8cm]{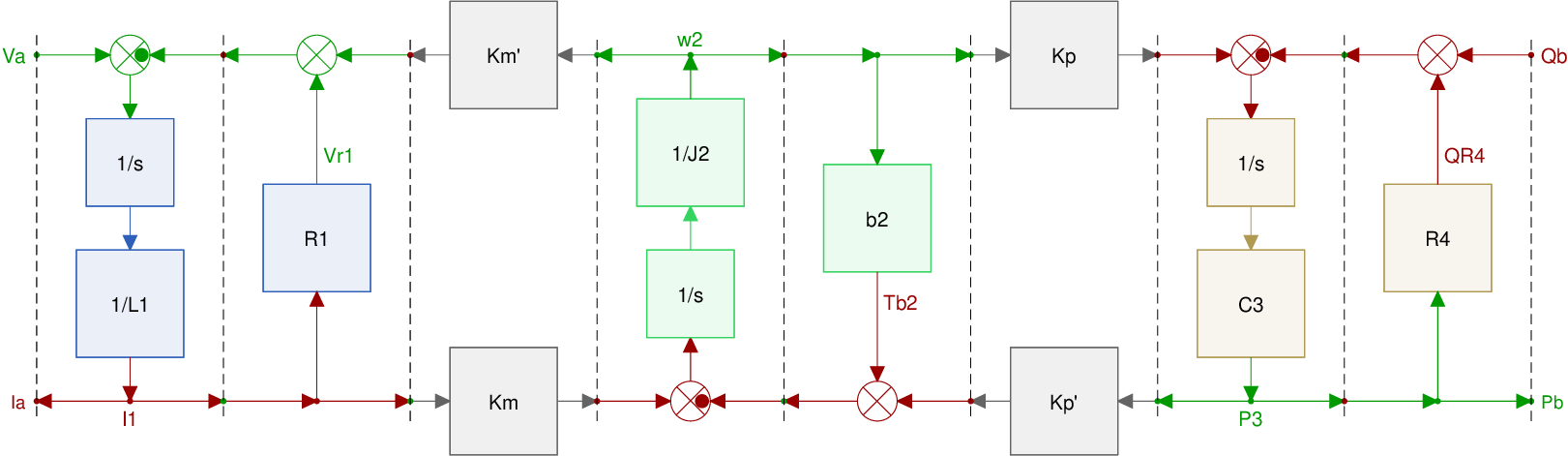}}
    \rput(-7.3,2.5){(d)}
     \rput[t](-1.5,-1.7){\tiny $\tau_{12}$}
     \rput[b](3.7,1.7){\tiny $Q_{23}$}
     \psellipse[linestyle=dashed,linecolor=red](-4.3,0)(0.75,2.5)
     \psellipse[linestyle=dashed,linecolor=red]( 0.85,0)(0.75,2.5)
     \psellipse[linestyle=dashed,linecolor=red]( 6.05,0)(0.75,2.5)
\end{pspicture}
  \end{center}
  \caption{DC motor supplying an hydraulic pump taken as a third case study. a) Simscape schematic, including the definition of the positive direction of the input and state variables representing {\bf Step 1} of the modeling procedure. b) {\bf Step 2} of the modeling procedure. c) {\bf Step 3} of the modeling procedure, while {\bf Step 4} is not necessary for the considered system. d) {\bf Step 5} and {\bf Step 6} of the modeling procedure, resulting in the final POG block scheme of the system.
  %
  }\label{MultiDom_Circuit_1}
    \vspace{-3mm}
  \end{figure}

The case study in Figure~\ref{MultiDom_Circuit_1}(a) involves three different energetic domains: electrical, mechanical and hydraulic.
The input variables are the source voltage $V_a$ of the DC motor and the volume flow rate $Q_b$,
while the state variables
are the current {$I_1$},
the angular speed {$\omega_2$}
and the pressure {$P_3$} within the hydraulic capacitor $C_3$.
By applying the FMPOG step-by-step procedure 
illustrated in Section~``Step-By-Step Modeling Of Physical Systems'',
the system in Figure~\ref{MultiDom_Circuit_1}(a) can be straightforwardly modeled following the intermediate steps shown in Figure~\ref{MultiDom_Circuit_1}(b)-(d). It is worth noticing that {\bf Step 4} is not necessary in this case, since there are no dynamic PEs that admit two possible configurations in the considered system.

\begin{Remar}
\label{conn_block_remar}
If one or more energy conversions take place within the system, the corresponding CBs need to be introduced into the POG scheme at {\bf Step 2}, as shown in Figure~\ref{MultiDom_Circuit_1}(b), and properly oriented after {\bf Step 3} and {\bf Step 4}, as shown in Figure~\ref{MultiDom_Circuit_1}(c)-(d).
The presence of a CB of {\it gyrator} type causes the {\mygreen across power line} and the {\myred through power line} to swap places. The swap does not happen if the CB is of {\it transformer} type.
\end{Remar}
The system in Figure~\ref{MultiDom_Circuit_1}(a) exhibits two energy conversions, between the electrical/mechanical rotational domains and between the mechanical rotational/hydraulic domains. According to Remark~\ref{conn_block_remar}, two CBs as that of Figure~\ref{basic_EBs_and_CBs}(d) are introduced in {\bf Step 2}. The CBs are initially empty, as shown in Figure~\ref{MultiDom_Circuit_1}(b), and their orientation is to be determined as follows. By applying {\bf Step 3}, the dynamic EBs shown in Figure~\ref{MultiDom_Circuit_1}(c) can be obtained, from which the proper orientation 
of the CBs can be determined
according to the EBs next to them.
The energy conversion coefficients $K_{12}$ and $K_{23}$ in the CBs of Figure~\ref{MultiDom_Circuit_1}(d) introduce the following relations:
\begin{equation}
\label{cbs_relat}
\tau_2=K_{12}\,I_1 \hspace{8mm} \mbox{and} \hspace{8mm}
Q_3=K_{23}\,\omega_2.
\end{equation}
The {\mygreen across} and {\myred through} power lines swap places in correspondence of the second CB in Figure~\ref{MultiDom_Circuit_1}(d). This well agrees with what is stated in Remark~\ref{conn_block_remar}, since the relation in \eqref{cbs_relat} on the right physically describes a gyrator, that is
the conversion of the across power variable $\omega_2$ (angular speed)
into the through power variable $Q_3$ (volume flow rate),
as shown in Table~\ref{Ambiti_Energetici}.

 \section{Simulation Comparison Against Simscape}\label{Comparison_Against_Simscape}
This section shows a comparison between the simulation results provided by the Simscape schematic shown in Figure~\ref{Electric_Circuit_1}(a) and the simulation results provided by the obtained POG model shown in Figure~\ref{Electric_Circuit_1}(d), using the parameters, inputs, and initial conditions reported in Table~\ref{simul_param}. 
 \begin{figure}[t!]
 \psfrag{V1 [V]}[][][0.85]{[V]}
 \psfrag{I2 [A]}[][][0.85]{[A]}
 \psfrag{I3 [A]}[][][0.85]{[A]}
 \psfrag{V4 [V]}[][][0.85]{[V]}
 \psfrag{V1 POG Model}[][][0.62]{$V_1$ POG Model}
 \psfrag{V1 Simscape}[][][0.62]{$V_1$ Simscape}
 \psfrag{I2 POG Model}[][][0.62]{$I_2$ POG Model}
 \psfrag{I2 Simscape}[][][0.62]{$I_2$ Simscape}
 \psfrag{I3 POG Model}[][][0.62]{$I_3$ POG Model}
 \psfrag{I3 Simscape}[][][0.62]{$I_3$ Simscape}
 \psfrag{V4 POG Model}[][][0.62]{$V_4$ POG Model}
 \psfrag{V4 Simscape}[][][0.62]{$V_4$ Simscape}
 \psfrag{(a)}[b][b][0.68]{(a)}
 \psfrag{(b)}[b][b][0.68]{(b)}
 \psfrag{(c)}[b][b][0.68]{(c)}
 \psfrag{(d)}[b][b][0.68]{(d)}
  \psfrag{Time [s]}[t][t][0.85]{Time [s]}

  \psfrag{Simulink ours (r) and Simscape (b)}[b][b][0.75]{}
\includegraphics[clip,width=1\columnwidth]{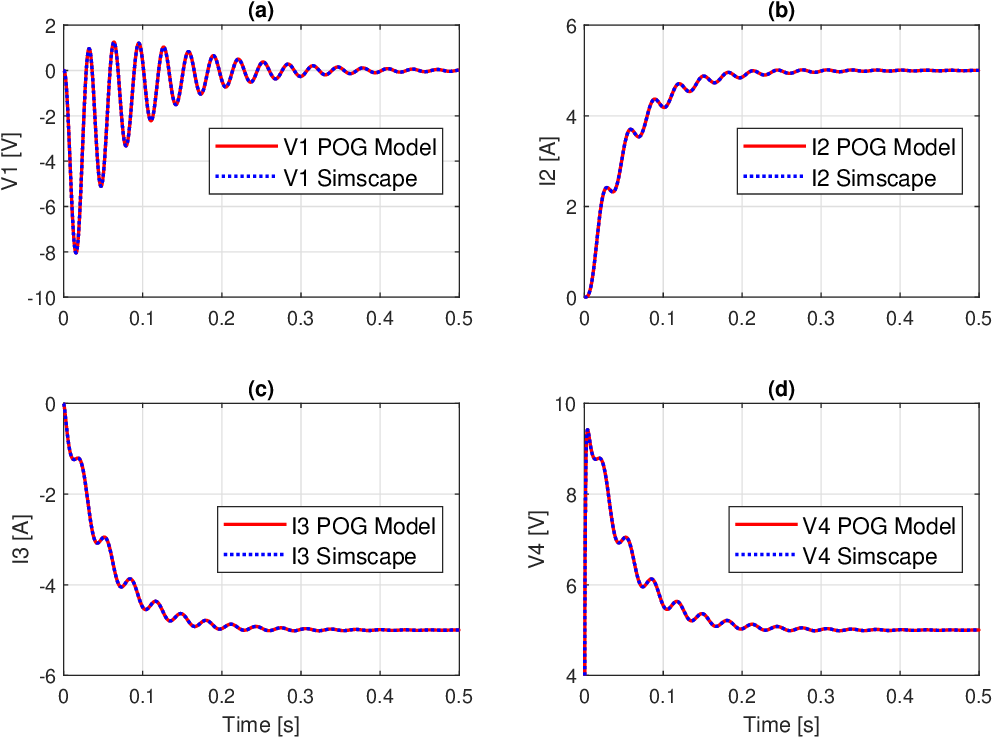}
 \caption{Comparison between the
 simulation results provided by the Simscape schematic of Figure~\ref{Electric_Circuit_1}(a) and the POG model of Figure~\ref{Electric_Circuit_1}(d). (a) The voltage $V_1$.  (b) The current $I_2$.  (c) The current $I_3$.  (a) The voltage $V_4$.}\label{Sim_Res}
    \vspace{-3mm}
\end{figure}
The obtained simulation results are reported in Figure~\ref{Sim_Res}:
the very good superposition between the results provided by the Simscape and the POG dynamic models clearly show
\begin{table}[t!]
  \centering \scriptsize
  \caption{Simulation parameters, inputs and initial conditions for the simulation of the Simscape schematic of Figure~\ref{Electric_Circuit_1}(a) and of the POG model of Figure~\ref{Electric_Circuit_1}(d).
  }\label{simul_param}
  \vspace{2mm}
\begin{tabular}{|@{}c@{}|c|c|} \hline
$V_a=0$ [V], $\;$ $V_b=10$ [V] & $C_1=C_4=1$ [mF] & $L_2=L_3=50$ [mH] \\ \hline
$R_3=R_4=1$ [$\Omega$]  &
$V_{1_0}=V_{4_0}=0$ [V] & $I_{2_0}=I_{3_0}=0$ [A]
\\ \hline
\end{tabular}
\end{table}
%
%
the correctness of the proposed FMPOG step-by-step procedure.

\section{CONCLUSIONS}
This article has proposed the Fast Modeling Power-Oriented Graphs (FMPOG), 
a step-by-step methodical procedure that guides the users through the derivation of the POG block schemes and, ultimately, the state-space models of physical systems starting from their schematics. 

In order to present the FMPOG procedure, the first part of this article has been dedicated to the description of the fundamental principles and properties of the POG technique. The comparison of Power-Oriented Graphs with the two other main graphical modeling techniques in the literature, namely Bond Graph and Energetic Macroscopic Representation, has also been discussed. This has highlighted the convenient properties of the POG technique: its ease of use for non-expert users thanks to its user-friendly symbolism, which generates block schemes directly implementable in the Simulink environment, the capability to derive the state-space model directly
from the POG block scheme, and its predisposition to model reduction through the use of congruent transformations. To help the readers to familiarize themselves with the POG technique, its application to various case studies of different complexity and in different energetic domains has also been addressed.


The proposed FMPOG procedure has the quality of being fully systematic and applicable to
different physical systems in different energetic domains. Simulators allow the
user to simulate the system schematic, but typically do not provide the user
with the system dynamic model. On the other hand,
the proposed FMPOG procedure allows for the direct reading of the system dynamic model from the POG block scheme, which is a functionality of
great interest from a control point of view,
in both graphical and analytical forms. 

In order to demonstrate the versatility of the FMPOG procedure, we applied it to the modeling of three physical systems in the electrical,
hydraulic, and electro-mechanical-hydraulic domains, respectively. 
 Furthermore, a freely available
 Matlab/Simulink program implementing the FMPOG procedure is provided in the given repository to make it accessible to the entire community.
 
We believe that the FMPOG procedure, developed within the framework of the POG modeling technique, is a powerful tool for engineers who need to quickly and automatically find the mathematical model of complex physical systems. In contrast, the derivation of these models may be more challenging and error-prone using standard, less systematic methods.

\section{ACKNOWLEDGMENT}
This work was partly supported by the University of Modena and Reggio Emilia
through the action FARD (Finanziamento Ateneo Ricerca Dipartimentale) 2023/2024, and funded under the National Recovery and Resilience Plan (NRRP), Mission 04 Component 2 Investment 1.5 – NextGenerationEU, Call for tender n. 3277 dated 30/12/2021
Award Number:  0001052 dated 23/06/2022.





\bibliographystyle{IEEEtran}

\bibliography{references}
\clearpage


\clearpage

\clearpage


\clearpage

\clearpage

\clearpage


\clearpage

\clearpage

\clearpage


\clearpage

\clearpage
\clearpage
\clearpage

  \clearpage

  \clearpage

  \clearpage

\clearpage

 \setcounter{equation}{0}


 \setcounter{equation}{0}
\section{Author Information}

\begin{IEEEbiography}{Davide Tebaldi}
received the bachelor's degree in electronics engineering in 2015 and the master’s degree in electronics engineering (cum laude) in 2018 from the University of Modena and Reggio Emilia, Italy. He received the Ph.D. degree in information and communication technologies in 2022 from the University of Modena and Reggio Emilia, discussing a thesis entitled Mathematical Modeling Control and Simulation of Hybrid Electric Vehicles. From October 2021 to January 2022, he was a visiting scholar at the Center for Automotive Research, The Ohio State University, Ohio, USA. From March 2022 to February 2023, he held a post-doc position at the University of Modena and Reggio Emilia. From March 2023, he has been an Assistant Professor at the University of Modena and Reggio Emilia. His research interests include the energetic modeling, control, energy efficiency analysis and simulation of mechatronic systems, with main application in the automotive, electrical machines and power electronics fields. Another branch of his research interests concerns the collaborative robotics field, with specific reference to the energy efficiency improvement and to the human-robot safe interaction.
\end{IEEEbiography}

\begin{IEEEbiography}{Roberto Zanasi}
 received the degree in electrical engineering (cum laude) from the University
of Bologna in 1986. He received the Ph.D. degree
in systems engineering in 1992. From 1994 to 1998,
he was a Researcher in automatic controls with the
Department of Electronics, Computer and System
Science, University of Bologna. From 1998 to 2004, he was an Associate Professor of Automatic Control with the ``Enzo
Ferrari'' Engineering Department, University of Modena and Reggio Emilia. From 2004, he has been a Full Professor of Automatic Control with the ``Enzo
Ferrari'' Engineering Department, University of Modena and Reggio Emilia.
He held the position of Visiting Scientist with the IRIMS of Moscow in 1991, at
the MIT of Boston in 1992, and at the Universit\'e Catholique de Louvain in 1995.
He has authored or coauthored many technical and scientific publications, and five books on automatic controls, simulation and digital control systems. His
research interests include: mathematical modeling, power-oriented graphs, simulation, automotive, control of multi-phase electrical motors, variable-structure
systems, sliding mode control with integral action, robotics, linear and nonlinear
control.
\end{IEEEbiography}

\endarticle

\end{document}